\documentclass[a4paper,11pt,nofootinbib]{book}
\pdfoutput=1 
\usepackage[margin=1.0in]{geometry}
\usepackage[utf8]{inputenc}

\usepackage{enumitem}
\usepackage{hyperref}
\usepackage{fancyhdr}
\usepackage{youngtab}
\usepackage[titletoc]{appendix}
\usepackage{amsfonts}
\usepackage{amsmath}

\usepackage[pages=some]{background}

\usepackage{amsbsy}
\usepackage{graphicx}
\usepackage{amssymb}
\usepackage{amsthm}
\usepackage{bm}
\usepackage{comment}
\usepackage{epsfig}
\usepackage{latexsym}
\usepackage{pdflscape}
\usepackage{cancel} 
\usepackage{color}
\usepackage{cleveref}
\usepackage{bm,nicefrac}

\usepackage{lmodern}
\usepackage[T1]{fontenc}

\usepackage[backend=bibtex,style=numeric-comp,sorting=none]{biblatex}
\makeatletter
\def\@fpheader{\relax}
\makeatother

\DeclareSymbolFont{AMSa}{U}{msa}{m}{n}
\DeclareSymbolFont{AMSb}{U}{msb}{m}{n}
\DeclareMathSymbol{\fieldR}{\mathalpha}{AMSb}{"52}

\usepackage{color}
    \definecolor{darkgreen}{rgb}{0,0.5,0}
    \definecolor{darkblue}{rgb}{0,0,0.6}
    \definecolor{purple}{rgb}{0.4,.2,0.7}

\newcommand{\myspace}{\vspace{0.2cm} \\}
\newcommand{\lag}{\mathcal{L}}
\newcommand{\del}{\partial}
\newcommand{\identity}{\mathbb{I}}
\newcommand{\trace}{\text{Tr}}
\newcommand{\sutwo}{\text{SU(2) }}
\newcommand{\chiral}{$SU(2)_L\times SU(2)_R$ }
\newcommand{\chior}{n_{\chi}}
\newcommand{\bra}[1]{\langle #1 |}
\newcommand{\ket}[1]{| #1 \rangle}
\newcommand{\matelem}[2]{\bra{#2}\mathcal{A}\ket{#1}}
\newcommand{\matelemgen}[3]{\bra{#2} #3 \ket{#1}}
\newcommand{\ih}{-\frac{i}{2}}
\newcommand{\quadx}{\hspace{0.2cm}}
\newcommand{\redcross}{\color{red}-}
\newcommand{\doublepole}{{\color{red}{\bigotimes}}}
\newcommand{\doublerow}[2]{\hspace{0.3cm} \begin{matrix} #1 \\ #2 \end{matrix} \hspace{0.3cm}}
\newcommand{\wpwm}{W^{+}W^{-}}

\newcommand{\masspl}{M_{\text{pl}}}

\begin{document}
\pagenumbering{gobble}

\begin{titlepage}


\begin{center}
{\fontfamily{qtm}\selectfont
\Large
\textbf{PhD Thesis}
\vspace*{1cm}

{\fontsize{32pt}{40pt}\selectfont \textbf{Anomalous Higgs Couplings as a Window to New Physics}\par }

\vspace*{0.5cm}

\vspace*{0.58cm}
\huge
\textbf{Iñigo Asiáin Jiménez}
}

\vspace*{2cm}
\includegraphics[width=0.5\textwidth]{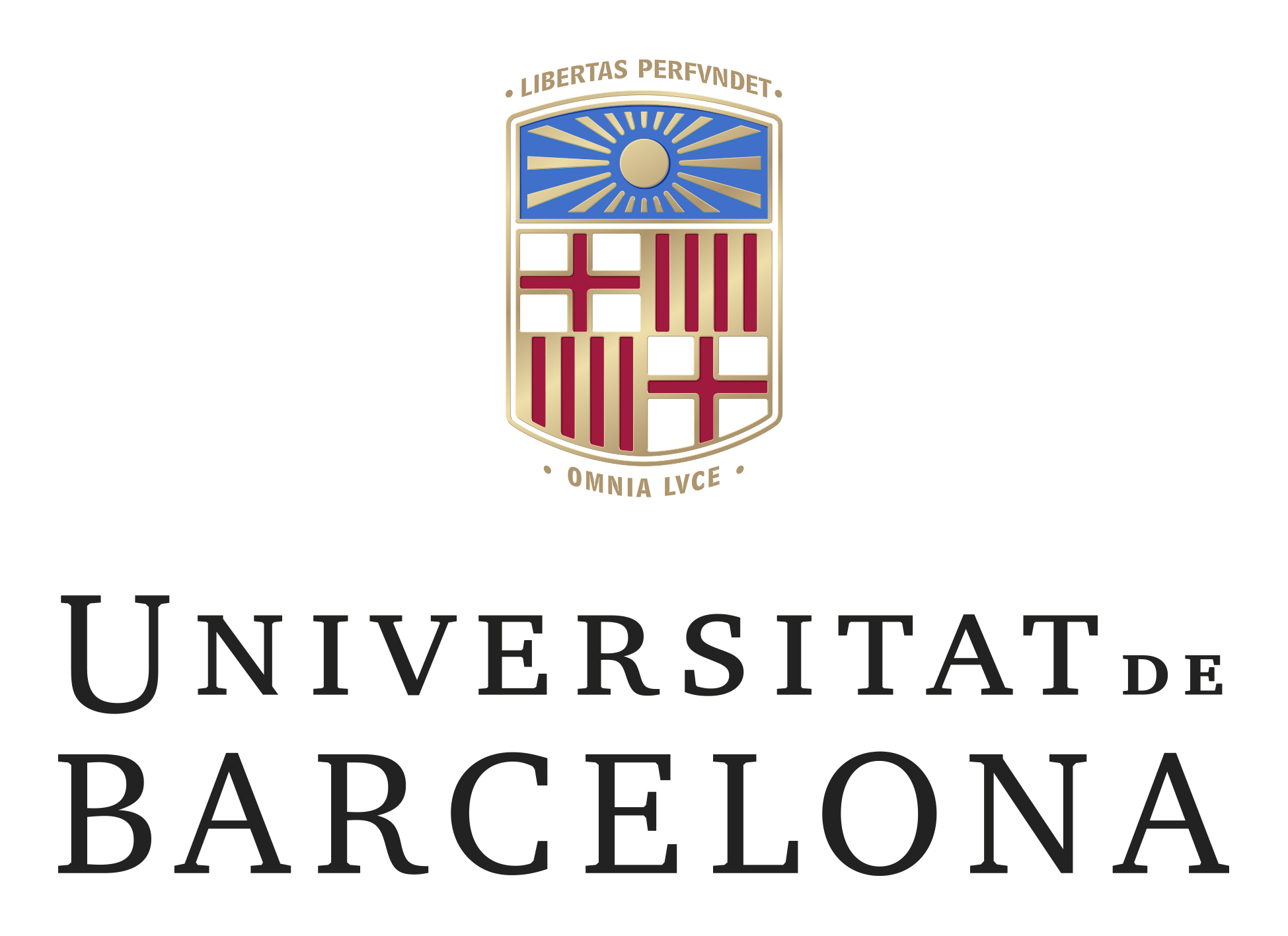}

\end{center}
\end{titlepage}
\pagecolor{white}
\thispagestyle{empty}
\clearpage
\thispagestyle{empty}
\hfill
\clearpage
\begin{center}
\Huge
\textbf{Anomalous Higgs Couplings as a Window to New Physics}\\
\vspace*{5cm}

\normalsize
Memòria presentada per optar al grau de doctor per la \\
\vspace{1cm}
Universitat de Barcelona\\
\vspace*{2cm}
Programa de doctorat en Física\\

\vspace*{5cm}
Autor: 
I\~nigo Asiáin Jiménez\\
Directors:
Domènec Espriu i Federico Mescia\\
Tutor:
Joan Soto Riera\\

\vspace*{1cm}
\includegraphics[width=0.5\textwidth]{images/logo-UB.jpeg}

\end{center}
\clearpage
\thispagestyle{empty}
\hfill
\clearpage



	

\pagestyle{fancy}
\fancyhf{}
\renewcommand{\headrulewidth}{0pt}
\cfoot{\thepage}

\newpage
\pagenumbering{roman}
\setcounter{page}{1}
\Large
\textbf{Agradecimientos}
\normalsize
\vspace*{2cm}

\noindent
¡Menudo viaje!\myspace
Esta tesis es el resultado de los últimos años de trabajo que he realizado en el departamento de Física Quàntica i Astrofísica (FQA) de la Universitat de Barcelona. Un departamento que realmente he sentido como mi casa, y donde ha sido un placer desarrollar mi labor tanto investigadora como docente.\myspace
En primer lugar, esta tesis no habría sido posible sin mis supervisores, el Dr. Espriu y el Dr. Mescia. Agradezco a Domènec y Federico todas las horas de trabajo en conjunto con enorme paciencia y comprensión, sobre todo en aquellos momentos cuando he estado más perdido. Que este proyecto haya salido adelante no se debe únicamente a vuestra calidad profesional, fuera de toda duda, sino también a la calidad personal que habéis demostrado. Agradezco cada uno de los consejos y los distintos puntos de vista que me habéis trasladado y que, seguro, podré aportar donde sea que vaya ahora mi camino.\myspace
También me gustaría agradecer en este trabajo al proyecto PID2022-136224NB-C21 que me ha apoyado económicamente para asistir a las distintas escuelas, congresos, conferencias y estancias y distintos eventos que han sido cruciales para la elaboración de esta tesis, y personalmente a la Dra. González García por todas las gestiones necesarias.\myspace
I would also like to acknowledge my collaborators Dr. Gr\"{o}ber and Lorenzo Tiberi for their warm welcome at the University of Padova and the excellent work we accomplished together. También a mi colaborador Antonio Dobado de la Universidad Complutense de Madrid, con el que hemos desarrollado el último capítulo de esta tesis y con el que he aprendido muchísimo sobre técnicas de unitarización.\myspace
Si algo he descubierto durante estos años de tesis es mi pasión por la docencia. Agradezco enormemente al FQA que me haya dado un lugar para desarrollar esta tarea con total libertad. A Muntsa Guilleumas, como anterior secretaria de docencia, por la confianza depositada y por contar siempre conmigo, y a todo el profesorado con el que he compartido alguna asignatura. Un agradecimiento en particular a Bruno Julià, por el intenso trabajo compartido y porque siempre ha tenido en cuenta mis opiniones. ¡Ha sido un placer formar parte del equipo de compu!\myspace
Al estudiantado de Física UB con el que he compartido horas infinitas de pizarra, tutorías y códigos fortran. Siento esto último.\myspace
A todo el grupo de estudiantes de doctorado. Sería una lista enorme, pero ya sabéis quiénes sois. Por todas las cenas, salidas y apoyo emocional compartidos.\myspace
Agradezco también a todas mis compañeras y compañeros con los que he realizado la actividad de Divulgación \textit{Quarks, els maons de la matèria}, por el enorme trabajo, los buenos ratos y las buenas experiencias vividas bien sea divulgando sobre el modelo de quarks o haciendo de niñeros. Un agradecimiento en particular a Àngels Ramos y a mi exvecina de despacho Assumpta Parreño por todo el trabajo como coordinadora del programa de doctorado y por todos los ratos de charlas estos años.\myspace
Y no todo en el departamento iba a ser trabajo, claro. A mi equipo de los viernes Vincent, Juan, María, Mariona, Miguel Àngel, Sergi, Nadine, Robert, Kim y Antonio. No se me ocurre mejor manera que dar carpetazo a la semana y dar la bienvenida al finde que con unas alhambras en la cafetería.\myspace
No puedo dejar de agradecer a mi grupo de inicio en el doctorado; Paula, Iván, Diego, Cristian, Carlos, Marc, Aniol y Miguel, la gente con la que compartí las dudas, los (muchos) problemas con las becas y sus tortuosos procedimientos.\myspace
Agradezco enormemente a mis grandes amigos Javi, Sergio, Jasper y Jordi; la gente más lista y más guapa de toda la facultad. Sí, soy completamente objetivo. Este viaje sin vosotros habría merecido menos la pena y, desde luego, habría sido mucho más aburrido. Habéis sido mi mayor descubrimiento, mi núcleo duro, mi zona de seguridad. Sois gente brillante y dedicada y cualquier departamento de física será afortunado de teneros entre sus filas. Aunque nuestros caminos ahora se hayan separado, estaré siempre ahí para celebrar vuestros logros laborales y, sobre todo, personales. Sois enormes y, para mí, eternos.\myspace
Por supuesto, sería muy injusto no extender este agradecimiento a vuestras respectivas parejas; María, Ana, Alej y Claire, que son igual de maravillosas. Todas las birras, cafés, bravas (de estas sólo algunas de ellas, hay otras que son motivo de denuncia), fiestas de Gràcia,... En definitiva, todos los ratos compartidos con vosotras, que han contribuido a que este camino sea menos duro.\myspace
También a Blanca, Laia, Robert, Marc, Zaida, con los que he descubierto que hay algo mucho más complicado que la física; los bolos.\myspace
En Madrid fue la primera etapa en la que me empecé a relacionar con la física. Allí es donde hice la carrera y donde guardo grandes recuerdos. De esta época, a nivel personal, no me puedo olvidar de Isa, Naia e Iribarren---y por consiguiente de \textit{La Oreja de Van Gogh} por crear un himno que vociferamos con pésimo gusto, apabullante falta de talento y muchas risas por los bares de Zaragoza---y de Adrián y Almu.\myspace
Después de Madrid empezó mi aventura en la UB con el máster. También fue una época de conocer gente nueva con la que compartí muchos intereses y momentos. De esa época me llevo, y agradezco en esta tesis también, a Rafa y Mikel, que a día de hoy siguen siendo gente con la que puedo contar siempre.\myspace
Mi llegada a la universidad y a la física en particular estoy seguro de que no se entiende sin mi Pamplona natal y todos los círculos que allí he tenido en mi vida y tengo a día de hoy. A mi tropa infalible; Amaia, Andrea, Bea, Eneko, Gara, Iribarren, Laura, Rubén y Silvia, y a Marta, Raquel y Pablo.\myspace
A mi familia, que ha estado ahí desde el principio y sé que siempre estará. A mi padre y mi madre, por todo el esfuerzo, el amor y la libertad con la que me han criado y educado, apoyando siempre mis decisiones aunque no las compartan, y siempre guiándome. A mi \textit{sis}, mi compañera desde que tengo memoria. A mís tías Eva y Sonia y mis primos Javier y Valeria. A mi querida abuela, mi \textit{yaya}, que es núcleo y pilar fundamental de toda esta estructura, sin la cual no sería capaz de entender la palabra familia. Y a mis abuelos que ya no están, porque todo lo que hago y soy lleva también vuestra huella.\myspace
Por último a Tomek y Mara, que sois no sólo mi casa sino mi hogar. No tengo ni idea de qué va a ser de mi vida a partir de ahora pero tengo claro que quiero recorrer cualquiera que sea mi camino con vosotros. Kocham was. 

\newpage
\Large
\textbf{Resumen}
\normalsize
\myspace

\textit{Título:}
\textbf{\underline{Acoplamientos anómalos del Higgs como una ventana a Nueva Física}}
\phantom{x}

\textit{Palabras clave:}
Ruptura espontánea de la simetría, simetría quiral, bosón de Higgs, bosones de Goldstone, bosones gauge, unitarización, resonancias, partículas elementales, LHC
\phantom{x}

\vspace{2ex}

A lo largo de esta tesis, hemos explorado el potencial que las teorías efectivas combinadas con técnicas de unitarización pueden ofrecer para estudiar aspectos desconocidos de la física más allá del Modelo Estándar (física BSM). En particular, nuestro principal interés se centra en el origen potencialmente dinámico del sector electrodébil de la ruptura espontánea de simetría.\myspace
Las teorías efectivas pueden llevar a un comportamiento no unitario con amplitudes que crecen rápidamente con la energía. Este comportamiento no físico requiere de técnicas de unitarización de las amplitudes antes de compararlas con los resultados experimentales. En la primera parte de esta disertación nos centramos en el scattering longitudinal $WW$, un proceso que, aunque subdominante en el LHC, resulta ser una buena ventana para estudiar física del Higgs. Realizamos un cálculo a un loop en el marco de la \textit{Teoría Efectiva para el Higgs} (del inglés, HEFT) de todos los procesos $2\to 2$ relevantes para el proceso de unitarización, determinando los contratérminos necesarios en el esquema on-shell, y estudiamos por primera vez cómo la inclusión de modos transversos de los bosones gauge modifica las masas y anchuras previamente calculadas de las resonancias dinámicas que surgen del proceso de unitarización en los canales vector-isovector y escalar-isoscalar. En conjunto, proporcionamos las herramientas técnicas necesarias para estudiar los acoplamientos a baja energía en la teoría efectiva del Higgs bajo los requisitos de unitariedad y causalidad.\myspace
Si bien la fenomenología de las resonancias vectoriales está razonablemente entendida dentro del marco de las reglas de suma de Weinberg y los estudios de unitarización, las resonancias escalares están mucho menos restringidas y, lo que es más importante, su correcta interpretación depende de acoplamientos efectivos a baja energía de HEFT diferentes de los de las resonancias vectoriales, y que son difíciles de restringir experimentalmente. Más específicamente, las técnicas de unitarización combinadas con el requisito de causalidad nos permiten establecer límites no triviales sobre las auto-interacciones del Higgs. Esto se debe a la necesidad de considerar canales acoplados en el caso escalar durante el proceso de unitarización. Como subproducto, podemos obtener información relevante sobre el sector del Higgs a partir de los procesos elásticos $WW \to WW$ sin necesidad de considerar la producción de dos Higgses.\myspace
En la última parte de la tesis, aplicaremos toda la maquinaria desarrollada en los capítulos anteriores para unitarizar las amplitudes de una teoría cuántica de la interacción gravitatoria, entendida ésta como una teoría efectiva.\myspace
Esta tesis está estructurada en seis capítulos y un capítulo final con las conclusiones. El contenido de estos capítulos incluye todo el trabajo original de nuestro grupo recogido en las publicaciones de \textit{Physical Review D} en las Refs.~\cite{Asiain:2021lch, paper2_doi, Asiain:2023myt}, así como de un artículo que se encuentra en fase final de preparación relativo al capítulo final.\myspace
En el \textbf{Capítulo 1}, repasamos la construcción del Modelo Estándar, haciendo especial hincapié en los temas relacionados con la ruptura espontánea de la simetría electrodébil y el sector de ruptura espontánea de la simetría (EWSBS). Este capítulo también incluye el estado actual de la investigación sobre el Higgs y las perspectivas futuras a nivel experimental.\myspace
El \textbf{Capítulo 2} sirve como introducción a las teorías efectivas desde un punto de vista general y proporciona una descripción detallada del HEFT, que se usará a lo largo del trabajo. Además, se presentan todas las amplitudes de procesos $2 \to 2$ relevantes para el estudio a nivel de un loop, incluyendo los contratérminos necesarios para hacerlas finitas.\myspace
En el \textbf{Capítulo 3}, repasamos el análisis de ondas parciales y las técnicas de unitarización más utilizadas en el campo, destacando la importancia de dos métodos en particular debido a sus propiedades analíticas: el Método de la Amplitud Inversa (IAM) y el de la matriz K mejorado (IK-matrix). Se presentan tanto la versión de canal simple como la de canales acoplados de estos métodos, siendo esta última, como ya se ha mencionado, de gran relevancia para este trabajo.\myspace
El \textbf{Capítulo 4}, que está basado en nuestra Ref.~\cite{Asiain:2021lch}, presenta los resultados obtenidos en el estudio de resonancias vectoriales. Específicamente, se muestran las ondas parciales en la proyección vectorial, junto con su comportamiento no físico---no unitario---a altas energías. Se proporciona información detallada sobre las resonancias vectoriales encontradas durante el proceso de unitarización de IAM en su versión de canal simple, que es la única relevante para las ondas vectoriales.\myspace
El \textbf{Capítulo 5} repite el análisis realizado previamente con las resonancias vectoriales pero centrado en la proyección escalar para el estudio de las resonancias escalares. En particular, se destacan las diferencias entre el uso del método de canal simple y el de canales acoplados, enfatizando la importancia de este último para una descripción correcta de las resonancias que emergen en este canal. Se presenta también una metodología sistemática para derivar límites sobre los parámetros de la teoría efectiva estudiando las propiedades de las resonancias escalares emergentes, o la ausencia de ellas. Este capítulo finaliza con unos resultados tentativos sobre el $H(650)$, una posible resonancia escalar de interés experimental reciente, reproduciendo sus propiedades mediante nuestro análisis de canales acoplados en el HEFT. Los resultados de este capítulo se pueden encontrar también en nuestros trabajos originales de las Refs.~\cite{paper2_doi, Asiain:2023myt}.\myspace
Por último, el \textbf{Capítulo 6} cierra esta tesis y se centra en la aplicación conjunta de teorías efectivas y técnicas de unitarización en un marco distinto al del Modelo Estándar. Específicamente, estudiamos el espectro de resonancias escalares que aparece en la teoría de gravedad de Einstein-Hilbert cuando, entendida como el orden más bajo de una teoría efectiva, se añaden operadores a siguiente orden en la expansión.

\clearpage
\hfill
\clearpage


\newpage
\Large
\textbf{Resum}
\normalsize
\myspace

\textit{Títol:}
\textbf{\underline{Acoblaments anòmals del Higgs com a finestra a Nova Física}}
\phantom{x}

\textit{Paraules clau:}
Trencament espontani de la simetria, simetria quiral, bosó de Higgs, bosons de Goldstone, bosons gauge, unitarització, ressonàncies, partícules elementals, LHC
\phantom{x}

\vspace{2ex}

Al llarg d'aquesta tesi, hem explorat el potencial que les teories efectives combinades amb tècniques d'unitarització poden oferir per estudiar aspectes desconeguts de la física més enllà del Model Estàndard (física BSM). En particular, el nostre principal interès se centra en l'origen potencialment dinàmic del sector electrofeble del trencament espontani de simetria.\myspace  
Les teories efectives poden portar a un comportament no unitari amb amplituds que creixen ràpidament amb l'energia. Aquest comportament no físic requereix de tècniques d'unitarització de les amplituds abans de comparar-les amb els resultats experimentals. En aquesta dissertació ens centrem en el scattering longitudinal $WW$, un procés que, tot i ser subdominant al LHC, resulta ser una bona finestra per estudiar la física del Higgs. Realitzem un càlcul a un loop en el marc del HEFT de tots els processos rellevants, determinem els contratermes necessaris en l'esquema on-shell, i estudiem per primera vegada com la inclusió completa dels modos transversos dels bosons gauge modifica les masses i amplades prèviament calculades de les ressonàncies dinàmiques que sorgeixen del procés d’unitarització en els canals vector-isovector i escalar-isoescalar. En conjunt, proporcionem les eines tècniques necessàries per estudiar els acoblaments a baixa energia en la teoria efectiva de l'Higgs sota els requisits d'unitarietat i causalitat.\myspace  
Tot i que la fenomenologia de les ressonàncies vectorials està raonablement compresa dins del marc de les regles de suma de Weinberg i els estudis d'unitarització, les ressonàncies escalars estan molt menys restringides i, el que és més important, depenen dels acoblaments efectius a baixa energia de HEFT, diferents dels de les ressonàncies vectorials, els quals són difícils de restringir experimentalment. Més específicament, les tècniques d'unitarització combinades amb el requisit de causalitat ens permeten establir límits no trivials sobre les auto-interaccions de l'Higgs. Això es deu a la necessitat de considerar canals acoblats en el cas escalar durant el procés d’unitarització. Com a subproducte, podem obtenir informació rellevant sobre el sector de l'Higgs a partir dels processos elàstics $WW \to WW$ sense necessitat de considerar la producció de dos Higgs.\myspace  
A la darrera part de la tesi, aplicarem tota la maquinària desenvolupada en els capítols anteriors per unitaritzar les amplituds d'una teoria quàntica de la interacció gravitatoria, entesa com una teoria efectiva.\myspace
Aquesta tesi està estructurada en sis capítols i un capítol final amb les conclusions. El contingut d’aquests capítols inclou tota la feina original del nostre grup recollida a les publicacions de \textit{Physical Review D} a les Refs.~\cite{Asiain:2021lch, paper2_doi, Asiain:2023myt}, així com un article que es troba en fase final de preparació relatiu al capítol final.\myspace
\textbf{Capítol 1}, repassem la construcció del Model Estàndard, fent especial èmfasi en els temes relacionats amb la ruptura espontània de la simetria electrofeble i el sector de trencament espontani de la simetria (EWSBS). Aquest capítol també inclou l'estat actual de la recerca sobre el Higgs i les perspectives futures a nivell experimental.\myspace  
El \textbf{Capítol 2} serveix com a introducció a les teories efectives des d’un punt de vista general i proporciona una descripció detallada del HEFT, que s’utilitzarà al llarg del treball. A més, es presenten totes les amplituds dels processos $2 \to 2$ rellevants per a l’estudi a nivell d’un loop, incloent-hi els contratermes necessaris per fer-les finites.\myspace  
En el \textbf{Capítol 3}, repassem l’anàlisi d’ones parcials i les tècniques d’unitarització més utilitzades en el camp, destacant la importància de dos mètodes en particular per les seves propietats analítiques: el Mètode de l’Amplitud Inversa (IAM) i el de la matriu K millorada (IK-matrix). Es presenten tant la versió de canal simple com la de canals acoblats d’aquests mètodes, sent aquesta última, com ja s’ha esmentat, de gran rellevància per a aquest treball.\myspace  
El \textbf{Capítol 4}, que està basat en la nostra Ref.~\cite{Asiain:2021lch}, presenta els resultats obtinguts en l’estudi de ressonàncies vectorials. Específicament, es mostren les ones parcials en la projecció vectorial, juntament amb el seu comportament no físic---no unitari---a altes energies. Es proporciona informació detallada sobre les ressonàncies vectorials trobades durant el procés d’unitarització d’IAM en la seva versió de canal simple, que és l’única rellevant per a les ones vectorials.\myspace  
El \textbf{Capítol 5} repeteix l’anàlisi realitzat prèviament amb les ressonàncies vectorials, però centrat en la projecció escalar per a l’estudi de les ressonàncies escalars. En particular, es destaquen les diferències entre l’ús del mètode de canal simple i el de canals acoblats, tot emfatitzant la importància d’aquest últim per a una descripció correcta de les ressonàncies que emergeixen en aquest canal. Es presenta també una metodologia sistemàtica per derivar límits sobre els paràmetres de la teoria efectiva estudiant les propietats de les ressonàncies escalars emergents, o l’absència d’aquestes. Aquest capítol finalitza amb uns resultats provisionals sobre l’$H(650)$, una possible ressonància escalar d’interès experimental recent, reproduint-ne les propietats mitjançant la nostra anàlisi de canals acoblats en el HEFT. Els resultats d’aquest capítol es poden trobar també en els nostres treballs originals de les Refs.~\cite{paper2_doi, Asiain:2023myt}.\myspace  
Finalment, el \textbf{Capítol 6} tanca aquesta tesi i se centra en l’aplicació conjunta de teories efectives i tècniques d’unitarització en un marc diferent del Model Estàndard. Específicament, estudiem l’espectre de ressonàncies escalars que apareix en la teoria de gravetat d’Einstein-Hilbert quan, entesa com l’ordre més baix d’una teoria efectiva, s’hi afegeixen operadors al següent ordre en l’expansió.
\clearpage
\hfill
\clearpage


\newpage
\Large
\textbf{Abstract}
\normalsize
\myspace

\textit{Title:}
\textbf{\underline{Anomalous Higgs couplings as a window to New Physics}}
\phantom{x}

\phantom{x}

\vspace{2ex}

Throughout this thesis, we have explored the potential of effective theories combined with unitarization techniques to study unknown aspects of physics beyond the Standard Model (BSM physics). In particular, our main interest lies in the potentially dynamical origin of the weak sector of spontaneous symmetry breaking.\myspace
Effective theories can lead to non-unitary behavior with rapidly growing amplitudes. This unphysical behavior necessitates the unitarization of the amplitudes before comparing them with experimental results. In the first part of this dissertation, we focus on longitudinal $WW$ scattering, a process that, although subdominant at the LHC, turns out to be a good window for studying Higgs physics. We perform a one-loop calculation in the HEFT framework of all relevant processes, determine the necessary counterterms in the on-shell scheme, and study for the first time how the full inclusion of the transverse modes of the gauge bosons modifies the previously computed masses and widths of the dynamical resonances arising from the unitarization process in the vector-isovector and scalar-isoscalar channels. Altogether, we provide the technical tools needed to study the low-energy couplings in the Higgs effective theory under the requirements of unitarity and causality.\myspace
While the phenomenology of vector resonances is reasonably understood within the framework of Weinberg sum rules and unitarization studies, scalar resonances are far less constrained and, more importantly, depend on HEFT low-energy effective couplings distinct from those of vector resonances. These scalar couplings are difficult to constrain experimentally. More specifically, unitarization techniques combined with the requirement of causality allow us to set nontrivial bounds on Higgs self-interactions. This is due to the necessity of considering coupled channels in the scalar case during the unitarization process. As a byproduct, we can extract relevant information about the Higgs sector from $WW \to WW$ elastic processes without requiring two-Higgs production.\myspace
In the final part of the thesis, we will apply all the machinery developed in the previous chapters to unitarize the amplitudes of a quantum theory of gravitational interaction, understood as an effective theory.\myspace
This thesis is structured into six chapters and a final chapter with conclusions. The content of these chapters includes all the original work of our group, as published in \textit{Physical Review D} in Refs.~\cite{Asiain:2021lch, paper2_doi, Asiain:2023myt}, as well as an article currently in its final stage of preparation related to the final chapter.\myspace
\textbf{Chapter 1} reviews the construction of the Standard Model, with a special emphasis on topics related to spontaneous electroweak symmetry breaking and the Electroweak Symmetry Breaking Sector (EWSBS). This chapter also includes the current state of research on the Higgs and future experimental prospects.\myspace  
\textbf{Chapter 2} serves as an introduction to effective field theories from a general perspective and provides a detailed description of HEFT, which will be used throughout this work. Additionally, all the $2 \to 2$ process amplitudes relevant for the one-loop level study are presented, including the necessary counterterms to make them finite.\myspace  
\textbf{Chapter 3} reviews the partial wave analysis and the most commonly used unitarization techniques in the field, highlighting the importance of two methods in particular due to their analytical properties: the Inverse Amplitude Method (IAM) and the Improved K-matrix (IK-matrix). Both the single-channel and coupled-channel versions of these methods are presented, with the latter being, as already mentioned, highly relevant for this work.\myspace  
\textbf{Chapter 4}, based on our Ref.~\cite{Asiain:2021lch}, presents the results obtained in the study of vector resonances. Specifically, the partial waves in the vector projection are shown, along with their unphysical—non-unitary—behavior at high energies. Detailed information is provided on the vector resonances found during the unitarization process using the single-channel IAM, which is the only relevant one for vector waves.\myspace  
\textbf{Chapter 5} repeats the analysis previously performed for vector resonances but focuses on the scalar projection for the study of scalar resonances. In particular, the differences between using the single-channel and coupled-channel methods are highlighted, emphasizing the importance of the latter for a correct description of the resonances emerging in this channel. A systematic methodology is also presented to derive bounds on the parameters of the effective theory by studying the properties of the emerging scalar resonances or their absence. This chapter concludes with tentative results on the $H(650)$, a possible scalar resonance of recent experimental interest, reproducing its properties through our coupled-channel analysis in HEFT. The results of this chapter can also be found in our original works in Refs.~\cite{paper2_doi, Asiain:2023myt}.\myspace  
Finally, \textbf{Chapter 6} concludes this thesis and focuses on the combined application of effective theories and unitarization techniques in a framework different from the Standard Model. Specifically, we study the spectrum of scalar resonances that appears in Einstein-Hilbert gravity when, understood as the lowest-order term of an effective theory, higher-order operators are added in the expansion.


\clearpage
\thispagestyle{empty}
\hfill
\clearpage

\pagenumbering{arabic}
\setcounter{page}{1}
\tableofcontents

\renewcommand{\headrulewidth}{1pt}

\hypersetup{colorlinks=true, linkcolor=blue, citecolor=blue}


\lhead{Chapter 1}
\rhead{Introduction}

\chapter{Introduction}
\label{chp:Introduction}
Unraveling the mysteries of nature is undoubtedly one of the most exciting and ambitious tasks that human kind can undertake. Understanding the rules of how the natural world operates in its entirety, from the cosmological scales where galaxies collide to the microscopic ones where electron scattering takes place, and passing through biological tissues, is the greatest treasure we can create as a society. There is a Greek proverb that says that \textit{a society grows when old people plant trees whose shade they know they will never sit in}; the legacy of everything we are learning now, built upon what previous generations left us, will pave the way for future generations. In this endless cycle we grow as a whole, as a part of something bigger, and science stands as the backbone of this growth, generating popular culture around consensus. Hence, the beautiful journey towards knowledge is not only about the present; the infinite wheel will never stop as long as we know where we come from, where we are, and we have at least an intuition of where we should be heading.\myspace  
This dissertation explores the unknown aspects of the electroweak symmetry breaking sector (EWSBS) and, in particular, the properties of the Higgs boson, a hot topic in the field of particle physics. It focuses particularly on its self-interactions, which are fundamental to understanding the phenomena of symmetry breaking in the Standard Model (SM). Additionally, these self-interactions may provide insights into the origin of the Higgs boson which remains, as for today, an open question. Our starting point is the assumption of a high-energy strongly interacting theory, whose effects at the electroweak (EW) scale manifest as the EWSBS in the SM. We will combine the tools of effective field theories, which provide simplified models to describe complex interactions at a certain \textit{low}-energy scale, and unitarization techniques, which ensure that our predictions adhere to the fundamental principle of unitarity in quantum mechanics. This approach allows us to make robust and reliable predictions about the behavior of the EWSBS at experimentally reachable energies today.\myspace
Let us start from the beginning: where we come from. In the case of fundamental physics, one of the questions that has long captivated the community, prompting great efforts to answer, is related to the ultimate composition of matter: what we are made of at the most fundamental level. Even in ancient Greece, philosophers speculated about the idea of indivisible matter, the building blocks of nature. However, it was not until the end of 19th century that the first ideas, that many years later would give rise to the current atomic model, were proposed. It is certainly outstanding that in just about a century, we have progressed from describing the atom as a solid structure with embedded electrons in its surface to discovering the Higgs boson in 2012, the last missing piece that completes the puzzle of the Standard Model of particle physics.\myspace
\section{The Standard Model of particle physics}
The Standard Model (SM) of particle physics is one of the most sucesful theoretical frameworks ever created as it has been proven to be extremely predictive in the experiments. This theory combines the principles of quantum mechanics and special relativity to list the fundamental\footnote{they are fundamental in today's understanding} particles and to describe how they interact among each other. The SM is what we know as a \textit{Quantum Field Theory} (QFT).
\subsection{SM matter content and the fundamental forces}
\begin{figure}[t]
\begin{center}
	\includegraphics[width=0.45\textwidth]{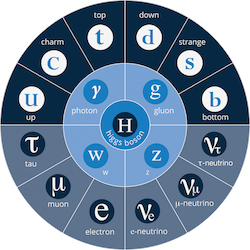} 
\end{center}
\vspace{-5mm}
\caption{\label{fig:SM_content} \small Complete matter content of the SM. The outer circle represents the flavor sector, featuring quarks in the upper half and leptons in the lower half. The middle circle contains the gauge sector with the so-called \textit{intermediante vector bosons}, and the Higgs particle is placed in the innermost circle. This picture is taken from Ref.~\cite{SMcircle}.}
\vspace{2mm}
\end{figure}

Within the SM, two fundamental building blocks are identified: \textit{leptons}, positioned in the outermost lower semicircle of Fig~\ref{fig:SM_content}, and \textit{quarks}, located in the upper half. Together, they receive the name of \textit{flavor sector} and they are all fermions, meaning they have a spin of $1/2$. The matter content happens to be organized in three families with equal properties, each containing two quarks; one positively charged and one negatively charged, along with one negatively charged lepton and its associated neutrino. The components of the three families are gathered in Table~\ref{tab:SM_families} together with their SM quantum numbers that will shortly be presented. There are no differences among the three families except for increasing masses, the first family being the lightest one. With this, a muon is nothing but a heavier copy of an electron, sharing the rest of properties and quantum numbers, and the same happens in the quark sector for the quarks $u$ and $c$. As for today, there is no fundamental reason for this symmetry to be present in the theory.
\myspace
While leptons are \textit{seen}\footnote{neutrinos cannot be seen directly but we know they are present to reconcile experiment with the laws of conservation.} in the experiment, no single quark has been ever observed. This is due to the other component of the SM: forces. Particles do no interact among themselves following the same rules; their intrisic properties differ and so do their interactions. In group-theory language, the SM is symmetric under the local, also called \textit{gauge}, transformations $SU(3)_c\times SU(2)_L\times U(1)_Y$, the SM group.\myspace
The first element in the SM group, $SU(3)_c$, encapsulates the theoretical framework describing the forces among the color sector that generally receive the name of strong interactions\footnote{not to be confused with nuclear strong force which is a manifestation of the strong force at nuclear lengths}, \textit{Quantum Chromodynamics} (QCD) in QFT nomenclature. The only matter fields that feel strong interactions are the quarks, as they possess a \textit{color charge}, typically referred to as $r,g\text{ and }b$. This unique characteristic places them in a special position within the SM, as they experience a phenomenon called \textit{confinement}, which means they cannot be isolated at low energies. Instead, quarks rapidly combine to form composite particles known as hadrons (such as protons and neutrons) through a process called \textit{hadronization}. Consequently, as stated before, quarks cannot be observed directly in experiments as asymptotic states; we can only detect the hadrons they form which are colorless (or \textit{white}) and lack a color charge. However, the theoretical structure of strong forces predicts that at high energies QCD becomes weakly coupled and quarks could be observed as free particles (see ALICE and RHIC experiments in Refs.~\cite{aliceexp,rhicexp}). In contrast, leptons are colorless and do not experience this force.\myspace
The $SU(2)_L\times U(1)_Y$ part of the SM group enfold the electroweak interactions; the unification of relativistic electromagnetic interactions and the weak forces at the EW scale. S.Weinberg and A.Salam independetly developed in 1967 and 1968, respectively, the framework for understanding electromagnetic and weak interactions as two facets of the same fundamental interaction. This framework is known as the Weinberg-Salam model, based in part on previous work by S. Glashow\footnote{The three authors were awarded the Nobel Prize in Physics in 1979, following the confirmation of the existance of neutral current interactions---mediated by $Z$---a key prediction of the electroweak theory. See Ref.~\cite{HASERT1973138}.}. They used the concept of \textit{weak isospin}, parallel to the idea of nuclear strong isospin originally introduced by Heisenberg in Ref.~\cite{Heisenberg+1979+229+238}, a quantum number that captures the fact that weak interactions maximally violate parity, meaning they distinguish between left-handed and right-handed particles. Left-handed particles, both quarks and leptons, have total weak isospin $I=1/2$ and third component $I_3=\pm 1/2$ depending on their position when arranged into the doublets of the fundamental representation of $SU(2)$. On the other side, right-handed charged particles are singlets of $SU(2)_L$ so they do not transform under that symmetry group. Within SM, rigth-handed neutrinos are not included. For this reason, the SM is said to be \textit{chiral} in the sense that it distniguishes between the left-handed and right-handed components of every particle.
\begin{table}[t]
\centering
\begin{tabular}{|c|c|c|c|c|c|}
\hline
 & \parbox[c][1.2cm][c]{2cm}{\centering Family \\ \vspace{1mm} 1 \quad 2 \quad 3} & Q  & \parbox[c][1cm][c]{3cm}{\centering I \\ \vspace{1mm} left \quad right} & \parbox[c][1cm][c]{3cm}{\centering $\text{I}_3$ \\ \vspace{1mm} left \quad right} & Color  \\
\hline
Leptons & \parbox[c][1cm][c]{2cm}{\centering $\nu_e$\quad$\nu_\mu$\quad$\nu_\tau$\\ \vspace{1mm} $\,e$\quad$\,\mu$\quad$\,\tau$}  & \parbox[c][1cm][c]{2cm}{\centering $0$\\ \vspace{1mm} \centering $-1$} & $1/2$\quad\parbox[c][1cm][c]{1cm}{\centering $\text{-}$ \\ \vspace{1mm}$0$} & \parbox[c][1cm][c]{1cm}{\centering $1/2$ \\ \vspace{1mm}$-1/2$}\quad\parbox[c][1cm][c]{1cm}{\centering $\text{-}$ \\ \vspace{1mm}$0$} & $\text{-}$  \\
\hline
 Quarks & \parbox[c][1cm][c]{2cm}{\centering $u$\quad$c$\quad$t$\\ \vspace{1mm} $d$\quad$s$\quad$b$} & \parbox[c][1cm][c]{2cm}{\centering $+2/3$\\ \vspace{1mm} \centering $-1/3$}  & $1/2$\quad\parbox[c][1cm][c]{1cm}{\centering $0$ \\ \vspace{1mm}$0$} & \parbox[c][1cm][c]{1cm}{\centering $1/2$ \\ \vspace{1mm}$-1/2$}\quad\parbox[c][1cm][c]{1cm}{\centering $0$ \\ \vspace{1mm}$0$} & $\text{r, g, b}$  \\
\hline
\end{tabular}
\caption{\small Quantum numbers under SM symmetries for the three families of matter content of the SM: leptons and quarks.}
\label{tab:SM_families}
\end{table}
\myspace
When imposing the theory to be symmetric under local transformations by means of the covariant derivative, the gauge sector naturally arises, giving a clear interpretation of the interactions among the matter sector. The conception of two particles interacting requires a mediation of a spin-$1$ particle exchanging, called gauge boson or \textit{carrier}. This nontrivial requirement ensures the renormalizabilty of all local interactions, allowing the infinites that appear in QFT calculations to be controlled and treated meaningfully. The gauge bosons are depicted in the middle circle of Fig~\ref{fig:SM_content} which shows: the photon, $W^{\pm}$ and $Z$ as carriers of the electroweak interactions and the gluon being the mediator of the strong force.\myspace 
In a Lagrangian density language, customary for QFT, the interactions described above that are invariant under the SM group are:
\begin{equation}\label{eq_int_lagbeforessb}
\begin{aligned}
\mathcal{L}_{\text{SM}} = & -\frac{1}{4} G^a_{\mu\nu} G^{a\mu\nu} - \frac{1}{4} W^i_{\mu\nu} W^{i\mu\nu} - \frac{1}{4} B_{\mu\nu} B^{\mu\nu} \\
& +  \overline{Q}_L i \cancel{D} Q_L + \overline{u}_R i \cancel{D} u_R + \overline{d}_R i \cancel{D} d_R + \overline{L}_L i \cancel{D} L_L + \overline{e}_R i \cancel{D} e_R ,
\end{aligned}
\end{equation}
where we define the field-strength
\begin{equation}\label{eq_int_buildingblocks}
\begin{aligned}
G^a_{\mu\nu} &= \partial_\mu G^a_\nu - \partial_\nu G^a_\mu + g_s f^{abc} G^b_\mu G^c_\nu \\
W^i_{\mu\nu} &= \partial_\mu W^i_\nu - \partial_\nu W^i_\mu + g \epsilon^{ijk} W^j_\mu W^k_\nu \\
B_{\mu\nu} &= \partial_\mu B_\nu - \partial_\nu B_\mu, \\
\end{aligned}
\end{equation}
and the covariant derivative
\begin{equation}
\begin{aligned}
\cancel{D}&=\gamma^{\mu}D_{\mu}\\
D_{\mu}&=\partial_{\mu}-ig_sG_{\mu}^a\frac{\lambda^a}{2}-igW_{\mu}^i\frac{\sigma^i}{2}-ig^{\prime}YB_{\mu},
\end{aligned}
\end{equation}
where $\lambda$ and $\sigma$ are the set of Gell-Mann and Pauli matrices generating the transformations of the $SU(3)_c$ and $SU(2)_L$ group, respectively.\myspace  
There are eight vector fields, $G^a_\mu$ with $a=1,...,8$, associated to the $SU(3)_c$ group describing strong interactions and, in total, there are four associated to the electroweak force; three of which come from the $SU(2)_L$ part, $W^i_{\mu}$ with $i=1,2,3$, and the remaining one, $B_{\mu}$, from $U(1)_Y$.\myspace
It also should be notice how the tensor structures $G_{\mu\nu}^a$ and $W_{\mu\nu}^i$ in Eq.~(\ref{eq_int_buildingblocks}) force the vector fields that belong to the same symmetry group, sharing the gauge coupling. This translates into the fact that all these components interact the very same way as they mix among one another via gauge transformations. We call $g_s$ the gauge coupling for the strong interactions $SU(3)_c$, $g$ the gauge coupling of the $SU(2)_L$ part of the electroweak interactions and $g^{\prime}$ the gauge coupling for $U(1)_Y$.\myspace
In Eq~({\ref{eq_int_lagbeforessb}), $Q_L$ and $L_L$ are the quark and lepton doublets, respectively, and $u_R$, $d_R$ and $e_R$ are the type-$u$, type-$d$ and type-$e$ right-handed component, respectively. The chiral nature of the SM is made explicit in this separation of the fermionic spinors into their left (L) and right (R) components.\myspace
However, there is a big problem emerging from the gauge principle that guides the construction of the SM. While gauge invariance requires massless fermions and mediators --see the abscense of mass terms in Eq.~(\ref{eq_int_lagbeforessb})-- the flavor sector and part of the gauge sector, in particular the $W^{\pm}$ and $Z$, have to be massive in order for the theory to agree with data. The opposite situation would give rise to a long-range weak force. The problem to solve is now clear and reads as the need for a mechanism that gives mass to the whole spectrum of elementary particles while preserving gauge invariance and, even more, the renormalizability of the theory. 
A first attempt to solve this issue was made by Salam and Weinberg, who proposed \textit{spontaneous symmetry breaking}, which, when applied to the case of an $SU(2)\times U(1)$ symmetry, would lead to the \textit{Higgs mechanism}.
\subsection{Spontaneous Symmetry Breaking and the Goldstone Theorem}
The idea behind spontaneous symmetry breaking (SSB) is as follows. Let us consider the case of a theory with a single massive scalar. The Lagrangian density describing the most simple local interactions that are symmetric under parity $\left(\phi\to -\phi\right)$ reads
\begin{equation}\label{eq_int_lagtoy}
\lag=\frac{1}{2}\partial^{\mu}\phi\partial_{\mu}\phi-\frac{1}{2}m^2\phi^2-\frac{\lambda}{4}\phi^4\equiv \frac{1}{2}\partial^{\mu}\phi\partial_{\mu}\phi-V\left(\phi\right).
\end{equation}
The parabolic potential above presents one minimum at $\phi=0$ and the system remains perfectly stable around that point. For this situation, it is pretty much the end of the story; the theory is renormalizable and nothing noteworthy happens. However, let's assume for a moment that we flipped the sign of the mass term in Eq.~(\ref{eq_int_lagtoy}). Up to a constant, the new Lagrangian is
\begin{equation}\label{eq_int_lagtoy_broken}
\lag=\frac{1}{2}\partial^{\mu}\phi\partial_{\mu}\phi-\frac{\lambda}{4}\left(\phi^2-\frac{m^2}{\lambda}\right)^2\equiv \frac{1}{2}\partial^{\mu}\phi\partial_{\mu}\phi-V\left(\phi\right).
\end{equation}
This modified potential, that for obvious reasons receives the name of \textit{mexican hat} potential, is shown in the right pannel of Fig~\ref{fig:SSB} and exhibits three relative extrema; one remains at $\phi=0$, which becomes a maximum, and the two new, that are located at $\phi=\pm\sqrt{m^2/\lambda}$, are symmetric minima. Perturbing now around $\phi=0$ is not a good idea if we want to avoid tachyonic states, i.e., particles with imaginary masses, and manifestly unstable vacuum solutions. Instead, what we can do is select one out of the two degenerate minima and apply all the perturbation theory machinery around it to get a perfectly sensible theory. It is precisely the arbitrary choice of the ground state among the minima that makes the theory said to be \textit{spontaneously broken}. 
\begin{figure}[t]
\begin{center}
	\includegraphics[width=0.45\textwidth]{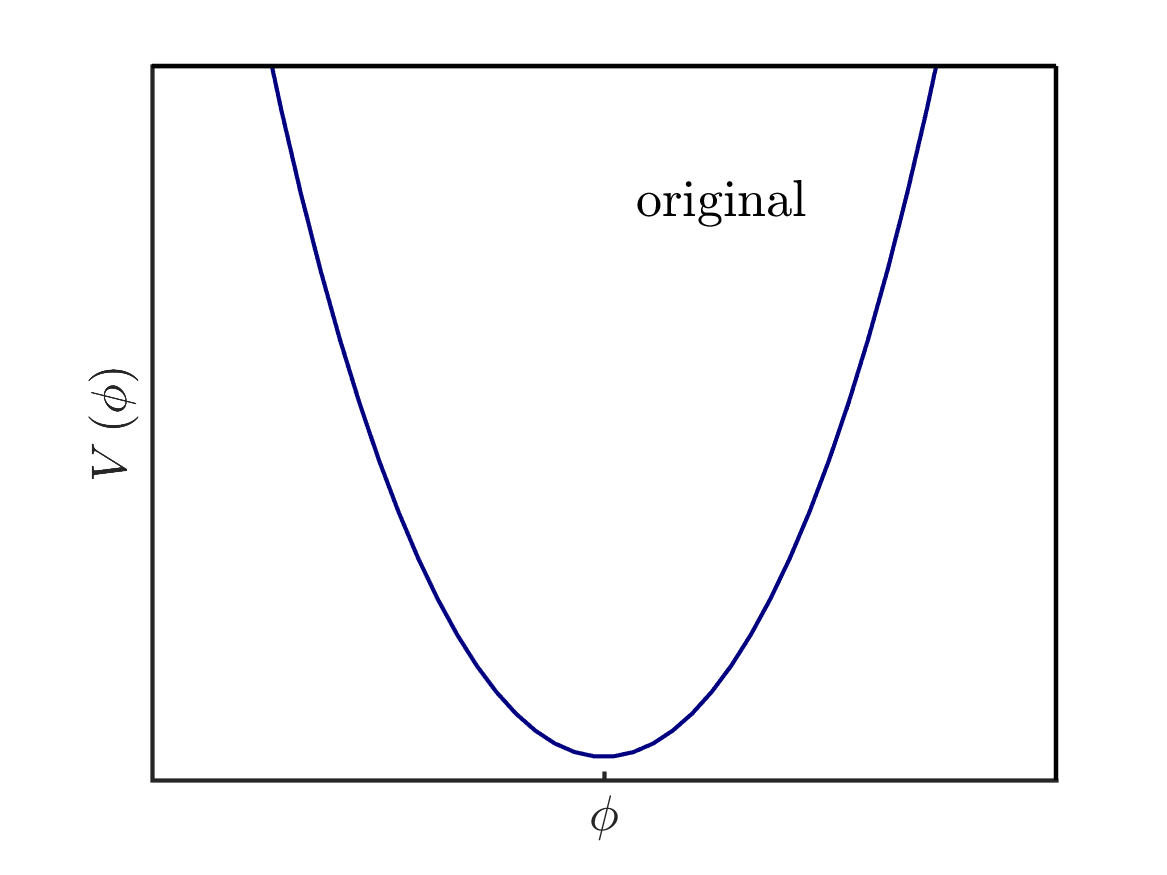} 
	\includegraphics[width=0.45\textwidth]{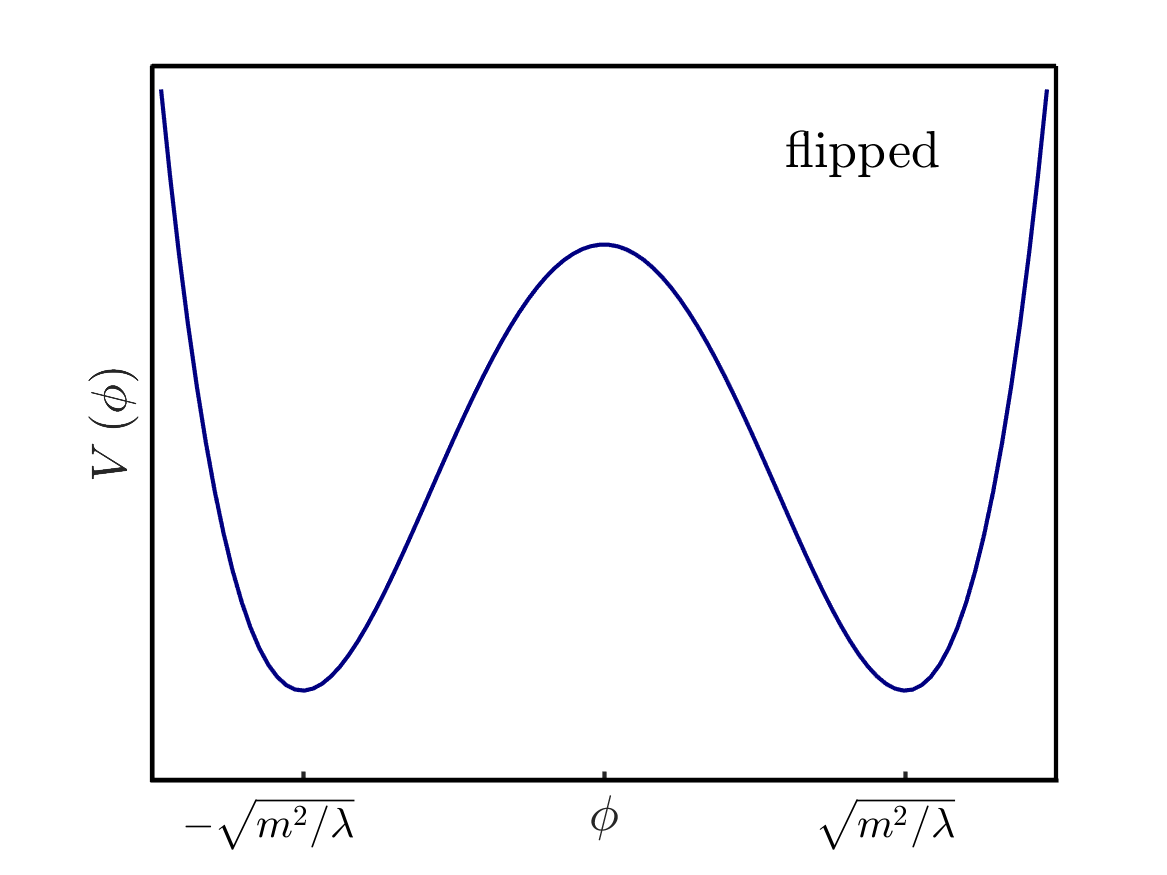} 
\end{center}
\vspace{-5mm}
\caption{\label{fig:SSB} \small Potential of a theory with one self-interacting scalar (left) before and (right) after \textit{spontaneous symmetry breaing} of the vacuum of the theory by flipping the sign of the mass term.}
\vspace{2mm}
\end{figure} 
\myspace
Now, we can define a new field $\phi^{\prime}$ whose expected value in the ground state, referred as \textit{vaccum expectation value}, or simply vev, and denoted by $\langle 0|\phi^{\prime}|0\rangle$, is equal to zero:
\begin{equation}\label{eq_int_field_shifted}
\phi^{\prime}=\phi-\sqrt{m^2/\lambda}.
\end{equation}
In terms of the shifted field, the Lagrangian in Eq.~(\ref{eq_int_lagtoy_broken}) reads 
\begin{equation}\label{eq_int_lag_shifted}
\lag=\frac{1}{2}\partial^{\mu}\phi^{\prime}\partial_{\mu}\phi^{\prime}-m^2\phi^{\prime\,2}-\sqrt{m^2\,\lambda}\phi^{\prime\,3}-\frac{\lambda}{4}\phi^{\prime\,4}.
\end{equation}
This theory, whose original discrete $\mathbb{Z}_2$ symmetry is now hidden, is still described by two parameters, enhancing renormalizability, and it characterizes a physical scalar with mass $m_s=\sqrt{2}m$.\myspace
When this idea is applied to continuous symmetries, an interesting situation shows up. If $\phi$ is now promoted to be a multiplet of single fields, the Lagrangian is invariant under the general infinitesimal transformation
\begin{equation}\label{eq_int_transformation}
\delta\phi=i\epsilon_a T^a\phi,
\end{equation}
for $a=1,...,N$, where $\epsilon_a$ is an infinitesimal parameter controling the extent of the transformation and $T^a$ are the set of generators of some generic symmetry group. By playing the same game as in the case of discrete symmetries, one can write a theory with a shifted field taking values around the vev, $v$, as in Eq.~(\ref{eq_int_field_shifted}), that apparently lacks the original symmetry for \textit{some} of the generators: $T^av\neq 0$ for \textit{some} $a$. These directions in flavor space constitute a subgroup of broken directions and it can be shown, see Ref.~\cite{PhysRev.127.965}, that for every broken generator, a massless scalar field will appear in the spectrum. At the end of the day, the total number of degrees of freedom is, of course, conserved as the broken charges \textit{disappearing} into the vacuum manifest themselves as massless and spinless fields with the same quantum numbers as the ground state. This is the Goldstone Theorem and the massless scalars are called the Goldstones.\myspace
The spontaneous breaking of the chiral symmetry by the vacuum of strong interactions is paradigmatic in the context of particle physics. Chiral symmetry is a global $SU(2)_L\times SU(2)_R$ symmetry that possesses the QCD Lagrangian when one takes as light degrees of freedom massless quarks from the first family, \textit{i.e.}, $u$ and $d$. The ground state of this simplified version of the theory has non-zero vev, corresponding to a spinless quark condensate, and the theory after SSB is only invariant under the diagonal subgroup of transformations satisfying $L=R$. Thus, the symmetry breaking pattern $SU(2)_L\times SU(2)_R\to SU(2)_{L+R}$ triggers the appearance of three Goldstones, the precise reduction in dimensionality\footnote{The number of generators of $SU(N)$ corresponds to $N^2-1$.} of the system before and after the symmetry breaking, and generate mass terms for the lightest baryons in the effective QCD Lagrangian. These three Goldstones are identified with the pions, which, although light compared to the mass of the proton\footnote{The mass of the pions is $m_{\pi}\sim 150\text{MeV}< \Lambda_{QCD}.$}, are massive, in clear contradiction to what the Goldstone Theorem states. This can be understood when one recovers the masses of the quarks $u$ and $d$, that softly break chiral symmetry explicitely; the light masses of the pions are the way for theory to remember that chiral symmetry was never exact but approximate. For this reason, pions receive the name of \textit{pseudo}-Goldstone bosons.
%
%
\subsection{Electroweak Symmetry Breaking for the SM}
Having established the foundational concepts of SSB and the implications of the Goldstone theorem in the preceding section, we now focus specifically on its application within the context of electroweak symmetry in the Standard Model.\myspace
What we expect from a symmetry-breaking-like mechanism is that, by introducing a scalar field that spontaneously breaks the vacuum of the electroweak theory, masses for the fermions and the gauge sector of the SM will be generated while respecting the principles of gauge invariance. This scalar field, along with the Goldstone bosons produced through the breaking process, forms the scalar sector. While the necessity of such a mechanism is clear, the parametrization of the breaking is not an evident question. This subsection will be devoted to this matter.\myspace
Before SSB, the Lagrangian density that describes the dynamics of the scalar sector is
\begin{equation}\label{eq_in_Higgslagrangian}
\lag_{\Phi} = (\partial^\mu \Phi)^\dagger (\partial_\mu \Phi) - \mu^2 \Phi^\dagger \Phi - \lambda (\Phi^\dagger \Phi)^2,
\end{equation}
where $\Phi$ transforms under \sutwo and hence contains four real scalar degrees of freedom. This Lagrangian exhibits a global $SU(2)_L\times SU(2)_R$ symmetry called electroweak chiral symmetry. However, how the Goldstones and the Higgs field transform separately under EW chiral transformations, is a choice that depends on the specific physics one wishes to describe.\myspace
In particular, there are two ways of realizing the EW chiral symmetry, depending on the assumptions about the scalar sector: linear and non-linear realizations. In the linear realization, which is the way that the symmetry is realized in the SM, the Higgs and the Goldstone bosons are embedded in an \sutwo doublet that transforms linearly under chiral transformations. On the other hand, the non-linear realization assumes that the Goldstone bosons are included in a triplet and transform non-linearly while the Higgs is treated separately as a chiral singlet, i.e., does not transform under the chiral symmetry.\myspace
The properties and interpreation of both approaches are described in the following subsections, taking as an illuminating example the $\sigma$-model, firstly proposed back in the 60s by Gell-Mann and Levy in Ref.~\cite{Gell-Mann:1960mvl}, in the context of low-energy QCD. 
\subsubsection{\centering The Linear $\sigma$-Model and the Linear Realization of the Symmetry}
%
Let us now turn to the $\sigma$-model in its linear version. Neglecting for the moment the gauge sector and the fermions, we can describe the dynamics of a scalar sector containing four degrees of freedom, $\sigma$ and $\pi^a$ with $a=1,2,3$, with the following Lagrangian density, similar to Eq.~(\ref{eq_in_Higgslagrangian}),
\begin{equation}\label{eq_in_linearsigma}
\lag_{\sigma}^{\text{lin}}=\frac{1}{2}\del_\mu\sigma\del^\mu\sigma+\frac{1}{2}\del_\mu\pi^a\del^\mu\pi^a-\frac{\mu^2}{2}\left(\sigma^2+\pi^2\right)-\frac{\lambda}{4}\left(\sigma^2+\pi^2\right)^2,
\end{equation}
where in the last two terms, one needs to understand $\pi^2=\pi^a\pi^a=(\pi^1)^2+(\pi^2)^2+(\pi^3)^2$ and not as the second component of the trio.\myspace
The Lagrangian in Eq.~(\ref{eq_in_linearsigma}) is invariant under \chiral chiral symmetry, although this might not be immediately obvious. To see this, one can construct a bi-doublet
\begin{equation}\label{eq_in_bidoubletlin}
\Sigma=\sigma\identity+i\sigma^a\pi^a \quad\to\quad \frac{1}{2}\text{Tr}\left(\Sigma^{\dagger}\Sigma\right)=\sigma^2+\pi^2,
\end{equation}
where $\sigma^a$ are the set of Pauli matrices generating the transformations of \sutwo. The structure above transforms as $\Sigma\to U_L\Sigma U^{\dagger}_R$ with $U_{L,R}=\exp\left(-i\varepsilon_{L,R}^a\sigma^a/2\right)$ being arbitrary \sutwo matrices.\myspace
The Lagrangian of the linear $\sigma$-model can be expressed as
\begin{equation}\label{eq_in_linearsigma_Sigma}
\lag_{\Sigma}^{\text{lin}}=\frac{1}{4}\trace\left(\del_\mu\Sigma^{\dagger}\del^{\mu}\Sigma\right)-\frac{\mu^2}{4}\trace\left(\Sigma^{\dagger}\Sigma\right)-\frac{\lambda}{16}\trace\left(\Sigma^{\dagger}\Sigma\right)^2,
\end{equation}
which now exhibits chiral invariance explicitely. The scalar fields transform as follows:
\begin{equation}
\begin{aligned}
\sigma&\to\sigma+\frac{1}{2}\pi^a(\varepsilon^a_L-\varepsilon^a_R)\\
\pi^a&\to\pi^a-\frac{1}{2}(\varepsilon^a_L-\varepsilon^a_R)\sigma-\frac{1}{2}\epsilon^{ajk}\pi^j(\varepsilon^k_L+\varepsilon^k_R),
\end{aligned}
\end{equation}
where $\epsilon^{ajk}$ is the fully-antisymmetric Levi-Civitta tensor. There are two particular cases to consider. First, the case of an isospin rotation $\varepsilon^a_L=\varepsilon^a_R$, which makes $\sigma$ transform as a singlet and the three $\pi^a$ together as an \sutwo triplet. Secondly, in the case of an axial rotation  $\varepsilon^a_L=-\varepsilon^a_R$, all the components of the scalar sector mix among each other.\myspace
As it happened in the case of having one scalar, the vacuum of the linear $\sigma$-model in Eq.~(\ref{eq_in_linearsigma}) presents nothing but a trivial minimum at $\sigma=\pi^i=0$, corresponding to a system of four degenerate scalar excitations around the origin, and it remains perfectly invariant under the chiral symmetry. However, if we flipped the sign of the mass term, or simply assumed that $\mu^2<0$, we can still make perfect sense of the theory. The true vaccum of that flipped theory is any point in the scalar manifold such that
\begin{equation}\label{eq_in_vevofSigma}
\frac{1}{2}\langle\trace\left(\Sigma^{\dagger}\Sigma\right)\rangle=\sigma^2+\pi^2=-\frac{\mu^2}{\lambda}\equiv f^2.
\end{equation}
One choice among all the possible vacua is the one given by
\begin{equation}\label{eq_in_vaccum_linear}
\langle\sigma\rangle = f,\qquad \langle\pi^a\rangle =0, 
\end{equation}
that breaks the chiral symmetry sponteneously.\myspace
Of course, the vacuum in Eq.~(\ref{eq_in_vaccum_linear}) is just one of the options, but it undoubtedly represents the simplest choice satisfying Eq.~(\ref{eq_in_vevofSigma}). Any other physically viable selection would lead to more complicated expressions without altering the physical interpretation.\myspace
In terms of the shifted fields $\sigma^{\prime}=\sigma-f$ and $\pi^a$, that take values close to the minimum of the potential, the linear $\sigma$-model after SSB can be written as
\begin{equation}\label{eq_in_linearsigma_broken}
\lag_{\sigma}^{\text{lin-SSB}}=\frac{1}{2}\del_\mu\sigma^{\prime}\del^\mu\sigma^{\prime}+\frac{1}{2}\del_\mu\pi^a\del^\mu\pi^a-\lambda f^2\sigma^{\prime 2}-\lambda f\sigma^{\prime}\left(\sigma^{\prime 2}+\pi^2\right)-\frac{\lambda}{4}\left(\sigma^{\prime 2}+\pi^2\right)^2,
\end{equation}
which describes the dynamics of a massive $\sigma^{\prime}$ particle with $m_{\sigma^{\prime}}=\sqrt{2\lambda}f$ and three massless $\pi^a$.\myspace
The vaccum of the linear $\sigma$-model after SSB in Eq.~(\ref{eq_in_linearsigma_broken}) remains invariant only under the diagonal, or vector, subgroup $L=R$ triggering the global symmetry breaking pattern $SU(2)_L\times SU(2)_R\to SU(2)_{V}$. The corresponding reduction in dimensionality when the chiral group is broken down to the vector group is precisely three; the number of massless particles that we have found in the broken phase. In other words, the $\pi^a$ are the Goldstones associated to the global SSB of the chiral symmetry.\myspace
In this simplified model, any Dirac fermionic field $\psi$ acquires mass from the SSB mechanism via a Lagrangian that contains interactions with the scalar sector,
\begin{equation}\label{eq_in_lagpsi}
\lag_{\psi}^{\text{lin}}=i\bar{\psi}\cancel{\partial}\psi+g\bar{\psi}\left(\sigma+i\sigma^a\pi^a\gamma^5\right)=i\bar{\psi}\cancel{\partial}\psi+g\bar{\psi}_L\Sigma\psi_R+g\bar{\psi}_R\Sigma^{\dagger}\psi_L,
\end{equation}
where the decomposition $\psi=\psi_L+\psi_R$ has been used.\myspace
The kinetic piece preserves the full chiral symmetry as the chiral components of the fermion transform as $\psi_{L,R}\to U_{L,R}\psi_{L,R}$. After SSB, the interaction terms among the fermions and the spontaneously-broken scalar sector generate contributions such as
\begin{equation}
\lag_{\psi}^{\text{lin}}\xrightarrow{\text{SSB}}\lag_{\psi}^{\text{lin-SSB}}\supset gf\bar{\psi}\psi,
\end{equation}
which is a mass term for the fermionic field. This term describes a fermion with mass $m_{\psi}=gf$, proportional to the vev of the scalar sector.
\subsubsection{\centering The Non-Linear $\sigma$-Model and the Non-Linear Realization of the Symmetry}
In the previous section, we showed that in the broken phase, the scalars in the spectrum, $\sigma$ and $\pi^a$, transform as a singlet and a triplet, respectively, under the vector subgroup. The fact that $\sigma$ is a singlet was of particular interest historically when the $\sigma$-model can be related to the electroweak sector of the SM. This is because, if $\sigma$ in the $\sigma$-model is identified with the Higgs boson of the SM, it could be decoupled from the spectrum. This scenario was intriguing at a time when the Higgs boson had not yet been discovered, as one possibility was that it might have a mass significantly larger than the EW scale, thus not substantially influencing low-energy dynamics. However, by simply integrating out $\sigma$, one would break the \chiral symmetry. This raises the question: is it possible to have a theory that can be described solely by $\pi^a$ while maintaining full chiral invariance? The answer is affirmative, though it requires accepting non-linear transformations for $\pi^a$. In this case, the symmetry is said to be non-linearly realized.\myspace 
The Goldstone contribution is parametrized using the exponential representation:
\begin{equation}\label{eq_in_exponential}
U = \exp\left(\frac{i\sigma^a\zeta^a}{f}\right) = \cos\left(\frac{\zeta}{f}\right)\identity + i \frac{\sigma^a\zeta^a}{\zeta}\sin\left(\frac{\zeta}{f}\right),
\quad U^{\dagger}U = \identity,
\end{equation}
which is customary in non-linear studies, where $\zeta = \sqrt{\zeta^a\zeta^a}$. There are alternative parametrizations, such as the spherical parametrization in Ref.~\cite{Dobado:2019fxe}, but they will ultimately lead to the same physical results as they consist in field redefintions (see Refs.~\cite{Kamefuchi:1961sb,CHISHOLM1961469}).\myspace
With the discovery of the Higgs boson at the LHC, we have a phenomenological reason for incorporating a scalar singlet into the low-energy spectrum of the non-linear $\sigma$-model. This inclusion does not introduce any additional caveats to the original question, as it does not transform under chiral symmetry. In non-linear variables, we denote the singlet as $\varphi$ and the set of three Goldstones as $\zeta^a$. In this non-linear framework, we redefine the bi-doublet as
\begin{equation}\label{eq_in_bidoubletnonlin}
\Sigma = \sigma + i \sigma^a\pi^a \equiv \varphi U = \varphi \exp\left(\frac{i\sigma^a\zeta^a}{f}\right).
\end{equation}
Since $\varphi$ is a chiral singlet, $U$ must transform under chiral transformations in the same way as $\Sigma$:
\begin{equation}
U \to U_L U U_R.
\end{equation}
From Eq.~(\ref{eq_in_exponential}) and Eq.~(\ref{eq_in_bidoubletnonlin}), we can relate the fields in both realizations
\begin{equation}\label{eq_in_fieldredefinition}
\begin{aligned}
\sigma&=\varphi\cos\left(\frac{\zeta}{f}\right)\\
\pi^a&=\varphi\frac{\zeta^a}{\zeta}\sin\left(\frac{\zeta}{f}\right),
\end{aligned}
\end{equation}
where non-linearity is evident.\myspace
From the transformations above and the properties of Pauli matrices, the relation $\varphi^2=\sigma^2+\pi^2=\frac{1}{2}\trace\left(\Sigma^{\dagger}\Sigma\right)$ should be noticed and, once again, one can select after SSB, when $\mu^2$ changes sign, a vacuum satisfying
\begin{equation}\label{eq_in_vaccum}
\langle\varphi\rangle = f,\qquad \langle\zeta^a\rangle =0,
\end{equation}
where the definition of $f$ remains unchanged with respect to the linear case in Eq.~({\ref{eq_in_vevofSigma}}).\myspace
Once again, as was the case with the linear $\sigma$ model, the choice of this vacuum represents the simplest option but is not the only one that minimizes the potential.\myspace
Now, the agrangian in Eq.~(\ref{eq_in_linearsigma}), can be written in non-linear shifted variables $\varphi^{\prime}=\varphi-f$ and $\zeta^a$ as
\begin{equation}\label{eq_in_nonlinearlag}
\lag_{\varphi}^{\text{non-lin}}=\frac{1}{2}\partial_{\mu}\varphi^{\prime}\partial^{\mu}\varphi^{\prime}+\frac{\left(\varphi^{\prime}+f\right)^2}{4}\trace\left(\partial_{\mu}U^{\dagger}\partial^{\mu}U\right)-\lambda f^2\varphi^{\prime 2}-\lambda f \varphi^{\prime 3}-\frac{\lambda}{4}\varphi^{\prime 4},
\end{equation}
which represents a massive scalar $\varphi^{\prime}$ with mass $m_{\varphi^{\prime}}=\sqrt{2\lambda}f$ and a set of three massless $\zeta^a$. The Lagrangian in Eq.~(\ref{eq_in_nonlinearlag}) is the non-linear $\sigma$-model.\myspace
One important different between the linear realization of the symmetry and the non-linear one is that, in the latter the Goldstones interact among themselves and the singlet with derivatives.\myspace
From the Lagrangian above it can also be seen that the Goldstone sector is completely factorized out from the $\varphi$ field, which makes integrating-out the latter a trivial task in contrast to the linear case. A non-linear effective theory that only contains the pions as dynamical states can be built by simply removing $\varphi$ from the Lagrangian in Eq.~(\ref{eq_in_nonlinearlag}). This way of proceeding gives a leading order contribution
\begin{equation}\label{eq_in_nonlineargoldstoneEFT}
\lag^{\text{EFT non-lin}}=\frac{f^2}{4}\trace\left(\partial_{\mu}U^{\dagger}\partial^{\mu}U\right),
\end{equation}
that produces, by expanding $U$ up to a desired order in the $o\left(\zeta/f\right)$ expansion, the kinetc term and interactions with an arbitrary number of Goldstone insertions.\myspace
Again, and as it happened in the linear case, any fermionic field $\psi$ aquires mass via spontaneous symmetry breaking mechanism. If we insert the bi-doublet $\Sigma$ written in terms of the non-linear variables, Eq.~(\ref{eq_in_bidoubletnonlin}), in the fermionic Lagrangian in Eq.~(\ref{eq_in_lagpsi}), we obtain, after SSB,
\begin{equation}
\lag_{\psi}^{\text{non-lin}}=g\varphi\left(\bar{\psi}_L U\psi_R+\bar{\psi}_R U^{\dagger}\psi_L\right)\xrightarrow{\text{SSB}}\lag_{\psi}^{\text{non-lin-SSB}} \supset gf\bar{\psi}\psi,
\end{equation}
since at first order in Goldstone insertions $U=\identity$.\myspace
Formally, the effective theory above has little, if anything, to do with two-flavor QCD, even though they share the symmetry breaking pattern for the vaccum $SU(2)_L\times SU(2)_R\to SU(2)_V$ in the chiral limit. However, the low-energy dynamics, i.e., processes involving pions as external states, can be accurately described below an energy scale related to the mass of $\varphi$, the lightest state that had been integrated out.\myspace
In the previous sections, we have presented a model consisting of four scalars whose interactions are constructed solely based on the conditions imposed by the conservation of chiral symmetry and its subsequent spontaneous breaking. This is the $\sigma$-model. After the symmetry breaking, one of the scalars becomes massive while the other three remain massless, that are identifyed with the Goldstones. Additionally, mass terms for fermionic fields are also produced. All these are ingredients that we undoubtedly want in a model of symmetry breaking for the electroweak vacuum in the SM.\myspace
We have also presented both the linear and nonlinear realizations of the complete chiral symmetry by redefining the scalar fields in the $\sigma$-model. However, it is important to emphasize that the difference between both realizations is merely a change of variables and, as such, should not affect the physical results. It is a matter of reordering diagram contriburions that makes physical quantities invariant. More about this will be said in the next chapter.
\subsection{The Higgs Mechanism}
From the previous section, it is clear that SM requires a symmetry-breaking-like mechanism to provide mass to elementary particles. The solution proposed for the SM case is the \textit{Higgs mechanism}, developed in the 1960s by several physicists, including Peter Higgs, François Englert, and Robert Brout; hence, from the initials, it is sometimes referred to as the BEH mechanism. The original paper by Higgs is found in Ref.~\cite{PhysRevLett.13.508}.\myspace 
Inspired by the $\sigma$-model presented above, one can construct a bi-doublet, containing four scalars, that couples to the SM content such that, after SSB, it produces mass terms for the fermions. In the SM, the EW symmetry is realized lineraly, as this approach is more suitable for weakly coupled theories. By gauging the electroweak chiral symmetry present in the scalar sector of the SM, we also hope that the mechanism dotates the weak gauge bosons with mass. This is particularly imortant as the SM is fundamentally built upon the idea of gauge invariance---a feature that was neither present nor needeed in the $\sigma$-model, which exploits the theory's invariances under global symmetries.\myspace
From now on, we will be referring to electroweak chiral symmetry as simply chiral symmetry but it should not be confused with the chiral symmetry of the massless QCD Lagrangian or the $\sigma$-model discussed previously.\myspace
We start from the scalar sector of the SM that is described by the Lagrangian in Eq.~(\ref{eq_in_Higgslagrangian}), which presents a full chiral symmetry \chiral. Gauging this chiral symmetry consists in introducing the corresponding gauge fields via the covariant derivative, which ensures the invariance of the theory under gauge transformations of the group $SU(2)_L\times U(1)_Y$. The covariant derivative is given by 
\begin{equation}\label{eq_in_covariantderivative}
D_{\mu}=\partial_{\mu}-igW_{\mu}^aT^a-ig^{\prime}YB_{\mu},
\end{equation}
where $a=1,2,3$ runs over weak isospin indices and $W_{\mu}^a$ and $B_{\mu}$ are the gauge fields for $SU(2)$ and $U(1)$, respectively.\myspace
By doing this, the entire left component of the chiral symmetry group is gauged with gauge coupling $g$, while only the abelian hypercharge $U(1)_Y$, subgroup of the right component, respects the gauge symmetry, with gauge coupling constant $g^{\prime}$. Consequently, $g^{\prime}$ explicitly breaks chiral invariance. To what extent this breaking occurs is important to assess whether the global symmetry can be retained, at least approximately, as a good symmetry of the theory.\myspace
The scalar content in the SM responsible for spontaneous symmetry breaking of the electroweak vacuum, is introduced via the Higgs field $\Phi$, that transforms as a spinor of $SU(2)_L$ and has hypercharge $Y=1/2$. In the fundamental representation:
\begin{equation}\label{eq_in_higgs_doublet}
\Phi=\begin{pmatrix}
\phi^{+}\\
\phi^0
\end{pmatrix}=\frac{1}{\sqrt{2}} \begin{pmatrix}
\phi_1+i\phi_2\\
\phi_3+i\phi_4
\end{pmatrix},
\end{equation}
where $\phi_i$ with $i={1,2,3,4}$ are real scalar fields.\myspace
By performing an appropriate rotation in the global $SU(2)\times U(1)$ space, the breaking of the EW vaccum in Eq.~(\ref{eq_in_Higgslagrangian}) can be oriented in a specific direction of our choice, such that $\langle \phi_{1,2,4}\rangle=0$ and $\langle \phi_3\rangle=v$, where $v$ is real and positive, leaving:
\begin{equation}
\langle\Phi\rangle=\begin{pmatrix}
0\\
v
\end{pmatrix}.
\end{equation}
After this rotation, the theory can be written in terms of the shifted fields $\phi_3^{\prime}=h+v$ and $\phi_{1,2,4}$ and the Lagrangian is only invariant under $U(1)$ rotations, triggering the symmetry breaking pattern $SU(2)\times U(1)\to U(1)$. This results in a mass term for $h$, while the three Goldstones from the unbroken theory remain massless.\myspace
However, a special situation arises in comparison to the previous case. Unlike the $\sigma$-model, the Higgs mechanism must include gauge invariance, not just global. The local $SU(2)$ symmetry provides extra freedom that allows us to perform field transformations to rotate away the Goldstones. By doing so, we select a gauge called \textit{unitary gauge}, where
\begin{equation}\label{eq_in_higgsfields}
\begin{aligned}
&\phi_1(x)=\phi_2(x)=\phi_4(x)=0\\
&\phi_3(x)=h(x)+v.
\end{aligned}
\end{equation}
The Higgs particle, $h$, is associated with the sole surviving degree of freedom.\myspace
Based on the $\sigma$-model studied previously, it is now clear how fermionic fields, including the electrons and quarks within the SM, acquire mass via the Higgs mechanism. The interaction terms between the Higgs field and the flavor sector, embedded in the corresponding $SU(2)_L \times U(1)_Y$ structures, are encapsulated in the Yukawa sector, described by the following Lagrangian density:
\begin{equation}\label{eq_in_yukawalagrangian}
\lag_{Yuk}=Y_u\bar{Q}_L \Tilde{\Phi} u_R+Y_d\bar{Q}_L\Phi d_R+Y_e\bar{L}_L\Phi e_R+\text{h.c.}, \quad \Tilde{\Phi}=i\sigma^2\Phi^{*},
\end{equation}
where the coefficients accompanying the operators in the expanded Lagrangian are the elements of the Yukawa matrices $Y_{u,d,e}$, whose indices belong to flavor space. Each element of the Yukawa matrices is known as a Yukawa coupling, or simply Yukawa. After spontaneous symmetry breaking, the Lagrangian in Eq.~(\ref{eq_in_yukawalagrangian}) generates mass terms for the fermionic fields that were previously forbidden by gauge invariance. The resulting masses for each field are given by:
\begin{equation}\label{eq_in_yukawamasses}
m_u \propto Y_u v,\quad m_d \propto Y_d v,\quad m_e \propto Y_e v.
\end{equation}
The Higgs mechanism clearly and trivially provides mass to the matter content of the SM through interactions with the Higgs field, as expected. However, the process of endowing the gauge sector with mass is more nuanced. A subtlety arises in Eq.~(\ref{eq_in_higgsfields}), which is closely related to the question of why the gauge sector remains massless. Before SSB, the theory contains four scalar fields; after SSB and gauge rotations, only one scalar field remains—the Higgs particle, $h$. Does this imply that the Higgs mechanism violates the conservation of degrees of freedom? The answer, as expected, is no.\myspace
Let us now come back to the point where we gauged the full chiral symmetry of the scalar sector. Substituting the partial derivative by the covariant one in Eq.~(\ref{eq_in_covariantderivative}), the kinetic term for the Higgs field after SSB reads
{\small
\begin{equation}\label{eq_in_covariantderivativeHiggs}
\left(D^{\mu}\Phi\right)^{\dagger}D_{\mu}\Phi\supset \frac{1}{2}\partial_{\mu}h\partial^{\mu}h+\frac{1}{8}g^2v^2\left(W^1_{\mu}W^{1\mu}+W^2_{\mu}W^{2\mu}\right)+\frac{1}{8}\begin{pmatrix}W^3_{\mu} & B_{\mu}\end{pmatrix}\begin{pmatrix}
g^2 & gg^{\prime}\\
gg^{\prime} & g^{\prime 2}
\end{pmatrix}
\begin{pmatrix}
W^{3\mu}\\
B^{\mu}
\end{pmatrix},
\end{equation}
}
which contains the kinetic term for the Higgs and mass terms for the guage bosons once the last term is diagonalized. Now, we can define the following fields in the charged basis
\begin{equation}
W^{\pm}_{\mu} = \frac{W^1_{\mu} \mp i W^2_{\mu}}{\sqrt{2}}, \quad
Z^0_{\mu} = \frac{g W^3_{\mu} - g^{\prime} B_{\mu}}{\sqrt{g^2 + g^{\prime\,2}}}, \quad
A^0_{\mu} = \frac{g^{\prime} W^3_{\mu} + g B_{\mu}}{\sqrt{g^2 + g^{\prime\,2}}},
\end{equation}
with physical\footnote{In this context, we refer to physical mass as the parameters of the quadratic Lagrangian with the abscense of cross terms, such as the last term of Eq.~(\ref{eq_in_covariantderivativeHiggs}).} masses
\begin{equation}
m_{W^{\pm}}=\frac{1}{2}gv, \quad m_{Z}=\frac{1}{2}\sqrt{g^2+g^{\prime 2}}v,\quad m_{A}=0.
\end{equation}
The situation now is clear: the three Goldstones, which were removed from the spectrum thorough gauge transformations, manifest themselves as the longitudinal components of the gauge fields $W^{\pm}$ and $Z$, that were massless prior to SSB due to gauge invariance, and have now become massive. In jargon, it is said that the Goldstones have been \textit{eaten} by the gauge fields. Thus, the number of degrees of freedom is conserved.\myspace
This is an enormous success from both a theoretical and phenomenological standpoint. Not only do the weak gauge bosons acquire mass, which is essential for reproducing experimental evidence, but the photon, associated with the gauge field $A_{\mu}$, remains massless. This masslessness is related to the residual $U(1)$ symmetry after SSB. Thus, the pattern of spontaneous gauge symmetry breaking in the SM is $SU(2)_L \times U(1)_Y \to U(1)_{\text{em}}$.\myspace
In addition to this successful symmetry breaking, another significant constituent of the SM after SSB is the predicted relationship between the masses of the $W$ and $Z$ bosons. If one defines the Weinberg angle as the rotation angle for diagonalizing the Lagrangian piece in Eq.~(\ref{eq_in_covariantderivativeHiggs}), one obtains
\begin{equation}\label{eq_in_cos_weinberg}
\cos\theta_W = \frac{g}{\sqrt{g^2 + g^{\prime 2}}}, \qquad \sin\theta_W = \frac{g^{\prime}}{\sqrt{g^2 + g^{\prime 2}}},
\end{equation}
where $g$ and $g^{\prime}$ are fixed by gauge invariance, and the masses of the guage boson are predicted to satisfy the relation
\begin{equation}\label{eq_in_mw_mz_relation}
m_W = m_Z \cos\theta_W \quad \to \rho\equiv \frac{m_W^2}{m_Z^2\cos^2\theta_W}=1
\end{equation}
which is independent of the value of the vev. This relation underscores a fundamental aspect of electroweak theory parametrized by the $\rho$ parameter and it predicts that the $Z$ boson is heavier than the $W$ boson. The well-known $\rho$-parameter has been measured with a really good accuracy, finding very good agreement with the SM prediction.\myspace
The mass relationship in Eq.~(\ref{eq_in_mw_mz_relation}) is intimately connected to a symmetry known as \textit{custodial symmetry}. Custodial symmetry is an exact symmetry of the pure Higgs sector (Higgs boson + Goldstones) that is described by the following Lagrangian density:
\begin{equation}\label{eq_in_higgssectorlag}
\lag_{\Phi}=\partial_\mu\Phi^\dagger\partial^\mu \Phi+m^2\Phi^\dagger \Phi-\frac{\lambda}{2}\left(\Phi^\dagger\Phi\right)^2,
\end{equation}
where $\Phi$ is the Higgs field defined in Eq.~(\ref{eq_in_higgs_doublet}).\myspace
The Lagrangian above can alternatively be written in terms of the field $\phi$ as  
\begin{equation}
\lag_\phi = \frac{1}{2} \left(\partial^\mu \phi\right) \cdot \left(\partial_\mu \phi\right) + \frac{m^2}{2} \phi \cdot \phi - \frac{\lambda}{8} \left(\phi \cdot \phi\right)^2,
\end{equation}  
where  
\begin{equation}
\phi = \begin{pmatrix}
\phi_1 \\ \phi_2 \\ \phi_3 \\ \phi_4
\end{pmatrix}.
\end{equation}\myspace
$\lag_\phi$ is invariant under transformations $\phi \to O\phi$, with $O \in O(4)$. In terms of this new field, and after symmetry breaking by the Higgs vev with $\phi_4 = v + h$ and $\langle\phi \cdot \phi\rangle = v^2$, three massless Goldstones $\phi_i = \varphi_i$ with $i = \{1, 2, 3\}$ appear as a byproduct. These transform under rotations of the $O(3)$ group.\myspace
It is precisely the symmetry under the $O(3)$ group that ensures the relation in Eq.~(\ref{eq_in_mw_mz_relation}) is exactly satisfied at tree level.\myspace
In the SM, custodial symmetry is an approximate global $SU(2)_R$ symmetry which ensures that the tree-level relation in Eq.~(\ref{eq_in_mw_mz_relation}) holds. It is an approximate symmetry as it is explicitly broken by the $U(1)_Y$ gauge interaction ($g^{\prime}\neq 0$) and by differences in the masses of the fermions; the Yukawa couplings. However, the breaking has been tested in the experiment to be relatively small, which is why is a good aproximation. This symmetry helps to protect the mass ratio of the $W$ and $Z$ bosons from large radiative corrections at the loop level, making it an essential feature for the consistency of the electroweak theory.\myspace
With all these ingredients, the Standard Model offers a theoretical framework to understand both theoretically and from the phenomenological perspective, three of the four fundamental forces of nature: electromagnetism and weak interactions, which happen to be part of the same interaction, and strong interactions. The only force it does not account for is gravity.
%


\section{Current status of the Higgs boson}
From the previous section, it is clear that the discovery of the Higgs boson is an enormously significant milestone in particle physics. So much so that it earned two of its proponents, Peter Higgs and François Englert, the Nobel Prize in Physics in 2013. It is also worth mentioning Robert Brout, Englert's collaborator, who could not receive the Nobel Prize as he had passed away in 2011.\myspace
Is that it, then? With the discovery of the Higgs boson, the last fundamental particle needed to complete the Standard Model, is the story of particle physics over? As one might expect, the answer is no.\myspace
Since the LHC's Run 2, when the center-of-mass energy reached 13 TeV, all measurements of the Higgs boson's couplings to gauge bosons and third-generation charged fermions leave little doubt that the particle discovered in 2012 is indeed the Standard Model Higgs boson, rather than some "other" Higgs. However, many theoretical and experimental questions about its nature remain unresolved.\myspace
One major issue is that the Higgs mechanism does not predict a value for the Higgs boson’s mass, which has been experimentally determined to be $m_h = 125.10 \pm 0.14 \, \text{GeV}$. This characteristic is closely tied to a longstanding conceptual problem within the Standard Model; the Higgs naturalness problem, also known as the hierarchy problem, where the notion of naturalness is usually understood according to the definition given by t'Hooft in Ref.~\cite{tHooft:1979rat}. The essence of this problem lies in the fact that the Higgs boson's mass is highly sensitive to the contributions of the heaviest particles in the Standard Model, resulting in large radiative corrections that require extreme fine-tuning to cancel out.\myspace
Another unresolved question is that identifying the Higgs boson as the Standard Model particle does not necessarily confirm that its potential matches the one predicted by the Higgs mechanism---a mechanism that, we must recall, is an \textit{ad hoc} artefact proposed on simplicity and renormalizability grounds. This minimal potential driving the electroweak symmetry breaking includes the Higgs self-interactions, parametrized by $\lambda$, and the Higgs boson mass, which is related to these self-interactions via $m_h = \sqrt{2\lambda}v$. However, the theory provides no stringent constraints on these parameters and they remain largely unknown up to now.\myspace
Additionally, there are proposals suggesting the Higgs boson might have a composite nature rather than being a fundamental particle. This possibility arises within frameworks where the EWSBS is dynamically generated through the spontaneous symmetry breaking of a vacuum of a strongly interacting theory. In several scenarios, the Higgs boson would emerge as a fourth Nambu-Goldstone boson, alongside the three others that, below the electroweak scale, become the longitudinal components of the massive gauge bosons. Examples of such breaking patterns are $SU(3)/SU(2)$ or $SO(5)/SO(4)$, see Ref.~\cite{Espriu:2017mlq}.\myspace
Thus, even after the discovery of the Higgs boson, many open questions remain. Addressing these challenges has driven significant theoretical and experimental efforts across the particle physics community worldwide. Well-founded answers are expected to emerge through the collaborative work between these two fields of research. Theorists refine their ideas based on the latest experimental results, while experimental physicists focus on addressing unresolved problems highlighted by theoretical predictions. Ultimately, the success of this symbiotic effort hinges on validation through experimentation.\myspace
The most important Higgs boson production channels at the LHC are (in decreasing order of cross section) gluon-gluon fusion (ggF)---mediated by a loop of virtual top quarks---, vector boson fusion (VBF), associated productions with a gauge boson (VH), and then associated productions with a pair of quarks $t\bar{t}$ ($t\bar{t}$H) or a single top quark ($tHq$). Numerical information on the cross sections computed for $pp$ collisions in all these processes is collected in Figure~(\ref{fig_in_xsHiggs}). In all cases, a Higgs boson mass of $m_H = 125$GeV is assumed. The bands for each process indicate the uncertainties in the calculation. This figure is not of our authorship and can be found in Ref.~\cite{LHCHiggsWG}.
\begin{figure}[t]
\begin{center}
	\includegraphics[width=0.45\textwidth]{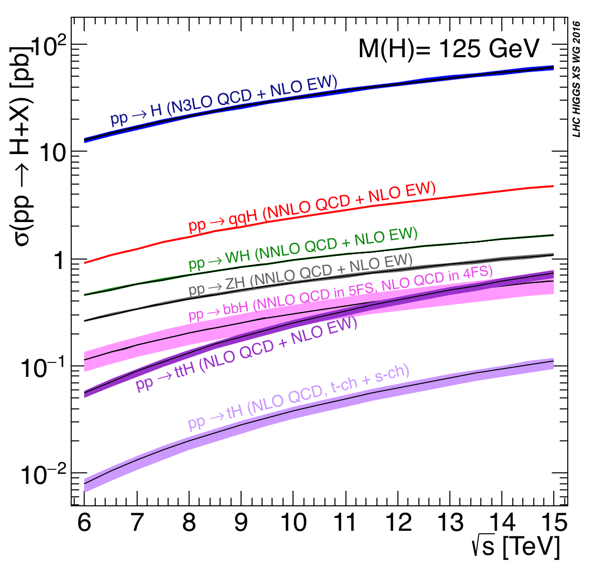} 
\end{center}
\vspace{-5mm}
\caption{\label{fig_in_xsHiggs} \small Figure showing the production cross sections of a Higgs boson at the LHC, computed for the processes gluon-gluon fusion ($pp \to H$), vector boson fusion ($pp \to qqH$), associated productions with a gauge boson ($pp \to VH$), with a pair of quarks $t\bar{t}$ and $b\bar{b}$ ($pp \to bbH/ttH$), and with a single quark $t$ ($pp \to tH$), assuming a Higgs boson mass of $m_H = 125~\mathrm{GeV}$. The labels for each process indicate the precision level of the calculations achieved. This figure is not of our authorship and can be found in Ref.~\cite{LHCHiggsWG}.}
\vspace{2mm}
\end{figure}\myspace
Other interesting events for studying the origin of the EWSBS at the LHC involve the production of not just a single Higgs boson but \textit{two Higgs bosons} in the final state at the subprocess level—a process known as \textit{double Higgs production} or \textit{di-Higgs production}. These events are particularly significant because the couplings involved are closely tied to the Higgs potential, especially the three-Higgs vertex or Higgs trilinear coupling. Measuring this trilinear coupling experimentally would impose strong constraints on the Higgs potential's shape, testing whether it matches the predictions of the Standard Model. Additionally, the study of this channel is relevant for testing BSM models with extra heavy scalar resonances, as the effective theory emerging after integrating out such heavy states typically leads to modifications of the trilinear and quartic self-couplings of the Higgs. For an example of this scenario starting from the 2HDM as the full model, see Ref.~\cite{Dawson:2023ebe}\myspace
However, the cross section of this process is much smaller than that of single Higgs production, due to a destructive interference among the diagrams involved ,which makes it, to this day, a challenging task to obtain precise measurements of the Higgs self-couplings from these processes. It is expected that the next phase of the LHC, known as the high-luminosity phase or simply the \textit{high-lumi phase}, will act as a true Higgs boson factory. This means the large-scale production of Higgs bosons, providing much more statistical data on processes involving the Higgs.
\section{The future of Higgs boson research}
Continuing the discussion from the previous section, it is worth asking: what are the future plans in the field of Higgs boson research?\myspace
Regarding the construction of new experiments, two main proposals are currently under consideration, both thought to be a \textit{Higgs factory}. The first option involves upgrading the LHC by building a larger circular accelerator with higher center-of-mass energy for proton collisions. This project, known as the Future Circular Collider (FCC), would consist of a 91km-long ring located beneath CERN, compared to the current 27 km circumference of the LHC.\myspace
The implementation of this project is envisioned in two phases, similar to the evolution from the LEP to the LHC. The first phase, expected to be running 15 years starting in the 2040s, would involve an \textit{electron-positron collider} called the FCC-ee (electron-positron). This machine would collide electrons and positrons at energies ranging from 90~GeV to 365~GeV, reaching the threshold for top-antitop pair production. At these energy ranges, the FCC-ee would serve for sure as a Higgs factory, as well as enabling precision studies in EW physics, flavor physics, top quark physics, and QCD.\myspace
Once the FCC-ee phase is complete, the plan would be to reuse the existing infrastructure, replacing the electron and positron beams with proton beams circulating in the same ring. This second stage would be called FCC-hh, where 'hh' refers to hadron-hadron (not Higgs-Higgs). The FCC-hh is expected to start around the 2070s, with the goal of reaching a center-of-mass energy of 100~TeV---approximately seven times the maximum energy of today’s LHC.\myspace
The objectives of this ambitious project are diverse. On the one hand, the FCC is not only a tool for searching for \textit{new physics} by discovering previously unknown states (currently inaccessible due to energy limitations in collisions) but also an instrument for enhancing our understanding of known physics and achieving unparalleled levels of precision. For instance, the FCC would allow extremely precise measurements of the mass of the $W$ boson and the Yukawa couplings between the Higgs boson and matter particles, which are directly related to their masses. In particular, it would be highly adequate for obtaining highly precise measurements of the top quark mass.\myspace
In summary, this ambitious project pursues two fundamental goals: precision and exploration. These objectives are deeply interconnected. While no one can guarantee the existence of new physics accessible at current energy scales, it is possible that we have yet to detect it due to a lack of precision. A paradigmatic example of this issue is dark matter, whose interactions with ordinary matter are extremely weak.\myspace
However, the future of this project depends on several factors. It is expected that by 2025, the feasibility assessment will be completed, with a detailed analysis of its political, financial and technical implications. The final decision regarding the construction of the FCC could be made around 2027 or 2028. This information can be found in the official CERN webpage in Ref.~\cite{cernwebpage}. In the meantime, we await an official confirmation.\myspace
Another interesting alternative is a linear $e^+e^-$ accelerator. Because electrons and positrons are point-like particles, the signals in this type of experiment would be much cleaner than those obtained from proton collisions, as protons are composite particles made up of quarks and gluons with complex QCD interactions. Additionally, the linear nature of these accelerators allows them to achieve higher energies by overcoming the synchrotron radiation problem that lighter particles face in circular accelerators.\myspace
At present, and at the time of writing this thesis, there are two notable projects in this direction: CLIC and ILC.\myspace
On one hand, there is the CLIC, short for \textit{Compact Linear Collider}. This is a proposed linear accelerator to be based at CERN. Its primary objective is similar to that of the FCC project: to obtain precise information about Higgs interactions, explore the properties of the electroweak sector, and search for new physics.\myspace
The project is planned in three phases, targeting different center-of-mass energies: $380$~GeV, $1.5$~TeV, and ultimately $3$~TeV. Construction of the CLIC would be expected to begin in 2026, with the first measurements anticipated by 2035. The experiment is projected to have an operational lifespan of 25 to 30 years.\myspace
On the other hand, there is the ILC, short for the \textit{International Linear Collider}, an international initiative proposed to be constructed in Japan. Similar to the CLIC, it aims to serve as a high-precision tool for studying the Higgs boson and probing physics beyond the Standard Model. Initially designed to operate at a center-of-mass energy of $250$~GeV, the ILC is planned to undergo an upgrade to 1~TeV in the future, enabling detailed investigations of Higgs self-interactions and other electroweak processes.\myspace
The project is led by the \textit{International Committee for Future Accelerators (ICFA)}, but the Government of Japan has yet to decide on its approval.\myspace
Additionally, the ILC could serve as an alternative to the FCC at CERN if the latter is not approved. These discussions are currently taking place within the framework of the European Strategy for Particle Physics; see Ref.~\cite{ESPP}.\myspace
However, we must not consider the LHC as obsolete. On the contrary, the ongoing data collection of Run 3, scheduled to last until June 2026, promises significant advancements. During this phase, the number of collisions is expected to surpass those of the two previous runs combined, thanks to the implemented upgrades.\myspace
During Run 3 of the LHC, researchers will focus on more precise measurements of how the Higgs boson interacts with matter and force particles, exploring potential decays into new particles, such as those linked to dark matter. They will also investigate the Higgs-top quark interaction and search for signs of new physics through direct searches or precision studies of known particles. Key goals include testing lepton flavour universality and studying quark-gluon plasma (QGP), a primordial state of matter, through heavy-ion collisions and, for the first time, oxygen collisions.\myspace
Looking ahead, the High-Luminosity LHC (HL$-$LHC) phase, set to begin after the Run 3 shutdown, marks a major advancement in the LHC's capabilities. Scheduled to start in June 2030, this phase—referred to as Run 4—will increase the collider's luminosity by a factor of 10, enabling the collection of a vastly larger dataset. This will allow to study rare processes with unprecedented precision, further refine measurements of the Higgs boson’s properties, probe the nature of electroweak symmetry breaking, and deepen searches for physics beyond the Standard Model. Additionally, the HL-LHC is designed to improve sensitivity to phenomena such as heavy neutrinos, exotic decays, and potential new particles, offering insights into enduring mysteries like the nature of dark matter and the matter-antimatter asymmetry in the universe.\myspace
I hope that with this introduction of my thesis dissertation, it is clear my vision of where we come from, where we are and where we are heading. It is time for us to plant the trees whose shade we know we will (may) never sit in. 

\thispagestyle{empty}

\lhead{Chapter 2}
\rhead{Effective Theories}

\chapter{Effective Theories}
\label{chp:effective_theories}

As we discussed in the introduction, the SM is a highly successful theoretical framework, as it fundamentally explains most of the known physics governing the realm of particles with impressive precision, as demonstrated by experiments over the years. However, despite this remarkable success, we often encounter situations where calculations become extremely complex or we need to understand phenomena that the SM does not directly address. This is especially true when the energy scales involved are very different; no one would ever care about gluons being exchanged inside the proton to study a block sliding down a plane, even though both the block and the plane are basically made out of protons (and neutrons). The essence of \textit{Effective Field Theories} (EFTs) lies in the assumption that detailed knowledge of a theory across all scales is unnecessary when studying processes at a specific energy. By appropriately treating energy scales as either negligible or large compared to the characteristic energy of the process under study, reliable results should be obtained, albeit with some associated margin of error.\myspace
One prominent example is the Fermi theory of weak interactions, which effectively describes processes such as beta decay at energies much lower than the W boson mass. The Electroweak Effective Theory (EWET) allows us to derive the Fermi constant, $G_F$, from the SM by eliminating the heavy W boson effects, leading to a four-fermion interaction. Another example which is of great importance for this dissertation is \textit{Chiral Perturbation Theory} (ChPT), which is an effective frame containing the pions as light degrees of freedom that describes the dynamics of low-energy QCD without having to consider the complexities of the full theory at higher energies.\myspace
The formal idea behind EFTs is to eliminate \textit{irrelevant} degrees of freedom from the path integral at a given energy scale. In this context, irrelevant refers to degrees of freedom that do not contribute significantly at energy scales where the effects of heavy states are suppressed. The process is as folows:
\begin{equation}\label{eq_ET_pathintegral}
\begin{aligned}
Z[J_l,J_H]&=\int\mathcal{D}\phi_H\mathcal{D}\phi_l\exp\left[\int d^4x\left(\mathcal{L}(\phi_l,\phi_H)+J_l\phi_l+J_H\phi_H\right)\right]\\
&\to Z_{\text{EFT}}[J_l]\equiv Z[J_l,0]=\int\mathcal{D}\phi_l\exp\left[\int d^4x\left(\mathcal{L}_{\text{EFT}}(\phi_l)+J_l\phi_l\right)\right],
\end{aligned}
\end{equation}
where $\phi_l$ and $\phi_H$ represent the light and heavy degrees of freedom, respectively, in the full theory. The full S-matrix elements are obtained by differentiating $Z[J_l, J_H]$ with respect to the sources $J_l$ and $J_H$. At low energies, where the heavy state $\phi_H$ cannot be produced, one sets $J_H = 0$. Consequently, the S-matrix elements in the \textit{infrared} theory, or IR, only account for the dynamics of $\phi_l$, reflecting the fact that the effects of $\phi_H$ become negligible at these energy scales. The complete theory is also referred to as \textit{ultraviolet} or simply UV theory.\myspace
The process in Eq.~(\ref{eq_ET_pathintegral}) is called \textit{integrating out} the heavy degree of freedom and essentially involves averaging over all $\phi_H$ configurations. The resulting effective Lagrangian, $\lag_{EFT}$, can be quite complex, potentially even containing non-local interactions. These interactions are typically expanded into local ones when expressed as a power series in the small quantity $E/M_H$, where $E$ represents IR energy scales. In such cases, there is a clear separation of scales, as we expect in BSM physics.\myspace
Notably, no new states have been yet found in the energy gap between the top quark, the heaviest SM particle, and the TeV scale, which is currently being tested at the LHC. Therefore, it seems well justified to assume that either no additional light states participate in the low-energy dynamics, or that, even if they do exist, their interactions with the SM content are weak enough to be safely neglected at the EW scale.\myspace
With all this, a general effective Lagrangian can be written as
\begin{equation}\label{eq_ET_effectivelagrangian}
\lag_{\text{EFT}}=\lag_{d\leq 4}+\sum_{d=5}^\infty \frac{1}{\Lambda^{d-4}}C^{d,i}\mathcal{O}^{d,i},
\end{equation}
where $d$ orders the energy dimensions and the $i$ index runs over all possible local operators with a certain energy dimension that are allowed by QFT requirements and possible restrictions specific to the physics that one wants to investigate.\myspace
A complete set of non-redundant, invariant operators is referred to as a basis. The size of the basis at each order in the effective expansion depends on several factors. This is primarily because operators in the Lagrangian can be reshuffled---for example, through integration by parts or field redefinitions---resulting in a theory that makes the same predictions for physical observables, even if the number of operators at each order is not preserved. The case of integration by parts is particularly intuitive, as it represents nothing more than shifting the Lagrangian by a total derivative, which does not contribute when evaluated on-shell. For example, within the SMEFT framework, which will be presented in the next section, one can choose to work in the Warsaw basis, that contains 15 dimension-6 bosonic operators (see Ref.~\cite{Grzadkowski:2010es}), or the SILH basis (see Ref.~\cite{Contino:2013kra}), which contains 20 bosonic operators at the same order. The choice between these bases depends on the convenience for different scenarios.\myspace
The $d>4$ part of the effective Lagrangian in Eq.~(\ref{eq_ET_effectivelagrangian}) has three components. First of all we have $\Lambda$, that sets the high-energy scale up to which the effective theory is applicable. In Fermi theory, $\Lambda$ would be the mass of the $W$ boson and in low-energy QCD it would correspond to the mass of the proton.\myspace
The second component consists of the effective operators $\mathcal{O}$, which capture physics of the integrated-out heavy states, each having mass dimension $\Lambda^d$ with $d>4$. The expansion in effective operators has two important consequences. Firstly, it allows for the possibility that the effective theory does not have to adhere strictly to the unitarity requirements dictated by the quantum-mechanical principle of probability conservation for energies near the threshold of applicablity. This means that the theory can tolerate potential violations of unitarity, as long as it accurately reproduces the low-energy behavior of the system. Secondly, the mass dimension indicates that the effective theory is non-renormalizable in the traditional sense. The UV infinities that arise cannot be absorbed into a finite set of coefficients. Instead, in the framework of EFTs, renormalization is performed order by order, which requires, in principle, an infinite tower of effective operators to achieve a UV-finite theory. However, the Lagrangian in Eq.~(\ref{eq_ET_effectivelagrangian}) naively produces amplitudes that scale as powers of $E/\Lambda$, where $E$ is the energy of the process. This scaling implies that at energies much lower than $\Lambda$, higher-dimensional operators contribute less to the amplitudes, effectively suppressing their impact. Consequently, we can truncate the series at a certain order, retaining only the most significant terms and discarding higher-order terms that have a negligible effect. This truncation makes the effective theory \textit{renormalizable} order by order and allows it to be rendered UV-finite with a finite, though not necessarily small, set of effective operators. \myspace
The third and last component in Eq.~(\ref{eq_ET_effectivelagrangian}) consists of the coefficients $C$, known as Wilson coefficients. These dimensionless coefficients quantify how much each higher-order local operator $\mathcal{O}^d$ contributes to the effective observables. They essentially determine the magnitude or strength of the interaction described by each operator within the effective theory framework.\myspace
There are two approaches to constructing an effective theory based on our prior knowledge of the complete theory at high-energy scales. If we know the details of the UV theory, we apply the integration process described in Eq.~(\ref{eq_ET_pathintegral}) to predict the behavior of the system in the energy range below the last integrated state. This method is referred to as the \textit{top-bottom} approach. In this case, Wilson coefficients and operators can be explicitly determined through matching procedure at the energy scale that separates the high and low energy limits of the theory. However, it may happen that we do not have complete knowledge of the details of the ultraviolet theory but rather some features. In such cases, the RG flow equations do not allow us to construct the effective theory directly, and we must rely on our limited knowledge to build the basis of effective operators using symmetry arguments. This is the \textit{bottom-up} approach.\myspace
%
For studies related to Higgs physics and the EWSBS, the most commonly utilized EFTs are the \textit{Standard Model Effective Field Theory} (SMEFT), Refs.~\cite{Brivio:2017vri, RevModPhys.96.015006} and the \textit{Higgs Effective Field Theory} (HEFT), also referred to as Electroweak Chiral Lagrangian with a light Higgs, Refs.\cite{Dobado:1990zh,Herrero:1993nc}. They primarily differ in their assumptions regarding the Higgs field and its role in the realization of electroweak symmetry, whether the symmetry is realized linearly or non-linearly, and the power counting used for ordering the effective operators.\myspace
In particular, we aim to investigate the features of electroweak symmetry breaking, which is primarily governed by the scalar and gauge sectors of the SM. Together, they form what is known as the bosonic sector. Therefore, from this point on, fermionic components will not be included in the theory unless explicitly stated otherwise. Some numerical values to support this assumption are found in Ref.~\cite{Delgado:2017cls}, where the authors discussed the contribution of top quark loops with SM interactions to the total cross section, finding more than three orders of magnitude below the prediction of the pure bosonic sector.
\section{SMEFT or HEFT?}
In the effective linear Lagrangian (SMEFT), the Higgs boson and the Goldstones associated with spontaneous symmetry breaking are embedded in an $SU(2)$ doublet and transform linearly under the EW gauge symmetries. It is no surprise that the SMEFT is constructed analogously to the linear $\sigma$-model presented in Chapter \ref{chp:Introduction}, but incorporating the gauge symmetry. On one hand, the NLO tower of effective operators, which describes BSM physics, is organized using canonical counting, meaning it is sorted by mass dimensions. This canonical counting constrains the SMEFT Lagrangian to take the form shown in Eq.~(\ref{eq_ET_effectivelagrangian}), where inverse powers of the cut-off scale are explicitely introduced to ensure a dimensionless action. Thus, the expected size of the BSM effects are of order $\sim \left(E/\Lambda\right)^{(d-4)}$, where $d$ represents the canonical dimension and $\Lambda$ is the scale of new physics, which explains why this linear description is a more suitable approach for realizing weakly-coupled theories in the IR. On the other hand, the LO Lagrangian contains the unsupressed renormalizable interactions of the SM, i.e., $\lag_{d\leq 4}=\lag_{SM}$.\myspace
In contrast, HEFT assumes that the Goldstone bosons and the Higgs boson are independent participants in the realization of electroweak gauge symmetry, similar to the non-linear $\sigma$-model. The non-linear Goldstones are incorporated into the theory through a matrix $U$ (see Eq.~(\ref{eq_in_exponential})), which transforms linearly, while the Higgs boson is treated as a singlet and therefore does not contribute to symmetry aspects. HEFT is organized based on chiral counting, which essentially tracks the number of derivatives and/or soft mass scales in each invariant operator. This expansion in powers of the number of derivatives makes the HEFT particulary suited for describing the low-enery regime of a theory that contains strong interactions. As a result, HEFT predicts contributions that scale order-by-order as powers of $p/(4\pi v)$, where $v$ is the electroweak scale. More details on this will be discussed in the next section.\myspace
The operator bases of SMEFT and HEFT are fundamentally different, and expressing one in terms of the other is a complex and non-trivial task. In Figure~\ref{fig_ET_heftandsmeftbasis} (taken from Ref.\cite{Brivio:2016uid}), a schematic illustration shows how SMEFT operators $\{\mathcal{O}_i\}$ from various canonical orders are required to reconstruct a series of leading-order HEFT operators $\{\mathcal{P}_i\}$.
\begin{figure}[t]
\begin{center}
	\includegraphics[clip,width=12cm,height=5cm]{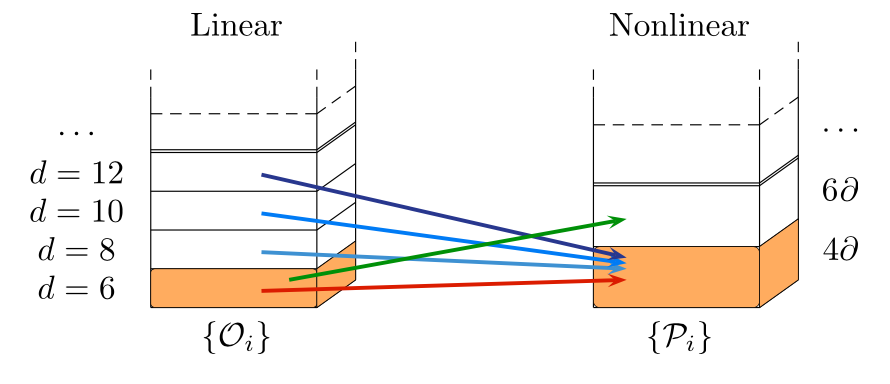} 
\end{center}
\vspace{-5mm}
\caption{\label{fig_ET_heftandsmeftbasis} \small Schematic representation showing the correspondence between some SMEFT and HEFT arbitrary operators. This illustration is taken from Ref.~\cite{Brivio:2016uid}}
\vspace{2mm}
\end{figure}\myspace
One way to determine whether a low-energy description of a system is better suited to HEFT or SMEFT is by examining the correlation, or lack thereof, of the anomalous couplings in each effective theory. For example, if SMEFT provides a better description, a linear correlation among the coefficients governing single-Higgs and double-Higgs production via VBS, is expected. This is because these interactions originate from the gauge sector coupled to an $SU(2)$ doublet containing both the Higgs field and the Goldstone bosons that transform linearly together. Conversely, if no such correlation is observed in the interactions, HEFT is more expected. From this perspective, it is clear that SMEFT---which explicitely contains the SM---is a particular case of HEFT and therefore is contained within it. SMEFT can be realized by an appropriate choice of HEFT parameters that satisfy certain conditions regarding the correlation (or decorrelation) among the interaction coefficients. See Refs.~\cite{Cohen:2020xca,Salas-Bernardez:2022hqv} for detailed information.\myspace
A different approach that has gained traction in recent years is a geometric interpretation of the four-scalar sector of the SM as coordinates on a four-dimensional scalar manifold, $\mathcal{M}$, with the Higgs boson $h \equiv \omega^4$, previously defined in Eq.~(\ref{eq_in_higgsfields}). In this framework, the physics of the scalar sector is understood in terms of the geometry of the field space. The distinction between SMEFT and HEFT (when HEFT is not a SMEFT as SMEFT is fully contained in the HEFT) is based on the linearization lemma (see Ref.~\cite{PhysRev.177.2239}).\myspace
According to this lemma, if there exists a fixed point---which remains unchanged under the action of a group $\mathcal{G}$---in the manifold $\mathcal{M}$ and this fixed point has a smooth neighborhood, then the theory admits a linear representation and thus a SMEFT description. Conversely, if there is no fixed point, or if the fixed point exists but is singular, a HEFT is required. This distinction helps clarify why some works associate linear realizations with analytical theories (around the neighborhood of the Higgs field before SSB $\Phi=0$, where electroweak symmetry is restored) and non-linear realizations with non-analytical theories. Recall that SMEFT is built upon a symmetry breaking pattern from a fully gauge-invariant vacuum, while HEFT is already formulated in the broken phase, with the Higgs boson treated as an excitation of the Higgs field around the non-zero vev. Thus, the fixed point $\Phi=0$ is not present in a HEFT description. All this geometrical discussion above can be found in the recent notes in Ref.~\cite{Alonso:2023upf}.\myspace
As stated in the Introduction, this work will focus on the search for resonances and how their predicted properties might provide insights—if any—into the anomalous couplings of the effective theory, particularly concerning Higgs self-interactions, which are considered a key avenue for investigating the nature of the Higgs boson. Following the simplified example of the $\sigma$-model, we emphasize that in both versions of the symmetry realizations, the linear and non-linear fields representing the Goldstones and the extra scalar are related by a non-linear transformation (see Eq.~(\ref{eq_in_fieldredefinition})). Since this transformation is merely a change of variables, the physics must remain invariant. Thus, the choice between SMEFT and HEFT depends exclusively on convenience in different scenarios. The discovery of additional resonances that have not yet been observed in experiments would strongly suggest that the UV completion of the effective theory involves strong interactions. This is the primary reason we have chosen to work within the HEFT rather than the SMEFT.
\section{The HEFT - The Electroweak Chiral Lagrangian with a light Higgs}
For the purposes of this dissertaion, we hope to have an effective theory that parametrizes our ignorance of the high-energy physics that originates the EWSBS and serves as a tool to perform calculations around an energy scale that can be tested at the LHC. We will work under the solely assumptions that this UV theory indeed exists, that is characterized by strong interactions and that the sector produced after the spontaneous breaking of the vacuum of the unknown strongly interacting theory, preserves the custodial symmetry, preventing the masses of the gauge bosons to grow with radiative corrections. The ligth degrees of freedom correspond to those of the EW sector of the SM.\myspace
This approach is completely model independent as it relies only on symmetry arguments so it would correspond to the bootom-up prescription presented before. The effective theory fulfilling all the above conditions is called a \textit{Higgs Effective Field Theory} (HEFT) or Electroweak Chiral Lagrangian (EChL) in a more old-fashioned nomenclature with the permission of the seniors that may be reading this work.\myspace
Formally, the HEFT is a gauged, non-linear EFT valid up to the TeV scale. Firstly, it is gauged because it incorporates the covariant derivative, ensuring invariance under local transformations, which is essential for an accurate low-energy description of the electroweak sector of the SM. Regardless of the symmetry-breaking pattern that governs the dynamics of the UV theory, the symmetries of the broken phase must align with those of the SM. Secondly, it is non-linear because the Goldstone bosons included in the spectrum, which drive the spontaneous symmetry breaking down to the custodial group producing masses for the gauge bosons, transform non-linearly under the symmetries of the Lagrangian. This aspect mirrors the case of the non-linear realization of the strong chiral symmetry in the $\sigma$-model, as discussed in Chapter \ref{chp:Introduction}. Additionally, for the purposes of this dissertation, the HEFT must include a scalar singlet, which plays the role of the Higgs boson, $h$.\myspace
The range of validity—or range of application—of the HEFT is at the order of the TeV scale. This means that, as the energy of the processes approaches this scale, perturbation theory begins to fail, and the description of the system becomes invalid because the parameter controlling the expansion is no longer small. Although this scale cannot be uniquely fixed, we can identify two scenarios where perturbation theory is expected to break down. First, loop contributions are typically suppressed by a factor of $1/4\pi$, so at energies around $4\pi v$, where $v$ is the electroweak scale, the loop expansion loses significance. When this occurs, the system enters a strong regime, with non-perturbative couplings of order 1, rendering the effective description invalid. Another possibility is the emergence of evidence of a previously integrated-out particle in the observables predicted by the effective theory below $4\pi v$. In such a case, energies exceeding the mass of this particle, $M_{\Lambda}$, would no longer permit a sensible effective description. Thus, we define the cutoff scale as $\Lambda_{HEFT} = \min(M_{\Lambda}, 4\pi v) \sim \text{TeV}$.\myspace
Now, we focus on the counting scheme that will guide the sorting of local operators involved in each process under study: the chiral counting. As explained in the non-linear $\sigma$-model when deriving the low-energy theory of Goldstone bosons after integrating out the $\varphi$ particle, an expansion in powers of momentum (or derivatives in configuration space) is appropriate. Unlike the linear case, commonly applied in weakly-coupled theories where the canonical dimension is useful, the non-linear chiral Lagrangian suggests a different approach. Here, because Goldstone boson insertions are always included through derivative interactions, a series expansion in powers of the momentum is more suitable to focus on the low-energy regime. An operator with two derivatives, which naively contributes to the amplitude as $\sim p^2$, where $p$ is the momentum transferred in the process, will be more significant at low momenta than an operator with four derivatives that contributes as $p^4$. The number of derivatives must be even to ensure Lorentz invariance. Thus, the chiral counting is primarily based on the counting of derivatives.\myspace
This counting of the (even) number of derivatives, each of them having chiral order $\chior=1$, is enough for a low-energy theory constructed solely using Goldstones. However, in our case, we must also include other particles as light degrees of freedom, the gauge bosons, whose dynamics are described by operator structures specific to gauge theories. These structures are defined to contribute differently to the chiral counting, of course in addition to the explicit counting of derivatives, and this must be accounted for to maintain full consistency. For example, soft scales of the HEFT Lagrangian, such as $M^2_W$ or $M^2_h$, the masses for the W gauge boson and the Higgs boson, respectively, must be assigned a chiral order of $\chior=2$ for the counting to be coherent at the leading order in the operator expansion. This requirement, in turn, dictates that the gauge coupling, $g$, and the trilinear self-coupling of the Higgs, $\lambda$, have chiral orders of 1 and 2, respectively, as derived from the tree-level relations $M_W^2=\frac{1}{4}g^2 v^2$ and $M_h^2=2\lambda v^2$. The pieces of the full Lagrangian that share chiral order of $\chior$ are represented by $\lag_{\chior}$.\myspace
With all the assumptions described in the preceeding paragraphs, we are now in disposition to present the building blocks and the Lagrangian itself that is useful for our purposes. First, the Goldstones are introduced via a matrix $U$, analogous to that in non-linear chiral theory in Eq.~(\ref{eq_in_exponential}) but now including the Goldstones of the electroweak theory,
\begin{equation}\label{eq_ET_UEW}
U=\exp\left(\frac{i\sigma^a\omega^a}{v}\right)\in SU(2)_L\times SU(2)_R/SU(2)_V,\\
\end{equation}
and the building blocks that dotates the HEFT with the gauge $SU(2)_L\times U(1)_Y$ invariance are the following:
\begin{equation}\label{eq_ET_buildingblocksHEFT}
\begin{aligned}
&\hat{W}_{\mu\nu}=\partial_{\mu}\hat{W}_{\nu}-\partial_{\nu}\hat{W}_{\mu}+i\left[\hat{W}_{\mu},\hat{W}_{\nu}\right], \quad B_{\mu\nu}=\partial_{\mu}\hat{B}_{\nu}-\partial_{\nu}\hat{B}_{\mu},\\
&\hat{W}_{\mu}=g\frac{W^a_{\mu}\sigma^a}{2}, \quad \hat{B}_{\mu}=g^{\prime}\frac{B_{\mu}\sigma^3}{2},\\
&D_{\mu}=\partial_{\mu}+i\hat{W}_\mu-i\hat{B}_{\mu}.
\end{aligned}
\end{equation}
In Eq.~(\ref{eq_ET_buildingblocksHEFT}), $\omega^a$ are the set of EW Goldstone bosons and, again, $\sigma^a$ three Pauli matrices generating the transformations of $SU(2)$. As usual, the kinetic terms for the gauge fields are built with the strength tensor fields defined above.\myspace
The masses of the weak gauge bosons are generated at the leading order in the effective operator expansion, similar to the non-linear $\sigma$-model in Eq.~({\ref{eq_in_nonlineargoldstoneEFT}}) but now with a covariant derivative gauging the theory. Thus, the leading order bosonic operators, with chiral order of $\chior=2$ hence contributing to $\mathcal{O}(p^2)$, $\lag_{\chior=2}$, is
\begin{equation}\label{eq_ET_leadingorderHEFT}
\lag_2=-\frac{1}{2g^2}\trace\left(\hat{W}_{\mu\nu}\hat{W}^{\mu\nu}\right)-\frac{1}{2g^{\prime 2}}\trace\left(\hat{B}_{\mu\nu}\hat{B}^{\mu\nu}\right)+\frac{v^2}{4}D_{\mu}U^{\dagger}D^{\mu}U.
\end{equation}
As an exercise, let us discuss the chiral dimension of each term in the Lagrangian above. The third term is the simplest case: in the absence of soft masses, the chiral dimension is directly determined by the number of derivatives, which is two. The first term also has a chiral order of two: each tensor structure consists of one derivative and one gauge coupling $g$, both assigned a chiral order of $\chior=1$, resulting in a total chiral order of two. Next, the two contracted field strength tensors have a combined chiral order of four, which is reduced to two due to the normalization factor $1/g^2$ in front. This reasoning similarly applies to the kinetic term for $B_{\mu}$.\myspace
Indeed, the Lagrangian $\lag_2$ described above represents the lowest order ---in chiral counting--- of the most general effective theory with nonlinear variables, which is invariant under the transformations of the local symmetry group $SU(2)_L \times U(1)_Y$. However, it does not account for a Higgs boson. This omission is expected because, as mentioned earlier, when the gauge symmetry is realized non-linearly, the Higgs boson is a singlet under the symmetry and therefore does not play a role in the fundamental group structure. Nevertheless, precisely because of this, the Higgs boson can be trivially incorporated into the Lagrangian. The most common approach is to include it through the most general polynomial function, known as the \textit{flare function} $\mathcal{F}(h)$, in a weak-field expansion:
\begin{equation}\label{eq_ET_L2}
\begin{aligned}
&\lag_2=-\frac{1}{2g^2}\trace\left(\hat{W}_{\mu\nu}\hat{W}^{\mu\nu}\right)-\frac{1}{2g^{\prime 2}}\trace\left(\hat{B}_{\mu\nu}\hat{B}^{\mu\nu}\right)+\frac{v^2}{4}\mathcal{F}(h)D_{\mu}U^{\dagger}D^{\mu}U+\frac{1}{2}\partial_{\mu}h\partial^{\mu}h-V(h)\\
&\mathcal{F}(h)=1+2a\frac{h}{v}+b\left(\frac{h}{v}\right)^2+\cdots,
\end{aligned}
\end{equation}
where, without loss of generality, the condition $\mathcal{F}(0)=1$ is chosen so that the kinetic term for the Goldstones remains canonically normalized. The rest of the terms in the expansion of $\mathcal{F}(h)$, generate arbitrary number of insertions of Higgs bosons in each Goldstone vertex with two derivatives.\myspace
There are special cases with specific parameters $a$ and $b$ that, due to their theoretical and/or phenomenological interest, have been widely studied over the years. The coefficients in the expansion of $\mathcal{F}(h)$ of order $\mathcal{O}\left(\frac{h}{v}\right)^3$ or higher are not included, as they produce vertices with at least five particles. The values $a=b=1$, of course, reproduce the SM limit, and thus, a linear realization of electroweak symmetry is recovered. Another case of great interest in past years was the limit $a=b=0$, which explicitly cancels the couplings of the Higgs with the gauge bosons and the Goldstones included in the covariant derivative term. To fully decouple the Higgs boson, $M_h$, and consequently its self-interactions, must tend to infinity.\myspace
Finally, a minimal composite Higgs model (MCHM) is achieved by choosing $a^2=1-\frac{v^2}{f^2}$ and $b=1-2\frac{v^2}{f^2}$. In these models, $f$ is the high-energy scale of the symmetry breaking that produces the EWSBS, and $\epsilon\equiv\frac{v^2}{f^2}$ is an interpolating parameter between the SM $\left(\epsilon\to 0, f\to\infty\right)$ and purely technicolor models $\epsilon\to 1$, based on the spontaneous breaking of strong chiral symmetry as a solution to the hierarchy problem in the electroweak sector. The case of purely technicolor models appears to have been ruled out by experimental data.\myspace
The MCHM is based on a symmetry-breaking pattern $SO(5) \to SO(4)$ in the ultraviolet, producing four Goldstones that act as the low-energy SM Higgs field. It is minimal in the sense that the breaking pattern is $G/H=SU(2)_L\times U(1)_Y$.\myspace
The potential of the Higgs, $V(h)$, which includes the mass of the Higgs boson along with its self-interactions, is introduced in the usual way, allowing for modifications with respect to the SM case:
\begin{equation}\label{eq_ET_higgspotential}
\begin{aligned}
&V(h)=\frac{1}{2}M_h^2h^2+\lambda_3 vh^3+\frac{\lambda_4}{4}h^4+\cdots \\
&\lambda_3=d_3\lambda_{SM}, \quad \lambda_4=d_4\lambda_{SM},
\end{aligned}
\end{equation}
where $d_{3}$ and $d_4$ are dimensionless quantities parametrizing deviations of the trilinear and quartic self-couplings of the Higgs, respectively, from their SM values, and $\lambda_{SM}=\frac{M_h^2}{2v^2}\sim 0.129$.\myspace
The full custodial symmetry, which is a requirement for the precision we aim for our study, is achieved by simply setting 
\begin{equation}\label{eq_ET_custodiallimit}
g^{\prime}=0, \quad c_W\equiv\cos\theta_W=1, \quad s_W\equiv\sin\theta_W=0.
\end{equation}
In this custodial limit, the three weak gauge bosons transform exactly as a triplet under the custodial group and share the same mass, the W mass $M_W=M_Z=\frac{1}{2}gv$. Thus, in the absence of $\hat{B}_{\mu}$ ---proportional to $g^{\prime}$---, pure electromagnetic interactions are turned off, an photons are explicitely removed from the Lagrangian vertices. The mixing term between $W^3_{\mu}$ and $A_\mu$ dissapears, eliminating the need to diagonalize the last term in Eq.~(\ref{eq_in_covariantderivativeHiggs}), with $\cos\theta_W=1$ and $\sin\theta_W=0$.\myspace
The \textit{next-to-leading} (NLO) HEFT Lagrangian, corresponding to chiral order of $\chior=4$, $\lag_4$, in the chiral expansion, describes processes up to $\mathcal{O}(p^4)$. At this order, there are a lot of more options for combining the above structures and obtain local operators that respect the EW gauge symmetry. However,   \myspace
Summing up, for the purposes of this thesis, the complete bosonic Lagrangian up to chiral order 4 that incorporates a Higgs, and including all CP-even operators that respect custodial symmetry up to $\mathcal{O}(g)$ is:
\begin{equation}\label{eq_ET_L4_custodial}
\begin{aligned}
\mathcal{L}_4 =&-i a_3\trace\left(\hat{W}_{\mu\nu}\left[V^{\mu},V^{\nu}\right]\right)
  +a_4 \left(\trace\left(V_{\mu}V_{\nu}\right)\right)^2
  +a_5 \left(\trace\left(V_{\mu}V^{\mu}\right)\right)^2+\frac{\gamma}{v^4}\left(\partial_{\mu}h\partial^{\mu}h\right)^2\\
  &+\frac{\delta}{v^2}\left(\partial_{\mu}h\partial^{\mu}h\right)\trace\left(D_{\mu}U^{\dagger}D^{\mu}U\right)
  +\frac{\eta}{v^2}\left(\partial_{\mu}h\partial_{\nu}h\right)\trace\left(D^{\mu}U^{\dagger}D^{\nu}U\right) \\
  & +i\chi\,\trace\left(\hat{W}_{\mu\nu}V^{\mu}\right)\partial^{\nu}\mathcal{G}(h),
\end{aligned}
\end{equation}
where the vector field $V_\mu=\left(D_{\mu}U\right)U^{\dagger}$, of chiral order of $\chior=1$, has been defined to write the Lagrangian more compactly. The flare function $\mathcal{G}(h)=1+b_1\left(\frac{h}{v}\right)+b_2\left(\frac{h}{v}\right)^2+\cdots$ couples the Higgs boson with at least one physical gauge boson. For the purposes of this thesis, we only need the coefficient $\zeta\equiv b_1\chi$, defined from the leading term in the expansion of $\partial^{\nu}\mathcal{G}(h)$, which generates a vertex with one Higgs boson.\myspace
A complete list of local operators relaxing the conditions mentioned above, can be found in Ref.~\cite{Alonso:2012px}. In particular, the authors include all bosonic operators up to order $\mathcal{O}(g^2,g^{\prime 2},gg^{\prime})$, without neglecting custodial symmetry-breaking sources. These terms are included through the structure $T=U\sigma^3 U^{\dagger}$, which explicitly breaks $SU(2)_R$, with $\sigma^3$ pointing in a preferred direction in the chiral space.\myspace
In general, all first-order corrections, both those that respect custodial invariance and those that do not, are described within the so-called Appelquist-Longhitano-Feruglio (ALF) basis, introduced in Refs.~\cite{PhysRevD.22.200,Longhitano:1980iz,Longhitano:1980tm,Feruglio:1992wf} in the context of the Higgsless chiral Lagrangian. This basis remains useful for describing the HEFT since, once again, the Higgs boson is trivially introduced in the theory.\myspace
The Lagrangians $\lag_2$ and $\lag_4$ in Eqs.~(\ref{eq_ET_L2}) and (\ref{eq_ET_L4_custodial}), respectively, are expressed using an on-shell basis. This means that the equations of motion (e.o.m.) for the various fields have been used to reduce the number of possible operators that can be constructed at NLO. For example, using the e.o.m. for the Higgs and the Goldstones ---omitting leptonic contributions---:
\begin{equation}\label{eq_ET_eoms}
\Box\omega=-\frac{2a}{v}\partial_{\mu}\omega\partial^{\mu}h+\cdots, \quad \Box h=-V'(h)+\left(\frac{a}{v}+\frac{b}{v^2}h\right)\partial_{\mu}\omega\partial^{\mu}\omega + \cdots,
\end{equation}
where the d'Alembertian is defined $\Box=\partial_\mu\partial^{\mu}$, operators such as 
\begin{equation}\label{eq_ET_aboxbox}
a_{\Box\Box}\frac{\Box h\Box h}{v^2},\quad a_{\Box VV}\frac{\Box h}{v} \trace\left(V_\mu V^\mu\right),
\end{equation}
which are perfectly valid on symmetry grounds and can contribute to $\lag_4$, are shown to be redundant with operators of the lower-order Lagrangian $\lag_2$. In particular, making use of the e.o.m. for the Higgs, the first of these two contributes to redefinitions for $M_h^2,\,a,\, b,\,\lambda_3,\,\lambda_4$ and $a_5$ in the on-shell basis, while the second one modifies $a,\, b$ and $a_5$. More detailed information will be provided in the section devoted to the renormalization of the HEFT.\myspace
These on-shell Lagrangians, $\lag_{2,4}$, describe the classical dynamics of the bosonic sector, including both scalars and vectors. To address the quantization of the HEFT and account for the quantum fluctuations of the theory—--essential for the objectives of this work—--, we need to add a \textit{gauge-fixing} term and a \textit{Faddeev-Popov} term. In the custodial limit and using an arbitrary gauge, these are built using the following functions
\begin{equation}\label{eq_ET_GFandFP}
\begin{split}
&f_i=\partial_{\mu}W_i^{\mu}-\frac{gv\xi}{2}\omega_i+\ldots\quad{i=1,2,3}\\
  &\mathcal{L}_{GF}= -\frac{1}{2\xi}f_i^2, \qquad\mathcal{L}_{FP}=c_a^{\dagger}\frac{\delta f^{\prime}_a}{\delta \alpha_{b,L}}c_b
\end{split}
\end{equation}
where $\alpha_{b,L}$ are the parameters associated to the $SU(2)_L\times U(1)_{Y}$ transformation of the function $f$, $f^{\prime}$, and $\xi$ is the gauge parameter controlling the choice of the gauge. The fields $\left(c^{\dagger}_a\right)c_a$ are the usual Faddeev-Popov (anti-)ghosts of the non-abelian theory which are introduced for removing non-physical configurations from the path integral through the Faddeev-Popov determinant in Ref.~\cite{FADDEEV196729}. The explicit gauge and custodial transformations for these pieces are the following:
\begin{equation}\label{eq_ET_gauge_fixing_trans}
\begin{aligned}
&g_L=e^{i\frac{\alpha^a_L\sigma^a}{2}}, &\hat{W}_{\mu}^{\prime}=g_L\hat{W}_{\mu}g_L^{\dagger}-\frac{1}{g}g_L\partial_{\mu}g_L^{\dagger}, & &U^{\prime}=g_LU.
\end{aligned}
\end{equation}
The gauge-fixing (GF) term in Eq.~(\ref{eq_ET_GFandFP}) includes a $\xi$-dependent mass for the Goldstones and cancels the unwanted term mixing the propagator of Goldstones and gauge fields, generated in the last term of Eq.~(\ref{eq_ET_leadingorderHEFT}). Usual values for the gauge parameter are $\xi=1$ and $\xi=0$, known as Feynman and Landau gauges, respectively. In the Feynman gauge, useful for one-loop calculations, the Goldstones have a mass equal to that of their associated gauge boson, while in the Landau gauge, the Goldstones remain massless. Of course, the choice of the gauge should not affect physical results.\myspace
It can be seen from the Faddeev-Popov term that, in contrast to the linear case, no couplings among ghosts and the Higgs appear in the non-linear realization of the symmetry. This is another consequence arising from the fact that the Higgs remains a singlet under the custodial group and does not contribute to the realization of the symmetry. The non-linearity manifests as new gauge-dependent vertices that involve a higher number of Goldstones coupled to ghost-antighost pairs. This feature clearly indicates the strongly interacting nature of the theory.\myspace
After fixing the gauge, and thus losing gauge invariance, a residual symmetry remains: the BRST symmetry. This symmetry, without going into further details, ensures that physical states remain gauge-invariant, while non-physical states do not contribute to physical observables. The interested reader may find more information in Ref.~\cite{Becchi:1974md,Peskin:1995ev}.\myspace
With all this, we define the (gauge-fixed) HEFT Lagrangian
\begin{equation}\label{eq_ET_fullHEFT}
\lag=\lag_2+\lag_4+\lag_{GF}+\lag_{FP}.
\end{equation}
The SM can be smoothly recovered from the HEFT Lagrangian. Specifically, when $a = b = 1$ and the remaining coefficients in $\mathcal{F}(h)$ are set to zero, \(\mathcal{L}_2\) reduces to \(\mathcal{L}_{SM}\). All interactions in $\mathcal{L}_4$ are absent in the SM, so in this limit, they must vanish, i.e., $a_i = \delta = \gamma = \eta = 0$. In the following sections, we will use a general notation as follows: $\alpha_{p^2}$ refers to the set of coefficients in $\mathcal{L}_2$, and $\alpha_{p^4}$ refers to the couplings in $\mathcal{L}_4$, which vanish in the SM.\myspace
The extent to which these couplings—referred to as anomalous couplings—deviate from their corresponding SM values determines the presence (or absence) of New Physics underlying the EWSBS of the SM. To identify such deviations, precise measurements from experiments, such as those conducted at the LHC, play a crucial role. Below, we summarize the current experimental status on the key couplings of interest.\myspace
The best experimental bounds available for the chiral coupling $a$ have been measured by ATLAS in the subprocess $h\to WW$ at 95\% C.L. to be $0.97<a<1.13$ with the 13 TeV data, see Ref.~\cite{ATLAS:2019nkf}. Also, the first experimental bounds on the chiral parameter $b$ have been also set by CMS looking for di-Higgs decays into $b\bar{b}\gamma\gamma$ and $bb\bar{b}\bar{b}$ final states and can be bound in Ref.~\cite{CMS:2024fkb}. The result of these two analysis, that are based on the assumptions that the rest of Higgs coupling are the SM ones, are found to be $-0.5<b<2.7$ and $0.55<b<1.49$, respectively, excluding the latter a Higsless scenario $\left(b=0\right)$ with a significance of 3.8. As we see, there are still large experimental uncertainties regarding the Higgs couplings to vector bosons. These uncertainties affect operators of chiral dimension two and are accordingly expected to be the most relevant ones. \myspace
Regarding $d_3$, and recalling the parameterization $\lambda_3 = d_3 \lambda$, recent bounds have been obtained by ATLAS, see Ref.~\cite{ATLAS:2022jtk}, using 13 TeV data. A study combining double-Higgs final states ($b\bar{b}b\bar{b}$, $b\bar{b}\tau^+\tau^-$, $b\bar{b}\gamma\gamma$) with single Higgs decays to $\gamma\gamma$, $ZZ^\ast$, $WW^\ast$, $\tau^+\tau^-$, and $b\bar{b}$ yields the experimental constraint $-0.4 < d_3 < 6.3$ at 95\% C.L., assuming no deviations in other Higgs couplings to Standard Model content. In a more flexible scenario that allows for modifications of additional couplings in the fit, the bound shifts to $-1.4 < d_3 < 6.1$, allowing for a wider range of negative values for $d_3$.\myspace
Concerning the Higgs quartic self-coupling, there are not relevant bounds on $d_4$ (i.e. on possible departures from the SM relation $\lambda_4=\lambda = M_h^2/2 v^2$).\myspace
The chiral couplings $a_4$ and $a_5$ have received a lot of attention in the past because to a large extent they control the appearance of resonances in the vector-isovector and scalar-isoscalar channels, at least in the approximation where $g^{\prime}=g=0$ and contributions coming from derivative operators dominate. This limit excludes the propagation of transverse modes in the internal lines of the loops and is known as the \textit{Equivalence Theorem} (ET) limit, that will be presented below. Using only the 8 TeV data, in 2017 ATLAS (Ref.~\cite{ATLAS:2016nmw}) set the bounds $-0.024 < a_4 < 0.030$ and $-0.028 < a_5 < 0.033$. More recently, CMS (Ref.~\cite{CMS:2019uys}) using the 13 TeV data and only $4l$ decays from $WZ$ scattering was able to set the bounds $-0.0061<a_4 < 0.0063$, $-0.0094 <  a_5 < 0.0098$, about three times better. In Ref.~\cite{Sirunyan:2019der}, CMS studies $2j+2l$ decays from both $WW$ and $WZ$ scatterings to set the rather stringent bound $|a_5| < 0.0008$.\myspace
CMS does not provide results for $a_4$ and $a_5$ directly as the analysis relies on the SMEFT, where the Higgs is treated as a doublet and the operators contributing to the scattering of four $W$ are of dimension eight (unlike in the HEFT where they are of dimension four). The basis adopted is the one introduced in Ref.~\cite{Eboli:2006wa}, namely $f_{S,0}/\Lambda^4$ and $f_{S,1}/\Lambda^4$. However,  as was later noted in Refs.~\cite{Eboli:2016kko, Rauch:2016pai}, a third operator containing four derivatives of the Higgs doublet, with coefficient $f_{S,2}/\Lambda^4$, exists in the SMEFT and cannot be in general missed. In order to get $f_{S,0}/\Lambda^4, f_{S,1}/\Lambda^4$ and $f_{S,2}/\Lambda^4$ one needs to measure $WW$ and  $WZ$  final states. This was done in the $4l$ analysis of CMS in Ref.~\cite{CMS:2019uys}. However in Ref.~\cite{Sirunyan:2019der}, $WW$ and $WZ$ are combined together and  it is not possible to extract $f_{S,2}$ and $f_{S,0}$ separately. Note that only the sum of the operators corresponding to $f_{S,0}$ and $f_{S,2}$ is custodially invariant, but neither of them is. The sum matches the chiral operator multiplying $a_4$ in the HEFT, see Ref.~\cite{Rauch:2016pai}. Since a valid comparison requires custodially invariant quantities, we assume $f_{S,0}=f_{S,2}$ for matching $a_4$, while $a_5$ directly matches $f_{S,1}$, as given by the relations:
\begin{equation}\label{eq_ET_a4match}
a_4=\frac{v^4}{16}\left(\frac{f_{S,0}}{\Lambda^4}+\frac{f_{S,2}}{\Lambda^4}\right)=
\frac{v^4}{8} \dfrac{f_{S,0}}{\Lambda^4}\Big\vert_{f_{S,2}=f_{S,0}}, \quad a_5=\frac{v^4}{16},\frac{f_{S,1}}{\Lambda^4}.
\end{equation}
As for the coupling $a_3$, its range of uncertainty is quite large and its influence in BSM physics has not been fully evaluated. However, as will be discussed later, this coupling will be simply neglected for the most part of this dissertation due to its small impact when compared with the rest of the coefficients listed herein.\myspace
To our knowledge, there are  no experimental studies on the $\mathcal{O}(p^4)$ chiral parameter $\zeta$.\myspace
Lastly, the rest of the $\alpha_{p^4}$ couplings relevant for the present discusseration, namely $\delta$, $\eta$ and $\gamma$, remain unconstrained experimentally, but taking into account the fact that they are absent in the SM, we will allow these to have a maximum (absolute) value of $10^{-3}$.\myspace
All the information regarding the current experimental status in the paragraphs above is gathered in Table.~\ref{tab_ET_chiralparams}
\begin{table}[tb]
\begin{center}
\renewcommand{\arraystretch}{1.2}
\begin{tabular}{|c|c|c| }
\hline
Couplings & Ref. & Experiments\\ 
\hline \hline
$0.97<a<1.13$  & \cite{ATLAS:2019nkf}
& ATLAS (from $H\to VV$)  \\ \hline 
$0.55<b<1.49$ & \cite{CMS:2024fkb} & CMS (from $HH\to X$) \\ \hline
$-0.4\lambda<\lambda_3<6.3\lambda$ &\cite{ATLAS:2022jtk} & ATLAS (from $H\to X$ and $HH\to X$) \\ \hline
 $-0.0061<a_4 < 0.0063$&\cite{CMS:2019uys} & CMS (from $WZ\to 4l$) \\ \hline
$-0.0094<a_5 < 0.0098$ &  \cite{Sirunyan:2019der} & CMS (from $WZ/WW\to 2l2j$) \\  \hline
\end{tabular}
\caption{{\small
    Current experimental constraints on bosonic HEFT anomalous
    couplings at 95\% CL. See the text about the issue to extract the $a_4$ bound from the CMS 
analysis of \cite{Sirunyan:2019der}. $X$ stands for different combinations of $l^{+}l^{-}$, $b\bar{b}$ and $\gamma\gamma$ that can participate in the process of Higgs decays.}} \label{tab_ET_chiralparams}
\end{center}
\end{table}
\section{The Equivalence Theorem}
In the previous section, we introduced the HEFT as an effective theory that enables the calculation of all relevant amplitudes. However, even within this effective framework, calculating processes involving gauge bosons in the initial and/or final states at the one-loop level can be quite challenging. While this level of precision is now available in the literature (see Refs.~\cite{Herrero:2021iqt, Herrero:2022krh}), it becomes cumbersome for our purposes, as the number of independent chiral parameters increases dramatically. However, the custodial limit $g^{\prime}=0$, assumed in our work, significantly simplifies the calculations by removing pure electromagnetic interactions and avoiding a source of potential IR divergences caused by photon exchange. Moreover, to further improve the manageability of the one-loop calculation, we will rely on the Equivalence Theorem, at the cost of introducing a small error, the magnitude of which will be quantified in this section.\myspace
The Equivalence Theorem (ET) was first proved by Cornwall, Levin, Tiktopoulos and Vayonakis in Ref.~\cite{PhysRevD.10.1145}, and essentially states that, at energies much higher than the electroweak scale, the longitudinal polarization of weak gauge bosons can be replaced, at leading order, by the associated Goldstone bosons. A formal derivation of this theorem can be found in detail in any of the Refs.~\cite{PhysRevD.10.1145,CHANOWITZ1985379}, but for our purposes, it will suffice to highlight the relative significance of the different polarizations of the gauge bosons at high energies. The two types of polarizations and the 4-momentum of an incoming gauge boson are given by:
\begin{equation}
\epsilon_T^\mu=\frac{1}{\sqrt{2}}\left(0,\mp 1,-i,0\right), \quad \epsilon_L^\mu=\frac{1}{M}\left(p,0,0,E\right), \quad k^\mu=\left(E,0,0,p\right),
\end{equation}
where $k_\mu k^\mu=E^2-p^2=M^2$.\myspace
One can easily express the longitudinal polarization vector as:
\begin{equation}
\epsilon_L^{\mu} = \frac{k^\mu}{M} + \frac{E - p}{M} \left(-1, 0, 0, 1\right),
\end{equation}
which, at energies much higher than the vector boson mass, becomes:
\begin{equation}
\epsilon_L^{\mu} = \frac{k^\mu}{M} + \frac{1}{2} \frac{M}{E} \left(-1, 0, 0, 1\right) \equiv \frac{k^\mu}{M} + v^\mu,
\end{equation}
where $v^\mu$ scales as $\mathcal{O}\left(M/E\right)$. This implies that, as the energy of the process increases beyond the mass of the vector boson, the longitudinal polarization becomes increasingly parallel to the four-momentum. This simple calculation is complemented by the Ward identity, which ensures the conservation of gauge symmetry and implies that the unphysical degrees of freedom, such as the longitudinal polarization at high energies, do not contribute in the ultraviolate regime. At high energies, well above the symmetry breaking scale, the theory is characterized by a full gauge invariance, where gauge fields have only two physical polarization states, and the Goldstone bosons decouple.\myspace
Thus, the Equivalence Theorem (ET) tells us that
\begin{equation}\label{eq_ET_equivalencetheorem}
\mathcal{A}(W_L W_L \to W_L \cdots W_L) = (-i)^n \mathcal{A}(\omega \omega \to \omega \cdots \omega)+\mathcal{O}(M_W/E),
\end{equation}
where $n$ is an integer counting the number of gauge bosons—their longitudinal components—that have been replaced by their associated Goldstone bosons. The prefactor accounts for a normalization that corrects the discrepancy:
\begin{equation}
\frac{1}{M^2}k_\mu k^\mu = 1 \qquad \epsilon_{L,\mu} \epsilon_L^\mu = -1.
\end{equation}
There is a thorough discussion in Ref.~\cite{Espriu:1994ep}, where the authors studied the applicability of the ET in the on-shell scheme. In particular they found that the relation in Eq.~(\ref{eq_ET_equivalencetheorem}) may not be entirely precise, except for the crudest approximations. Firstly, it is missing an overall factor $\mathcal{C}^n$ that accounts for the renormalization constants of the external legs, including both the gauge field and the Goldstones. Secondly, not all the additional terms, such as those where not all the gauge fields have been substituted by Goldstones (e.g., $\mathcal{A}\left(W_L\omega\to\omega\omega\right)$) and are naivly categorized as $\mathcal{O}\left(M_W/E\right)$, can be safely discarded to obtain an accurate description, even at high energies.\myspace
Whether the simplest form of the Equivalence Theorem, as expressed in Eq.~(\ref{eq_ET_equivalencetheorem}), provides sufficient precision for our purposes will be clarified in the following sections.
\section{Relevant Processes}
This dissertation, as mentioned in the introduction, primarily focuses on the Higgs boson and, in particular, on how it couples to the rest of the particles of the Standard Model, assuming that the low-energy physics (near the spontaneous symmetry breaking scale) is described by a HEFT. In the first part of this section, we will present the processes of interest and the required level of precision in detail. Subsequently, we will address the renormalization of the effective theory at the one-loop level to eliminate ultraviolet divergences that lack physical significance. The results we find are also compared with the pre-existing literature. \myspace
One of the most prominent channels for probing deviations from SM predictions is WW scattering, which refers to $2 \to 2$ processes of the form $WW \to X$. For reasons that will be clear in future parts of this dissertation, we focus on the initial state with opposite-charged pair of Ws, this is $W^+W^-\to X$ where $X=ZZ$ or $hh$. These processes, commonly referred to as WW fusion, are present both at the LHC and in a -future- $e^+e^-$ collider, see for instance Refs.~\cite{Domenech:2022uud, Anisha:2024ryj} for a full list of the processes involved. The interactions involved in these processes, whether at tree level or one loop, are described by the anomalous chiral parameters under study in this work and are highly sensitive to modifications compared to the SM case. Additionally, there is significant interest in the $hhh$ final state, as HEFT couplings such as $\lambda_3$ and $\lambda_4$ can be probed in a future $e^+e^-$ collider; however, this will not be covered in this dissertation.\myspace
The type of WW-fusion processes in the previous paragraph, can be tested looking for non-resonant scenarios-ideal for EFTs- but also in resonant scenarios where one can look for explicit resonances in the s-channel. Naturally, this work will focus on the former, as no resonances, aside from those that may emerge from the unitarization process, are expected in the spectrum within the effective framework used.\myspace 
Before computing all the $2 \to 2$ amplitudes from the Lagrangian in Eq.~(\ref{eq_ET_fullHEFT}), recall that any massive external gauge boson has three polarization states: two transverse modes and one longitudinal component. The latter is closely tied to electroweak symmetry breaking through the boson's mass generation and will be of primary interest, as will become clear shortly. It is important to note that experiments do not distinguish between the different polarizations of the incoming particles involved in a process, and the polarization of any external gauge boson---whether incoming or outgoing---takes the form
\begin{equation}
\begin{aligned}
&\epsilon^\mu_{T=\pm}=\frac{1}{\sqrt{2}}\left(\mp\cos\theta\cos\phi+i\sin\phi,\mp\cos\theta\sin\phi-i\cos\phi,\pm\sin\theta\right)\\
&\epsilon^\mu_{L=0}=\frac{1}{M_W}\left(p,E\sin\theta\cos\phi,E\sin\theta\sin\phi,E\cos\theta\right),
\end{aligned}
\end{equation}
where $p$ is the module of the tri-momentum. The four-momentum of the external gauge boson reads
\begin{equation}\label{eq_ET_pols}
k^\mu=\left(E,\vec{p}\right)=\left(E,p\sin\theta\cos\phi,p\sin\theta\sin\phi,p\cos\theta\right).
\end{equation}
All the expressions above satisfy the conditions
\begin{equation}\label{eq_ET_moms}
k_\mu\epsilon^\mu_\lambda=0, \quad \epsilon_{\mu,\lambda}\epsilon^\mu_{\lambda^{\prime}}=-\delta_{\lambda\lambda^{\prime}}, \quad \lambda,\lambda^{\prime}=\left\{\pm,0\right\}
\end{equation}
when applied to a particular gauge boson.\myspace
However, not all helicity combinations in $ W_{\lambda_1}W_{\lambda_2} \to W_{\lambda_3}W_{\lambda_4} $ contribute equally to the amplitude. Ref.~\cite{Delgado:2017cls} illustrates this in their Figure~ 3, showing the cross-section of the process $WZ \to WZ $ for each polarization combination in the Standard Model limit, using Madgraph event generator. The results reveal that of the nine possible helicity combinations, three dominate by three orders of magnitude over the others. In descending order of significance, these combinations are $ TT \to TT $, $ LT \to LT $, and $ LL \to LL $, following the notation $\lambda_1 \lambda_2 \to \lambda_3 \lambda_4 $ for $ W_{\lambda_1}W_{\lambda_2} \to Z_{\lambda_3}Z_{\lambda_4} $, which will be used here and throughout this text in the context of helicity amplitudes.\myspace
We have verified that for the process of interest, $W^+_{\lambda_1}W^-_{\lambda_2} \to Z_{\lambda_3}Z_{\lambda_4}$, and in our custodial limit $g^{\prime}=0$, even for values of $a$ (the coupling between one Higgs boson and two gauge/Goldstone fields) that deviate from the SM limit, the fully transverse combination remains a significant contribution, as observed in the case of the crossed process studied in Ref.~\cite{Delgado:2017cls} where it was the dominant one. However, this combination is insensitive to modifications in the anomalous couplings. Moreover, in the high-energy regime of the EFT, near the validity threshold, the purely longitudinal and transverse scatterings become comparable in magnitude. This is shown in the left panel of Figure~\ref{fig:xs_WWWW/WWHH}, where the tree-level cross section for different helicity combinations of the elastic scattering process \(W_{\lambda_1}W_{\lambda_2} \to W_{\lambda_3}W_{\lambda_4}\) is depicted for energies slightly above the Higgs resonance and up to the validity of the EFT. The plot includes results for SM interactions (represented by straight lines) and for an anomalous value of the chiral parameter, specifically $a = 0.94$.
\begin{figure}[t]
\begin{center}
	\includegraphics[width=0.45\textwidth]{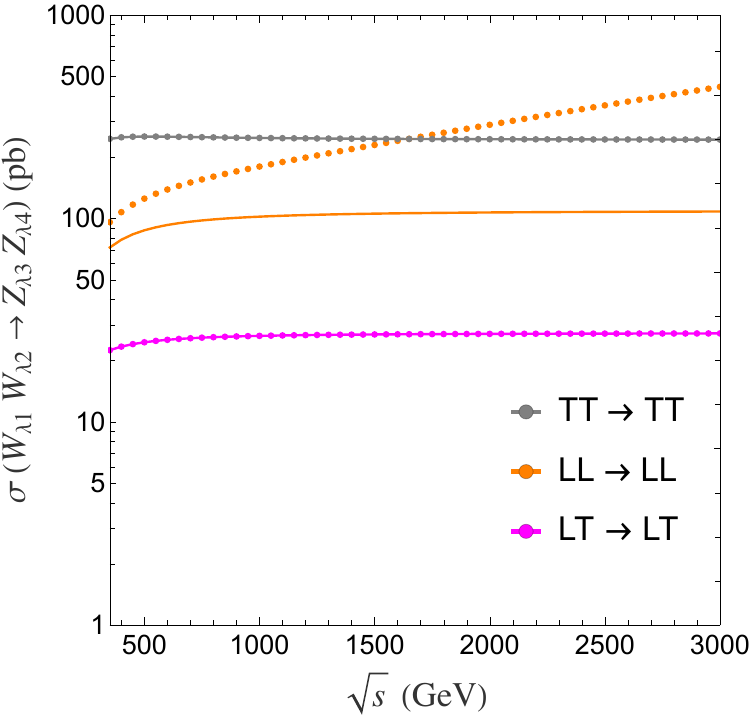} 
	\includegraphics[width=0.45\textwidth]{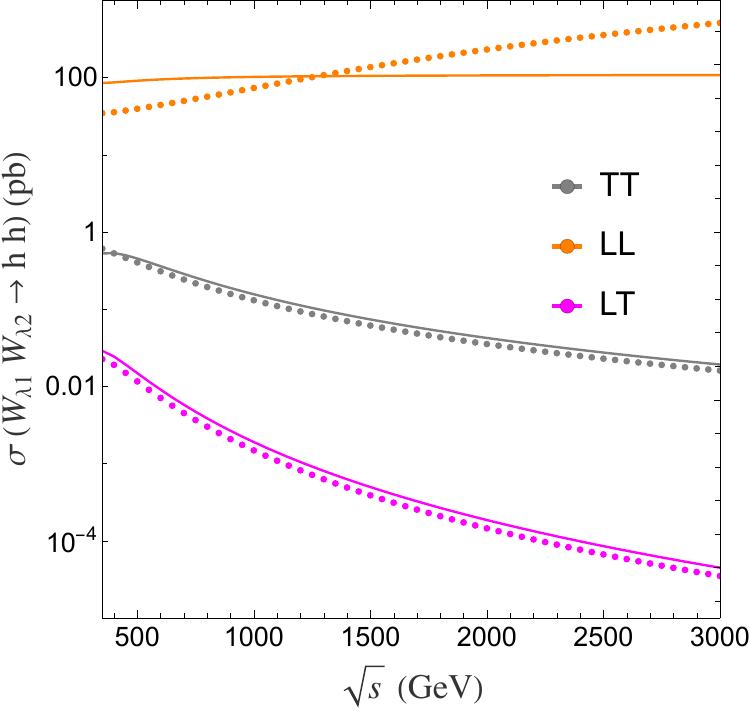} 
\end{center}
\vspace{-5mm}
\caption{\label{fig:xs_WWWW/WWHH} \small Cross section for different helicity combinations of the process (left) $W_{\lambda_1}W_{\lambda_2} \to Z_{\lambda_3}Z_{\lambda_4}$ and (right) $W_{\lambda_1}W_{\lambda_2} \to hh$. In both panels, the solid line represents the cross section in the SM limit, while the dots indicate the cross section with anomalous Higgs couplings, specifically for $a = 0.94$ and $ b = 1.05$, the latter contributing only to the $HH$ final state. The modifier of the trilinear coupling, $d_3$, also participating only in $WW\to HH$ is set to its SM value.}
\vspace{2mm}
\end{figure}\myspace
A similar situation arises when studying the process $W_{\lambda_1}W_{\lambda_2}\to hh$. In this case, the LL helicity amplitude dominates by far across the entire energy range relative to the other helicity channels. Importantly, the purely longitudinal scattering is again the only channel noticeably affected by modifications to the couplings $a$ and $b$ as it can be seen in the right pannel of Figure~\ref{fig:xs_WWWW/WWHH}. Particularly, we have selected $a=0.94$ and $b=1.05$. In the double-Higgs production process, the modifier of the Higgs triple self-coupling, $d_3$, affects all helicity combinations, as it appears in the $s$-channel Higgs boson exchange, independent of the tensor structure involved in the polarization vectors. For the plot above, it has been set to its SM value $d_3=1.$\myspace
This study suggests that the scattering of longitudinally polarized gauge bosons will be of primary interest, as it represents one of the dominant contributions and allows for probing anomalous Higgs couplings. The NLO contributions to the cross section have not been explicitly computed in Figure~\ref{fig:xs_WWWW/WWHH}, even though terms that scale as \(\mathcal{O}(p^4)\) are expected to dominate in the high-energy regime. This further supports our approach by referring to the equivalence theorem: at high energies, the purely longitudinal interaction behaves like its associated Goldstone boson. Thus, at NLO, with the tree-level interactions depending on $a_4$ and $a_5$, the scattering with transverse modes scales as $g^4$ and the longitudinal ones as $s^2/v^4$.\myspace
As mentioned previously, a full NLO calculation for the processes \(WZ \to WZ\) and \(WW \to hh\) has been published in Refs.~\cite{Herrero:2021iqt, Herrero:2022krh}. The authors computed all the Green functions for these processes up to one-loop level, including transverse modes in the internal lines and with \(g' \neq 0\). However, for our purposes, the results from these references are not practical, as the number of chiral parameters increases significantly, making them less suitable for phenomenological studies. Instead, we will use a more simplified approach.\myspace
Our calculation, up to \(\mathcal{O}(p^4)\) in the chiral expansion, will proceed as follows. First, the leading-order amplitude, which consists only of the tree-level contribution from Lagrangian in Eq.~(\ref{eq_ET_L2}), will be computed exactly in the custodial limit, i.e., with physical gauge bosons on the external legs but taking the limit \(g' = 0\). Second, the NLO contributions will be computed by separating the components: the NLO tree level, the real part of the one-loop correction, and the imaginary part of the one-loop correction.\myspace
The NLO tree-level calculation is also performed exactly from the custodially invariant Lagrangian of chiral order $n_\chi=4$ in Eq.~(\ref{eq_ET_L4_custodial}). Regarding the imaginary part of the one-loop contribution, it will likewise be calculated exactly by applying the Optical Theorem, which will be discussed in the following sections, hance turning it into a tree-level calculation with $\lag_2$. The only source of approximation in the calculation comes from the real part of the one-loop contribution, aside from the custodial limit. Based on the previous discussion regarding the contributions of different helicity amplitudes, the real part of the one-loop amplitude will be computed using the ET, where the external gauge bosons are replaced by their associated Goldstones, thereby restricting the analysis to purely longitudinal scattering. This explicit one-loop calculation will also be carried out using the leading-order Lagrangian $\mathcal{L}_2$ and performed at order $\mathcal{O}(g)$, allowing transverse modes of gauge bosons to propagate in the internal lines of the diagrams.
\subsection{Amplitudes}
In this section, we present the amplitudes for the different $2\to 2$ processes which are important for our study: $W^+_LW^-_L\to Z_LZ_L$, $W^+_LW^-_L\to hh$ and $hh\to hh$. When possible the explicit expressions will be given. This reduces basically to the tree-level amplitudes.
\subsubsection{\centering Tree level}
We show every tree-level amplitude decomposed in its different diagrams using the following notation: a superindex indicates the different processes labeled as $WW$ for $W^+W^-\to ZZ$, $Wh$ for $W^+W^-\to hh$ and $hh$ for $hh\to hh$. Also, each amplitude carries a  subindex ${xy}$ that represents a process with a particle $y$ propagating in the $x$ channel. In the case with $x=c$ and no $y$, $\mathcal{A}_{c}$, represents the contact interaction of the four external particles. For instance, the amplitude $\mathcal{A}^{WW}_{sh}$ represents a Higgs exchanged in the $s$-channel of $W^+W^-\to ZZ$ scattering.\myspace
For reasons that will become clear in the next chapter, where we discuss partial wave decomposition and the symmetries of the isospin amplitudes, we now present the aforementioned amplitudes with explicit dependence on the four-momenta of the external particles involved in the process.
\clearpage
\begin{center}
$\underline{W^+W^- \to ZZ}$
\end{center}
The tree-level amplitude includes contribution from the $\mathcal{O}(p^2)$ and $\mathcal{O}(p^4)$ Lagrangian
\begin{equation}\label{eq_ET_tree_WW}
\begin{split}
\mathcal{A}^{WW}_c=&g^2\left(\left((-2 a_3+a_4)g^2+1\right)\left(\left(\varepsilon_1\varepsilon_4\right) \left(\varepsilon_2\varepsilon_3\right)+ \left(\varepsilon_1 \varepsilon_3\right) \left(\varepsilon_2 \varepsilon_4\right)\right)\right.\\
&\left.+2 \left((2 a_3+a_5)g^2-1\right) (\varepsilon_1\varepsilon_2) \left(\varepsilon_3\varepsilon_4\right)\right)\\
\mathcal{A}^{WW}_{sh}=&-\frac{a^2g^2M_W^2\left(\varepsilon_1\varepsilon_2\right)\left(\varepsilon_3\varepsilon_4\right)}{(p_1+p_2)^2-M_h^2}+\frac{ag^4\zeta}{4((p_1+p_2)^2-M_h^2)}\left[2(\varepsilon_3\varepsilon_4)\left((p_1\varepsilon_2)(p_2\varepsilon_1) \right.\right.\\
&\left.\left. -(\varepsilon_1\varepsilon_2)(p_1+p_2)^2\right)+2(\varepsilon_1\varepsilon_2)(p_3\varepsilon_4)(p_4\varepsilon_3)\right]\\
\mathcal{A}^{WW}_{tW}=&-\frac{(1-2a_3g^2)g^2}{(p_1-p_3)^2-M_W^2} \left[-4\left((\varepsilon_1\varepsilon_2)(p_1\varepsilon_3)(p_2\varepsilon_4)+(\varepsilon_1\varepsilon_4)(p_1\varepsilon_3)(p_4\varepsilon_2)\right.\right.\\
&\left. +(\varepsilon_2\varepsilon_3)(p_3\varepsilon_1)(p_2\varepsilon_4)+(\varepsilon_3\varepsilon_4)(p_3\varepsilon_1)(p_4\varepsilon_2) \right)\\&+2\left( (\varepsilon_2\varepsilon_4)\left( (p_1\varepsilon_3)(p_2+p_4)\varepsilon_1+(p_3\varepsilon_1)(p_2+p_4)\varepsilon_3\right)\right.\\
&\left. +(\varepsilon_1\varepsilon_3)((p_2\varepsilon_4)(p_1+p_3)\varepsilon_2+(p_4\varepsilon_2)(p_1+p_3)\varepsilon_4)\right)\\
&\left.-(\varepsilon_1\varepsilon_3)(\varepsilon_2\varepsilon_4)((p_1+p_3)p_2+(p_2+p_4)p_1) \right]\\
\mathcal{A}^{WW}_{uW}=&\mathcal{A}^{WW}_{tW}(p_3\leftrightarrow p_4, \varepsilon_3 \leftrightarrow \varepsilon_4)
\end{split}
\end{equation}
where $\varepsilon_i$ is the abbreviation for $\varepsilon_L(p_i)$.\myspace
From the HEFT Lagrangian in Eq.~(\ref{eq_ET_fullHEFT}), there are also allowed reactions with Goldstones exchaning in the t and u channels, $\mathcal{A}^{WW}_{t\omega}$ and $\mathcal{A}^{WW}_{u\omega}$,but they contribute as $a_3^2$, thus, subdominant for our purposes.
\begin{center}
$\underline{W^+W^- \to hh}$
\end{center}
\begin{equation}\label{eq_ET_tree_Wh}
\begin{split}
\mathcal{A}^{Wh}_c=&\frac{g^2\,b}{2}(\varepsilon_1\varepsilon_2)-\frac{g^2\,\eta}{v^2}((\varepsilon_1 p_4)(\varepsilon_2 p_3)+(p_3\varepsilon_1)(\varepsilon_2 p_4))-\frac{2g^2\,\delta}{v^2}(p_3 p_4)(\varepsilon_1\varepsilon_2)\\
&+\frac{g^2\,\zeta}{v^2}((\varepsilon_1\varepsilon_2)(p_1+p_2)^2-2(p_1\varepsilon_2)(p_2\varepsilon_1))\\
\mathcal{A}^{Wh}_{sh}=&\frac{3d_3g^2M_h^2}{2((p_1+p_2)^2-M_h^2)}\left(a(\varepsilon_1\varepsilon_2)+\frac{\zeta}{v^2}((\varepsilon_1\varepsilon_2)(p_1+p_2)^2-2(p_1\varepsilon_2)(p_2\varepsilon_1))\right)\\
\mathcal{A}^{Wh}_{t\omega}=&\frac{2a^2g^2+a\zeta g^4}{2(p_1-p_3)^2}\left((p_3\varepsilon_1)(p_4\varepsilon_2)\right)\\
\mathcal{A}_{tW}=&\frac{a^2g^2M_W^2}{((p_1-p_3)^2-M_W^2)}\left(\varepsilon_1\varepsilon_2+\frac{(p_4\varepsilon_2)(\varepsilon_1 p_3)}{(p_1-p_3)^2}\right)+\frac{ag^4\zeta}{2((p_1-p_3)^2-M_W^2)}\left( 2M_h^2(\varepsilon_1\varepsilon_2)\right. \\
&\left. -(p_4\varepsilon_2)(p_2\varepsilon_1)-(\varepsilon_1p_3)(\varepsilon_2p_3)+M_W^2\frac{(p_4\varepsilon_2)(\varepsilon_1p_3)}{(p_1-p_3)^2}\right)\\
\mathcal{A}^{Wh}_{u\omega}=&\mathcal{A}^{Wh}_{t\omega}(p_3\leftrightarrow p_4)\\
\mathcal{A}^{Wh}_{uW}=&\mathcal{A}^{Wh}_{tW}(p_3\leftrightarrow p_4)\\
\end{split}
\end{equation}
\begin{center}
$\underline{hh \to hh}$
\end{center}
\begin{equation}\label{eq_ET_tree_hh}
\begin{split}
\mathcal{A}^{hh}_c=&\frac{8\gamma}{v^4}\left((p_1 p_4)(p_2 p_3)+(p_1 p_3)(p_2 p_4)+(p_1 p_2)(p_3 p_4)\right)-6\lambda_4\\
\mathcal{A}^{hh}_{sh}=&-\frac{36\lambda_3^2v^2}{(p_1+p_2)^2-M_h^2}\\
\mathcal{A}^{hh}_{th}=&\mathcal{A}^{hh}_{sh}(p_2\leftrightarrow -p_3)\\
\mathcal{A}^{hh}_{uh}=&\mathcal{A}^{hh}_{sh}(p_2\leftrightarrow -p_4)
\end{split}
\end{equation}
\subsubsection{\centering One Loop}
In this section we present the one loop calculation of the relevant amplitudes. 
The amplitudes cannot be expressed in terms of elementary functions as they are given by Passarino-Veltman integrals and they are quite cumbersome. For this reason we just show here the divergent parts (only present in the real part of the amplitude) and the explicit expression for the counterterms.\myspace
The calculation of quantum corrections for the processes requires gauge fixing and the inclusion of the Faddeev-Popov ghosts. The results presented below will be given in the Landau gauge $\xi=0$. Obviously, physical amplitudes should be independent of the gauge choice, but some renormalization constants do depend on the gauge election. In the gauge-fixing processes several differences are present in the HEFT (Refs.~\cite{Herrero:1993nc, Herrero:2020dtv}) with respect to the textbook SM. On one hand, the Higgs is a singlet so it does not play any role in the symmetry and thus it is not present either in the gauge fixing piece, or in the Faddeev-Popov one, i.e, there are no Higgs-ghost interactions. On the other hand, the gauge condition in Eq.~(\ref{eq_ET_GFandFP}) translates into ghost-antighost pairs coupled to an arbitrary number of Goldstones insertions with a strength depending on the gauge parameter.\myspace
As mentioned before, the calculation of the  $2\to 2$ amplitudes we are interested in is relatively involved, even at the one-loop level. Recall that we will be interested both in the divergent part (to determine counterterms) but also in the much more involved finite part. This is particularly so because there are several free parameters 
that have to be considered when one moves away from the SM. For this reason it has become customary starting with the work of Ref.~\cite{Espriu:2012ih} to split the one-loop calculation into two parts. The imaginary part is to be computed exactly from the tree-level results in Eqs.~(\ref{eq_ET_tree_WW})-(\ref{eq_ET_tree_hh}) using the optical theorem, including only the $\mathcal{O}(p^2)$ pieces. The real part is computed making use of the equivalence theorem, replacing the longitudinal vector bosons in the external legs with the corresponding Goldstone bosons. However, the full set of polarizations (including of course transverse modes) will be kept internally inside the loops in the present study. We emphasize that this procedure is done only for efficiency reasons and there is no fundamental reason to do so.\myspace
At this point one concern could be whether gauge invariance is preserved by doing this splitting. The answer is obviously in the affirmative in the following sense. The ET is derived from gauge invariance by requiring that in and out states fulfill the gauge condition. Retaking the discussion in the preceding  section, a precise implementation of the ET tells us that corrections to the leading term (i.e. the one where the longitudinal gauge boson amplitude is approximated by the corresponding Goldstone boson scattering) are given by a succession of subleading contributions each one lower with respect to the previous by a power of momenta. More practically, the ET relies on the splitting of the polarization vector $\epsilon^\mu_L = k^\mu/M_W + v^\mu$. Here $v^\mu$ is of order $M_W/E$. Substituting the splitting into our amplitude leads to corrections with  higher and higher powers of $E$ in the denominator. When summed up they all reproduce the original $W_L$ amplitude. The reader can see Ref.~\cite{Espriu:1994ep} for details. We note that the ET is used here for the one-loop correction only, not for the tree-level contribution - different orders of $\hbar$. The one-loop correction is of $O(s^2)$ and the corrections implied by the ET might change the $O(s)$ contribution, but the latter ---tree level--- is calculated exactly without appealing to the ET. Therefore gauge invariance is respected even if the splitting is itself not gauge invariant.
\begin{center}
$\underline{\text{Real part}}$
\end{center}
The ET states that at high energies compared to the electroweak scale, the longitudinal projection of the vector boson can be substituted by the associated Goldstone boson allowing an error 
\begin{equation}\label{eq: et_pol}
\varepsilon^{\mu}_L(k)=\frac{k^{\mu}}{M_W}+\mathcal{O}\left(\frac{M_W}{\sqrt{s}}\right).
\end{equation}
This error assumed at the TeV scale, the cutoff of our theory, is then, nominally, lower than 10\% but actually much lower because $M_W$ can appear only quadratically.\myspace
The calculation carried out in Ref.~\cite{Espriu:2013fia} just allowed the longitudinal part of the gauge bosons running inside the loops but for this study a full $\mathcal{O}(g)$ calculation is performed and the number of diagrams that needs to be taken into account scales to more that 1500. This calculation has been done with the help of FeynArts~\cite{Hahn:2000kx}, FeynCalc~\cite{Shtabovenko:2020gxv} and FeynHelpers~\cite{Shtabovenko:2016whf} Mathematica packages. These routines are able to evaluate the one-loop integrals in the Passarino-Veltman notation introduced in Ref.~\cite{Passarino:1978jh} and extract just the divergent part of the diagrams when is required.\myspace
The $2\to 2$ amplitudes obtained from these packages are expressed in terms of the usual Mandelstam variables, which are defined as follows:
\begin{equation}\label{eq_ET_mandelstam}
\begin{aligned}
&s=\left(p_1+p_2\right)^2, \quad  t=\left(p_1-p_3\right)^2, \quad u=(p_1-p_4)^2, \\
& s+t+u=\sum_{i=1}^4 m_i^2, \qquad p_1+p_2=p_3+p_4,
\end{aligned}
\end{equation}
where $i=\{1,2\}\left(\{3,4\}\right)$ are the incoming (outgoing) particles.\myspace
After use of the ET we have to consider the (real part of) the following processes.
\begin{enumerate}
\item \textbf{$\omega^+\omega^-\to zz$}\myspace
From the isospin point of view, this is the fundamental amplitude for elastic $\omega\omega$ scattering. This will be clear in Chapter. III.In this process
294 1PI diagrams participate at one-loop level. The divergences that appear need to be absorbed by redefinitions of
coefficients of the tree-level amplitude up to NLO. When the $W_L$ are replaced by the $\omega$, following the equivalence theorem, the amplitude tree-level amplitude reads
{\small
\begin{equation}\label{eq_ET_tree_ww}
\begin{split}
  \mathcal{A}^{\omega\omega}_{tree}=& -\frac{s(M_h^2-s(1-a^2))}{(s-M_h^2)v^2}+\frac{4}{v^4}(a_4(t^2+u^2)+2a_5s^2)+
  \left[\frac{g^2}{4}\frac{u-s}{t-M_W^2}\left(1+\frac{8a_3t}{v^2}\right) \right.\\
  &\left.+ u\Longleftrightarrow t \right],
\end{split}
\end{equation}}
with the infinitesimal substitutions
\begin{equation}\label{eq_ET_counters_wwww}
\begin{split}
&M_h^2\to M_h^2+\delta M_h^2, \qquad M_W^2\to M_W^2+\delta M_W^2, \qquad v^2\to v^2+\delta v^2,\\
&a\to a+\delta a, \qquad a_4\to a_4+\delta a_4, \qquad a_5\to a_5+\delta a_5, \qquad a_3\to a_3+\delta a_3.
\end{split}
\end{equation}
Besides, a redefinition of the Goldstone fields in the Lagrangian needs to be used for the divergent corrections of the external legs
\begin{equation}\label{eq_ET_Z_w}
\{\omega^{\pm},z\}\to \sqrt{Z_{\omega^{\pm},z}}\{\omega^{\pm},z\} \approx (1+\frac{1}{2}\delta Z_{\omega^{\pm},z} )\{\omega^{\pm},z\}.
\end{equation}
\item \textbf{$\omega^+\omega^-\to hh$}\myspace
This scattering requires computing 505 one-loop 1PI diagrams . The tree-level amplitude is
\begin{equation}\label{eq_ET_tree_wh}
\begin{split}
  \mathcal{A}^{\omega h}_{tree}=&-b\frac{s}{v^2}-\frac{6a\lambda_3 s}{s-M_h^2}
  -\left[\frac{g^2}{4(t-M_W^2)}\left(2a^2s+\frac{a^2}{t}(t-M_h^2)^2\right)+t \Longleftrightarrow u\right]\\
&-\frac{1}{v^2}\left[\frac{\zeta ag^2}{2(t-M_W^2)}\left(t(s-u)+
M_h^4\right)+ t\Longleftrightarrow u\right]-\frac{1}{v^2}\left[\frac{a^2}{t}(t-M_h^2)^2\right.\\
&\left.+t\Longleftrightarrow u\right]+\frac{1}{v^4}\left(2\delta\, s(s-M_h^2)+\eta\left((t-M_h^2)^2+(u-M_h^2)\right)\right).
\end{split}
\end{equation}
To get rid of the divergences of this process, the following substitutions for the couplings are needed
\begin{equation}\label{eq_ET_counters_wwhh}
\begin{split}
&M_h^2\to M_h^2+\delta M_h^2, \qquad M_W^2\to M_W^2+\delta M_W^2, \qquad v^2\to v^2+\delta v^2, \qquad a\to a+\delta a\\
  &b\to b+\delta b, \qquad \lambda_3\to \lambda_3+\delta \lambda_3, \qquad \delta\to \delta+\delta\, \delta,
  \qquad \eta\to \eta+\delta\eta, \qquad \zeta\to\zeta+\delta\zeta 
\end{split}
\end{equation}
Now, apart from Eq.~(\ref{eq_ET_Z_w}), we will also need the redefinition of the classical Higgs field 
\begin{equation}\label{eq_ET_Z_h}
h\to \sqrt{Z_h}h \approx (1+\frac{1}{2}\delta Z_h)h.
\end{equation}
\item \textbf{$hh\to hh$}\myspace
This process at the one-loop level contains 654 1PI diagrams and the divergences must be canceled from the
parameters of the amplitude in Eq.~(\ref{eq_ET_tree_hh}) once the usual Mandelstam definitions have been applied,\myspace
\begin{equation}\label{eq_ET_tree_hh_2}
\begin{split}
\mathcal{A}_{tree}^{hh}=&-6\lambda_4-36\lambda_3^2v^2\left(\frac{1}{s-M_h^2}+\frac{1}{t-M_h^2}+\frac{1}{u-M_h^2}\right)+\frac{8\gamma}{v^4}\left(\left(\frac{s}{2}-M_h^2\right)^2\right.\\
&\left.+\left(\frac{t}{2}-M_h^2\right)^2+\left(\frac{u}{2}-M_h^2\right)^2 \right).
\end{split}
\end{equation}
The universal counterterms
\begin{equation}
\begin{split}
  &M_h^2\to M_h^2+\delta M_h^2, \quad v^2\to v^2+\delta v^2, \quad \lambda_3\to \lambda_3+\delta \lambda_3,\\
  &\qquad \quad \lambda_4\to \lambda_4+\delta \lambda_4, \quad \gamma\to \gamma+\delta \gamma
\end{split}
\end{equation}
are required for absorbing the divergences, plus the Higgs redefinition in Eq.~(\ref{eq_ET_Z_h}).
\end{enumerate}

\begin{center}
$\underline{\text{Imaginary part: the optical theorem}}$
\end{center}
The imaginary part of the NLO amplitude is obtained exactly using the optical theorem. The fact that some states can go on shell in the process forces the presence of a physical cut in the analytic structure of an amplitude that depends on the variable $s$ promoted to a complex quantity. This amplitude is obtained after the analytical continuation to the whole complex plane of the Feynman amplitude depending on the centre of mass energy, a real variable. \myspace
Given a physical amplitude $\mathcal{A}(s)$, once we know the discontinuity of the complex amplitude across the physical cut with the usual Cutkosky rules, we find
\begin{equation}\label{eq: OT}
\text{Im}\,\mathcal{A}(s)=\sigma (s)|\mathcal{A}(s)|^2,
\end{equation}
where $\sigma (s)=\sqrt{1-\frac{(M_1+M_2)^2}{s}}$ is the two-body phase space. This allows us to compute the imaginary part of any amplitude at the one-loop level from the tree-level result.\myspace
As an example, if we are interested in computing the full amplitude in the process $W_L^+W_L^-\to Z_LZ_L$
\begin{equation}\label{eq_ET_full_amp}
\mathcal{A}(W_L^+W_L^-\to Z_LZ_L)=\mathcal{A}_{tree}^{(2)}+\mathcal{A}_{tree}^{(4)}+\mathcal{A}_{loop}^{(4)},
\end{equation}
where $\mathcal{A}^{(2)}_{tree}+\mathcal{A}_{tree}^{(4)}$ is the amplitude in Eq.~(\ref{eq_ET_tree_WW}) and $\mathcal{A}_{loop}^{(4)}$ 
is the full one-loop amplitude
\begin{equation}\label{eq_ET_exampleA4}
\mathcal{A}^{(4)}_{loop}=\text{Re} \left[\mathcal{A}^{(4)}_{loop}(\omega^+\omega^-\to zz)\right]+i \sigma (s) |\mathcal{A}_{tree}^{(2)}|^2 .
\end{equation}
This procedure is not necessary but speeds up the calculation.
\subsubsection{\centering The massless limit}
Another potential advantage of using the ET has to do with the convenience to make direct contact with some existing analytical results that are only available in the ET limit and for $g=0$. In this limit, the expressions become analytically tractable, because a full one-loop calculation in the custodial $g^\prime\neq 0$ limit involves expressions that are much more difficult to handle and do not have a simple analytical continuation to the $s$-complex plane, required when searching for resonances.\myspace
For further reference, we give below the expression for the tree and one-loop results in the limit $M_W=M_h=0$ from Ref.~\cite{Delgado:2013hxa}. They are useful to identify physical poles in the full-fledged calculation that, as said, is not amenable to analytical continuation to the appropriate Riemann sheet. In the limit where all the particles are massless and hence the SM values $g$ and $\lambda_{SM}$ are set to zero, the amplitudes with $W$s in the external states vanish, and if we want to have some analytical expressions for the tree level for our unitarity study, we are forced to go to the ET and place Goldstone bosons in the external legs.\myspace
The authors in Ref.~\cite{Delgado:2013hxa} worked with the chiral parameters $\alpha$ and $\beta$, instead of $a$ and $b$, since they introduced a vacuum-tilt extra free parameter, $\xi=\sqrt{\frac{v}{f}}$, interpolating between composite models ($v=f$) and
the SM limit ($f\to\infty$) where the new resonant states completely decouple from the theory (this vacuum-tilt parameter should not be confused with the  gauge parameter). In fact, with the massless and naive custodial limit $g=g^{\prime}=0$ in Ref.~\cite{Delgado:2013hxa} [also known as \textit{naive Equivalence Theorem} (nET)], all the gauge dependence disappears and all the amplitudes are trivially gauge invariant. This parametrization makes contact with ours with the redefinitions $a=\alpha \sqrt{\xi}$ and $b=\beta\xi$. In our framework, only the electroweak scale is used for the Higgs mechanism and weighs both Goldstone and Higgs fields, so we rewrite their amplitudes in the particular case: $\xi=1$, $a=\alpha$ and $b=\beta$.\myspace
The expressions are shown below and, in contrast to the full calculation previously described, due to the simplicity of the formulas, we do not split the full amplitude in the different channels:

\begin{enumerate}
\item \textbf{$\omega\omega\to \omega\omega$ Massless limit:}\myspace
\begin{equation}\label{eq_ET_ww_tree_net}
\mathcal{A}^{tree}=\left(1-a^2\right)\frac{s}{v^2}+\frac{4}{v^4}\left(2a_5s^2+a_4(t^2+u^2)\right)
\end{equation}
\begin{equation}\label{eq_ET_ww_loop_net}
\mathcal{A}^{loop}=\frac{1}{576\pi^2 v^4}\left[f^W(s,t,u)s^2+(1-a^2)^2(g(s,t,u)t^2+g(s,u,t)u^2)\right]
\end{equation}
with the definitions
\begin{equation}\label{eq_ET_fandgfunction}
\begin{split}
f^W(s,t,u)&=20-40a^2+56a^4-72a^2b+36b^2+\Delta(12-24a^2+30a^4-36a^2b+18b^2)\\
&+(-18+36a^2-36a^4+36a^2b-18b^2)\log\left(\frac{-s}{\mu^2}\right)\\
&+3(1-a^2)^2\left[\log\left(\frac{-t}{\mu^2}\right)+\log\left(\frac{-u}{\mu^2}\right)\right]\\
\\
g(s,t,u)&=26+12\Delta-9\log\left(\frac{-t}{\mu^2}\right)-3\log\left(\frac{-u}{\mu^2}\right)
\end{split}
\end{equation}
\item \textbf{$\omega\omega\to hh$ Massless limit:}\myspace
\begin{equation}\label{eq_ET_wh_tree_net}
\mathcal{M}^{tree}=(a^2-b)\frac{s}{v^2}+\frac{2\delta}{v^4}s^2+\frac{\eta}{v^4}(t^2+u^2)
\end{equation}
\begin{equation}\label{eq_ET_wh_loop_net}
\mathcal{M}^{loop}=\frac{a^2-b}{576\pi^2v^2}\left[f^{WH}(s,t,u)\frac{s^2}{v^2}+\frac{a^2-b}{v^2}(g(s,t,u)t^2+g(s,u,t)u^2)\right]
\end{equation}
where
\begin{equation}
\begin{split}
f^{WH}(s,t,u)=&-8(-9+11a^2-2b)-6\Delta(-6+7a^2-b)-36(1-a^2)\log\left(\frac{-s}{\mu^2}\right)\\
&+3(a^2-b)\left(\log\left(\frac{-t}{\mu^2}\right)+\log\left(\frac{-u}{\mu^2}\right)\right)
\end{split}
\end{equation}
and the function $g(s,t,u)$ is the same as in the elastic case.
\item \textbf{$hh\to hh$ Massless limit:}\myspace
\begin{equation}\label{eq_ET_hh_tree_net}
\mathcal{T}^{tree}=\frac{2\gamma}{v^4}(s^2+t^2+u^2)
\end{equation}
Notice that this process has no $\mathcal{O}(p^2)$ contribution since in the massless Higgs limit, the triple self-coupling of the Higgs vanishes and there is no diagram contributing to the process,
\begin{equation}\label{eq_ET_hh_loop_net}
\mathcal{T}^{loop}=\frac{3(a^2-b)^2}{32\pi^2 v^4}\left(f^H(s)s^2+f^H(t)t^2+f^H(u)u^2\right)
\end{equation}
with
\begin{equation}
f^H(s)=2+\Delta-\log\left(\frac{-s}{\mu^2}\right)
\end{equation}

\end{enumerate}
\subsection{Renormalization scheme and Counterterms}
The divergences eventually appearing in all these processes at the one-loop level have to be absorbed by redefining the parameters appearing at tree level. Namely,
\begin{equation}\label{eq: renor_couplings}
\begin{split}
&v^2\rightarrow v^2+\delta v_{\text{div}}^2+\delta \bar{v^2}, \quad \{h,\omega\} \rightarrow Z_{h,\omega}\{h,W,\omega\}, 
\quad M^2_{h,W}\rightarrow M^2_{h,W}+\delta M^2_{h,W}, \\ 
&\lambda_{3,4}\rightarrow \lambda_{3,4}+\delta \lambda_{3,4}, \quad a\rightarrow a+\delta a, \quad b\rightarrow b+\delta b, \quad a_i \rightarrow a_i+\delta a_i,\\ 
&\delta \rightarrow \delta + \delta \delta, \quad \eta \rightarrow \eta + \delta \eta, \quad \gamma \rightarrow \gamma 
+ \delta \gamma, \quad \zeta \rightarrow \zeta + \delta \zeta
\end{split}
\end{equation}
where we recall that $\zeta \equiv\chi b_1$.\myspace
Even though the gauge coupling $g$ appears in some of the previous formulae, the relation $M_W= gv/2$ is assumed to all orders and the renormalization of $g$ is fixed by the ones of $v$ and $M_W$. On the contrary, we cannot assume the SM relation $M_h^2 = 2 v^2 \lambda$ because this already assumes the persistence of the SM Higgs potential ---something that we want to eventually test. It is for this reason that we keep separate notations $\lambda_3$ and $\lambda_4$ for the three- and four-point Higgs vertices.\myspace
In general all counterterms have both a divergent and a finite part, determined by the renormalization conditions. However, for reasons that will be clear later, we have split the counterterms for $v^2$ explicitly into divergent and finite pieces.\myspace
As we will see subsequently, we will determine all counterterms for processes involving only Goldstone bosons, whose calculation is substantially simpler than using vector bosons. This is enough to get all the necessary counterterms. The corresponding tree level amplitudes for the Goldstones are those in Eqs.~(\ref{eq_ET_tree_ww}), (\ref{eq_ET_tree_wh}) and (\ref{eq_ET_tree_hh_2}).
\subsubsection{Auxiliary processes: $h\to\omega\omega$, $h\to hh$ and $h\to W\omega$}
In this subsection we collect a series of $1\to 2$ tree-level processes that are useful to uniquely determine  the counterterms. One of the processes ($h \to hh$) cannot take place on shell, but it has to be rendered finite through the renormalization procedure. They  are
\begin{enumerate}
\item \textbf{$h\to\omega\omega$ process}.\\
The tree-level amplitude of this decay up to NLO is (with $p_h$  the Higgs 4-moment) 
\begin{equation}\label{eq_ET_decay_hww}
\mathcal{A}^{h\to \omega\omega}_{\text{tree}}=-\frac{a p_h^2}{v}
\end{equation}
which leads to the on-shell renormalization condition
\begin{equation}\label{eq_ET_count_hww}
\frac{M_h^2}{2v^3}\left( a\delta v^2-2v^2\delta a\right)+\text{div}\left(\mathcal{A}^{h\to\omega\omega}_{\text{1-loop}}\right)=0.
\end{equation}
From (\ref{eq_ET_decay_hww}) and with the substitutions that will be specified later, we find the relation between $\delta a$ and $\delta v^2$, being the counterterms associated to the chiral parameter $a$ and to the vev, respectively. Note that obviously $p_h$ does no get a counterterm even though on shell $p_h^2 = M_h^2$. 
\item\textbf{$h\to hh$ process}.\\
At tree level, the corresponding amplitude reads
\begin{equation}\label{eq_ET_decay_hhh}
\mathcal{A}^{h\to hh}_{\text{tree}}=-6\lambda_3v\,.
\end{equation}
From the cancellation of the divergences of this process at one loop, we get  a relation between $\delta \lambda_3$ and $\delta v^2$
\begin{equation}\label{eq_ET_count_hhh}
-\frac{3}{2v^3}\left(d_3 M_h^2\delta v^2+4v^4\delta \lambda_3 \right)+\text{div}\left(\mathcal{A}^{h\to hh}_{\text{1-loop}}\right)=0.
\end{equation}
Note that this (off-shell) process cannot be modified by using the equation of motion for $h$.

\item \textbf{$h\to W\omega$ process}
\begin{equation}
\mathcal{A}^{h\to W\omega}_{\text{tree}}= ig\left(a+\frac{M_W^2\zeta}{v^2}\right) \varepsilon_W p_h\,.
\end{equation}
From the cancellation of the divergences of this process at one loop level and with the assumption that the relation $M_W=\frac{1}{2}gv$ is satisfied at every order, we obtain a relation among $\delta v^2,\, \delta M_W^2,\, \delta a$ and $\delta\zeta$.
\begin{equation}\label{eq_ET_count_hWw}
  -i\left(a M_W^2\delta v^2-2M_W^4\delta\zeta-av^2\delta M_W^2-2 M_W^2 v^2 \delta a\right)\frac{\varepsilon_W p_h}{M_W v^3} 
  +\text{div}\left(\mathcal{A}^{h\to W\omega}_{\text{1-loop}}\right)=0.
\end{equation}
\end{enumerate}
The real absorptive part has both finite and divergent parts. The divergences are reabsorbed in the amplitudes via new parameters from redefinitions of couplings and fields of the bare theory in Eq.~(\ref{eq_ET_fullHEFT})  given in the previous subsections.\myspace
The counterterms of our theory are not uniquely defined and depend on the choice of physical inputs to define the finite part of the amplitude. In this study the so-called on-shell (OS) scheme (see e.g. Ref.~\cite{Grozin:2005yg}) has been used. It states that the physical mass is placed in the pole of the renormalized propagator with residue 1. This means 
\begin{equation}\label{eq_ET_renor_prop}
\begin{split}
  \text{Re}\left[\Pi_{h,W_T}(q^2=M_{h,W_T}^2)-\delta M_{h,W_T}^2\right]=0, \quad \text{Re}\left[\frac{d\Pi_{h,\omega,W}}{dq^2}(q^2=M^2_{h,W,\omega})
    +\delta Z_{h,W,\omega}\right]=0
\end{split}
\end{equation}
where $\Pi(q^2)$ is the one-loop correction to the respective propagator. The OS, first used in the context of LEP physics, has the advantage that many relevant radiative corrections involve only two-point functions. This is obvious for the masses and wave function renormalization. After the splittings $\delta M_{h,W}^2 = \delta \overline{M}_{h,W}^2+\delta M_{h,W,\text{div}}^2$ and $\delta Z_{h,\omega}=\delta \bar{Z}_{h,\omega}+\delta Z_{h,\omega,\text{div}}$ we obtain 
\begin{equation}\label{eq_ET_counter_masses_Zs}
\begin{split}
&\delta M_{h,\text{div}}^2=\frac{\Delta}{32 \pi^2 v^2}\left(3\left[6\left(2a^2+b\right)M_W^4-6a^2M_W^2M_h^2+\left(3d_3^2+d_4+a^2\right)M_h^4 \right]\right),\\
&\delta M_{W,\text{div}}^2=\frac{\Delta}{48 \pi^2 v^2}\left(M_W^2\left[3\left(b-a^2\right)M_h^2+\left(-69+10a^2\right)M_W^2\right]\right),\\
&\delta Z_{h,\text{div}}=\frac{\Delta}{16 \pi^2 v^2}\left(3 a^2\left(3 M_W^2-M_h^2 \right)\right),\\
&\delta Z_{\omega,\text{div}}=\frac{\Delta}{16 \pi^2 v^2}\left(\left(b-a^2\right)M_h^2+3 \left(a^2+2\right) M_W^2\right)
\end{split}
\end{equation}
where $\Delta \equiv \frac{1}{\epsilon}+\log(4\pi)+\gamma_E$ and the dimensionality is set to $4+2\epsilon$.\myspace
The one-loop level propagator mixing between the gauge boson and its associated Goldstone is protected by the gauge fixing condition in Eq.~(\ref{eq_ET_GFandFP}) and no extra counterterms will be needed for this. In the absence of electromagnetic interactions assuming an exact custodial symmetry, no $Z-\gamma$ mixing in the gauge propagator can occur either.\myspace
Besides, the condition of vanishing tadpole is assumed. There is an extra counterterm $\delta T$ that cancels the Higgs tadpole contribution at one loop satisfying the usual relation (see Ref.~\cite{Espriu:2013fia})
\begin{equation}\label{eq_ET_tadpole_counter}
\delta T= -v\left(\delta M_h^2-2v^2\delta\lambda-2\lambda\delta v^2\right)=-\mathcal{A}_{\text{tad}}^h
\end{equation}
With our parametrization for the Higgs potential, $\lambda$ does not appear in any of the processes but its counterterm can be determined using Eq.~(\ref{eq_ET_tadpole_counter}) once $\delta M_h^2$ and $\delta v^2$ are obtained.\myspace
The matrix field $U$ defined in Eq.~(\ref{eq_ET_UEW}) containing the Goldstones in the HEFT should retain its unitarity and hence it cannot receive any multiplicative renormalization. Perturbatively, the redefinitions of the $n-th$ term
of the expansion of $U$ are:
\begin{equation}\label{eq_ET_expansion_U_term}
  \frac{1}{n!}\left(i\frac{\omega}{v}\right)^n \to \frac{1}{n!}\left(i\frac{\omega}{v}\right)^n
  +\frac{1}{2(n-1)!}\left(\delta Z_{\omega}-\frac{\delta v^2}{v^2}\right)\left(i\frac{\omega}{v}\right)^n.
\end{equation}
It turns out that to absorb the one-loop divergences, the counterterms for the Goldstone fields ($\sqrt{Z_\omega}$) and the vev ($\sqrt{\delta v^2}$) are equal so they cancel each other at every order in the expansion. The {\em finite} part of $\sqrt{\delta v^2}$ is fixed, at every order, by the condition
\begin{equation}\label{eq_ET_relation_Z_v}
\delta Z_{\omega}=\frac{\delta v^2}{v^2}\,.
\end{equation}
The counterterms of the HEFT whose renormalization is not determined by the OS scheme conditions, are obtained in the $\overline{MS}$ scheme. Since this is a mass
independent scheme, the counterterms corresponding to operators of dimension four (such as $a_3$, $a_4$, etc. ) are independent of $M_W$.\myspace
The complete list of counterterms allowing us to get rid of the divergences of the three amplitudes in the previous subsections is 
\begin{equation}\label{eq_ET_counter_coupling}
\begin{split}
&\delta v^2_{\text{div}}=\frac{\Delta}{16\pi^2}\left((b-a^2)M_h^2+3(a^2+2)M_W^2\right),  \quad \delta T_{\text{div}}=-\frac{\Delta}{32\pi^2 v}3\left(d_3M_h^4+6aM_W^4\right),\\
  &\delta a=\frac{\Delta}{32 \pi ^2 v^2}\left(6\,a \left(-2 a^2+b+1\right)M_W^2+(5a^3-a(2+3b)-3d_3(a^2-b))M_h^2\right), \\
&\delta b=\frac{\Delta}{32 \pi ^2 v^2}\left(6  \left(3 a^4-6 a^2 b+b (b+2)\right)M_W^2 \right. \\
&\qquad\left. -\left(21a^4-a^2(8+19b)+b(4+2b)+6ad_3(1+2b-3a^2)-3d_4(b-a^2)\right)M_h^2\right), \\
&\delta \lambda_{\text{div}}= \frac{\Delta}{64 \pi ^2 v^4}\left(\left(5a^2-2 b+3\left(d_3(3d_3-1)+d_4\right)\right)M_h^4 -12 \left(2 a^2+1\right) M_W^2 M_h^2\right.\\
&\qquad\left.+18  (a (2a-1)+b) M_W^4\right), \\
&\delta \lambda_3= \frac{\Delta}{64 \pi ^2 v^4}\left(36a b M_W^4+6 (3a^3-3ab-d_3(5a^2+1))M_W^2 M_h^2  \right. \\
&\qquad\left. +(-9a^3+3ab+d_3(10a^2-b)+9d_3d_4)M_h^4 \right), \\
&\delta \lambda_4=\frac{\Delta}{64\pi^2v^4}\left(36b^2M_W^4-12(a^2-b)(8a^2-2b-9ad_3)M_W^2M_h^2 \right.\\
&\qquad\left.+(96a^4+4b^2-d_3(114a^3-42ab)+9d_4^2+a^2(-64b+27d_3^2+12d_4))M_h^4 \right) ,\\
  &\delta a_3=-\frac{\Delta}{384\pi^2}\left(1-a^2\right),\quad \delta a_4=-\frac{\Delta}{192 \pi ^2}\left(1-a^2\right)^2, \\
 &\delta a_5=-\frac{\Delta}{768 \pi ^2}\left(5 a^4-2 a^2 (3b+2)+3 b^2+2\right),\\
  &\delta \gamma=-\frac{\Delta}{64\pi^2}3(b-a^2)^2, \quad \delta \delta = -\frac{\Delta}{192\pi^2} (b-a^2)(7a^2-b-6),
  \quad \delta \eta=-\frac{\Delta}{48\pi^2} (b-a^2)^2, \\
&\delta \zeta=\frac{\Delta}{96\pi^2}a(b-a^2)\,.
\end{split}
\end{equation}
For completeness we include the counterterm for $\delta g^2$ even though this is not an independent input of the theory anymore in the renormalization scheme used here.
\begin{equation}\label{eq_ET_counter_g}
\delta g^2=g^2\left(\frac{\delta M_W^2}{M_W^2}+\frac{\delta v^2}{v^2}\right)=\frac{\Delta}{12\pi^2v^4}M_W^2\left((-51+19a^2)M_W^2+6(b-a^2)M_h^2\right)
\end{equation}
Notice that our prescription is different from the usual one where one requires the renormalized $Z,\gamma$ two-point function to vanish at zero momentum. This condition cannot be implemented without electromagnetism, obviously. However the different result
for $\delta g$ is of no consequence in the on-shell scheme.
\subsection{Cross checks and comparison with previous results}
All these counterterms in Eq.~(\ref{eq_ET_counter_coupling}) have the correct SM limit. When $a=b=d_3=d_4=1$, all the parameters that are not present in the SM vanish, and we are left with $\delta v^2_{div},\, \delta \lambda_{div},\,\delta \lambda_3$ and $\delta \lambda_4$, that indeed renormalize the SM at one loop. In the SM limit, $\delta \lambda, \delta \lambda_3$ and $\delta \lambda_4$ have been checked to be exactly equal, as it should since they all derive from the unique SM Higgs potential coupling $\lambda$ present in the tadpole, triple and quartic self-couplings. In particular
\begin{equation}
\delta \lambda_{\text{div},\text{SM}}=\delta \lambda_{3,\text{SM}}=\delta \lambda_{4,\text{SM}}=\frac{\Delta}{16\pi^2v^4}3(3M_W^4-3M_W^2M_h^2+M_h^4)\,.
\end{equation}
As explained before, it can be seen just by direct comparison with Eq.~(\ref{eq_ET_counter_masses_Zs}), that the relation in Eq.~(\ref{eq_ET_relation_Z_v}) is satisfied.\myspace
We can also compare our counterterms with the results previously reported in the literature. As mentioned before, the authors in Ref.~\cite{Espriu:2013fia} made
a complete study of the elastic $\omega\omega$ scattering at one-loop level, allowing only longitudinal modes in the internal lines. That is, they set $g=0$ for the whole process and therefore they set the value $M_W=0$ for the vector boson mass. Our results
in Eq.~(\ref{eq_ET_counter_masses_Zs}) and Eq.~(\ref{eq_ET_counter_coupling}) have been checked with those relevant for the process in Ref.~\cite{Espriu:2013fia} in the limit $M_W=0$.\myspace
A cruder approximation was taken in Ref.~\cite{Delgado:2013hxa} where they studied all the processes including the $I=0$ final states, i.e., $WW$ and $hh$, but, besides setting $g=0$ and neglecting physical vector bosons in the loops, the authors took the limit $M_h=0$. In this limit where the self-interactions of the Higgs are absent, there is no need for redefinitions of $a,b$ and $v$ to absorb the one-loop divergences and
we are left with $a_4,a_5,\gamma,\delta,\eta$. Our results agree with theirs in the limit $M_W=M_h=0$, so the inclusion of the transverse gauge modes does not modify these counterterms.\myspace
We also find agreement with the results of Ref.~\cite{Gavela:2014uta}, where the authors carried out the renormalization of the off-shell Green functions of the three processes studied in this work for the purely scalar sector of the custodial preserving HEFT with a light Higgs in the limit $g=0, g^{\prime}=0$ (i.e. $M_W=0$). For the comparison with our on-shell calculation, we have made use of the equations of motion for the Higgs and the Goldstone fields, omitting the leptonic contribution
\begin{equation}\label{eq_ET_eoms}
\Box\omega=-\frac{2a}{v}\partial_{\mu}\omega\partial^{\mu}h+ \cdots\quad ,\quad \Box h=-V^{\prime}(h)+\left(\frac{a}{v}+\frac{b}{v^2}h\right)\partial_{\mu}\omega\partial^{\mu}\omega+\cdots
\end{equation}
leading to the following redefinitions of the electroweak and chiral parameters.
\begin{equation}\label{eq_ET_redefinitions_belen}
\begin{split}
&M_h^2=\widetilde{M}_h^2-2c_{\Box\,H}\frac{\widetilde{M}_h^4}{v^2},\quad a=a_C+\frac{2\widetilde{M}_h^2}{v^2}\left(c_7-a_C\,c_{\Box\,H}\right),\\
&b=b_C+2\mu_3^v\left(c_7-a_C\,c_{\Box H}\right)+\frac{\widetilde{M}_h^2}{v^2}\left(8\,a_7-8a_C\,a_{\Box H}-4b_C\,c_{\Box H}+a_C\,c_{\Delta H}\right), \\
&\lambda_3=\frac{\mu_3^v}{3!}-\frac{\widetilde{M}_h^2}{v^2}\mu_3^vc_{\Box H}+\frac{\widetilde{M}_h^4}{v^4}\left(\frac{1}{2}c_{\Delta H}-2a_{\Box H}\right),\\
&\lambda_4=\frac{\widetilde{\lambda}}{3!}-\left(\mu_3^v\right)^2c_{\Box H}+\frac{\widetilde{M}_h^2}{v^2}\left(\mu_3^v(\frac{5}{3}c_{\Delta H}-8a_{\Box H})-\frac{4}{3}\widetilde{\lambda}c_{\Box H}\right)+\frac{4\widetilde{M}_h^4}{v^4}\left(\frac{2}{3}a_{\Delta H}-b_{\Box H}\right),\\
&a_4=c_{11},\quad a_5=c_6-\frac{a_C}{2}c_7+\frac{a_C^2}{4}c_{\Box\,H}, \quad \delta=-c_{20}+\frac{1}{2}a_Cc_{\Delta H}, \\
&\eta=-c_8+2a_Cc_{10}-4a_C^2c_9, \quad \gamma=c_{DH} \\
\end{split}
\end{equation}
where $\mu_3^v\equiv \frac{\mu_3}{v}$ and all the quantities of the form $\widetilde{X}$ represent the parameters from their Lagrangian that have a direct counterpart in ours.\myspace
It is worth commenting on the relevance of off-shell calculations. First we see at once that the number of parameters simply explodes; trying to do phenomenology is in practice nearly impossible. Secondly, there is a large arbitrariness in using totally or partially the equations of motion so some degree of arbitrariness is unavoidable. Finally, we have to remember that off-shell couplings in an effective theory are devoid of any physical meaning. A glance at the l.h.s. and the r.h.s. of the previous equivalences should suffice to convince oneself that this is not the way to go.\myspace
We have also compared our results with the more recent study in Ref.~\cite{Herrero:2021iqt} where the authors performed a full (off-shell) renormalization of the one-loop Green functions involved in $W Z$ scattering (all polarizations considered) including custodially nonpreserving operators too. When we restrict our set of counterterms to those relevant for $W_L W_L$ scattering and take into account that custodially nonpreserving contributions are omitted, we find our results compatible with those of Ref.~\cite{Herrero:2021iqt} with two differences originating from the inclusion by these authors of two $\mathcal{O}(p^4)$ operators,
\begin{equation}
  a_{\Box\Box}\frac{\Box h\Box h}{v^2}, \qquad a_{\Box VV}\frac{\Box h}{v}\text{Tr}\left(V_{\mu}V^{\mu}\right),
\end{equation}
already showed in Eq.~(\ref{eq_ET_aboxbox}).\myspace
These operators can be reduced by using the equation of motion of the Higgs field  at leading order in Eq.~(\ref{eq_ET_eoms}), and are for our purposes redundant. The first of these two operators enters directly in the renormalization of the propagator of the Higgs so, even with the same OS renormalization condition as ours, $\delta Z_h$ and $\delta M_h^2$ differ in a consistent way. After using the e.o.m., both operators are actually redundant and change the coefficients $M_h$,$\,a$ and $a_5$ following Eq.~(\ref{eq_ET_redefinitions_belen}) with $a_{\Box\Box}=c_{\Box H}$ and $a_{\Box VV}=c_7$. As mentioned before, from this reference we are able to compare only those counterterms participating in our elastic $W_LW_L$ scattering within the custodial limit ($g^{\prime}=0,\, M_Z=M_W$) and in the Landau gauge ($\xi=0$): $v^2, M_h^2,M_W^2,a,\,a_3,a_4,a_5$ and $\zeta$ (called $a_{d2}$ in their notation).\myspace
$\delta T_{div}$ is also compatible up to a different sign in the renormalization condition for the Higgs tadpole. We do not find agreement, though, in the counterterm associated to the $SU(2)_L$ coupling $g$, coming from the fact that in our case it is a 
derived quantity as shown in Eq.~(\ref{eq_ET_counter_g}).\myspace
This work was later extended by the same authors in Ref.~\cite{Herrero:2022krh}, where they performed the one-loop renormalization of the process $WW \to hh$ under the same conditions as in their previous work---off-shell, in $R_\xi$ gauges, and including operators that do not preserve custodial symmetry. Their results coincide with ours when taking the custodial limit $g^\prime=0$ and in the Landau gauge.\myspace
One can check easily that all our additional or `anomalous' counterterms do vanish in the SM limit, while this is not in general the case for the off-shell calculations in Ref.~\cite{Gavela:2014uta} and Refs.~\cite{Herrero:2021iqt,Herrero:2022krh}.\myspace
The authors in Ref.~\cite{Buchalla:2013rka} also obtained the divergences of the HEFT local operators with the heat kernel formalism for the path integral. To this purpose, they needed to make redefinitions of the quantum fields, in particular the Higgs field, so they were present in the canonical normalization. These redefinitions alter the UV divergences of some operators with respect to those in our Lagrangian. All the $\mathcal{O}(p^4)$ divergences, not affected by field renormalizations, have been checked to coincide after some reparametrization of the chiral couplings.
\subsection{Amplitudes at high energies}
In the beginning of this section, we introduced the foundational ideas behind effective theories, emphasizing the specific properties they must satisfy. Notably, when formulating an effective field theory, one shall forgo acceptable high-energy behavior near the cutoff, as long as the low-energy regime is accurately reproduced. Whether the theory behaves correctly in the UV depends on the specific interactions, the values of the couplings, and the particular processes under study. However, what we want to emphasize here is that there is no requirement for the amplitudes of the effective theory to behave well---i.e., to remain unitary---in the UV. In this section, we quantify this trade-off by illustrating the UV behavior of our amplitudes.\myspace
In Figure~\ref{fig_ET_real_WWWW_a} we plot the modulus computed at the tree plus one-loop level for the process $W_L W_L \to Z_L Z_L$ for various values of the parameter $a$ at a fixed scattering angle $\cos\theta=0.3$. Departures from the SM value $a=1$ result in a clear bad high-energy behavior.  
\begin{figure}
\centering
\includegraphics[clip,width=12cm,height=9cm]{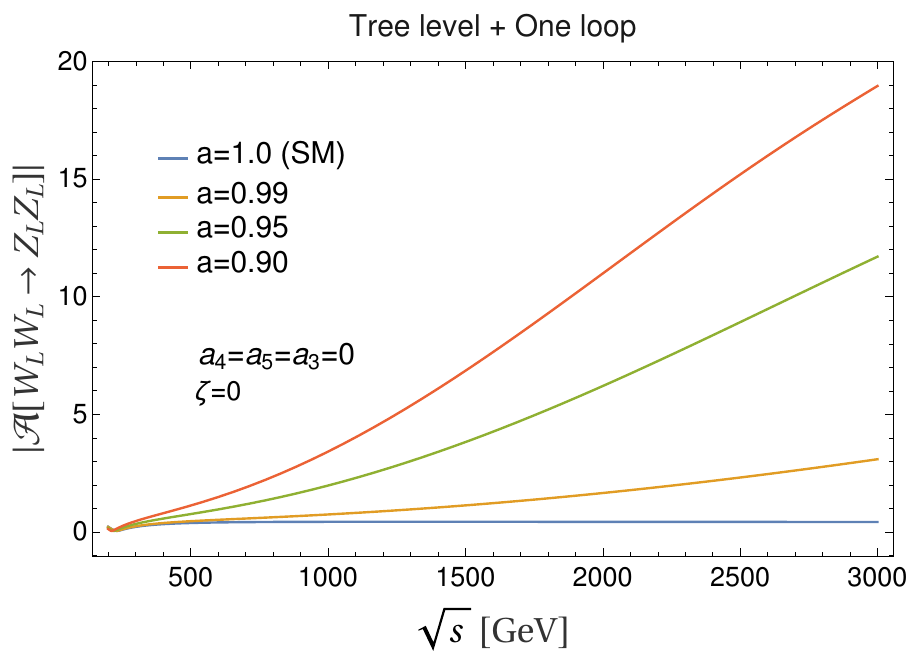}
\caption{\small{Plot of the modulus of the elastic vector boson scattering (VBS) amplitude in longitudinal polarization ($W_LW_L\to Z_L Z_L$) versus the center of mass energy $\sqrt{s}$ for some values of the chiral parameter $a$ at a fixed scattering angle $\cos\theta=0.3$. It can be seen how small departures for the SM value ($a=1$) leads to a quick violation of unitarity within the HEFT regime of validity. All the $\mathcal{O}(p^4)$ couplings contributing to the process ($a_3,\,a_4,\, a_5,\,\zeta$) are set to zero}}
\label{fig_ET_real_WWWW_a}
\end{figure}
This same behavior is seen in the remaining $2\to 2$ processes. For instance the modulus of the amplitude for the process $W_L W_L \to hh$ is depicted in the left panel of Figure~ \ref{fig_ET_real_WWHH} for the same values of the parameter $a$ parameterizing the Higgs-vector boson coupling in the HEFT.
\begin{figure}
\centering
\includegraphics[width=0.45\textwidth]{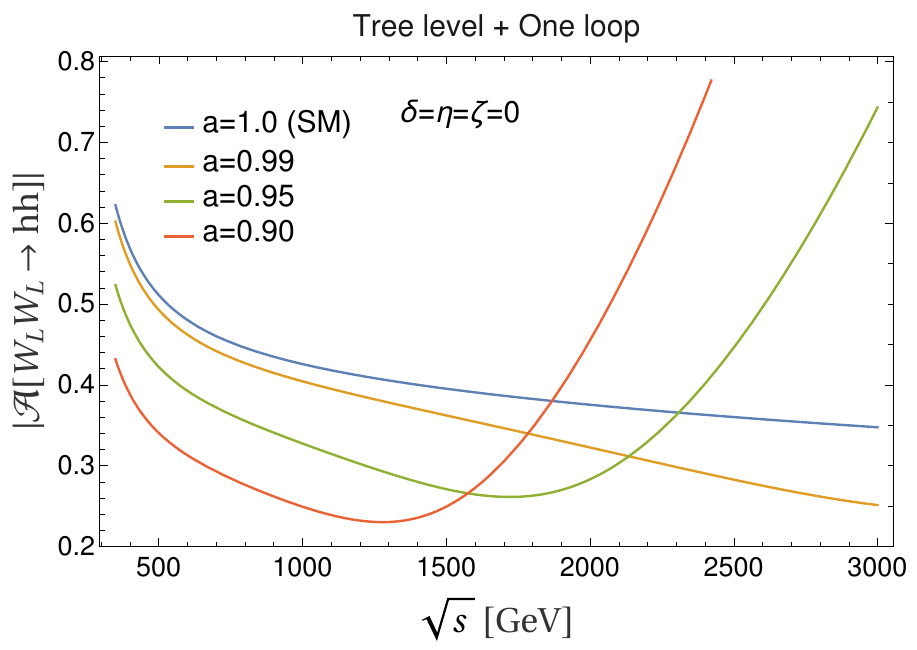} %
\includegraphics[width=0.45\textwidth]{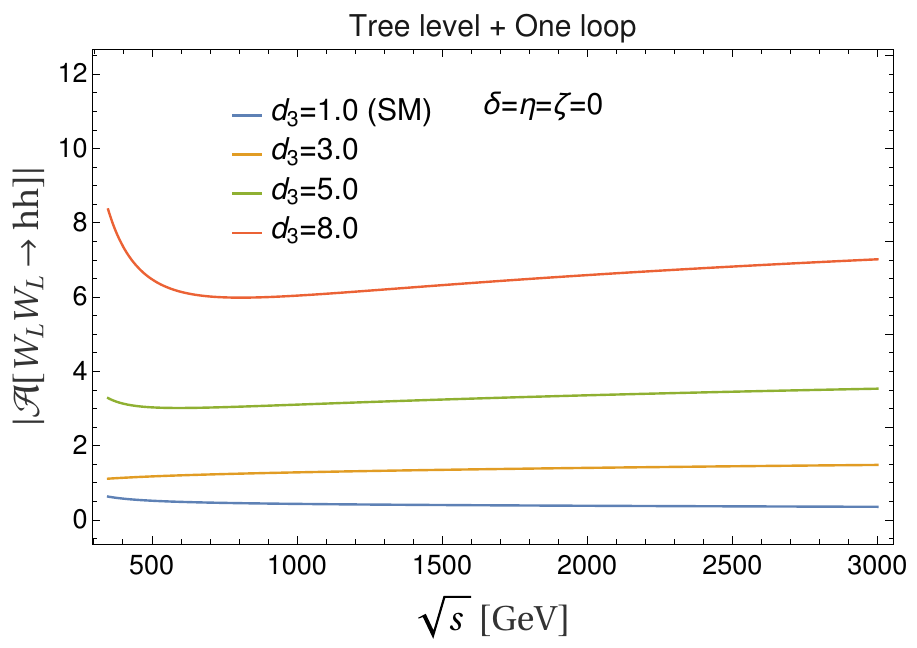} %
\caption{\small{Plot of the modulus of the $W_LW_L\to hh$ amplitude versus the center of mass energy $\sqrt{s}$ for some values of the chiral parameter $a$ (left) and the trilinear Higgs coupling $\lambda_3=d_3\lambda$ (right) at a fixed scattering angle $\cos\theta=0.3$. It can be seen how departures from the SM limit do not lead to an obvious bad UV behavior of the amplitude in the second case. Although it could seem that in the left panel there is no bad high energy limit for the case $a=0.99$, the reality is that the amplitude acquires an unphysical behaviour for a scale just above the cut-off of the theory (around 4 TeV), a region not shown in the plot. In each Figure~ the remaining parameters do not vary and are set to the corresponding SM values.}}%
\label{fig_ET_real_WWHH}%
\end{figure}\myspace
On the contrary, this same modulus of the $W_L W_L \to hh$ amplitude (Figure~ \ref{fig_ET_real_WWHH}) shows a milder dependence on the parameter $\lambda_3$ of the Higgs potential. For the SM value $a=1$, modifying $\lambda_3$ does not show obvious signs of bad high-energy behavior. At this level, this derives from the fact that this coupling is momentum independent. Note that this coupling is very poorly constrained so the overall uncertainty of the amplitude is accordingly large.\myspace
Higher-loop calculations will only worsen the high-energy behavior. It is thus clear that, except for tiny deviations from the SM, as soon as one enters the multi TeV region, the perturbative treatment is unreliable. Therefore, checking for constraints on the anomalous couplings present in the HEFT by just looking at growing cross sections is risky but may be justified (if the deviations are small) or plain wrong (if the anomalous coupling constants deviate significantly from their SM values). It is clear that physical amplitudes --even beyond the SM-- are necessarily unitary, meaning that in the HEFT higher-loops contributions have to be somehow summed up to render a reasonable high-energy behavior, which of course will be different from the SM one, but still in accordance with the general principles of field theory. We conclude that unitarization is necessary to compare the predictions of the HEFT with those of the SM \textit{vis-\`a-vis} the experiments at very high energies, particularly when we are close to the HEFT UV cutoff.\myspace
Although the poor high-energy behavior is directly observed in Figure~\ref{fig_ET_real_WWWW_a} and Figure~\ref{fig_ET_real_WWHH}, it is difficult to precisely quantify the energy scale at which unitarity is violated. We can, however, estimate an approximate range. In the next chapter, several unitarization methods will be introduced. These methods are described using partial-wave analysis, as this framework is well-suited for implementing the unitarity condition, which, as discussed, is crucial for obtaining amplitudes that lead to physically meaningful observables.


\thispagestyle{empty}

\lhead{Chapter 3}
\rhead{Unitarization}

\chapter{Unitarization Techniques}
\label{chp:unitarization}

In the last part of the previous chapter, we emphasized that, within the framework of effective field theories, one generally need not be concerned about amplitudes being well-behaved in the UV. Here, "well-behaved" means that the amplitudes exhibit controlled growth. In fact, the amplitudes relevant to our discussion lose this controlled behavior whenever the couplings in the Lagrangian deviate from their SM values. This is especially evident in the case of elastic scattering \(W_LW_L \to Z_LZ_L\), depicted in Figure~\ref{fig_ET_real_WWWW_a}, where even a 1\% variation in the coupling \(a\)---which governs the interaction between a single Higgs boson and a pair of gauge bosons---from its SM value of \(a = 1\) results in significant growth near the validity threshold of the theory. Larger deviations result in unphysical enhancements at even lower energies, where we expect the effective field theory to work reasonably well.\myspace
However, the effective amplitudes pose problems when making physical predictions in the high-energy regime---in our case, at the TeV scale. These uncontrolled growths in the amplitudes would lead to a significant overestimation of events.\myspace
Imagine we have an amplitude $\mathcal{A}=\mathcal{A}(s)$, and there exists a physical condition that requires $|\mathcal{A}(s)|<\mathcal{A}_0$ for all permitted values of $s$. Amplitudes that do not meet this condition are considered, in general, unphysical and, therefore, cannot represent a viable physical process in a theory with a UV completion. For example, the amplitude shown in Figure~\ref{fig_un_example} is unphysical for values of $s$ greater than $s_0$.
\begin{figure}
\centering
\includegraphics[clip,width=12cm,height=8cm]{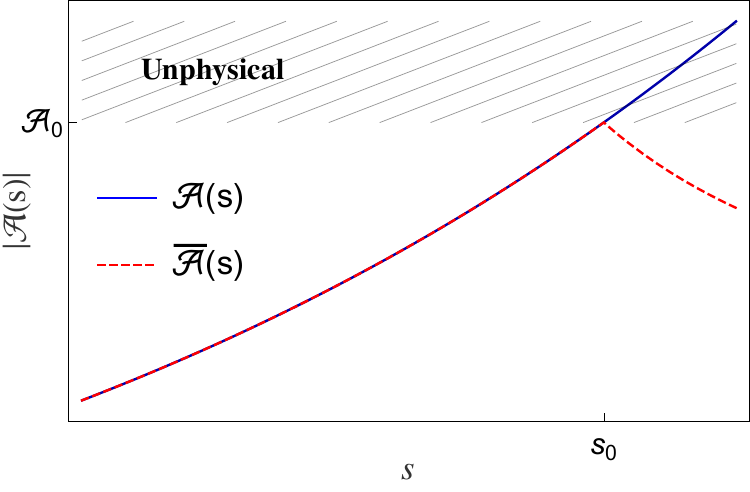}
\caption{\small{Illustration of a method to enforce a physical condition on an amplitude. The solid blue line represents an amplitude that violates the condition $|\mathcal{A}(s)|<\mathcal{A}_0$ for some values of $s < s_0$. The dashed red line depicts the physical amplitude $\overline{\mathcal{A}}(s)$ obtained using the method described in Eq.~(\ref{eq_un_example}).}}
\label{fig_un_example}
\end{figure}\myspace
To ensure that $\mathcal{A}(s)$ satisfies the physical condition, we can modify it using a multiplicative factor. Specifically, we could define a new amplitude $\overline{\mathcal{A}}(s)$ as follows:
\begin{equation}\label{eq_un_example}
\overline{\mathcal{A}}(s) = g(s) \mathcal{A}(s) = \left\{\begin{array}{lcc}
\mathcal{A}(s) & \text{if} & s < s_0 \\ \\
\mathcal{A}(s)\left(\frac{s}{s_0}\right)^{-3} & \text{if} & s \geq s_0
\end{array}\right. .
\end{equation}
This \textit{ad-hoc} modification ensures that $\overline{\mathcal{A}}(s)$ remains within the acceptable bounds for all $s$.\myspace
However, the justification for this modification of the amplitude raises several questions. Can this adjusted amplitude $\overline{\mathcal{A}}(s)$ truly represent a physical process? The transition applied in Eq.~(\ref{eq_un_example}), while ensuring that $\overline{\mathcal{A}}(s)$ remains within the physical region that we have defined, may not be fully justified if it lacks a clear connection to the underlying UV theory. It is essential to assess whether this modification is compatible with the properties of the full theory. Without such justification, the modified amplitude could potentially violate essential theoretical constraints or introduce inconsistencies in the theory's structure.\myspace
What are the main properties that any physical amplitude should have?
\section{General properties of the $2\to 2$ amplitudes}
It is unsurprising that we require the amplitudes in our theory to be analytic functions of the dynamical variables. When the variable $s$ is extended to take values in the complex plane, this invites a natural connection between the analytic properties of the amplitude and the underlying physical principles. In particular, the analyticity of the scattering amplitude is intimately tied to causality, as seen through dispersion relations, which are rooted in the requirement that no information or signal can propagate faster than the speed of light.\myspace
In terms of scattering amplitudes, this translates to the condition that the amplitude, as a function of the Mandelstam variable $s$-promoted to a complex value-, must be analytic in regions where no physical processes—such as particle production or intermediate states—can occur. The presence of singularities, i.e., poles and branch cuts, corresponds to physical states (e.g., resonances or thresholds for particle production), while the analytic behavior away from these singularities ensures that the amplitude is consistent with causal propagation.\myspace
More rigorously, this connection is formalized through the Kramers-Kronig relations, which relate the real and imaginary parts of the amplitude in a causal manner assuming the usual analiticity conditions. In this case, an amplitude $\mathcal{A}(s)$ satisfies the following relation:
\begin{equation}\label{eq_un_realandimaginary}
\text{Re}\, \mathcal{A}(s) = \frac{1}{\pi} \mathcal{P} \int_{s_{\text{threshold}}}^{\infty} \frac{\text{Im}\, \mathcal{A}(s')}{s' - s} \, ds',
\end{equation}
where $\mathcal{P}$ denotes the principal value of the integral. This expresion can be related to the full amplitude $\mathcal{A}(s)$ via the identity
\begin{equation}
\frac{1}{s^\prime-s-i\epsilon}=\mathcal{P}\frac{1}{s^\prime-s}+i\pi\delta(s^\prime-s)
\end{equation}
to give
\begin{equation}\label{eq_un_ampandimaginary}
\mathcal{A}(s)=\frac{1}{\pi}\int_{s_{\text{threshold}}}^\infty ds^\prime\frac{\text{Im}\mathcal{A}(s^\prime)}{s^\prime-s-i\epsilon},
\end{equation}
known as dispertion relation.\myspace
However, Eq.~(\ref{eq_un_ampandimaginary}) also presents a problem, as it requires knowledge of the imaginary part of the amplitude at large $s^\prime$ energy scales to determine the real contribution at low energies. The problem is softened by the use of substractions. The \textit{substracted} dispertion relation for $\left(\mathcal{A}(s)-\mathcal{A}(0)\right)/s$ is given by
\begin{equation}
\frac{\left(\mathcal{A}(s)-\mathcal{A}(0)\right)}{s}=\frac{1}{\pi}\int_{s_{\text{threshold}}}^\infty \frac{ds^\prime}{s^\prime-s-i\epsilon}\text{Im}\left[\frac{\mathcal{A}(s^\prime)-\mathcal{A}(0)}{s^\prime}\right],
\end{equation}
and taking into account that $\text{Im}\mathcal{A}(0)=0$,
\begin{equation}\label{eq_un_asubstracted}
\mathcal{A}(s)=\mathcal{A}(0)+\frac{s}{\pi}\int_{s_{\text{threshold}}}^\infty \frac{ds^\prime}{s^\prime}\frac{\text{Im}\mathcal{A}(s^\prime)}{s^\prime-s-i\epsilon}.
\end{equation}
The integral of the substracted amplitude in Eq.~(\ref{eq_un_asubstracted}) contains an extra power of $s^\prime$ in the denominator, which softens the contribution of the high-energy regime.\myspace
Consecutive substractions can be applied to further lessen the dependence of large $s^\prime$ by systematically adding inverse powers of $s^\prime$ in the integral and substraction constants $\mathcal{A}(0),\mathcal{A}^\prime(0),\mathcal{A}^{\prime\prime}(0),\cdots$.\myspace
The result for the n-substracted dispertion relation at some $s_0$ belonging to the IR regime is given by
\begin{equation}\label{eq_un_nsubstracted}
\begin{aligned}
\mathcal{A}(s)=&\sum_{n}\frac{\mathcal{A}^{(n)}(s_0)}{n!}(s-s_0)^n+\frac{(s-s_0)^n}{\pi}\int_{LC}ds^\prime\frac{\text{Im}\mathcal{A}(s^\prime)}{(s^\prime-s-i\epsilon)(s^\prime-s_0)^n}\\
&+\frac{(s-s_0)^n}{\pi}\int_{RC}ds^\prime\frac{\text{Im}\mathcal{A}(s^\prime)}{(s^\prime-s-i\epsilon)(s^\prime-s_0)^n},
\end{aligned}
\end{equation}
where the usual notation has been applied to identify $\mathcal{A}^{(n)}(s)$ as the n-th derivative of $\mathcal{A}(s)$. The description of the amplitude in the form of Eq.~(\ref{eq_un_nsubstracted}) will be valid as long as the series of derivatives exist in the selected subtraction points. We shall take $s_0=0$ from now on unless the contrary is specified.\myspace
The symbols RC and LC in Eq.~(\ref{eq_un_nsubstracted}) represent the so-called right-hand cut and left-hand cut, respectively. The kinematics of a $2\to2$ process requires a discontinuity along the real axis for $s > s_0$, extending up to $s \to \infty$, where $s_0$ is determined by the masses of the external particles. On the other hand, due to crossing symmetry, the left-hand cut (or simply left cut) appears on the negative semi-axis, starting at $s < 0$ and extending to $s \to -\infty$. This cut arises because, for the particular case of the $u-$channel, the crossed process exhibits a physical cut for $u > u_0$,  which, in terms of the variable $s$, enforces the presence of the left cut in our amplitude. The LC is also referred to as unphysical cut, since it involves unphysical values of the energy in the $s-$channel.\myspace
Both Eqs.~(\ref{eq_un_realandimaginary}) and (\ref{eq_un_ampandimaginary}) show that the real part of the amplitude at any given point is determined by an integral over its imaginary part. The imaginary part of the amplitude, which encodes information about intermediate physical states via the optical theorem, is non-zero only above certain thresholds, indicating the onset of particle production.\myspace
The unitarity condition requires that the sum of the probabilities of all possible processes at a certain energy equals unity. This is ensured by the definition of the S-matrix:
\begin{equation}\label{eq_un_smatrix}
S=1+i \mathcal{A}.
\end{equation}
In Eq.~(\ref{eq_un_smatrix}), the term \(1\) represents the identity matrix, corresponding to the possibility of a non-interacting reaction, while \(\mathcal{A}\) denotes the interacting part, i.e., the transition matrix between initial and final states, which are not necessarily the same. Imposing $S^\dagger S=1$, then
\begin{equation}\label{eq_un_discontinuity}
i\left(\mathcal{A}-\mathcal{A}^\dagger\right)=-2\text{Im}\,\mathcal{A}=-\mathcal{A}\mathcal{A}^\dagger,
\end{equation}
which is often referred to as the discontinuity condition.\myspace
From the unitarity condition of the $S-$matrix elements, an important result can be derived (see, for example, Refs.~\cite{book:tesisrafa}), which relates the imaginary part of an amplitude to its modulus squared:
\begin{equation}\label{eq_un_opticaltheorem}
\text{Im}\left[\mathcal{A}(a(p_1,p_2)\to b(p_3,p_4))\right]=\sum_f\frac{1}{64\pi^2K}\sqrt{1-\frac{4m^2_f}{s}}\int d\Omega\left[\mathcal{A}(a\to f)\right]\left[\mathcal{A}(f\to b)\right]^\ast,
\end{equation}
where $f$ sums over all possible intermediate states, and $a$ and $b$ denote the initial and final states of a unitary process, respectively. The factor $K$ accounts for the potential indistinguishability of the $a$ and $b$ states, ensuring that two-particle states are not double-counted.\myspace
An important consequence of Eq.~(\ref{eq_un_opticaltheorem}) is the Optical Theorem, which relates the imaginary part of a forward scattering amplitude to the total cross section of the process. Here, "forward" refers to scattering in the direction $\cos\theta=1$, or equivalently, when $t=0$. The standard form of the Optical Theorem is obtained by adding the corresponding kinematical factors into Eq.~(\ref{eq_un_opticaltheorem}), yielding:
\begin{equation}
\text{Im}\left[\mathcal{A}(a(p_1,p_2)\to b(p_1,p_2))\right](s,t=0)=\frac{s}{2}\sqrt{1-\frac{4m^2}{s}}\sigma_{tot}(a(p_1,p_2)\to f).
\end{equation}
Causality also imposes constraints on the high-energy behavior of the amplitude. For instance, the Froissart bound (see Ref.~\cite{PhysRev.123.1053}), which stems from both causality and unitarity, limits the growth of total cross-sections at high energies, further constraining the analytic properties of the amplitude at large $s$. In particular, any valid amplitude should obey
\begin{equation}\label{eq_un_amplitudedecomposed}
\text{Im}\,\mathcal{A}\left(s\to\infty,t=0\right)\leq \text{const}\cdot\,s\log^2\frac{s}{s_0} 
\end{equation}
All together, these conditions form a coherent structure that shapes the behavior of scattering amplitudes in our theory.
\subsection{Partial Wave Analysis}
The language of partial waves is particularly suitable for our study. A partial wave represents the projection of our amplitudes onto a specific channel with fixed angular momentum. Not only do the bound states and resonances we are interested in occur in specific partial waves, but the mathematical implementation of the unitarity condition also becomes much simpler in this framework.\myspace
The partial wave for any helicity combination and fixed angular momentum $J$ is defined as
\begin{equation}\label{eq_un_pw_wigner}
t_J^{\lambda_1\lambda_2\lambda_3\lambda_4}(s)=\frac{1}{32K\pi}\int_{-1}^{1}d\left(\cos\theta\right) W_J^{\lambda,\lambda^\prime}\left(\cos\theta\right)\mathcal{A}\left(s,\cos\theta\right),
\end{equation}
where, $K$ is a constant, taking the value $K=2$ if the particles participating in the process are identical, and $K=1$ otherwise. In the expression above we have also defined $\lambda=\lambda_1-\lambda_2$ and $\lambda^\prime=\lambda_3-\lambda_4$, with the Wigner functions denoted by $W_J^{\lambda\lambda^\prime}$.\myspace
However, we emphasize that we are only interested in the purely longitudinal channel for every process, so for us, $\lambda_i=0$, which implies $\lambda=\lambda^\prime=0$. In this case the Wigner functions reduce to the Legendre polynomials $P_J\left(\cos\theta\right)$, which depend solely on the angular momentum. Thus, our partial waves become:
\begin{equation}\label{eq_un_pw}
t_J(s)=\frac{1}{32K\pi}\int_{-1}^{1}d\left(\cos\theta\right) P_J\left(\cos\theta\right)\mathcal{A}\left(s,\cos\theta\right).
\end{equation}
In the centre-of-mass frame and for the case $m_1=m_2$ and $m_3=m_4$ (which applies to the processes we are interested in) the scattering angle $\theta$ is defined as
\begin{equation}\label{eq_un_costheta}
\cos\theta=\frac{1}{\sqrt{1-\frac{4m_1^2}{s}}\sqrt{1-\frac{4m_3^2}{s}}}\left(1+\frac{2\left(t-m_1^2-m_3^2\right)}{s}\right).
\end{equation}
All explicit dependence on the Mandelstam variable $u$ dissapears using the last relation in Eq.~(\ref{eq_ET_mandelstam}).\myspace
Inverting Eq.~(\ref{eq_un_pw}) using the orthonormal property of the Legendre polynomials,
\begin{equation}\label{eq_un_pwdecomposition}
\mathcal{A}\left(s,\cos\theta\right)=16\pi K\sum_{J=0}^\infty t_J(s)\left(2J+1\right)P_J(\cos\theta),
\end{equation}
one can explicitly see the meaning of partial waves as functions that decompose an amplitude into its fixed angular-momentum components.\myspace
The unitarity condition for partial waves, as previously discussed, has a simple form. However, deriving how Eq.~(\ref{eq_un_opticaltheorem}) can be expressed in terms of partial waves, is not straightforward. This requires accounting for relations among the initial, final, and intermediate state particles of the process, as well as the orthogonality condition for the Legendre polynomials. While we will not present the full derivation here, the interested reader can refer to Ref.~\cite{book:tesisrafa} for detailed steps. The result for the elastic case is:
\begin{equation}\label{eq_un_opticaltheorempw}
\text{Im}\left[t_{IJ}(s)\right]=\sqrt{1-\frac{4m^2}{s}}|t_{IJ}|^2\equiv \sigma(s)|t_{IJ}|^2\approx |t_{IJ}|^2,
\end{equation}
where the last step holds as $\sigma (s)\lessapprox 1$ for all values of $s$ in the physical region and in the TeV scale that is of our interest.\myspace
The unitarity condition in Eq.~(\ref{eq_un_opticaltheorempw}) can be rewritten, far from the production threshold, as
\begin{equation}\label{eq_un_relationimmodulus}
\text{Re}\left[t_{IJ}\right]^2+\left(\text{Im}\left[t_{IJ}\right]-\frac{1}{2}\right)^2=\frac{1}{4},
\end{equation}
which implies that the representation of the imaginary versus real part for a unitary partial wave must lie on a circumference of radius $\frac{1}{2}$ centered at $(0,1/2)$. For the case of our study focused on $WW$ scattering, any distortion of this pictorial unitarity condition, would come for values of the energy close to twice of the $W$ mass, which are of course of no interest for our study of the TeV region.\myspace
An equivalent unitarity condition is found noting that Eq.~(\ref{eq_un_opticaltheorempw}) can be solved explicitely to give
\begin{equation}\label{eq_un_pwdelta}
t_{IJ}(s)=\frac{i}{2\sigma(s)}\left(1-e^{2i\delta (s)}\right),
\end{equation}
where $\delta (s)$ is the \textit{phase shift}, which will play a crucial role in the following chapters of this dissertation.\myspace
From this, the unitarity condition follows straightforwardly as
\begin{equation}\label{eq_un_unitaritycondition}
|t_{IJ}|=\frac{1}{\sigma (s)}|\sin(\delta)|\leq \frac{1}{\sigma(s)}.
\end{equation}
For the case of an inelastic scattering, there is an extra contribution to Eq.~(\ref{eq_un_opticaltheorempw}), and of course it has the form of the modulus squared of some other processes with the corresponding phase space that can occur in the same isospin channel, according to the idea of the optical theorem. If we call this extra term $\Delta$, the unitarity condition changes to
\begin{equation}\label{eq_un_opticaltheorempw_inelastic}
\text{Im}\left[t_{IJ}(s)\right]=\sqrt{1-\frac{4m^2}{s}}|t_{IJ}|^2+\Delta,
\end{equation}
and again it can be solved explicitely by simply adding an inelasticity parameter, $\eta$,  to the partial wave in Eq.~(\ref{eq_un_pwdelta}), giving
\begin{equation}\label{eq_un_pwdelta_inelastic}
t_{IJ}(s)=\frac{i}{2}\left(1-\eta (s)e^{2i\delta (s)}\right),
\end{equation}
which satisfies $\eta (s)\leq 1$. By comparison one finds that $\eta^2=\frac{1}{2}-4\sigma (s)\Delta$.
\subsubsection{Construction of our partial waves}
Let us now particularize to the full $2\to 2$ $WW$ scattering, that is the case of interest for out analysis.\myspace
The amplitudes we derived in the previous chapter ignore purely electromagnetic interactions by setting $g^\prime=0$. The absence of these corrections corresponds to a theory that exactly preserves custodial symmetry, which provides an additional invariance that allows us to relate different amplitudes of the form $W^a_LW^b_L\to W_L^cW_L^d$. In particular, from Bose and crossing symmetries,
\begin{equation}\label{eq_un_isos_general}
\mathcal{A}^{abcd}=\delta^{ab}\delta^{cd}\mathcal{A}\left(p^a,p^b,p^c,p^d\right)
+\delta^{ac}\delta^{bd}\mathcal{A}\left(p^a,-p^c,-p^b,p^d\right)+\delta^{ad}\delta^{bc}\mathcal{A}\left(p^a,-p^d,p^c,-p^b\right),
\end{equation}
which allows us to write the following amplitudes in the charged basis
\begin{equation}\label{eq_un_isos_I1}
\begin{split}
\mathcal{A}^{+-00}&=\mathcal{A}(p^a,p^b,p^c,p^d)\\
\mathcal{A}^{+-+-}&=\mathcal{A}(p^a,p^b,p^c,p^d)+\mathcal{A}(p^a,-p^c,-p^b,p^d)\\
\mathcal{A}^{++++}&=\mathcal{A}(p^a,-p^c,-p^b,p^d)+\mathcal{A}(p^a,-p^d,p^c,-p^b),
\end{split}
\end{equation}
where we have used the definitions $\ket{W^{\pm}} =\frac{1}{\sqrt{2}}\left(\ket{W^1}\pm i \ket{W^2}\right)$ and $\ket{Z}\equiv\ket{W^0}=\ket{W^3}$. As an example, let us derive the first relation above:
\begin{equation*}
\begin{aligned}
\mathcal{A}^{+-00}&=\matelem{W^+W^-}{ZZ}=\frac{1}{2}\matelem{(W^1+i W^2)(W^1-iW^2)}{W^3W^3}\\
&=\frac{1}{2}\left[\matelem{W^1W^1}{W^3W^3}+\matelem{W^2W^2}{W^3W^3}+\cancel{i\matelem{W^2W^1}{W^3W^3}}\right.\\
&\left.-\cancel{i\matelem{W^1W^2}{W^3W^3}}\right]=\mathcal{A}(p^a,p^b,p^c,p^d),
\end{aligned}
\end{equation*}
where the last two terms vanish according to Eq.~(\ref{eq_un_isos_general}).\myspace
The rest of the relations are derived in the same, though more tortuous, manner.\myspace
The relation in Eq.~(\ref{eq_un_isos_general}), built from custodial invariance, implies that every amplitude with vector bosons as asymptotic states can be obtained by crossings from $W^{+}W^{-}\to ZZ$. This is why, in the previous section, this amplitude was the only one computed and was designated as the \textit{fundamental} amplitude. \myspace
When longitudinally-polarized gauge bosons are involved, crossing symmetry must be implemented via the momenta and not via Mandelstam variables because the polarization vectors do not transform covariantly (see e.g., the discussion in Ref.~\cite{Espriu:2012ih}). In the case where Goldstone bosons appear as external legs---as assumed in the real part of the one-loop calculation by virtue of the ET--- the situation is simpler. In the absence of polarization vectors, the (weak) isospin symmetry becomes manifest at the level of the Mandelstam variables, allowing Eq.~(\ref{eq_un_isos_general}) to be written as
\begin{equation}\label{eq_un_isos_general_golds}
\mathcal{A}^{abcd}=\delta^{ab}\delta^{cd}\mathcal{A}\left(s,t,u\right)
+\delta^{ac}\delta^{bd}\mathcal{A}\left(t,s,u\right)+\delta^{ad}\delta^{bc}\mathcal{A}\left(u,t,s\right).
\end{equation}
The relations in Eq.~(\ref{eq_un_isos_I1}) are derived accordingly in the framework of the ET.\myspace
From now on, unless otherwise specified, isospin will refer to weak isospin rather than the strong isospin associated with the custodial-symmetry-preserving limit of effective QCD.\myspace
In the isospin limit, every one-particle initial or final state is represented by the quantum state $\ket{I,I_3}$, where $I$ denotes the total weak isospin and $I_3$ its third component. For the gauge bosons and Goldstones, $I=1$, and $I_3=0,\pm 1$, which varies for each member of the isomultiplet, according to the standard definitions. These states are give by:
\begin{equation}\label{eq_un_isospinfields}
\ket{1,1}=-W^+, \quad \ket{1,0}=Z, \quad \ket{1,-1}=W^-.
\end{equation}
Using the definitions in Eq.~(\ref{eq_un_isospinfields}), we can now construct states with fixed isospin projections for $I=0,1,2$ corresponding to isoscalar, isovector and isotensor states, respectively. These states are formed using the appropriate Clebsch-Gordan coefficients, and the resulting isospin projections $T_I$ are given by:
\begin{equation}\label{eq_un_Ts_WW}
\begin{aligned}
T_0&=3\mathcal{A}^{+-00}+\mathcal{A}^{++++}\\
T_1&=2\mathcal{A}^{+-+-}-2\mathcal{A}^{+-00}-\mathcal{A}^{++++}\\
T_2&=\mathcal{A}^{++++}.
\end{aligned}
\end{equation}
As an example, we derive $T_0$, which is defined in the basis $\ket{I,I_3}$ as 
\begin{equation}\label{eq_un_definitionT0}
T_0\equiv\bra{0,0}\mathcal{A}\ket{0,0}.
\end{equation}
The decomposition into third components with Clebsch-Gordan coefficients gives
\begin{equation}
\begin{aligned}
\ket{0,0}&=\frac{1}{\sqrt{3}}\left(\ket{1,1}\times\ket{1,-1}-\ket{1,0}\times\ket{1,0}+\ket{1,-1}\times\ket{1,1}\right)\\
&=-\frac{1}{\sqrt{3}}\ket{W^+W^-+ZZ+W^-W^+}=-\frac{1}{\sqrt{3}}\ket{W^1W^1+W^2W^2+W^3W^3},
\end{aligned}
\end{equation}
where we use the definitions from Eq.~(\ref{eq_un_isospinfields}) when moving from the first to the second line.\myspace
Thus,
\begin{equation}
\begin{aligned}
T_0&=\frac{1}{3}\bra{W^1W^1+W^2W^2+W^3W^3}\mathcal{A}\ket{W^1W^1+W^2W^2+W^3W^3}\\
&=3\mathcal{A}(p^a,p^b,p^c,p^d)+\mathcal{A}(p^a,-p^c,-p^b,p^d)+\mathcal{A}(p^a,-p^d,p^c,-p^b)\\
&=3\mathcal{A}^{+-00}+\mathcal{A}^{++++}.
\end{aligned}
\end{equation}\myspace
The other two processes of interest in our study involve a two-Higgs final state, which, being the Higgs boson a singlet in this framework, is naturally an $I=0$ state. Consequently, for $W_L^aW^b_L\to hh$ the only possible fixed-isospin amplitude that can be constructed is the $I=0$ amplitude
\begin{equation}\label{eq_un_Ts_Wh}
\mathcal{A}(W_L^a W_L^b \to hh)=\mathcal{A}^{ab}(p^a,p^b,p_{h,1},p_{h,2}),\qquad
T_{Wh,0}=\sqrt{3}\mathcal{A}^{+-}.
\end{equation}
The prefactor $\sqrt{3}$ comes from the Clebsh-Gordan coefficient that projects to the $\ket{00}$ isospin state.\myspace
Lastly, for the elastic scattering of Higgsses
\begin{equation}\label{eq_un_Ts_hh}
\mathcal{A}(hh\to hh)=\mathcal{A}(p_{h,1},p_{h,2},p_{h,3},p_{h,4})= T_{hh,0}.
\end{equation}
Now that we have constructed amplitudes with fixed-isospin, we can define the corresponding partial wave with fixed angular momentum and isospin, $t_{IJ}$ by replacing $\mathcal{A}(s,\cos\theta)$ in Eq.~(\ref{eq_un_pw}) for the amplitude with a fixed isospin that can be found for each process in Eqs.~(\ref{eq_un_Ts_WW}), (\ref{eq_un_Ts_Wh}) and (\ref{eq_un_Ts_hh}):
\begin{equation}\label{eq_un_pwIJ}
t_J(s)\to t_{IJ}(s)=\frac{1}{32K\pi}\int_{-1}^{1}d\left(\cos\theta\right) P_J\left(\cos\theta\right) T_I\left(s,\cos\theta\right).
\end{equation}
From this point forward, we will refer to the expression in Eq.~(\ref{eq_un_pwIJ}) as partial wave, superseding the form given in Eq.~(\ref{eq_un_pw}).\myspace
The way we compute the fixed-isospin amplitudes using Feynman diagrams from the Lagrangian in Eq.~(\ref{eq_ET_fullHEFT}), leads to a perturbative expansion of our partial waves the form
\begin{equation}\label{eq_un_pwexpansion}
t_{IJ}(s)\approx t_{IJ}^{(2)}+t_{IJ}^{(4)}+\cdots,
\end{equation}
where the superscripts $(2), (4),\cdots$ indicate the chiral order. Each term $t_{IJ}^{(n)}$ is obtained by applying Eq.~(\ref{eq_un_pwIJ}) to the fixed-isospin amplitude $T_I^{(n)}$ for specific values of $I$ and $J$.\myspace
The unitarity condition in Eq.~(\ref{eq_un_unitaritycondition}) was derived from the very principles of quantum field theory by imposing the unitarity of the S-matrix. While the full result must satisfy this unitarity condition exactly, the expanded partial wave in Eq.~(\ref{eq_un_pwexpansion}), derived from effective amplitudes, satisfies the unitarity condition only perturbatively, a concept referred to as \textit{perturbative unitarity}.\myspace
In any case, unitarity should not be confused with perturbative unitarity, which relates the real and imaginary parts up to a given order in perturbation theory. Perturbative unitarity merely ensures the internal consistency of a calculation in field theory, even if the theory itself is not unitary, and is therefore automatic and of no relevance to our discussion.\myspace
Regarding unitarity, if we were computing the amplitudes in the UV theory, the study of full unitarity would yield little-if anything- insight, as the condition would be automatically satsified by construction, regardless of the values of the coefficients of the theory. However, we will still check full unitarity at the level of the unitarized waves as a sanity check for the unitarization method.\myspace
For elastic scattering, the conditions from perturbative unitarity on the two first orders of the partial wave expansion are given by
\begin{equation}\label{eq_un_perturbativeunitarity}
\begin{aligned}
&\text{Im}\left[t^{(2)}\right]=0\\
&\text{Im}\left[t^{(4)}\right]=\sum_i \sigma |t^{i,(2)}|^2,
\end{aligned}
\end{equation}
where $i$ runs over all possible combinations allowed by the particular elastic process. Thus, perturbative unitarity allows us to compute the imaginary part of the NLO wave in terms of the tree level calculation that we recall is obtained exactly using physical gauge bosons in teh external states.\myspace
Among the infinite $J$-terms required in principle to fully reproduce the amplitude in Eq.~(\ref{eq_un_pwdecomposition}), we assume that, for each isospin channel, the first term in the series suffices. This assumption holds because we are focused on the resonant region, where the unitarized amplitudes may exhibit a peak that dominates at specific $J$ values. In particular, we consider the cases $I=J=0$, corresponding to the isoscalar-scalar wave; $I=J=1$, corresponding to the isovector-vector wave; and $I=0, J=2$, corresponding to the isoscalar-tensor wave, corresponding to the first terms allowed in the series expansion.\myspace
We recall the three processes of interest for this dissertation, where we omit electric charges and the subscripts indicating the longitudinal polarization for the gauge bosons: $WW\to ZZ$ ($WW$), $WW\to hh$ ($Wh$) and $hh\to hh$ ($hh$). All three processes have a projection in the isoscalar-scalar channel, while the elastic scattering of gauge bosons ($WW$) is the only one that can contribute to the two remaining waves, namely the isovector-vector and the isoscalar-tensor.\myspace
With all the preceeding information we construct $t_{00}^{WW}$ up to NLO as
\begin{equation}\label{eq_un_t00wwfull}
\begin{aligned}
&t_{00}^{WW,(2)}=\frac{1}{64\pi}\int_{-1}^1 d\left(\cos\theta\right) T_0^{WW,(2)}\left(s,\cos\theta\right)\\
&\text{Re}\left[t_{00}^{WW,(4)}\right]=\frac{1}{64\pi}\int_{-1}^1 d\left(\cos\theta\right) T_{0}^{WW,(4)}\left(s,\cos\theta\right)\\
&\text{Im}\left[t_{00}^{WW,(4)}\right]=\sigma (s)|t_{00}^{WW,(2)}|^2+\sigma_H (s)|t_{00}^{Wh,(2)}|^2,
\end{aligned}
\end{equation}
the $t_{11}^{WW}$
\begin{equation}\label{eq_un_t11wwfull}
\begin{aligned}
&t_{11}^{WW,(2)}=\frac{1}{64\pi}\int_{-1}^1 d\left(\cos\theta\right)\cos\theta\,T_1^{WW,(2)}\left(s,\cos\theta\right)\\
&\text{Re}\left[t_{11}^{WW,(4)}\right]=\frac{1}{64\pi}\int_{-1}^1 d\left(\cos\theta\right)\cos\theta\,T_{1}^{WW,(4)}\left(s,\cos\theta\right)\\
&\text{Im}\left[t_{11}^{WW,(4)}\right]=\sigma (s)|t_{11}^{WW,(2)}|^2
\end{aligned}
\end{equation}
and the $t_{20}^{WW}$
\begin{equation}\label{eq_un_t20wwfull}
\begin{aligned}
&t_{20}^{WW,(2)}=\frac{1}{64\pi}\int_{-1}^1 d\left(\cos\theta\right) T_2^{WW,(2)}\left(s,\cos\theta\right)\\
&\text{Re}\left[t_{20}^{WW,(4)}\right]=\frac{1}{64\pi}\int_{-1}^1 d\left(\cos\theta\right) T_{2}^{WW,(4)}\left(s,\cos\theta\right)\\
&\text{Im}\left[t_{20}^{WW,(4)}\right]=\sigma (s)|t_{20}^{WW,(2)}|^2.
\end{aligned}
\end{equation}
In this expressions, the Legendre polynomials $P_0(\cos\theta)=0$ and $P_1(\cos\theta)=\cos\theta$ have been used. As discussed in the previous chapter and exemplified in Eq.~(\ref{eq_ET_exampleA4}), under our ET approximation for the loop calculation, one must consider $T_I^{WW,(4)}=\text{Re}\left[T_{I,loop}^{\omega\omega,(4)}\right]+T_{I,tree}^{WW,(4)}$.\myspace
Similarly, and for the crossed channels $WW\to hh$ and $hh\to hh$ that can only exist in the $IJ=00$ channel, the corresponding partial waves $t_{IJ}^{\{Wh,hh\}}$ are
\begin{equation}\label{eq_un_t00whfull}
\begin{aligned}
&t_{00}^{Wh,(2)}=\frac{1}{64\pi}\int_{-1}^1 d\left(\cos\theta\right) T_0^{Wh,(2)}\left(s,\cos\theta\right)\\
&\text{Re}\left[t_{00}^{Wh,(4)}\right]=\frac{1}{64\pi}\int_{-1}^1 d\left(\cos\theta\right) T_{0}^{Wh,(4)}\left(s,\cos\theta\right)\\
&\text{Im}\left[t_{00}^{Wh,(4)}\right]=t_{00}^{WW,(2)}t_{00}^{Wh,(2)\,\ast}+t_{00}^{Wh,(2)}t_{00}^{hh,(2)\,\ast},
\end{aligned}
\end{equation}
and
\begin{equation}\label{eq_un_t00hhfull}
\begin{aligned}
&t_{00}^{hh,(2)}=\frac{1}{64\pi}\int_{-1}^1 d\left(\cos\theta\right) T_0^{hh,(2)}\left(s,\cos\theta\right)\\
&\text{Re}\left[t_{00}^{hh,(4)}\right]=\frac{1}{64\pi}\int_{-1}^1 d\left(\cos\theta\right) T_{0}^{hh,(4)}\left(s,\cos\theta\right)\\
&\text{Im}\left[t_{00}^{hh,(4)}\right]=\sigma (s)|t_{00}^{Wh,(2)}|^2+\sigma_H (s)|t_{00}^{hh,(2)}|^2.
\end{aligned}
\end{equation}
Again, in the expressions above, $T_{0}^{\{Wh,hh\},(4)}=\text{Re}\left[T_{0,loop}^{\{\omega h,hh\},(4)}\right]+T_{0,tree}^{\{Wh,hh\},(4)}$.\myspace
\subsection{Partial waves at high energies}
The bad-energy behavior can also be seen at the partial wave level before unitarization. As an example, we consider the modulus of the vector-isovector contribution up to the NLO in the expansion of Eq.~(\ref{eq_un_pwexpansion}), the set of expressions in Eq.~(\ref{eq_un_t11wwfull}). This is illustrated in Figure~ \ref{fig_un_t11_loopvstree} for a small departure from the SM values via the parameter $a_4=10^{-4}$ (the rest of the parameters are set to their corresponding SM values). In that same figure (in the right axis), the contribution in percentage (in absolute value) to the one-loop partial wave is shown with respect to the full tree level [$\mathcal{O}(p^2)$+$\mathcal{O}(p^4)$] with the following definition
\begin{equation}\label{eq_un_treevsloop}
\Delta_{\text{1-loop}}=100 \cdot \Bigg | \frac{|t_{11}^{tree+loop}|-|t^{tree}_{11}|}{|t_{11}^{tree}|}\Bigg |.
\end{equation}
\begin{figure}
\centering
\includegraphics[clip,width=12cm,height=8.5cm]{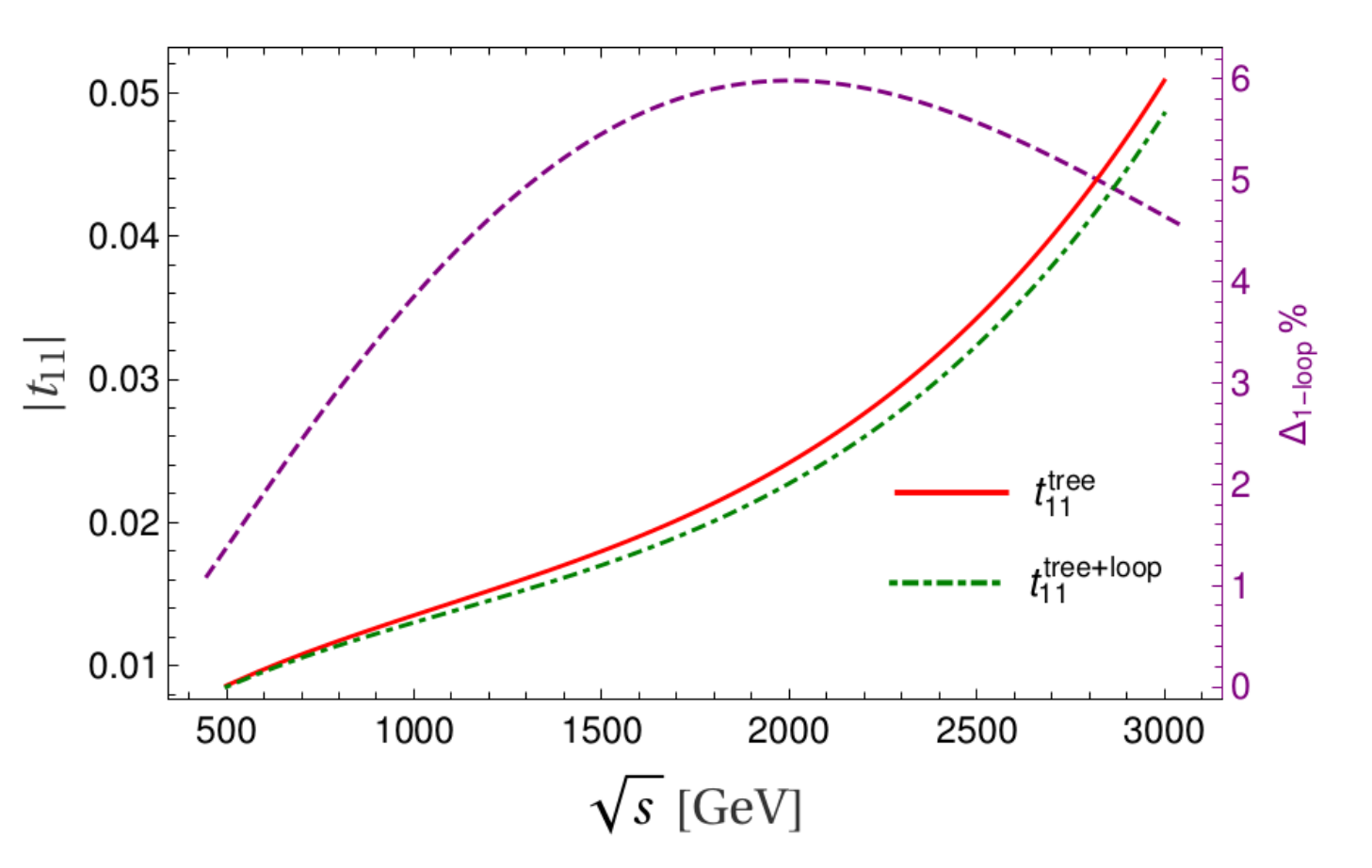}
\caption{\small{(Left axis) Plot of the modulus of the vector-isovector partial wave at tree level (solid red line) and tree + one loop level (dot-dashed green line) versus the center of mass energy $\sqrt{s}$. (Right axis) Plot of the percentage represented  by the one-loop contribution (purple dashed line), $\Delta_{1-loop}$ in Eq.~(\ref{eq_un_treevsloop}) versus the center of mass energy $\sqrt{s}$ in absolute value. The curves are depicted for $a_4=10^{-4}$ and the rest of the parameters are set to their SM values.}}
\label{fig_un_t11_loopvstree}
\end{figure}\noindent
The contribution of the one-loop level to the full partial wave turns out to be negative as it can be seen in the figure (the green dot-dashed line representing the full amplitude is always below the tree level contribution in solid red), reaching  a maximum value  $\sim 6\%$ around the $2$ TeV region. Here the tree level contribution includes, as mentioned, both the $O(p^2)$ and the $O(p^4)$ pieces. As was already noticed in previous works (see Ref.~\cite{Espriu:2013fia}), as soon as one departs 
from the SM the latter quickly dominates the real part of the $O(p^4)$ contribution.\myspace
The fact that the next-to-leading order calculation is dominated by the tree-level contribution with anomalous values of $a_4$ and $a_5$—--which, we recall, is an exact calculation--—allows us some flexibility in computing the real part at one loop, the only part calculated approximately. Thus, we believe the use of the ET is justified.\myspace
We can now ask ourselves about the precision of such an approximation and the extent to which it affects the predictions of perturbative unitarity. As we have seen, the \textit{perturbative unitarity} condition relates the imaginary part of the NLO amplitude to its LO modulus  (see Eq.~(\ref{eq_un_t00wwfull})).
\begin{equation}\label{eq_un_perturbativeunitarity2}
\text{Im}\,t_{00}^{WW,(4)}=\sigma_W|t_{00}^{WW,(2)}|^2+\sigma_H|t_{00}^{Wh,(2)}|^2.
\end{equation}
Our test consists in computing the left-hand side of (\ref{eq_un_perturbativeunitarity2}) within the ET, via a one-loop calculation,
and checks whether the relation (\ref{eq_un_perturbativeunitarity2}), with the rhs determined using physical $W$'s, stands and what level of agreement is obtained. For this, let us define the error, in a percentage, assumed by the ET with the quantity
\begin{equation}\label{eq_un_error_OT}
\Delta_{\text{Full-ET}}=\frac{\bigg |\text{Im}\,t_{00}^{\omega\omega,(4)}-\left(\sigma_W|t_{00}^{WW,(2)}|^2+\sigma_H|t_{00}^{Wh,(2)}|^2\right)\bigg |}{\sigma_W|t_{00}^{WW,(2)}|^2+\sigma_H|t_{00}^{Wh,(2)}|^2}\cdot 100.
\end{equation}
This quantity includes, by construction, couplings from the leading-order Lagrangian (\ref{eq_ET_L2}), and it is completely independent of any $\mathcal{O}(p^4)$ parameter since our calculation is made up to $\mathcal{O}(p^4)$ and they just enter at tree level and consequently do not produce any imaginary part for the left-hand side of Eq.~(\ref{eq_un_perturbativeunitarity2}).
\begin{figure}
\centering
\includegraphics[clip,width=13cm,height=9cm]{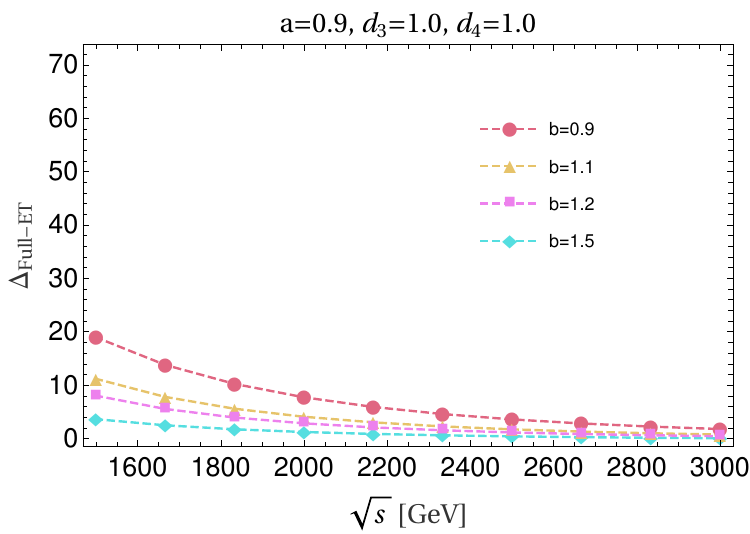}
\caption{\small{In this plot we show in percentage the quantity $\Delta_{\text{Full-ET}}$ defined in (\ref{eq_un_error_OT}) plotted with respect to the centre of mass energy. In this figure we set $a=0.9, d_3=1.0$ and $d_4=1.0$ and show different values of $b$. This figure is independent of $\mathcal{O}(p^4)$ parameters. This shows that for large values of $s$ the imaginary part computed via the ET agrees at the 10\% level with the one determined (exactly) via the optical theorem. As explained in the text, although the discrepancy may look large at low values of $s$, the amplitude is very small there and does not contribute significantly to the position and width of possible resonances. See also  Figure~\ref{fig_un_perturbative_unitarity}.}}
\label{fig_un_error_OT}
\end{figure}\myspace
Figure \ref{fig_un_error_OT} shows $\Delta_{\text{Full-ET}}$ for the BSM interaction of a Higgs to two gauge bosons, $a=0.9$, the self-couplings of the Higgs set to their
standard values, and various $b$ values. The behavior is that expected for the ET: the longitudinal components of the electroweak gauge bosons are well represented by the associated Goldstone boson for high energies compared to the gauge boson masses. This is independent of the value of $b$ in the plot.\myspace
What we also observe in Figure~\ref{fig_un_error_OT} is that for values of $b$ close to $a^2=0.9^2=0.81$ the error grows. This "failure" of the ET can be understood by going to the completely massless limit, useful for $\sqrt{s}\gg M_{\omega,h}$, where the leading-order $\omega\omega$ and $\omega h$ amplitudes are proportional to $(1-a^2)$ and $(a^2-b)$ [Eqs.~(\ref{eq_ET_ww_tree_net}) and (\ref{eq_ET_wh_tree_net})], respectively. For a fixed value of $a$, close to the SM which cancels the $\omega\omega$ massless amplitude, the closer $b$ is to $0.81$, the worse the comparison with the full calculation is. This happens because the right-hand side of (\ref{eq_un_perturbativeunitarity2}) approaches zero, being proportional to $b-a^2$, and cancels the leading part of the denominator of $\Delta_{\text{Full-ET}}$. Generally speaking, if one considers the SM parameters $b=a^2=1$, we do not find good agreement and more terms in the ET expansion, such as $\mathcal{O}(g^2)\,\, W_L\omega\to\omega\omega$, would be needed at low $s$ values as explained in Ref.~\cite{Espriu:1994ep}.\myspace
However, the apparent failure just described is actually a mirage because we are dealing with partial waves that are very small numerically. To see this,  we show in Figure~\ref{fig_un_perturbative_unitarity} a check of perturbative unitarity for a benchmark point away from the SM by a $10\%$, except for the Higgs self-couplings which remain set in their SM values.
\begin{figure}
\centering
\includegraphics[clip,width=13cm,height=9cm]{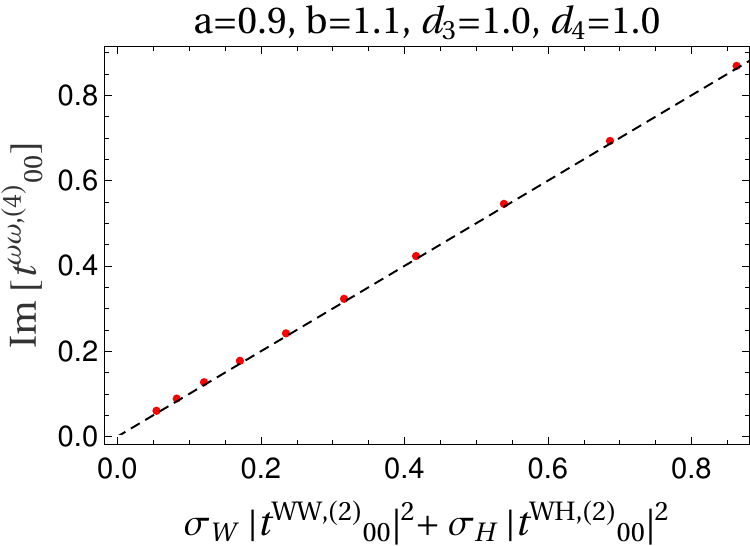}
\caption{\small{Figure showing perturbative unitarity (\ref{eq_un_perturbativeunitarity2}) for the chiral couplings specified in the title. The data shown here are 10 red points equally spaced along the energy range $1500-3000$ GeV, within the validity region of the theory. The dashed line is the bisector of the first quadrant and the points that lie over it satisfy perturbative unitarity exactly.}}
\label{fig_un_perturbative_unitarity}
\end{figure}\myspace
It could seem that Figure~\ref{fig_un_perturbative_unitarity} enters into contradiction with the previous Figure~ \ref{fig_un_error_OT} that shows a worse agreement for the same benchmark point, while it is almost unnoticeable for the low-energy points in the former. This situation, that is reproduced with any choice of parameters, is explained by the fact that, by construction, the chiral amplitudes are much smaller at low energies than in the high-energy regime where this uncontrolled growth leads precisely to the violation of unitarity. Having said that, we can conclude that it is safe for us to make use of the ET for the one-loop level along the whole range of energies since the differences with the full calculation are on one hand negligible in the high-energy
regime and, on the other hand, the low-energy contributions are much smaller when one is very close to the SM values. Away from the SM limit (which is of course our main focus), the agreement is good everywhere.\myspace
With the partial waves in Eqs.(\ref{eq_un_t00whfull})-(\ref{eq_un_t00hhfull}) we are in position to implement the unitarization methods in the simplest manner. What we expect from this process is the emergence of resonances in at least some of the different channels of the unitarized amplitudes. If these resonances do appear in the spectrum of the unitarized theory, we interpret them as heavy states belonging to the high-energy scale, which somehow becomes accessible through unitarization. After integrating out the heavy state from the UV theory, HEFT amplitudes are generated in the infrarred, and the violation of unitarity in these effective amplitudes serves as a remnant of the high-energy scale. Conversely, these heavy resonances unitarize the UV theory, much like the Higgs boson unitarizes the longitudinal scattering of gauge bosons in the SM.
\section{Some Unitarization Methods}
Thus, the loss of unitarity in the amplitudes requires unitarization methods, of which there are several. In this section we list some of them, with a special emphasis on the K-matrix (and its improved version, the IK-matrix) and the Inverse Amplitude Method (IAM), as they are cornerstones due to their analytical properties.\myspace
Following the idea of enforcing unitarity on an amplitude, as we did at the beginning of this chapter, several methods are available. All of them are summarized in Refs.~\cite{Rauch:2016pai, PhysRevD.91.096007, Garcia-Garcia:2019oig}.\myspace
The first method, which is not a method \textit{per se}, involves applying a cut-off to the effective amplitude to avoid the unitarity-violating region; that is, the amplitude is redefined only for those energies where it remains physical. In the example of Figure~\ref{fig_un_example}, this would correspond to defining the amplitude as $\mathcal{A}_{\text{cut-off}}(s) = \mathcal{A}(s) / s < s_0$.\myspace 
Other methods do not directly discard part of the amplitude; instead they modify it by multiplying it by an energy-dependent factor that suppresses its problematic behavior. Specifically, we encounter the Form Factor (FF) method and the Kink method. While the multiplicative factor in the FF method suppresses the amplitude smoothly, the Kink method does so abruptly using a step function:  
\begin{equation}
\mathcal{A}_{\text{FF}}(s) = \mathcal{A}(s) \left(1 + \frac{s}{\Lambda^2}\right)^{-\alpha}, \quad 
\mathcal{A}_{\text{Kink}}(s) = \mathcal{A}(s) \begin{cases} 
1 & \text{if } s < \Lambda^2 \\ 
\left(\frac{s}{\Lambda^2}\right)^{-\alpha} & \text{if } s \geq \Lambda^2 
\end{cases}.
\end{equation}\myspace
The specific value of $\alpha$ depends on each particular amplitude. Comparing this with the example introduced at the beginning of this chapter, it becomes evident that the method we used was the Kink method.
The next method we introduce is the \(N/D\) method. Although it is not used for the unitarization of the amplitudes appearing in this thesis, it is the first on our list that is constructed based on the analytic properties of scattering amplitudes. This method, whose name stands for "numerator over denominator," is based on the ansatz for the elastic amplitude  
\begin{equation}
t(s) = \frac{N(s)}{D(s)},
\end{equation}  
where the numerator function only has left-hand cuts (LC), and the denominator function only has right-hand cuts (RC), ensuring the expected analytic structure. These definitions impose that in the RC, \(\text{Im} N = 0\), and in the LC, \(\text{Im} D = 0\), with \(\text{Im} N = D \text{Im} t\). Furthermore, the elastic unitarity condition also imposes that \(\text{Im} D = -N\) in the RC. Together, these relations lead to a system of coupled equations:  
\begin{equation}
D(s) = 1 - \frac{s}{\pi} \int_0^\infty \frac{ds^\prime \, N(s^\prime)}{s^\prime \left( s^\prime - s - i\epsilon \right)},  
\quad  
N(s) = \frac{s}{\pi} \int_{-\infty}^0 \frac{ds^\prime \, D(s^\prime) \text{Im} t(s^\prime)}{s^\prime \left( s^\prime - s - i\epsilon \right)},
\end{equation}  
which give unitary partial waves by construction. Depending on the ultraviolet behavior of the partial waves to be unitarized, successive subtractions can be applied, as in Eq.~(\ref{eq_un_nsubstracted}). A detailed discussion of this method and its application in the context of the chiral waves for EWSBS can be found in Ref.~\cite{Delgado:2015kxa}.  

\section{The Inverse Amplitude Method}
This section is dedicated to explaining the method used throughout this work in relation to the unitarization of the chiral amplitudes. Although there is no fundamental reason to adopt this specific method, it holds historical significance. In particular, it echoes the prediction of the rho meson (see Ref.~\cite{Oller:2000ma}) and its properties thorugh an unitarized analysis of pion-pion scattering, computed effectively in the low-energy regime of QCD---the strong chiral theory--- which is well described up to about $\sqrt{s}\simeq 1.2$ GeV. The rho meson, $\rho (770)$, emerges in the unitarized vector wave, along with its companions $K^\ast (892)$ and $\phi (1020)$. In the scalar channel, the unitarization procedure reproduces the $\sigma$, $f_0 (980)$, and $a_0 (980)$ resonances, with the latter two requiring the use of the so-called coupled-channel formalism, which will also be discussed in this section.
\subsection{IAM Single channel}
The essencial point for deriving the IAM comes from noting that the full unitarity condition, as expressed in Eq.~(\ref{eq_un_opticaltheorempw}), also constrains
\begin{equation}\label{eq_un_imaginaryinverse}
\text{Im}\left[\frac{1}{t_{IJ}(s)}\right]=-\sigma(s).
\end{equation}
This implies that the imaginary part of the inverse partial wave that satisfies the unitarity condition exactly-not just perturbatively-is determined purely by kinematics.\myspace
Thus, the question that now arises is straightforward: given complete knowledge of the imaginary part, can we exploit the analytical properties of the amplitudes to extract information about the real part so the full unitary partial wave is completely determined? The answer is affirmative and follows from the relation in Eq.~(\ref{eq_un_realandimaginary}), where the real part can be computed as an integral over the imaginary part in the physical region.\myspace
For the inverse amplitude, it is customary to define the function
\begin{equation}\label{eq_un_functionG}
\mathcal{G}(s)=\frac{\left(t^{(2)}\right)^2}{t(s)},
\end{equation}
where $t^{(2)}$ is the first order in the chiral expansion of a partial wave (the subindices $IJ$ have been removed for clarity).\myspace
%
We now construct the three-times substracted function $\mathcal{G}(s)$---see Eq.~(\ref{eq_un_nsubstracted})---:
\begin{equation}\label{eq_un_functiongSubs3}
\mathcal{G}(s)=G_0+G_1\,s+G_2\,s^2+\frac{s^3}{\pi}\int_{LC}ds^\prime\frac{\text{Im}\,\mathcal{G}(s^\prime)}{s^{\prime\,3}(s^\prime-s-i\epsilon)}+\frac{s^3}{\pi}\int_{RC}ds^\prime\frac{\text{Im}\,\mathcal{G}(s^\prime)}{s^{\prime\,3}(s^\prime-s-i\epsilon)},
\end{equation}
which is sufficient to soften the behavior of the waves of interest at $s\to \infty$.\myspace
The three substractions constants $G_0,\,G_1$ and $G_2$ of the function $\mathcal{G}(s)$ in Eq.~(\ref{eq_un_functionG}) are determined perturbatively from the chiral waves:
\begin{equation}
\begin{aligned}
&t^{(2)}(s)=a_0+a_1\,s\\
&t^{(4)}(s)=b_0+b_1\,s+b_2\,s^2+\frac{s^3}{\pi}\int_{LC}ds^\prime\frac{\text{Im}\,t^{(4)}(s^\prime)}{s^{\prime\,3}(s^\prime-s-i\epsilon)}+\frac{s^3}{\pi}\int_{RC}ds^\prime\frac{\text{Im}\,t^{(4)}(s^\prime)}{s^{\prime\,3}(s^\prime-s-i\epsilon)}.
\end{aligned}
\end{equation}
Thus, the dispertion relation for $\mathcal{G}(s)$ can be rewritten as:
\begin{equation}\label{eq_un_Gexpanded}
\mathcal{G}(s)=t^{(2)}-t^{(4)}+\frac{s^3}{\pi}\int_{LC}ds^\prime\frac{\text{Im}\,\mathcal{G}(s^\prime)+\text{Im}\,t^{(4)}(s^\prime)}{s^{\prime\,3}(s^\prime-s-i\epsilon)}+\frac{s^3}{\pi}\int_{RC}ds^\prime\frac{\text{Im}\,\mathcal{G}(s^\prime)+\text{Im}\,t^{(4)}(s^\prime)}{s^{\prime\,3}(s^\prime-s-i\epsilon)}.
\end{equation}
The RC integral is treated exactly since, along the RC, the following relation holds exactly:
\begin{equation}\label{eq_un_relationsG}
\text{Im}\,\mathcal{G}(s)=\left(t^{(2)}\right)^2\text{Im}\left[\frac{1}{t}\right]=-\sigma(s)\left(t^{(2)}\right)^2=-\text{Im}\,t^{(4)},
\end{equation}
where in the step in the middle, Eq.~(\ref{eq_un_imaginaryinverse}) has been used.\myspace
Substituting Eq.~(\ref{eq_un_relationsG}) into Eq.~(\ref{eq_un_Gexpanded}), the integral over the RC vanishes exactly.\myspace
This yields:
\begin{equation}
\mathcal{G}(s)=\frac{\left(t^{(2)}\right)^2}{t}=t^{(2)}-t^{(4)}+\frac{s^3}{\pi}\int_{LC}ds^\prime\frac{\text{Im}\,\mathcal{G}(s^\prime)+\text{Im}\,t^{(4)}(s^\prime)}{s^{\prime\,3}(s^\prime-s-i\epsilon)}.
\end{equation}
If we approximate $\text{Im}\mathcal{G}(s)\approx-\text{Im}t^{(4)}(s)$ also along the LC, which is true perturbatively, the remaining integral over the LC vanishes too---this time approximately--- and the IAM amplitude at NLO is obtained;
\begin{equation}\label{eq_un_IAMamplitude}
t^{\text{IAM}}_{IJ}(s)\equiv t_{IJ}(s)\approx\frac{\left(t^{(2)}_{IJ}\right)^2}{t^{(2)}_{IJ}-t^{(4)}_{IJ}},
\end{equation}
where we have explicitely reintroduced the subscripts $IJ$ to indicate that this IAM wave is constructed specifically for a particular $IJ$ channel.\myspace
Some advantages of the IAM will be now discussed. First, the IAM wave in Eq.~(\ref{eq_un_IAMamplitude}) satisfies the unitarity condition exactly, provided that perturbative unitarity is also satisfied---a requirement that is guaranteed by construction. Explicitely, we have
\begin{equation}
\begin{aligned}
&\text{Im}\,t^{IAM}=\frac{\left(t^{(2)}\right)^2\text{Im}\,t^{(4)}}{\left(t^{(2)}-\text{Re}\,t^{(4)}\right)^2+\left(\text{Im}\,t^{(4)}\right)^2},\\
&\sigma\left|t^{IAM}\right|^2=\frac{\sigma\left(t^{(2)}\right)^4}{\left(t^{(2)}-\text{Re}\,t^{(4)}\right)^2+\left(\text{Im}\,t^{(4)}\right)^2}=\frac{\left(t^{(2)}\right)^2\text{Im}\,t^{(4)}}{\left(t^{(2)}-\text{Re}\,t^{(4)}\right)^2+\left(\text{Im}\,t^{(4)}\right)^2},
\end{aligned}
\end{equation}
where, in the last step, the relations of perturbative unitarity from Eq.~(\ref{eq_un_perturbativeunitarity}) has been applied in their single-channel version.\myspace
Secondly, the IAM wave accurately reproduces the chiral amplitudes for the low energy regime
\begin{equation}
t^{IAM}=\frac{\left(t^{(2)}\right)^2}{t^{(2)}-t^{(4)}}\approx t^{(2)}+t^{(4)}+\cdots
\end{equation}
at order $\mathcal{O}(s^2)$, as it should.\myspace
Another reason to employ the IAM  our analysis, as apposed to other methods, is that the resonances emerging as poles of the amplitudes, if present, occur when
\begin{equation}
t^{(2)}(s)-t^{(4)}(s)=0,
\end{equation}
and are interpreted naturally as dynamically generated resonances.\myspace
This can be seen by expanding the IAM wave in Eq.~(\ref{eq_un_IAMamplitude}) up to an arbitrary order and see that each term corresponds to consecutive insertions of bubble diagrams. For the case under study in this dissertartation, $WW\to WW$, these bubbles correspond to $WW/ZZ$ pairs forming an infinite chain $WW\to ZZ\to WW\to\cdots \to ZZ$. This infinite resummation of bubble diagrams to infinite loop-order, guarantees the unitary condition, and the resonances appeareing are understood as dynamical states emerging from the rescattering process. Figure~\ref{fig_un_IAM_resum} provides a pictorial representation of this resummation in the single-channel case, where the intermediate bubbles can only correspond to $WW$ pairs in an $I=1$ state. As we will see soon, for the case $I=0$ the coupled channel formalism is required, leading to a slightly different picture.
\begin{figure}
\centering
\includegraphics[clip,width=13cm,height=4.3cm]{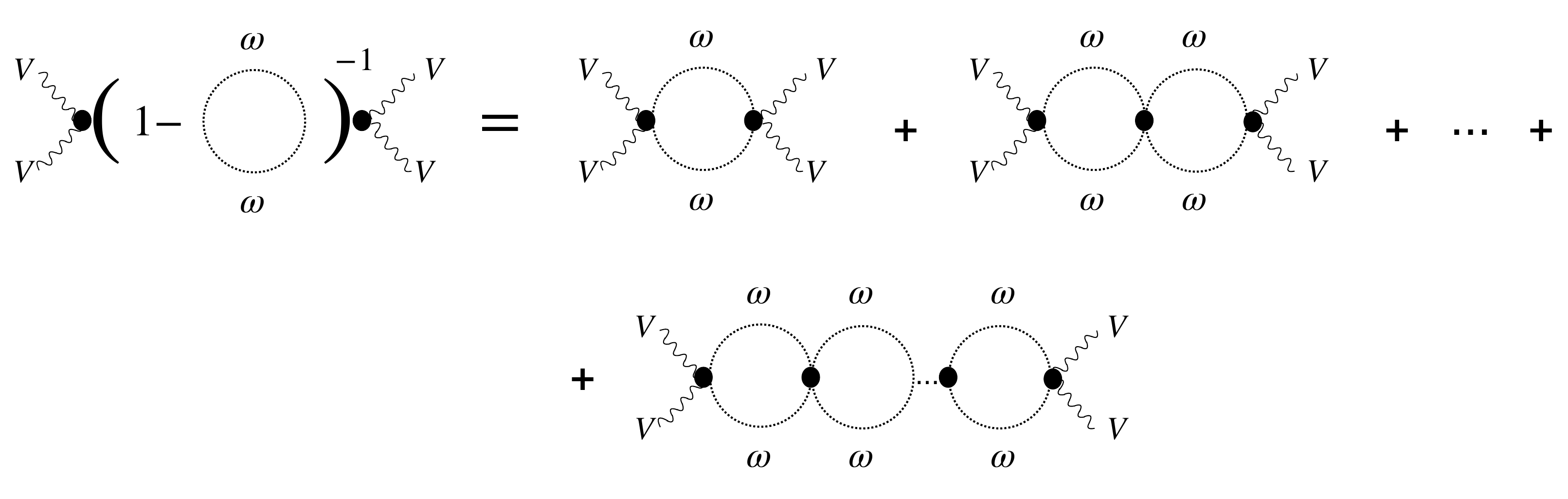}
\caption{\small{Schematic representation of the expansion of the IAM-wave in Eq.~(\ref{eq_un_IAMamplitude}), illustrating the interpretation of a vector dynamically generated resonance. The infinite series of bubbles representing the intermediate states of the process consists exclusively of $WW$ pairs in the $I=1$ projection, as they are the only possible combination compatible with a isovector-vector wave.}}
\label{fig_un_IAM_resum}
\end{figure}\myspace
A thorough revisiting of the IAM can be found in Ref.~\cite{Salas-Bernardez:2020hua}, where the authors study in depth the derivation of the inverse amplitude and the uncertanties involved at each step of the process.
\subsection{IAM Coupled channel formalism}
Accounting for the possible insertions of $I=0$ pairs in the intermediate states of the expansion of Eq.~(\ref{eq_un_IAMamplitude}) is not straightforward, necessitating the use of the coupled-channel formalism. Unlike the isovector-vector channel, where only $WW$ pairs are present, the isoscalar-scalar waves can include both $WW$ and $hh$ pairs in the $I=0$ projection. Therefore, it becomes essential to consider the crossed-channel process $WW\to hh$ and the elastic channel $hh\to hh$. A modification of Figure~\ref{fig_un_IAM_resum} to include the coupled-channels is shown in Figure~\ref{fig_un_IAM_resum_cc}.
\begin{figure}
\centering
\includegraphics[clip,width=13cm,height=4.3cm]{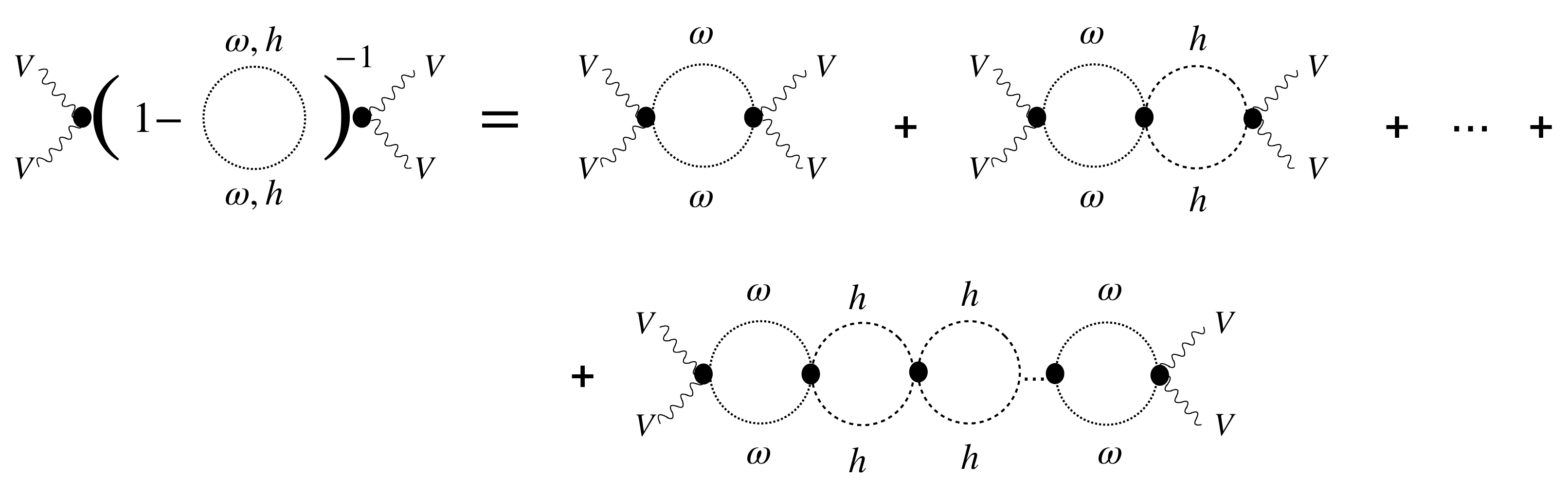}
\caption{\small{Schematic representation of the expansion of the IAM-wave in Eq.~(\ref{eq_un_IAMamplitude}), illustrating the interpretation of a scalar dynamically generated resonance. Unlike the case of the vector resonance, the infinite series of bubbles representing the intermediate states of the process can exist either pairs of $WW$ or $hh$ in the $I=0$ projection.}}
\label{fig_un_IAM_resum_cc}
\end{figure}\myspace
The extension of the IAM wave to the coupled-channel formalism is made by organising the $IJ$-waves into the matrices
\begin{equation}\label{eq_un_wavematrix}
t_{00}=\begin{pmatrix}
t_{00}^{WW} & t_{00}^{Wh}\\
t_{00}^{Wh} & t_{00}^{hh}\\
\end{pmatrix},
\end{equation}
that is the fundamental structure that will be rendered unitary.\myspace
For the case $b=a^2$ and in the high-energy limit where the mass of the Higgs can be neglected, the off-diagonal elements (what we call the crossed channel) of Eq.~(\ref{eq_ET_wh_tree_net}) vanish in the ET limit when we set $g=0$, i.e., in the nET framework. This actually leads to the decoupling limit: there is no mixing among the different scalar channels whatsoever. However, this is not true as soon as we set $g\neq 0$, even if $b=a^2$, and the full coupling matrix needs to be considered.\myspace
Assuming that an expansion in chiral orders is also admited $t_{00}\approx t_{00}^{(2)}+t_{00}^{(4)}+\cdots$, as in the case of the single channel, it can be found, for example in Ref. \cite{Oller:1998hw}, that when cutting the expansion of the scalar wave at NLO [$\mathcal{O}(p^4)$], the multichannel IAM amplitude is just the generalization of the elastic case in matrix form
\begin{equation}\label{eq_un_tIJ_cc}
t_{00}^{IAM}=t_{00}^{(2)}\cdot\left(t_{00}^{(2)}-t_{00}^{(4)}\right)^{-1}\cdot t_{00}^{(2)}.
\end{equation}
From Eq.~(\ref{eq_un_tIJ_cc}), it can be seen how the scalar resonances, if present, are located as zeros of the determinant
\begin{equation}\label{eq_un_t00_determinant}
\text{det}\left(t_{00}^{(2)}(s_R)-t^{(4)}_{00}(s_R)\right)=0,
\end{equation}
which is the matrix version of the condition for the cancellation of the denominator in single-channel IAM unitarization.\myspace
The elements of the IAM matrix are all the unitary scalar waves participating in the process up to NLO: unitary $WW$ in the first diagonal entry, $Wh$ in the off-diagonal and $hh$ in the second diagonal element.\myspace
This IAM matrix, besides keeping the analytical properties on the right cut required for partial wave analysis, has a low-energy expansion that coincides with Eq.~(\ref{eq_un_pwexpansion}) and fulfills the exact unitarity condition
\begin{equation}\label{eq: exact_unitarity}
\text{Im}\,t_{00}^{\text{IAM}}=\sigma\left(t_{00}^{\text{IAM}}\right)^{\dagger}t_{00}^{\text{IAM}}.
\end{equation}
At this point, we find an ambiguity in the crossed channel of this expression: we have two kinds of particles with different masses, the gauge bosons and the Higgs, but yet we only include a unique phase space, that we choose to be the one with the $W$ boson mass, $\sigma=\sqrt{1-\frac{4M_W^2}{s}}$. This choice, of course, will be of no relevance at the high-energy regime that we want to explore where $M_W^2\approx M_h^2\ll s$ and $\sigma \approx 1$.
\section{$K$-matrix}
The second method that warrants a detailed study in our work is the K-matrix formalism. This method is based on expressing the S-matrix elements in a way that, by definition, satisfies the relation
\begin{equation}\label{eq_un_smatrixK}
S=\frac{1+iK/2}{1-iK/2},
\end{equation}
where $K^\dagger=K$.\myspace
From the definition of the S-matrix in Eq.~(\ref{eq_un_smatrixK}), we can derive the interaction part $iT=S-1$ as
\begin{equation}\label{eq_un_tmatrixK}
T=\frac{K}{1-iK/2}.
\end{equation}
This expression for $T$ ensures that the optical theorem is satisfied, as the S-matrix in Eq.~(\ref{eq_un_smatrixK}) mantains the unitarity condition. Inverting Eq.~(\ref{eq_un_tmatrixK}) one finds the K-matrix
\begin{equation}\label{eq_un_kmatrixK}
K=\frac{T}{1+iT/2},
\end{equation}
which satisfies Eq.~(\ref{eq_un_discontinuity}).\myspace
The process of constructing the $K$-matrix, guided by the principle of maintaining full unitarity, can be translated into the language of partial waves that we are interested in. This leads us to build the corresponding "$K$-wave" as follows:
\begin{equation}\label{eq_un_Kwave}
t^{K}_{IJ}=\frac{t_{IJ}}{1+it_{IJ}},
\end{equation}
that satisfies the unitarity condition for partial waves exactly, this is Eq.~(\ref{eq_un_opticaltheorempw}).\myspace
Unlike the IAM, the $K$-matrix method does not rely on a perturbative expansion of the waves, although such expansion will be used for amplitudes inherited from the effective framework. The truncation of the series expansion of the $S$-matrix in Eq.~(\ref{eq_un_smatrixK}), generally leads to approximate non-unitarity $S$-matrix. However, truncating instead a series in $K$, the corresponding expansion of $S$ satisfies unitarity exactly at every order (see Ref.~\cite{Delgado:2015kxa}). \myspace
When using the $K$-matrix method (in the improved version we will present here) to unitarize effective amplitudes derived from the Einstein-Hilbert action within the framework of $f(R)$ theories of gravity, we work at tree level, giving:
\begin{equation}
t^{K}_{IJ}=\frac{t^{(2)}_{IJ}}{1-it^{(2)}_{IJ}}.
\end{equation}
This formulation satisfies the unitarity condition exactly:
\begin{equation}
\text{Im}\left[t^{K}_{IJ}\right]=\left|t^{K}_{IJ}\right|^2=\frac{\left|t^{(2)}_{IJ}\right|^2}{\left(1+\left(t^{(2)}_{IJ}\right)^2\right)^2}.
\end{equation}\myspace
However, despite the exact fulfillment of the unitarity condition, the analytic properties of the resulting $K$-wave make the amplitude impractical. Specifically, $t^{K}_{IJ}$ lacks a right cut, preventing an analytic continuation to the second Riemann sheet for locating physical poles, as will be discussed in the following subsections. We therefore have constructed a partial wave that is well-defined in the physical region and satisfies the unitarity condition but it is unable to recover resonances. 
\subsection{Improved $K$-matrix Single channel}
To address this, we introduce the right cut manually, completing the necessary analytical structure with the function 
\begin{equation}\label{eq_un_functiong}
g(s)=\frac{1}{\pi}\left(C+\log\frac{-s}{\mu^2}\right), 
\end{equation} 
where $C$ is an arbitrary constant and $\mu$ an arbitrary scale. There is effectively only one arbitrary quantity since the constant $C$ can always be redefined to a running function $C(\mu)$, without loss of generality, to absorb the term $-\log\mu^2$. The choice to separate the constant and the logarithmic term is made solely for dimensional reasons.\myspace
The function $g(s)$ in Eq.~(\ref{eq_un_functiong}) defines explicitely a right cut and satisfies $\text{Im}[g(s)]=-1$, coming from the complex logarithm in the physical region. We then define the improved K-matrix (IK) method by substituting $-i\to g(s)$, yielding
\begin{equation}\label{eq_un_IKmatrix}
t^{IK}_{IJ}=\frac{t^{(2)}_{IJ}}{1+g(s)t^{(2)}_{IJ}}.
\end{equation}
The IK amplitude in Eq.~(\ref{eq_un_IKmatrix}) can be continued to the whole complex plane and the second Riemann sheet to look for resonances can be achieved by going \textit{beyond} the \textit{ad hoc} RC.
\subsection{Improved $K$-matrix Coupled channel formalism}
Just as we found that the IAM could be extended to include the case of coupled channels, the K-matrix method—particularly the improved K-matrix of our interest—can also accommodate resummations among states that share quantum numbers and thus mix with one another. The result of this extension again leads to an expression identical to the single-channel case, but now promoting single-process partial waves to matrices that include all relevant processes:
\begin{equation}\label{eq_un_IK_cc}
t^{IK}_{IJ}=\left(\mathbb{I}+G\cdot t^{(2)}_{IJ}\right)^{-1}\cdot t^{(2)}_{IJ}.
\end{equation}
Here, $t^{(2)}$ must be understood as a matrix containing all possible processes like in Eq.~(\ref{eq_un_tIJ_cc}), where the off-diagonal elements trigger the mixing between elastic channels.\myspace
In the coupled-channel formalism, the matrix $G$ in Eq.~(\ref{eq_un_IK_cc}) introduces the different thresholds at which new elastic channels can open up kinematically. For example, if we applied this unitarization method to $WW$ scattering projected onto the scalar channel, the first threshold would open at an energy $\sqrt{s}=2M_W$, and the second at $\sqrt{s}=2M_h$. In practice, however, the close proximity of these masses and the negligible difference relative to the effective theory's validity limit would lead us to take the limit $M_W = M_h$. This simplification streamlines the scheme for transitions between the different Riemann sheets and provides a good approximation.\myspace
However, when we address a simple case of quantum gravity at the end of this thesis, we will find that unitarizing the scalar channel involves contributions from graviton-graviton and $\varphi\varphi$ states, where $\varphi$ is a scalar field. The graviton is a massless particle, while $\varphi$ is necessarily massive. In this scenario, it would be inconsistent to disregard the mass difference, and the matrix $G$ must explicitly account for both cuts. Specifically, for this case:
\begin{equation}\label{eq_un_Gmatrix}
G(s) = \frac{1}{\pi} \begin{pmatrix}
\log\frac{-s}{\Lambda^2} & 0 \\
0 & g(s)
\end{pmatrix}, \quad g(s) = \log\frac{m^2}{\Lambda^2} + \sigma \log\frac{\sigma+1}{\sigma-1},
\end{equation}
where $\sigma$ is the two-body phase space for two particles of mass $m$ and $\Lambda$ an UV scale.\myspace
In Eq.~(\ref{eq_un_Gmatrix}), $\log(-s/\Lambda^2)$ describes the cut due to the presence of a massless particle, similarly to the Goldstones in chiral theory, and the function $g(s)$ describes the cut in the physical region arising from a particle with mass $m$. Of course, $t^{IK}_{IJ}$ in Eq.~(\ref{eq_un_IK_cc}), with this matrix $G$, fulfills all requirements from unitarity, since $\text{Im}\,g(s)=-\pi\sigma$ in the physical region.
\section{Resonances as Poles of the Amplitudes}
As we will see in the following chapters, resonances as poles in the S-matrix can—and do—appear in the unitarized theory, unlike in the effective framework, as they cannot be described by a series expansion in powers of the momenta. However, we need one way to parametrize the pole structure so the physical features of a resonance---mass and width---can be extracted from the amplitude.\myspace
In the experiment, we detect the presence of resonances as peaks emerging over the backgrounds in specific cross-section curves. A usual parametrization for the cross section in a region of a resonance is the through the Breit-Wigner formula
\begin{equation}
\sigma \propto \left|\mathcal{M}\right|^2\propto \left|\frac{1}{p^2-(M^2-iM\Gamma)}\right|^2,
\end{equation}
describing a resonance with mass $M$ and widht $\Gamma$, where it is implicit that $M\Gamma\leq M^2$.\myspace
Because poles in the cross section of a $2\to 2$ process are located at 
\begin{equation}\label{eq_un_poleposition}
s_R=M_R^2-i M_R\Gamma,
\end{equation}
it is sometimes said that the mass of a resonance is complex. This refers to its position in the complex plane when the Mandelstam variable $s$ is promoted to be a complex quantity. Needless to say, the physical mass measured in the experiment is a real value.\myspace
For a resonance, understood as a pole in the S-matrix according to Eq.~(\ref{eq_un_poleposition}), to have physical meaning, it must be located in the lower half of the complex $s$-plane, where the imaginary part is negative. A pole in the upper half of the plane would imply a negative resonance width, violating causality. Acausal poles, known as \textit{spurious resonances} or ghost-like particles, are viewed as artifacts of either unitarization methods failing to extend the low-energy amplitude's validity into the UV, or an inadequate parameterization of the strongly interacting theory in the infrarred. In either case, we exclude these scenarios by imposing bounds on the chiral-parameter space, ensuring the absence of spurious states.\myspace
However, such extensions of the partial waves, defined for real values of $s$ above possible thresholds, to the whole complex plane are not sufficient for the searching of physical poles. Poles represent physical states-and are therefore physical-only if they are found on a different branch of the multivalued partial waves, accesible by \textit{crossing}\footnote{Formally, this right-hand cut cannot be simply crossed but instead, we have to go \textit{beyond} it with the usual techniques.} the right cut. This becomes clear when considering the reflection principle, which states
\begin{equation}\label{eq_un_reflectionprinciple}
\mathcal{A}(s^\ast)=[\mathcal{A}(s)]^\ast.
\end{equation}
If a pole is located in the lower half-plane, even if the associated width is physically meaningful, the reflection principle would create a symmetric pole in the upper half-plane, resulting in a non-physical state that cannot be part of any consistent theory. The solution is to analytically continue the unitarized amplitudes to the so-called second Riemann sheet and search for poles there. By contraposition, the unitarized wave before crossing is defined in the first Riemann sheet. The second sheet is accessible by crossing the physical cut, where the reflection principle no longer applies. This is why the K-matrix wave in Eq.~(\ref{eq_un_Kwave}) is not suitable for defining physical poles, unlike the IAM or other methods that allow extensions to the second Riemann sheet, necessitating the improved version.\myspace
Thus, with the information above, once we have constructed the partial waves and, from them, the unitarized amplitudes, we perform an analytical continuation of the multivalued functions to the second Riemann sheet. The parameters of the resonance are then given by
\begin{equation}\label{eq_un_resonanceparameters}
M=\sqrt{\text{Re}\left[s_R\right]}, \qquad \Gamma=-\frac{1}{M}\text{Im}\left[s_R\right].
\end{equation}
\subsubsection{The parameters of the resonances in practice}
The relations in Eq.~(\ref{eq_un_resonanceparameters}) fully determine the characteristics of a resonance when described as a Breit Wigner pole in the cross section. However, one of the steps outlined in the previous section becomes challenging when dealing with complicated amplitudes. In particular, in our case, we do not have access to the analytical continuation to the second Riemann sheet for the various far-from-simple functions that appear at the one-loop level. This was not the case of earlier studies such as the ones in Refs.~\cite{Espriu:2013fia,Delgado:2014dxa,Arnan:2015csa}, where the only functions requiring continuation were complex logarithms-a well known expression that will be given explicitely later.\myspace
To overcome this situation, we focus on the real part of the amplitudes, this is the projection of the unitarized partial wave potentially containing the resonance onto the $s$-real axis, where physical values lie. When a resonance is found in a unitarized amplitude, we systematically determine the width using two methods, both of which we have confirmed yield nearly identical results: (1) by fitting the amplitude profile close to the resonant region to a Breit-Wigner form, with the width as the only free parameter—the mass is completely determined from the real part; and (2) by directly measuring the width from the partial wave profile.\myspace
Regarding the second method of  deriving the width directly from the partial wave profile, one point should be stressed out. When using the profile of the unitarized amplitude itself, the usual definition of the width as the difference between the two points at half-maximum height is not directly applicable and requires a correction factor of $\sqrt{3}$. This correction is unnecessary when measuring the width from the profile of the squared amplitude (modulus squared), which is proportional to the cross section.
\begin{figure}
\centering
\includegraphics[width=0.45\textwidth]{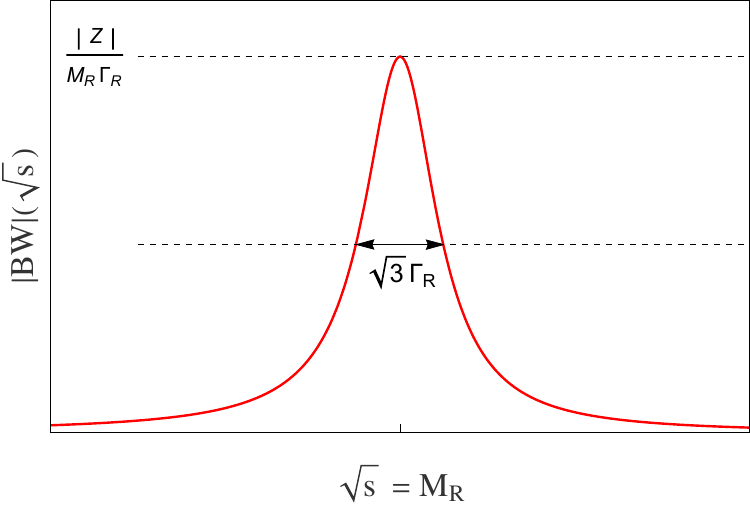} %
\includegraphics[width=0.45\textwidth]{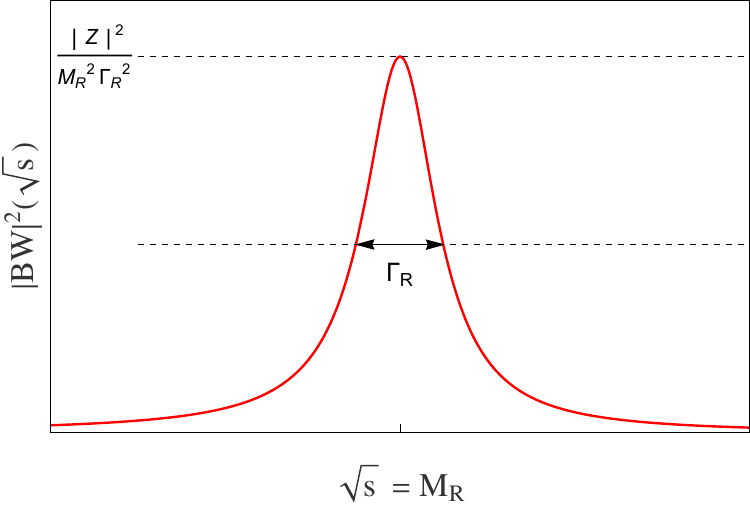} %
\caption{\small Plot of the (left) modulus of the Breit-Wigner and (right) modulus squared of the Breit-Wigner in a general parametrization as given in Eq.~(\ref{eq_un_bwparam}). It is explicitely shown that the distance (in $\sqrt{s}$ units) between the two points at half-maximum equals the resonance parameter $\Gamma$ only for the modulus squared for of the Breit-Wigner, whereas there is an additional factor of $\sqrt{3}$ for the modulus.}%
\label{fig_un_bw}%
\end{figure}\myspace
This becomes evident when comparing the explicit calculation of the distance between the two points of the profile at half-maximum height for both the modulus of the Breit-Wigner and modulus of the Breit-Wigner squared, depicted in Figure~(\ref{fig_un_bw}). The Breit-Wigner in a general parametrization is given by
\begin{equation}\label{eq_un_bwparam}
|BW|=\left| \frac{Z}{s-\left(M_R-\frac{i}{2}\Gamma_R\right)^2}\right|,
\end{equation}
being $\left|Z\right|$ the residue.\myspace
The two points $\sqrt{s}_{\pm}$ resulting from the intersection of the curve and the horizontal dashed line at half-maximum for each profile can be trivially computed from Eq.~(\ref{eq_un_bwparam}), yielding
\begin{equation}
\begin{aligned}
&\sqrt{s}^{\left|BW\right|}_{\pm}=M_R\sqrt{1\pm\frac{\sqrt{3}\Gamma_R}{M_R}-\frac{\Gamma_R^2}{4M_R^2}}\approx M_R\left(1\pm \frac{\sqrt{3}\Gamma_R}{2M_R}\right),\\
&\sqrt{s}^{\left|BW\right|^2}_{\pm}=M_R\sqrt{1\pm\frac{\Gamma_R}{M_R}-\frac{\Gamma_R^2}{4M_R^2}}\approx M_R\left(1\pm\frac{\Gamma_R}{2M_R}\right)
\end{aligned}
\end{equation}
in the limit $\Gamma_R\ll M_R$. Therefore, the distance (in energy) between these points is given by
\begin{equation}
\begin{aligned}
&d^{\left|BW\right|}\equiv\left|\sqrt{s}^{BW}_{+}-\sqrt{s}^{BW}_{-}\right|=\sqrt{3}\Gamma_R,\\
&d^{\left|BW\right|^2}\equiv\left|\sqrt{s}^{BW^2}_{+}-\sqrt{s}^{BW^2}_{-}\right|=\Gamma_R.
\end{aligned}
\end{equation}
Thus, in cases where we cannot perform the numerical evaluation over the entire complex plane and find the pole analytically, we opt to directly measure the width from the amplitude profile. This approach allows us to infer the behavior of the amplitude on the physical axis, using only the values that we have access to. However, this raises an important question: through this projection onto the physical axis, how can we be certain that the observed profile originates from the lower half of the complex plane and not the upper half, where the resonance would be non-physical?\myspace
To contour that difficulty we have three tools at our disposal: (1) comparison with the nET limit in order to determine if a pole represents a modification of a pole previously known to exist in the simplified model; (2) fitting the partial wave to Breit-Wigner resonances, leaving the sign of the width as a free variable; and (3) checking the behavior of the phase shift across the resonance. Of these three possibilities, tool 1 is not very informative especially for scalar resonances, because, as we will see, the modifications with respect the nET are large when coupled channels are involved. Tool 2 is quite useful, but tool 3 is the method of choice, particularly when combined with tool 2.\myspace
The phase of an amplitude that contains a physical resonance presents a shift from $\pi/2$ to $-\pi/2$ in the pole position. The derivative of the phase should always be positive as the expression $\Gamma\sim \left(\frac{\partial \delta(s)}{\delta \sqrt{s}}\right)^{-1}$ (see for example Ref.~\cite{Pelaez:1996wk}), where $\delta (s)$ represents the phase in Eq.~(\ref{eq_un_pwdelta}), can be derived analytically. In this way, we study the causal character of all resonances found. All cases are met: one resonance, two physical resonances, and also two resonances where only one of them happens to be physical.\myspace
\begin{figure}
\centering
\includegraphics[width=0.45\textwidth]{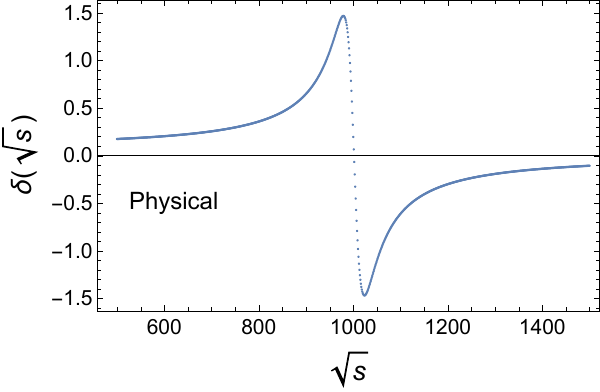} %
\includegraphics[width=0.45\textwidth]{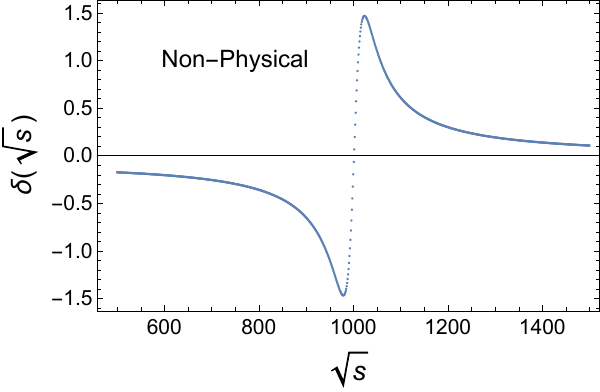} %
\caption{\small Phase shift associated with a (left) physical and (right) unphysical resonance. In the left panel, the phase shift transitions from $\pi/2$ to $-\pi/2$, with the derivative of $\delta (\sqrt{s})$ remaining positive throughout the entire energy range. In constrast, the resonance in the right panel exhibits non-physical behavior, with the phase shift occurring in the oposite direction.}%
\label{fig_un_phases}%
\end{figure}\noindent
As an example, Figure~\ref{fig_un_phases} shows the behavior of the phase shift associated with a resonance characterized by the parameters $M=1000$ and $\Gamma=40$ in some energy units. The resonance in the left panel is physical since it satisfies the condition discussed in the previous paragraph. In constrast, the resonance on the right panel exhibits an incorrect phase shift, resulting in a negative decay width and thus is unphysical.
\section{Argand plots for unitary partial waves}
In the previous sections, it has been shown that the unitarity condition has a straightforward expression for partial waves. As long as no inelasticities appear in the scattering, according to Eq.~(\ref{eq_un_relationimmodulus}), a unitary partial wave has a graphical representation as a circle with radius $1/2$, centered at $(0,1/2)$ in the $(\text{Re}\,t_{IJ}(s), \text{Im}\,t_{IJ}(s))$ plane. This representation, which displays the imaginary and real parts, is commonly referred to as an Argand plot.\myspace
With this graphical representation of partial waves, we identify different scenarios depicted in Figure~\ref{fig_un_scenarios_unit}. The first scenario is the absence of a resonance. In this case, illustrated in the left panel of Figure~\ref{fig_un_scenarios_unit}, the unitary partial wave asymptotically approaches its maximum modulus. When there is a resonance in the spectrum, the partial wave reaches this maximum value at an energy equal to the mass of the resonance, and then the modulus decreases again. This behavior is shown in the central panel of the Figure~\ref{fig_un_scenarios_unit}.
\begin{figure}
\centering
\includegraphics[width=0.30\textwidth]{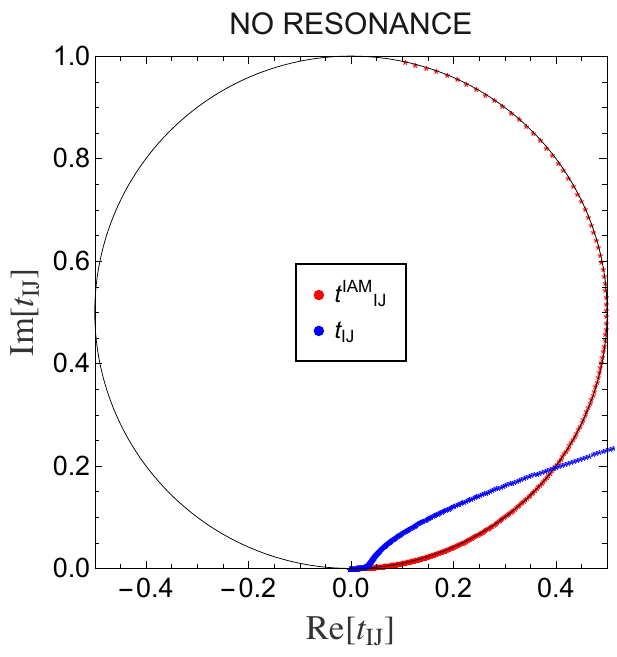} %
\includegraphics[width=0.30\textwidth]{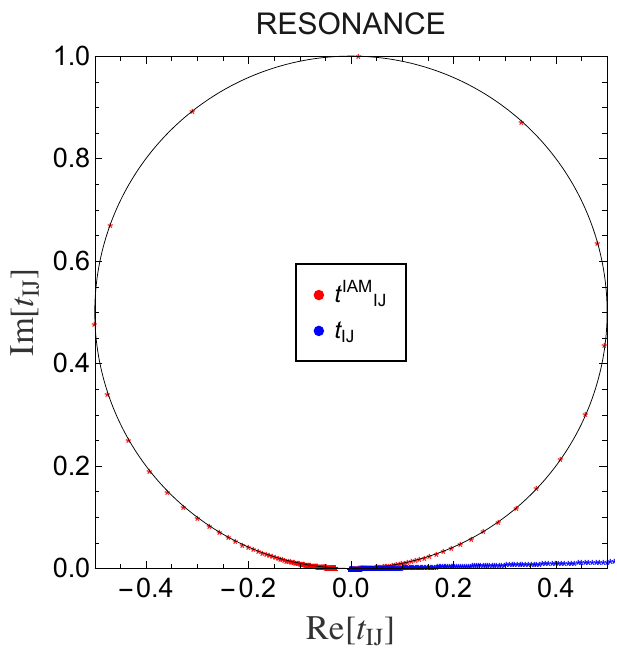}
\includegraphics[width=0.30\textwidth]{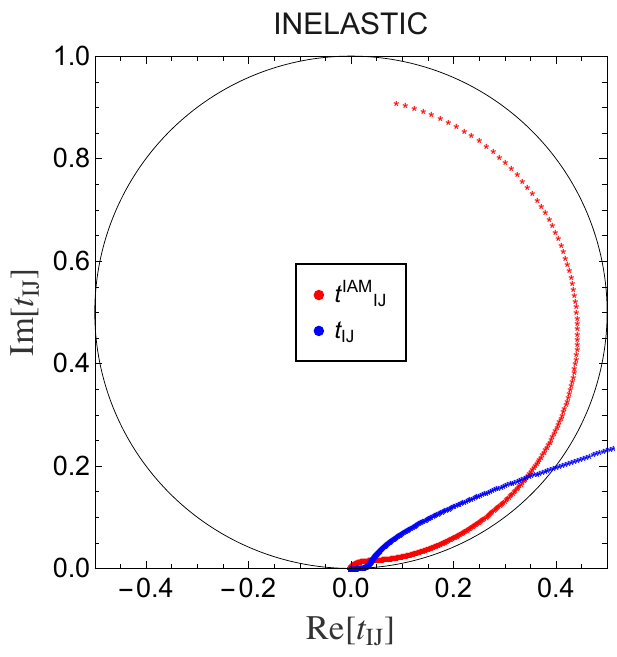} %
\caption{\small Different physical scenarios that can be observed in the Argand plots. In the left panel, a scenario with no resonance; the unitarized partial wave saturates at its maximum value. In the central panel, the case of a partial wave containing a resonance is represented, where the modulus decreases after reaching its maximum. Finally, in the right panel, the presence of an additional channel absorbing part of the amplitude.}
\label{fig_un_scenarios_unit}%
\end{figure}\noindent\myspace
As the energy increases, additional channels might open up, absorbing part of the total cross-section. In this case, the unitarity condition follows the form of Eq.~(\ref{eq_un_opticaltheorempw_inelastic}), with $\Delta>0$. This situation, represented in the right panel, results in an Argand diagram where the partial wave lies entirely within the circle but is separated from its boundary, as in the elastic case.\myspace
For VBS, the isovector-vector scenario is an example of this situation. The process projected in this channel is completely diagoanlized with the only possible process being elastic $WW\to WW$.
\section{IAM vs IK: a massless example}
To conclude this chapter, we apply the two most widely developed unitarization methods presented in the previous subsections, using them as examples in the elastic scattering process $WW \to WW$. To facilitate easy reproduction of the results for interested readers, we consider the amplitudes in the massless limit (Eq.~(\ref{eq_ET_ww_tree_net})-Eq.~(\ref{eq_ET_fandgfunction})) and consequently within the nET. Figure~\ref{fig_un_different_methods} presents the results of applying these methods to the partial waves—projected in the isovector-vector channel in the massless limit—using the IAM method (solid blue line) and the IK method (solid green line), alongside a comparison to the unitarized form (red dashed line). The chosen values are $a=0.9$, $b=a^2$, $a_4=5.5 \cdot 10^{-4}$, and $a_5=-2.5 \cdot 10^{-4}$, which we will refer to in the next chapter as $\text{BP2}^\prime$.\myspace
\begin{figure}
\centering
\includegraphics[clip,width=13cm,height=8cm]{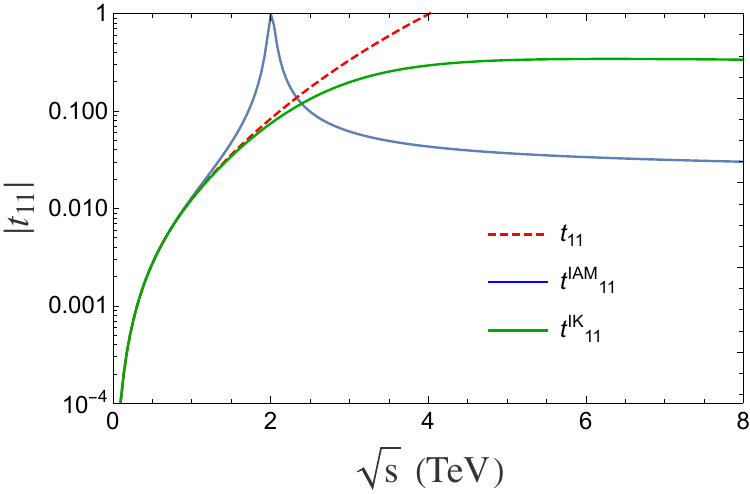}%
\caption{\small Result of applying the IAM (solid blue line) and the IK (solid green line) to the partial wave projected in the isovector-vector channel in the massless limit (dashed red line). Energy units are in TeV. Both unitarization methods produce manifestly different results even though they give unitary results in the whole energy range.}%
\label{fig_un_different_methods}%
\end{figure}\noindent
The results obtained with these two methods show a clear difference.\myspace
By analytical continuation of the only non-trivial function in the massless case, the logarithm, to the second Riemann sheet
\begin{equation}
\log^{II}(-s)=\log\left|s\right|+i\left(\arg s-\pi\right),
\end{equation}
the IAM predicts a pole at $s=4.05-i\,0.15$ TeV in the $s$-complex plane, defining an isovector-vector resonance characterized by $M=2012$ GeV and $\Gamma=73$ GeV. In Figure~\ref{fig_un_different_methods}, only the projection onto the real axis is shown and the resonant behevior is evident too. In contrast, the IK wave simply saturates at high energies, satisfying the unitarity condition but showing no indication of a resonant state whatsoever.\myspace
Does this imply that one of the two methods is inherently flawed and not useful for making predictions about the EWSBS? The answer is \textit{not always}, but in this case, the reasoning requires more than simply acknowledging that both methods are built on a series of approximations that may not be entirely reliable. \myspace
In constructing the IK method, it is customary to define the constant $C$ in the function $g(s)$ (see. Eq.~(\ref{eq_un_functiong})) as $C=\frac{B(\mu)}{D+E}$, with the NLO partial wave written as
\begin{equation}
t^{(4)}=s^2\left(B(\mu)+D\log\frac{s}{\mu^2}+E\log\frac{-s}{\mu^2}\right).
\end{equation}
This definition of $C$ is scale-independent, as all the $\mu$-dependence can be absorbed through the redefinitions of the coefficients in the bare Lagrangian that render the theory finite at one loop.\myspace
In the isovector-vector projection of elastic $\omega\omega$ scattering-recall the massless limit-(see for instance Ref.~{\cite{Delgado:2015kxa}}),
\begin{equation}
D + E \propto (b - a^2)^2,
\end{equation}
which obviously vanishes for $ b = a^2 $ and therefore prevents the definition of $C$, anc consequently $g(s)$, needed to apply the IK method.\myspace
A particular and noteworthy case is the Higgsless scenario, even though is essentially ruled out due to the discovery of the Higgs boson. This scenario is realized by setting $ a = b = 0 $, which clearly does not allow the construction of $g(s)$ for the IK method. Interestingly, in this limit, the associated electroweak chiral Lagrangian is fully equivalent to the chiral theory of QCD with two flavors when we substitute $v = 246 \, \text{GeV} \to f_\pi = 92 \, \text{MeV} $, the pion decay constant. In this low-energy QCD theory, the $ \rho $-meson should appear in the vector channel with a mass of about $ 770 \, \text{MeV} $. As noted, the IK will fail to reproduce it, at least in the parameter space where $ b = a^2 $. However, with an appropriate choice of chiral parameters $a_{4,5}^{\cancel{H}}$, we are able to place the $ \rho $-meson correctly using the IAM. In particular, with the choice $a_4^{\cancel{H}}=-2a_{5}^{\cancel{H}}=\frac{3}{192\pi^2}$-beyond $a=b=0$-and assigning zero to the other chiral parameters, the IAM predicts the $\rho$-meson mass at $0.79$ GeV, slightly above the experimental value of $0.775$ GeV. This discussion is detailed in Ref.~\cite{Delgado:2015kxa}.\myspace
Therefore, when studying the spectrum of vector resonances in $WW$ scattering, the IAM will be the preferred method. As previously mentioned, the IK unitarization method will be applied in the final chapter in the context of $f(R)$ effective theories for gravity, where a complete NLO calculation suitable for our purposes is neither available in the literature nor feasible with the tools at our disposal.

\thispagestyle{empty}

\lhead{Chapter 4}
\rhead{Vector Resonances}

\chapter{Vector Resonances}
\label{chp:vector}

With the machinery developed in previous chapters, we are now prepared to search for resonances in the parameter space defined by the bosonic chiral Lagrangian. Our approach seeks to determine how much information can be extracted about the Lagrangian parameters by examining the presence or absence of resonances in the spectrum. In particular, we aim to gather as much information as possible about the Higgs’s anomalous self-couplings—the deviations from their Standard Model values-, which would indicate new physics. As we will see throughout this chapter, studying the properties of vector resonances provides little to no information about these self-interactions carries out by non-standard $ d_3 $ and $ d_4 $. The scalar projection isthe way to go for this purpose.\myspace
Nonetheless, the study of vector resonances also yields interesting insights. For instance, the absence of coupled channels in the vector projection for $WW$ scattering simplifies the application of unitarization techniques, especially the IAM, which we will use here. The main difficulty is concentrated in the complete NLO calculation of the process, with the required precision.\myspace
In our case, we aim a level of precision that includes physical bosons in the loop’s internal lines, i.e., $g \neq 0$. This implies the first improvement with respect to previous works of the group (see the set of works in Refs.~(\cite{Espriu:2013fia, Espriu:2014jya})). It is worth noting that at the time of publication of our first work in Ref.~\cite{Asiain:2021lch}, a full set of one-loop counterterms was not yet available for all the $ 2 \to 2 $ processes participating in the isospin-spin structure of the EWSBS, specifically $ WW \to WW $, $ WW \to hh $, and $ hh \to hh $, for the case $g \neq 0$. Including transverse modes in the one-loop calculation inevitably requires considering the masses of both the electroweak bosons and the Higgs boson.\myspace
Coinciding in time, and as also noted in the first chapter, the authors of Ref.~\cite{Herrero:2021iqt} published a detailed study of the elastic process $ WZ \to WZ $ off-shell, introducing not only $ g \neq 0 $ but also $g^\prime \neq 0 $, and in an arbitrary gauge. This approach resolves the mass difference between $ M_W $ and $ M_Z $ and accordingly accounts for photons propagating in internal lines at both tree and one-loop levels. The process under study by the authors is, of course, a crossed version of our so-called fundamental amplitude, $ WW \to ZZ $, and therefore allows for cross-check at the level of counterterms, as these should be universal once a renormalization scheme has been chosen. As detailed in Chapter \ref{chp:effective_theories}, once the Landau gauge that we use is fixed and their expressions are taken to the custodial limit, the substitutions in Eq.~(\ref{eq_ET_redefinitions_belen}) with $ c_{\Box H} \to a_{\Box\Box} $ and $ c_7 \to a_{\Box VV} $, together with the equations of motion in Eq.~(\ref{eq_ET_eoms}), guarantee the agreement.\myspace
From the preceding comparison, further consistency checks also arise for simplified scenarios with \( g = 0 \) (see Refs.~\cite{Espriu:2013fia, Gavela:2014uta}) and \( \lambda_{SM} = 0 \) (see Ref.~\cite{Delgado:2014dxa}), which have been explicitly verified.\myspace
The starting point of this chapter, which contains original work of our group, is a complete set of finite amplitudes for relevant $2 \to 2$ processes in the bosonic sector of the EWSBS, up to one-loop precision. Let us recap here the list of assumptions made for our amplitudes. We take the custodial limit $ g^\prime = 0 $, meaning $M_Z=M_W$ and the absence of photons, and work in the Landau gauge ( $\xi = 0 $), where the Goldstones are massless. Furthermore, to simplify the calculation of the real part at one-loop order, we apply the equivalence theorem (see Chapter \ref{chp:effective_theories}), replacing longitudinally polarized external gauge bosons with their corresponding Goldstones. This approximation is valid as long as the process energy significantly exceeds the mass of the $W$—a reasonable assumption here, as our study probes new physics at the terascale, an order of magnitude above the electroweak scale. The imaginary part of the one-loop calculation is obtained exactly, with physical bosons as asymptotic states, through a tree-level calculation using the optical theorem.\myspace
The parameter space of the chiral Lagrangian for the complete elastic $WW$-scattering—the only one involved in the unitarization procedure in the vector projection—is relatively large. At leading order, only $ a $ and $ b $ are present; however, at next-to-leading order, additional parameters enter: $ a_4 $, $ a_5 $, $ a_3 $, and $ \zeta$ appear at the next order in the tree-level calculation, while $ a $, $ b $, and $ d_3 $ contribute to the real part of the loop. This set constitutes a seven-dimensional chiral parameter space. With this relatively high-dimensional parameter space, performing phenomenology becomes complicated. Therefore, beyond the existing experimental constraints on some of the chiral parameters (see Table~(\ref{tab_ET_chiralparams})), it would be very helpful to have other types of theoretical predictions to minimize the region to be explored. This is not an easy task, but we have, to our knowledge, two sum rules at our disposal.\myspace
In the context of HEFT, in Ref.~\cite{Urbano:2013aoa}, the following sum rule was derived
\begin{equation}\label{eq_v_sum_rule}
\frac{1-a^2}{v^2} = \frac{1}{6\pi}\int_0^\infty \frac{ds}{s}
\left(2\sigma_{I=0}(s)^{tot} + 3 \sigma_{I=1}(s)^{tot} -5
\sigma_{I=2}(s)^{tot}\right),
\end{equation}
where $\sigma_I^{tot}$ is the total cross section in the isospin channel $I$. This interesting result was derived making full use of the ET and setting $g=0$. Taking into account that unless there is an unlikely strong enhancement of the $I=2$ isospin channel, the rhs is positive definite, this would exclude values of the effective coupling $a$ greater than 1. As we will see there are no physical resonances in the isotensor-scalar channel. And, indeed, no satisfactory microscopic model has
been constructed with $a>1$ to our knowledge.\myspace
As we have seen, there are some deviations with respect to the ET predictions when using the proper longitudinal vector boson amplitudes and they affect the analytic properties of the amplitude. In Ref. \cite{Espriu:2014jya} it was seen that a complete calculation (as opposed to the simpler nET treatment) changes the previous result in several ways. For instance, it is not true that a given order in the chiral expansion corresponds to a definite power of $s$ ---a property that is used in order to derive Eq.~(\ref{eq_v_sum_rule}). Therefore,  when gauge transverse propagation is included, the order $s$ contribution will have corrections from all orders in perturbation theory. The contribution to the left-hand side of the integral obtained will then be of the form
\begin{equation}\label{eq_v_cont_sum_rule}
\frac{3 - a^2 + \mathcal{O}(g^2)}{v^2}\,.
\end{equation}
However, the right cut changes, too, when $g$ is taken to be nonzero due to $W$ propagation in the $t$ channel and this could compensate the modification on the lhs. Finally, as we have seen, crossing symmetry is not manifest in the Mandelstam variables when one moves away from the nET. This is again a necessary ingredient to derive Eq.~(\ref{eq_v_sum_rule}).\myspace
These subtleties, however, do not mean that the $a>1$ forbidden region is not present; it just means that proving this when the propagation of transverse modes is taken into account is not so easy. Indeed, Ref.~\cite{Espriu:2014jya}, it was seen that for $a>1$ the IAM led to pathologies in resonances appearing in various channels, including acausal resonances ---poles in the first Riemann sheet.\myspace
The reinforces the idea that an efficient way of setting bounds on the low-energy constants is provided by discarding those regions of parameter space in the effective theory, i.e., in the infrared, where resonances are acausal. The regions described by these effective theories do not have an ultraviolet completion.\myspace
While the phenomenology of vector resonances is well understood within the framework of Weinberg’s sum rules (see Ref.~\cite{Weinberg:1967kj}) for the SM, together with unitarization techniques, the impact of including higher orders in the calculation remains undetermined.\myspace
Our first step, therefore, is to test the relevance of the new operators not considered in previous studies. First, we examine the corrections from including transverse modes in the one-loop calculation, $ g \neq 0 $, and subsequently, the effects of new operators at chiral order 4 proportional to $ g $, accompanied by the coefficients $a_3 $ and $ \zeta $, which contribute to the NLO tree-level calculation.\myspace
From Eq. (\ref{eq_un_isos_I1}) and (\ref{eq_un_Ts_WW}), the fixed isospin amplitudes in the chiral expansion, $T_1^{(2)}$ and $T_1^{(4)}$ are obtained. $T_1^{(2)}$ using $\mathcal{A}_{tree}^{(2)}(p_1,p_2,p_3,p_4)$ and $T_1^{(4)}$ with
$\mathcal{A}_{tree}^{(4)}(p_1,p_2,p_3,p_4)+\\
\text{Re} \left[\mathcal{A}_{loop}(\omega^+\omega^-\to zz)\right](p_1,p_2,p_3,p_4)$. Using Eq.~(\ref{eq_un_pwIJ}) and Eq.~(\ref{eq_un_perturbativeunitarity}) perturbatively, we find the partial wave  for $I,J=1,1$
\begin{equation}\label{eq_ my_t11}
\begin{split}
t^{(2)}_{11}=&\frac{1}{64\pi}\int_{-1}^{1}d(\cos\theta)\cos\theta\,T^{(2)}_1(s,\cos\theta)\\
Re\left[t^{(4)}_{11}\right]=&\frac{1}{64\pi}\int_{-1}^{1}d(\cos\theta)\cos\theta\,T^{(4)}_1(s,\cos\theta)\\
Im\left[t_{11}^{(4)}\right]=&\sqrt{1-\frac{4M_W^2}{s}}|t_{11}^{(2)}|^2\\
\end{split}
\end{equation}
where the Legendre polynomial $P_1(\cos\theta)=\cos\theta$ has been used.\myspace
The vector-isovector resonances, if present, are located by searching for poles of the unitary IAM amplitude (\ref{eq_un_IAMamplitude}) i.e. looking for solutions of $t^{(2)}(s_R)-t^{(4)}(s_R)=0$ for specific complex value $s_R$.
\section{Inclusion of transverse gauge modes}
Let us first of all investigate how the proper inclusion of the transverse modes (i.e. $g\neq 0$) influence the results obtained in the extreme ET limit. Below we provide results for $g=0$ and $g=2M_W/v$. The benchmark points correspond to those used in Ref.~\cite{Delgado:2017cls}.\myspace
In Figure~\ref{fig_v_BPs}, the positions of all selected points within the $a_4$-$a_5$ plane are shown. The location of each resonance is primarily determined by these two chiral parameters, particularly by the combination $a_4 - 2a_5$, which appears when applying the IAM to amplitudes in the equivalence theorem limit, as discussed in Ref.~\cite{PhysRevD.55.4193}. Since these are resonances at the teraelectronvolt scale, we can therefore expect that their positions will not be drastically affected by the inclusion of $\mathcal{O}(g)$ corrections and the $a_4-a_5$ plane is a suitable place . This will be examined quantitatively in the following sections.
\begin{figure}
\centering
\includegraphics[clip,width=13cm,height=9cm]{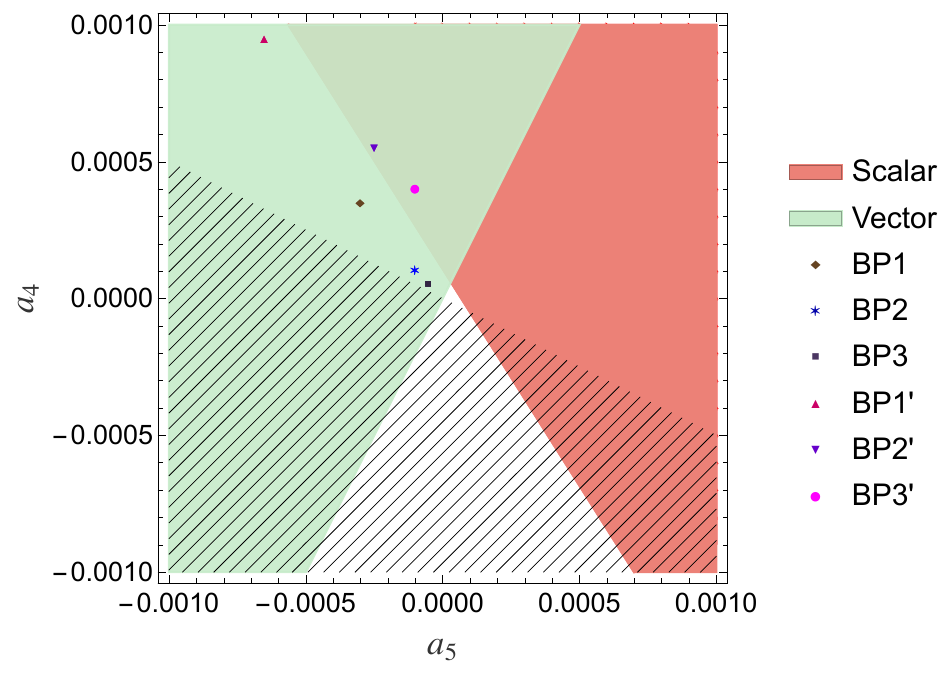} %
\caption{\small{Regions in the $a_4-a_5$ plane where scalar (red) and vector (green) resonant states appear. The striped area represents excluded paramter space by the presence of acausal isotensor resonances. The benchmark points used in this study are marked in the plot.}}
\label{fig_v_BPs}
\end{figure}\myspace
The results for these six benchmark points are gathered in Table \ref{table_v_benchmark_points}.
\begin{table}[h!]
\centering
\begin{tabular}{|c|c|c|c|c|c|}
\hline
    $\sqrt{s_V} \, (GeV)$ & $\quad g=0\quad $  &  $\quad g\neq 0\quad$   & $ \quad a \quad $   & $a_4\cdot 10^4$ & $a_5\cdot 10^4$ \\ \hline
BP1  & $\quad 1476\ih 14 \quad$ & $\quad 1503\ih 13 \quad$       & 1   &   3.5          & -3   \\ \hline
BP2  & $2039\ih 21$ & $2087\ih 20$        & 1   &   1            & -1   \\ \hline
BP3  & $2473\ih 27$ & $2540\ih 27$        & 1   &   0.5          & -0.5 \\ \hline 
BP1$^{\prime}$ & $1479\ih 42 $ & $1505\ih 44$        & 0.9 & 9.5            & -6.5  \\ \hline
BP2$^{\prime}$ & $1981\ih 97$ & $2025\ih 98$        & 0.9 & 5.5            & -2.5   \\ \hline
BP3$^{\prime}$ & $2481\ih 183$ & $2547\ih 183$        & 0.9 & 4              &  -1  \\ \hline
\end{tabular}
\caption{{\small
  Values for the location of the vector poles $\sqrt{s_V}=M_V-\frac{i}{2}\Gamma_V$ found in all the benchmark points of Ref.~\cite{Delgado:2017cls} once the transverse modes are included $(g\neq 0)$.}}\label{table_v_benchmark_points}
\end{table}\myspace
As it can be seen {\em ceteris paribus} the inclusion of the gauge boson masses systematically increases the masses of the resonances by a few per cent. The modifications in the widths are not significant. In these calculations $b=a^2$, and both $a_3$ and $\zeta$ have been set to zero.\myspace
As a check of the good unitarity behavior of the amplitudes obtained in the IAM and the validity of the approximations made we plot the partial wave for complex values of the kinematical variable $s$ in the $IJ=11$ channel. There are no threshold in this channel beyond the elastic channel and following what we learnt in the section devoted to the Argand plots in chapter \ref{chp:unitarization}, the results must lie accordingly in a circumference of radius 1/2 centered at $s= i/2$ when the imaginary versus the real parts are represented. This is shown in Figure~ \ref{fig: argand}. We also plot the results obtained for the  same $IJ=11$ channel in perturbation theory without resummation.  They obviously violate the unitarity bound. The plot correspond to the values of the benchmark point BP2$^{\prime}$ from Table \ref{table_v_benchmark_points}, corresponding to $a=0.9,\,a_4=5.5\cdot10^{-4}$ and $a_5=-2.5\cdot 10^{-4}$.
\begin{figure}
\centering
\includegraphics[clip,width=0.6\textwidth]{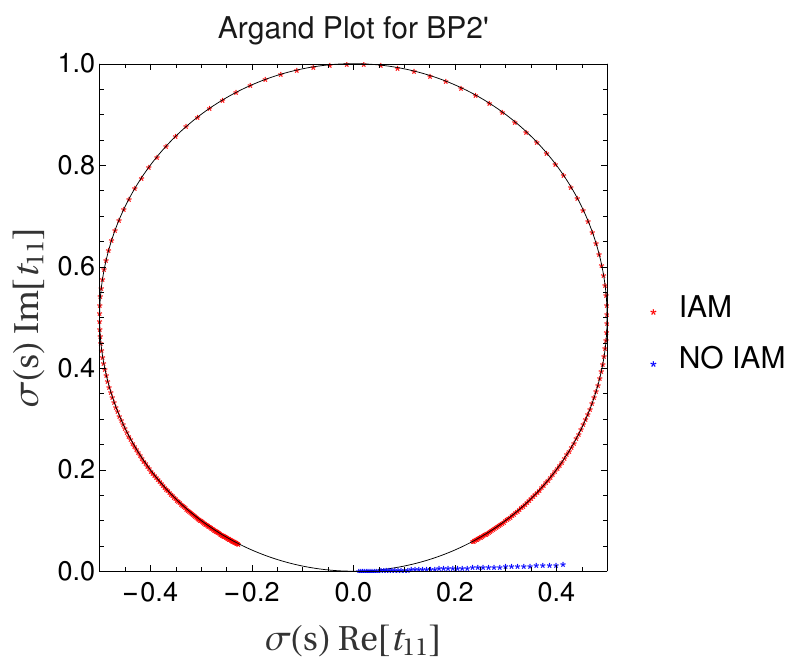}
\caption{\small{Argand plot showing the unitary VBS amplitude (red points) for the values of BP2$^{\prime}$ from Table \ref{table_v_benchmark_points}. Due to the elasticity of the process, the IAM amplitude lies exactly on the unitarity limit, i.e, the circumference of radius $1/2$ centered at $(0,1/2)$. The amplitude before applying the IAM is also present (blue points) and obviously lies entirely outside the unitarity condition.}}
\label{fig_v_argand}
\end{figure}
\section{Vector resonances at the Terascale}
Now we turn to the effect in the position of the poles with $g\neq 0$ gathered in Table.~(\ref{table_v_benchmark_points}) of the new operators that involve transverse modes. The new counterterms in the HEFT accompanying those are $a_3$ and $\zeta$. Let us see how their presence may affect the previous results.\myspace
For this analysis we focus in the four lightest resonances of Table \ref{table_v_benchmark_points}.  This is BP1, BP2, BP1$^{\prime}$ and BP2$^{\prime}$
\begin{table}[h!]
\begin{tabular}{|c|c|c|c|c|c|}
\hline
    $\sqrt{s_V} \, (GeV)$&$a_3=0$ & $a_3=0.1$  &  $a_3=-0.1$   & $ a_3=0.01 $   & $a_3=-0.01$ \\ \hline
BP1 & $\quadx 1503\ih 13 \quadx$ & $\quadx 1795\ih 11 \quadx$ & $\quadx 1215\ih 15 \quadx$        &  $ \quadx 1532\ih 13 \quadx$  &   $\quadx 1474\ih 13 \quadx$             \\ \hline
BP2 & $2087\ih 20$ & $2721\ih 15$ &   $1505\ih 23$      &   $2150\ih 19$ &  $2025\ih 21$              \\ \hline 
BP1$^{\prime}$& $1505\ih 44$& $1663\ih 46$ &  $1335\ih 43$       & $1520\ih 44$ &     $1488\ih 44$   \\ \hline
BP2$^{\prime}$&$2025\ih 98$ & $2278\ih 104$ &   $1752\ih 89$      & $2052\ih 98$ &      $1999\ih 97$  \\ \hline

\end{tabular}
\caption{{\small Values for the location of the vector poles $\sqrt{s_V}=M_V-\frac{i}{2}\Gamma_V$ found in all the benchmark 
points of Ref.~\cite{Delgado:2017cls} for different values of $a_3$ and $g\neq 0$. The chiral parameter $\zeta$ is set to zero.}}\label{table_v_benchmark_points_a3}
\end{table}
\begin{table}[h!]
\begin{tabular}{|c|c|c|c|c|c|}
\hline
    $\sqrt{s_V} \, (GeV)$ &$\zeta=0$ & $\zeta=0.1$  &  $\zeta=-0.1$   & $\zeta=0.01 $   & $\zeta=-0.01$ \\ \hline
BP1  & $\quadx 1503\ih 13 \quadx$ &   $\quadx 1637\ih 13 \quadx$      & $\quadx 1377\ih 14 \quadx$   & $\quadx 1516\ih 13 \quadx$   &  $\quadx 1489\ih 13 \quadx$          \\ \hline
BP2  & $2087\ih 20$ &     $2393\ih 18$    & $1809\ih 22$   &  $2117\ih 20$   &  $2058\ih 21$         \\ \hline 
BP1$^{\prime}$ & $1505\ih 44$ &  $1570\ih 46$       & $1439\ih 43$ & $1510\ih 45$  &  $1497\ih 45$   \\ \hline
BP2$^{\prime}$ & $2025\ih 98$ &   $2136\ih 100$       & $1915\ih 94$ & $2036\ih 98$  &  $2014\ih 97$   \\ \hline

\end{tabular}
\caption{{\small Values for location of the vector poles $\sqrt{s_V}=M_V-\frac{i}{2}\Gamma_V$ found in all the benchmark points of Ref.~\cite{Delgado:2017cls} for different values of $\zeta$ and $g\neq 0$. The chiral parameter $a_3$ is set to zero.}}\label{table_v_benchmark_points_zeta}
\end{table}
We see from the previous results that, of the two new parameters (not previously considered in unitarization analysis), $a_3$ is most relevant as it can be seen in Tables \ref{table_v_benchmark_points_a3} and \ref{table_v_benchmark_points_zeta}. Positive values of $a_3$ tend to increase the mass of the vector resonance and make it even narrower, making its detection harder. Negative values of $a_3$ work in the opposite direction. Although the bounds on $a_3$ allow it, the value $|a_3| = 0.1$ may be too large, and we also provide $M_V$ and $\Gamma_V$ for $|a_3| = 0.01$. If $a_3$ happened to be of the same order as the current bounds for $a_4$ and $a_5$, its effect would besubleading. The influence of $\zeta$ appears to be less than that of $a_3$ but the qualitative behavior remains.\myspace
This, as previously discussed in Chapter \ref{chp:effective_theories}, is not entirely surprising. By construction of the HEFT in a consistent chiral expansion, $g $, which remains fixed due to gauge invariance, is assigned a chiral order of $\chior=1$, as are derivatives. These lead to an uncontrolled growth in amplitudes as the energy increases. Consequently, operators accompanied by $a_4$ and $a_5$, which contain four derivatives, have the same chiral order in the expansion as those operators associated with $a_3$ and $\zeta$, which each contain three derivatives and a single factor of $g$.

\thispagestyle{empty}

\lhead{Chapter 5}
\rhead{Scalar Resonances}

\chapter{Scalar Resonances}
\label{chp:scalar}
Unlike the vector case discussed in the previous chapter, in the case of scalar resonances in $WW$ scattering, the elastic channel alone is not sufficient. The study of scalar resonances within the framework of unitarization techniques requires considering additional channels that contribute due to the mixing of quantum numbers. Specifically, the $WW$ states in $IJ=00$ share quantum numbers with $hh$, which contributes exclusively to the scalar partial wave. Thus, both the elastic processes $WW \to WW$ and $hh \to hh$, as well as the mixing between them $WW \to hh$, known as the crossed channel, must be taken into account along the unitarization process of scalar waves. This mixing offers an advantage: by searching for scalar resonances in VBS, we can access information about the Higgs self-couplings, which contribute at tree level in the elastic $hh$ channel and the crossed channel (see Eqs.~(\ref{eq_ET_tree_Wh}) and (\ref{eq_ET_tree_hh})), without the need to directly address di-Higgs final states in the experiment, which are highly challenging to measure.\myspace
The treatment of quantum number mixing in the unitarization process described in the previous paragraph is understood within the framework of the coupled-channel formalism, which is also detailed in the previous chapter.\myspace
With this coupled-channel formalism, more channels are available for the resummation of the intermediate and in the final states, making the resonances appearing in the scattering characteristically broader. They are short lived, compared to those found in single-channel, massive states. If these poles in the zeros of the determinant in Eq.~(\ref{eq_un_t00_determinant}) are to be interpreted as Breit-Wigner-like resonant states, we will be applying the broadly used criterion that the width satisfies  $\Gamma<\frac{M}{4}$, meaning this that the pole is located near the real axis as one can see from the definition of $s_R$ above. Otherwise, we would have found a simple enhancement of the scalar amplitude not to be interpreted as a physical pole with such an enhancement produced by the presence of a pole far from the real axis.\myspace
As in the vector case, we shall be looking for poles appearing in the second Riemann sheet where the Breit-Wigner interpretation leads to positive widths, required by causality arguments. If some pole appears in the first Riemann sheet, where imaginary parts are positive, it would be associated with a spurious resonance with negative width that cannot be present in a physical theory. Thus, we find here an empirical approach in order to discriminate {\em a priori} plausible sets of parameters in the HEFT.\myspace
As we did in the previous chapter, we begin with the finite amplitudes required to construct the partial waves in scalar projection, specifically for the processes $WW \to WW$, $WW \to hh$, and $hh \to hh$. Using these amplitudes, we construct the isoscalar $\left(I=0\right)$ amplitudes $T_0^{WW,Wh,hh}$ following Eqs.~(\ref{eq_un_isos_I1}) and (\ref{eq_un_Ts_WW}). We then project them onto $J=0$ using Eq.~(\ref{eq_un_pwIJ}). The explicit expressions for all channels are given in Eqs.~(\ref{eq_un_t00wwfull}), (\ref{eq_un_t00whfull}), and (\ref{eq_un_t00hhfull})
\begin{equation}\label{eq_s_scalarwaves}
\begin{aligned}
&t^{\{WW,Wh,hh\},(2)}_{00}=\frac{1}{64\pi}\int_{-1}^{1}d\left(\cos\theta\right)T_0^{\{WW,Wh,hh\},(2)},\\
&\text{Re}\left[t^{\{WW,Wh,hh\},(4)}\right]=\frac{1}{64\pi}\int_{-1}^{1}d\left(\cos\theta\right)T_0^{\{WW,Wh,hh\},(4)},\\
&\text{Im}\left[t^{\{WW,Wh,hh\},(4)}\right]=\left\{ \begin{array}{lcc} \sigma(s)\left|t_{00}^{WW,(2)}\right|^2+\sigma_H (s)\left|t_{00}^{Wh,(2)}\right|^2 & \text{for} & \text{WW}\vspace{0.3cm}\\t_{00}^{WW,(2)}t_{00}^{Wh,(2)\,\ast}+t_{00}^{Wh,(2)}t_{00}^{hh,(2)\,\ast} & \text{for} & \text{Wh}\vspace{0.3cm}\\
\sigma (s)\left|t_{00}^{Wh,(2)}\right|^2+\sigma_H (s)\left|t_{00}^{hh,(2)}\right|^2 & \text{for} & \text{hh} 
\end{array} \right.
\end{aligned},
\end{equation}
where $T_0^{\{WW,Wh,hh\},(4)}=\text{Re}\left[T_{0,\text{loop}}^{\{\omega\omega ,\omega h,hh\},(4)}\right]+T_{0,\text{tree}}^{\{WW,Wh,hh\},(4)}$ we have used that the Legendre polynomial $P_0\left(\cos\theta\right)=1$.\myspace
Previous works such as Refs. \cite{Delgado:2014dxa} and \cite{Arnan:2015csa} already searched for scalar resonances in $WW$ scattering following the procedure developed in the preceding section. These works relied on the ET (even at tree level) in the naive custodial limit, $g=g^{\prime}=0$, and the former assumed a completely massless scenario, so both of them could get exact analytical continuations of the partial waves to the second Riemann sheet to look for the resonant states. This is a step that we are not able to perform in our calculation, as the resulting expressions do not have an analytic treatment.
\section{Inclusion of transverse gauge modes}
The first task for our numerical analysis, following the ideas in Chapter \ref{chp:vector}, is to find modifications in the properties of the scalar resonances studied e.g., in Ref.~\cite{Arnan:2015csa} once one relaxes the $g=0$ approximation and includes gauge bosons in the external states at tree level and in internal lines of the one-loop calculation. In that study, the authors considered the relevant chiral parameter space giving scalar resonance masses in the range $1.8$ TeV$<M_S<2.2$ TeV. No coupled-channel formalism was used, and instead they assumed the decoupled-channel limit within the nET by setting $b=a^2$ for the particular case $b=a=1$. For some benchmark points in the mentioned region, we get the modifications on the location of the scalar points after allowing for transverse gauge propagation (see Table \ref{table_s_differences_arnan}).
\begin{table}[h!]
\centering
\begin{tabular}{|c|c|c|c|c|}
\hline
    $\sqrt{s_S} \, (GeV)$ &$a_4\cdot 10^4$ & $a_5\cdot 10^4$ & $g=0$ & $g\neq 0$\\ \hline
$ $  & $ \quad 1 \quad $ &     $\quad -0.2 \quad $    & $\quad 1805 \ih 130 \quad $ & $\quad 1856 \ih 125\quad$            \\ \hline
$ $  & $\quad 2 \quad$ &   $\quad -1  \quad$      & $\quad 2065 \ih 160 \quad$ & $\quad 2119 \ih 150 \quad$\\ \hline
$ $ & $\quad 3.5 \quad $ &  $\quad -2 \quad$       & $\quad 2175\ih 170 \quad$ & $\quad 2231 \ih 163 \quad$ \\ \hline
\end{tabular}
\caption{{\small Values for the location of the scalar poles $\sqrt{s_S}=M_S-\frac{i}{2}\Gamma_S$ for $g=0$ and $g\neq 0$ for some points in the $a_4-a_5$ plane and in the decoupling limit $b=a^2$ within the nET with $a=b=1$. The self-interactions of the Higgs are set to the SM values. Note that the coupling to other $I=0$ channels is ignored here for the purpose of assessing the effect of switching on the transverse modes.}}\label{table_s_differences_arnan}
\end{table}\myspace
From this table we extract similar conclusions as in the vector case. On one hand, the masses of the scalar resonances are pushed up by $2\%-3\%$ once the $SU(2)_L$ coupling is set to its SM value, the very same behavior as for $I=1$. On the other hand, we observe variations in the widths of around $4\%-6\%$, values much greater than in the vector case where the differences were almost unnoticeable. This gives us an idea of the significance of the propagation of transverse modes.\myspace
However, the above results for $M_S, \Gamma_S$ are only tentative because when one wants to make a full calculation beyond the nET and consider physical vector bosons in the external states, even in the case where $b=a^2$, there is no decoupling and one needs the coupled-channel formalism to get a proper description of the dynamics of the system in the $IJ=00$ channel. Let us now proceed to study how coupling the various relevant channels affects the results. The modifications will be
substantial in fact.\myspace
All the $\mathcal{O}(p^2)$ parameters are included in all the amplitudes of Eq.~(\ref{eq_s_scalarwaves}), but as one can see from the $\mathcal{L}_4$ Lagrangian in Eq.~(\ref{eq_ET_L4_custodial}), not every $\mathcal{O}(p^4)$ coupling affects all the channels. In particular, $WW$ depends on $a_4$, $a_5$, $a_3$, and $\zeta$; $Wh$ depends on  $\delta$, $\eta$, and $\zeta$; and the $hh$ elastic process depends only on $\gamma$. The operators accompanying these couplings could eventually dominate the corresponding amplitudes at high energies due to the presence of the four derivatives. However, not all these couplings contribute to the NLO scalar amplitude with the same strength. 
The aforementioned contribution is represented in Figure~ \ref{fig_s_relative} for values of the parameters of the expected (absolute) size of $10^{-3}$. 
\begin{figure}
\centering
\includegraphics[width=0.45\textwidth]{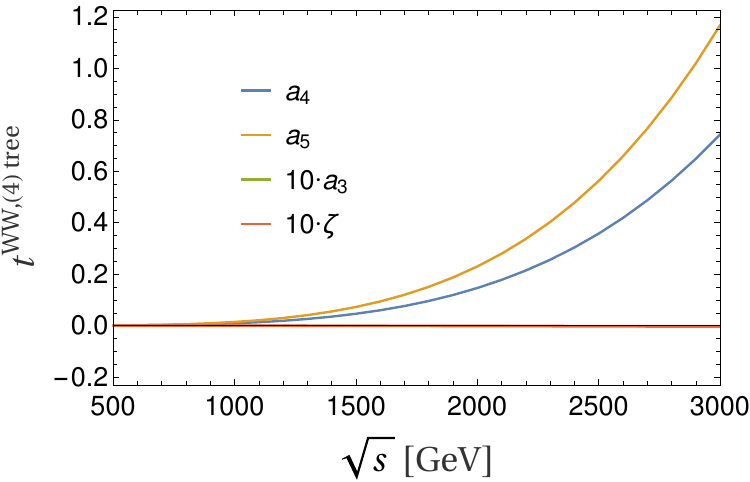} %
\includegraphics[width=0.45\textwidth]{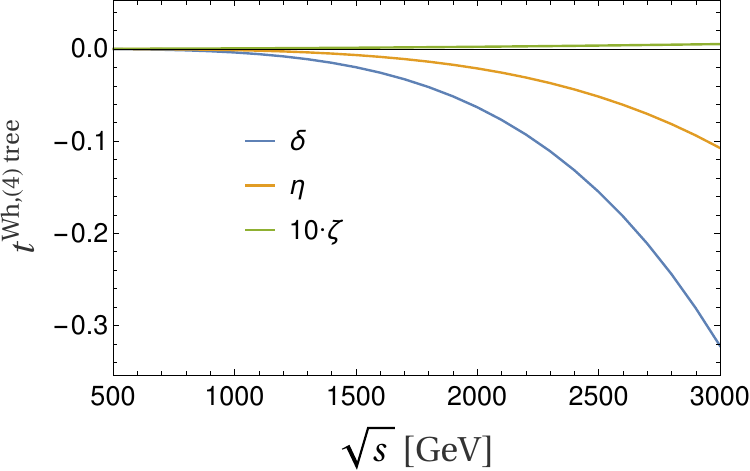} %
\caption{\small{Plot of the NLO tree-level scalar wave separated in the different chiral coupling contributions for (left axis) the elastic $WW$ and (right axis) the crossed channel $Wh$. All the values are chosen to be of the maximum expected size 
of $10^{-3}$.}}%
\label{fig_s_relative}
\end{figure}\myspace
From Figure~\ref{fig_s_relative}, we see an evident hierarchy among the different couplings: $a_4$ and $a_5$ contributions are much more relevant than those of $a_3$ and $\zeta$ in the elastic $WW$. For the crossed scattering, $Wh$, $\delta$, and $\eta$  contributions are much more important than the one of $\zeta$. This picture reinforces the conclusion that those operators surviving in the $g=0$ limit (the nET limit) are more relevant that the other ones. The reason why this happens lies in the fundamental structure of the HEFT. To be consistent in the chiral counting, both Higgs mass ($\sim \sqrt{\lambda_{SM}} v$) and EW gauge boson mass ($\sim gv$) must be understood as $\mathcal{O}(p)$ soft scales; therefore, a local operator with one gauge coupling plus three derivatives (like those accompanying $a_3$ and $\zeta$) is of chiral order $4$, just like one with four derivatives ($a_4, a_5,\cdots$), but the
latter dominates by far at high energies.\myspace
The behavior presented above agrees with what we found from vector-isovector resonances in Chapter~\ref{chp:vector} the pole position was almost completely determined by $a_4$ and $a_5$ with subleading effects after adding $a_3$ and $\zeta$, at least for values of $a,b, d_3$, and $d_4$ close to the SM values. This is why in the forthcoming analysis, in order to keep it as simple as possible, we will only consider the influence of $a_4$ and $a_5$ in determining the properties of resonances in the $IJ=11$ channel and neglect the role of $a_3$ and $\zeta$.
\section{Scalar resonances at the Terascale}
The space of parameters to analyze in the $IJ=00$ case is considerably larger than in the vector case and some sort of hierarchy is needed in order to proceed. One point to check is whether in the scalar case $a_4$ and $a_5$ dictates to a very good approximation the structure of resonances as it happens in the vector case (assuming for the time being that we stay close to the SM values $a=b=d_3=d_4=1$). To study this,  we will focus first on the benchmark points (BPs) in the $a_4-a_5$ plane defined in Table~\ref{table_s_BPs}. Other works have studied the spectrum of resonances in $WW$ scattering, in particular the group of the Ref.~\cite{Rosell:2020iub} that made use of Weinberg sum rules in Ref.~\cite{Weinberg:1967kj} derived from the $W^3B$ correlator, to set minimal bounds for the masses of vector resonances allowed by experimental constraints of the chiral parameters. For the region in the $a_4-a_5$ plane that we are interested in, they found that, for any scenario where an axial state is decoupled, the minimal mass for an experimentally allowed vector resonance is around $2$ TeV. We slightly relax that condition and require a parameter space where, if present, the vector resonances satisfy $M_V\gtrsim 1.8$ TeV. We choose the minimal mass for any observable scalar resonance to be the same value of $M_S\gtrsim 1.8$ TeV and assume that any lighter state should have already been seen in the experiment.\myspace
At this point, one should recall that only particular combinations of $a_4$ and $a_5$ appear in the various channels, namely, $5a_4+8a_5$ for $IJ=00$, $a_4-2a_5$ for $IJ=11$, and $2a_4+a_5$ for $IJ=20$, see Ref.~\cite{PhysRevD.55.4193}. In previous studies, it was found that isotensor resonances are always acausal and the corresponding region $2a_4+a_5<0$ is to be excluded from our considerations. Thus, we select BPs outside the region excluded by isotensor acausal resonances (see Ref.~\cite{Espriu:2014jya}) and within the vector-isovector and scalar-isoscalar space. In particular, we select one BP (BP1) that belongs to the region where both isovector and isoscalar resonances appear, satisfying the condition commented above regarding the vector resonance mass. The other two BPs (BP2 and BP3) lie in the purely scalar-isoscalar region. In Figure~\ref{fig_s_map}, we see the position of the BPs that we have just described within the $a_4-a_5$ plane where resonances are present when the coupled-channel formalism is applied.
\begin{figure}
\centering
\includegraphics[clip,width=13cm,height=9cm]{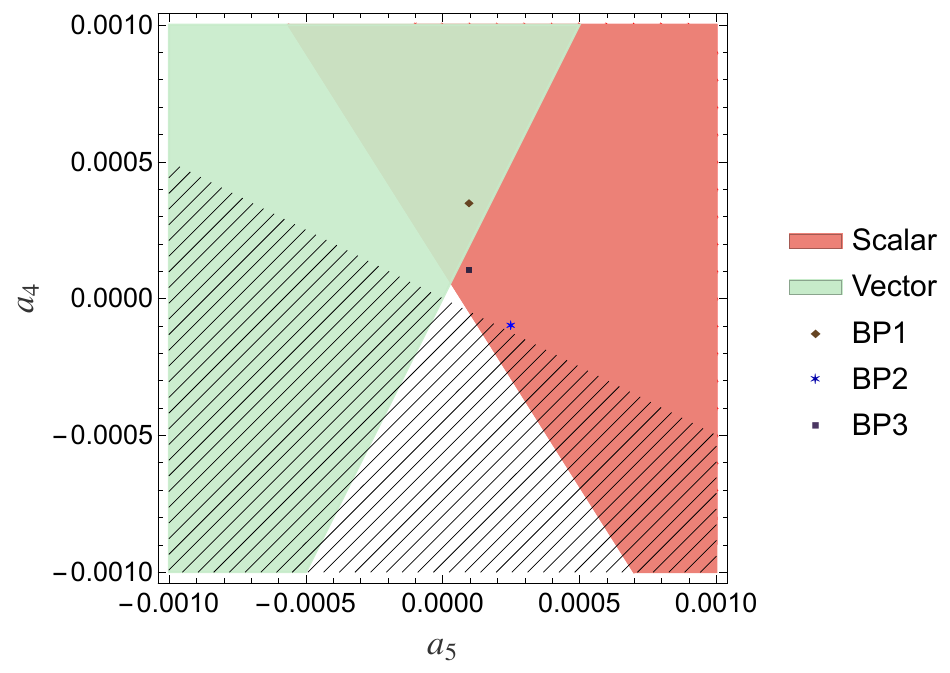} %
\caption{\small{Regions in the $a_4-a_5$ plane where scalar (red) and vector (green) resonant states appear. The striped area represents excluded paramter space by the presence of acausal isotensor resonances. The benchmark points used in this study are marked in the plot. One of them, BP1, lies in the region where both isoscalar and isovector states show up and the other two, BP2 and BP3, in the purely isoscalar sector.}}
\label{fig_s_map}
\end{figure}\myspace
These BPs are gathered in Table \ref{table_s_BPs} where we also include, even if they do not have a physical relevance, the values after applying the single channel formalism to the $WW$ scalar wave, obviating the crossed channel and the elastic $hh$ scattering. Both values are obtained with $g\neq 0$.
\begin{table}[h!]
\centering
\begin{tabular}{|c|c|c|c|c|c|}
\hline
$  $ & $ a_4\cdot 10^4 $ & $a_5\cdot 10^4$ &  $S.C. $ & $C.C. $ & $M_V\ih \Gamma_V$\\ \hline
$ BP1$ & $ 3.5  $ & $ 1 $   &  $ 1044\ih 50  $ & $ {\bf 1844\ih 487} $ & $2540\ih 27 $ \\ \hline
$ BP2$ & $ -1 $  & $ 2.5 $   &  $ 1219\ih 75 $ & ${\bf 2156\ih 637} $ & $\redcross$ \\ \hline
$ BP3$ & $ 1 $  & $1 $   &  $1269\ih 75 $ & ${\bf 2244\ih 675} $ & $\redcross$ \\ \hline
\end{tabular}
\caption{{\small Properties of the scalar resonances for the selected benchmark points in the $a_4-a_5$ plane, with the $\mathcal{O}(p^2)$ parameters set to their standard values, in both single-channel (S.C.) and coupled-channel (C.C.) formalism. We also include the values of the properties of vector resonances if present. The centerdots in red $\redcross$ indicates the absence of a zero in the determinant, Eq. (\ref{eq_un_t00_determinant}). The $\mathcal{O}(p^2)$ chiral parameters are set to their SM values. We see that coupling channels modifies very substantially masses and widths. Those poles not fulfilling the resonance condition are in boldface.}}\label{table_s_BPs}
\end{table}\myspace
The first thing that one notices is that when (correctly) considering coupled channels the results differ considerably from the ones one would obtain in the single channel and the resonance masses and widths visibly increase. Recall that here we are assuming $a=b=1$ where naively one would expect to have decoupling (this is the case in the nET). This is not so because we are setting $g\neq 0$. In fact some of the would-be resonances even dissapear as such by just becoming broad enhancements. Recall that conventionally a physical resonance must satisfy $\Gamma<M/4$ and this is not the case in many cases when applying the coupled-channel formalism. Obviously, coupled channels matter.\myspace
Finally, let us mention that in $pp$ collisions the scattering of vector bosons (VBS) is a subdominant process, but the production of $hh$ pairs via VBS is further suppressed with respect to the elastic channel $WW \to WW$. A relevant issue that will be studied below is the intensity of the coupling of the dynamical resonances to $hh$ final states. As we will see, they would be more visible in the elastic $WW\to WW$ channel and tend to couple weakly to final $hh$ pairs. To what extent this depends on the various couplings is an interesting question, too.\myspace
Now, we explicitly examine the high-energy behavior of the unitarized scalar wave within the coupled-channel extension of the IAM. In the coupled case, two elastic processes---$WW$ and $hh$---must satisfy full unitarity, which is visually represented in the Argand plot, as discussed in Chapter \ref{chp:unitarization}. However, this graphical condition does not apply to the crossed channel, $Wh$, which must still satisfy unitarity but does not require its points to lie within the circumference when the Argand plot is represented. We select benchmark point BP3 with $\gamma = 10^{-3}$, keeping the other chiral parameters at their standard values, as an example; however, this behavior holds for any other benchmark point as well.
\begin{figure}
\centering
\includegraphics[width=0.45\textwidth]{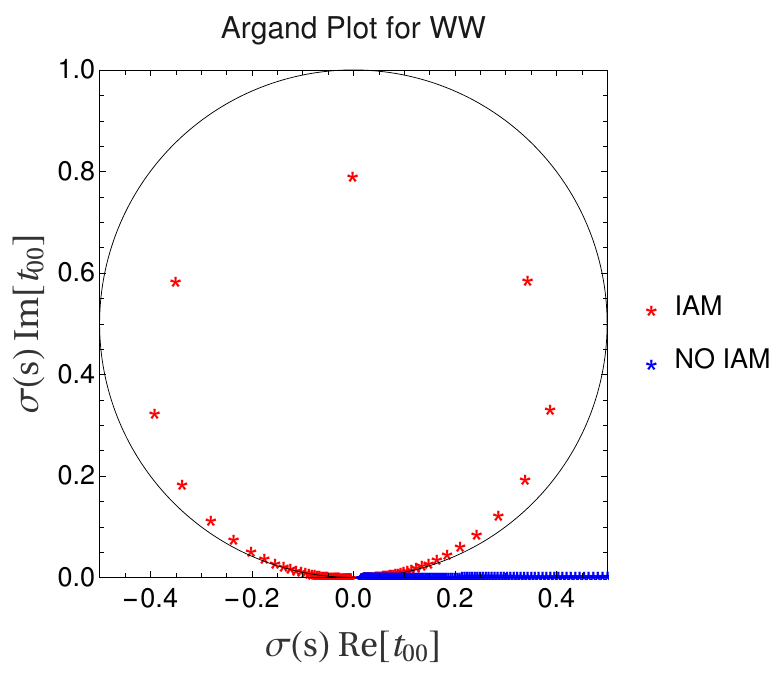} %
\includegraphics[width=0.45\textwidth]{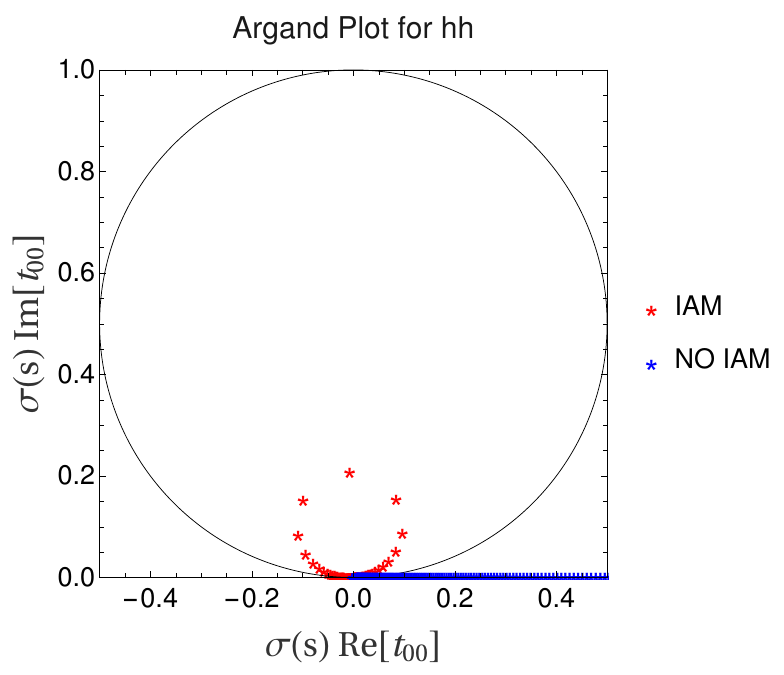} %
\caption{\small{Argand plot for the scalar wave of the (Left) elastic $WW$ and (Right) elastic $hh$ scattering for BP3 and the chiral coupling $\gamma=10^{-3}$. The rest of the parameters are set to their SM values. In red dots the unitarized amplitude satisfying the unitarity condition and in blue dots the non-unitary chiral amplitude from the Lagrangians in Eq.~(\ref{eq_ET_L2}) and Eq.~(\ref{eq_ET_L4_custodial}). The $Wh$ crossed channel alone needs not to satisfy this condition of lying on or inside the circumference.}}
\label{fig_s_argand_WW_and_HH}
\end{figure}\myspace
As we can see in Figure~ \ref{fig_s_argand_WW_and_HH}, no matter the point (every one of them corresponding to different energies), they all fall within the unitary circle. The fact that they do not lie exactly over the circumference is because, for the selection of parameters chosen for illustration, there is a big component of inelasticity in the process, i.e., the crossed channel cannot be neglected.\myspace
As previously  mentioned, bounds on the parameters of the HEFT from the study of unitarity and resonances can come in two ways. One is simply by experimentally falsifying a given set of parameters because they should give rise to resonances that are not seen in experiment. The other is giving rise to unphysical acausal resonances that lay on the wrong Riemann sheet.\myspace
To exploit all the potential of the analysis, we will group the parameters of the HEFT into two different sets. One of them contains all parameters that enter the $\mathcal{ O}(p^2)$ Lagrangian; namely $a,b,d_3$, and $d_4$. In the preliminary analysis previously presented, these values were all set to their SM values $a=b=d_3=d_4=1$.  The other set contains all the ${\cal O}(p^4)$ parameters: $a_4, a_5, \gamma, \eta$, and $\delta$. We shall assume that none of the parameters in the second group exceeds in absolute value $10^{-3}$. We will not include the chiral parameters $a_3$ and $\zeta$ as dsicussed in the previous chapter (based on our work in Ref.~\cite{Asiain:2021lch}), since they play only a marginal role in determining vector resonances.\myspace
In what follows, we will first study the influence of the relevant ${\cal O}(p^4)$ parameters while keeping the first set to their SM values. Later, we will repeat the analysis for values of $a,b$ that slightly differ from the SM, but still keeping $d_3=d_4=1$. Finally, we will study the influence of $d_3$ and $d_4$, but keeping the SM values $a=b=1$ to test the sensitivity of scalar resonances to the parameters in the Higgs potential.\myspace
For a given set of $\mathcal{O}(p^2)$ parameters ($a,b,d_3$, and $d_4$), once one fixes $a_4$ and $a_5$, the pole position in the elastic $WW$ channel is pretty much determined (up to small $a_3$ and $\zeta$ corrections that we neglect). To the extend that the elastic channel may dominate resonance production, we can treat the effect of the rest of parameters that participate in the mixing among the scalar channels as a perturbation. We have, for instance, searched for resonances in the case $a_4=a_5=0$ while varying the remaining ${\cal O}(p^4)$ terms with a negative result. The presence of resonances (both in the vector and scalar channels) is largely triggered by nonzero values of the chiral couplings $a_4$ and $a_5$.\myspace
However, not every set of low-energy parameters may correspond  to an effective  description of a strongly interacting theory. Therefore, we have to be able to discriminate which of the zeroes of Eq.~(\ref{eq_un_t00_determinant}) is a physical and which is not and also which resonances should have also been observed.\myspace
On one hand, we will be looking for resonant states that satisfy the condition $\Gamma < M/4$. If this is not fulfilled, we will be talking of an enhancement of the unitarized amplitude but never to be interpreted as a resonant state. In that case, even if Eq.~(\ref{eq_un_t00_determinant}) has a zero, the parameters $M$ and $\Gamma$ are not directly related the properties of a Breit-Wigner resonance. On the other hand, there are zeros that even satisfying the aforementioned condition cannot be taken as physical states since they have negative Breit-Wigner widths. These spurious states cannot be present in any physical theory. Analytically speaking, these zeros are found in the first Riemann sheet, above the physical cut in the complex $s$ plane.\myspace
Let us now proceed with the study in the case $a=b=d_3=d_4=1$.\myspace
To study the impact of the new parameters, we focus on the three BP points defined by specific values of $a_4$ and $a_5$ previously used. The new parameters are: $\gamma$, that enters in elastic $hh$, and $\delta$ and $\eta$ that carry out the mixing between the two elastic processes as one can see in Eq.~(\ref{eq_un_wavematrix}).\myspace
The effect of each $\mathcal{O}(p^4)$ anomalous coupling is reflected in the Tables \ref{table_s_differences_coupled_gamma}-\ref{table_s_differences_coupled_eta} below where we study the separate influence of each one for the above benchmark points. In the following analysis we keep the SM values for $a,b,d_3$, and $d_4$.
\begin{table}[h!]
\centering
\resizebox{16cm}{!}{
\begin{tabular}{|c|c|c|c|c|c|c|}
\hline
    $M_S\ih \Gamma_S$ & $\gamma=0$  & $\gamma=0.5\cdot 10^{-4}$ & $\gamma=1\cdot 10^{-4} $ & $\gamma=-0.5\cdot 10^{-4}$ & $\gamma=-1\cdot 10^{-4}$ & $\gamma=1\cdot 10^{-2} $ \\ \hline
 $ BP1 $ & \parbox[c][1.2cm][c]{2.2cm}{$\bf 1844\ih 487 $}    & $1668\ih 212 $   & $1594\ih 162 $ & $ \redcross $ &  $ \redcross$ & $1119\ih 50 $ \\ \hline
 $ BP2 $ &  \parbox[c][1.2cm][c]{2.2cm}{$\bf 2156\ih 637  $}    & $ 1881\ih 212  $  & $ 1781\ih 162 $ & $\redcross $ & $\redcross $  & $1269\ih 62 $   \\ \hline
 $ BP3 $ &  \parbox[c][1.2cm][c]{2.2cm}{$ \bf 2244\ih 675 $}    & $1931\ih 200   $  & $1831\ih 162  $ & $\redcross $ & $\redcross $ & $1319\ih 75 $   \\ \hline
\end{tabular}
}
\caption{{\small Pole position for the benchmark points in Table \ref{table_s_BPs} varying the $\mathcal{O}(p^4)$ parameter $\gamma$. The rest of the parameters are set to their SM values. Values in boldface indicate broad resonances that do not satisfy $\Gamma<M/4$.}}\label{table_s_differences_coupled_gamma}
\end{table}\myspace
From the previous table, we can see that the appearance of a nonzero $\gamma$  makes the profile of the zero narrower in such a way that we can even talk of a Breit-Wigner resonance with $\Gamma<M/4$. The values in boldface for $\gamma=0$ do not satisfy this condition. In the case of an extreme value of $\gamma$, a value we would not expect for naturalness reasons, we recover the single-channel approximation and, the coupled channel formalism is not necessary anymore. This is shown in the last column for a value of $\gamma=10^{-2}$, where very narrow resonances appear.
\begin{table}[h!]
\centering
\resizebox{16cm}{!}{
\begin{tabular}{|c|c|c|c|c|c|}
\hline
    $M_S\ih \Gamma_S$ & $\delta=0$  & $\delta=0.5\cdot 10^{-4}$ & $\delta=1\cdot 10^{-4} $ & $\delta=-0.5\cdot 10^{-4}$ & $\delta=-1\cdot 10^{-4}$  \\ \hline
 $ BP1 $ &  \parbox[c][1.2cm][c]{2.2cm}{$\bf 1844\ih 487 $}    & $ 1744\ih 362$   & $1669\ih 300   $   & $\bf 1994\ih 1100  $ & $ \doublepole $    \\ \hline
 $ BP2 $ &  \parbox[c][1.2cm][c]{2.2cm}{$\bf 2156\ih 637 $}    & $1981\ih 387   $  & $1869\ih 300  $   & $\bf 2644\ih \Gamma  $ & $\redcross $    \\ \hline
 $ BP3 $ &  \parbox[c][1.2cm][c]{2.2cm}{$\bf 2244\ih 675$}    & $2031\ih 400 $  & $1906\ih 287   $   & $\redcross $ & $\redcross $ \\ \hline
\end{tabular}
}
\caption{{\small Pole position for the benchmark points in Table \ref{table_s_BPs} varying the $\mathcal{O}(p^4)$ parameter $\delta$. The rest of the parameters are set to their SM values. Values in boldface indicate broad resonances that do not satisfy $\Gamma<M/4$.}}\label{table_s_differences_coupled_delta}
\end{table}
\begin{table}[h!]
\centering
\centering
\resizebox{16cm}{!}{
\begin{tabular}{|c|c|c|c|c|c|}
\hline
    $M_S\ih \Gamma_S$ & $\eta=0$  & $\eta=0.5\cdot 10^{-4}$ & $\eta=1\cdot 10^{-4} $ & $\eta=-0.5\cdot 10^{-4}$ & $\eta=-1\cdot 10^{-4}$  \\ \hline
 $ BP1 $ &   \parbox[c][1.2cm][c]{2.2cm}{$\bf 1844\ih 487$}    & $1806\ih 437  $  & $1769\ih 387 $ & $\bf 1881\ih 575$ & $\bf 1931\ih 712$     \\ \hline
 $ BP2 $ &   \parbox[c][1.2cm][c]{2.2cm}{$\bf 2156\ih 637$}    & $2094\ih 512  $  & $2031\ih 437 $ & $\bf 2256\ih 887 $ & $\bf 2394\ih \Gamma$      \\ \hline
 $ BP3 $ &   \parbox[c][1.2cm][c]{2.2cm}{$\bf 2244\ih 675 $}    & $2156\ih 537  $  & $2094\ih 450 $ & $\bf 2356\ih 925 $ & $\bf 2544\ih \Gamma $      \\ \hline
\end{tabular}
}
\caption{{\small Pole position for the benchmark points in Table \ref{table_s_BPs} varying the $\mathcal{O}(p^4)$ parameter $\eta$. The rest of the parameters are set to their SM values. Values in boldface indicate broad resonances that do not satisfy $\Gamma<M/4$.}}\label{table_s_differences_coupled_eta}
\end{table}\myspace
The symbol $\redcross$ represents the absence of a zero in the determinant of the IAM matrix. We have also introduced the symbol $\doublepole$ to indicate the situation where there are two poles in the unitarized amplitude but one is unphysical following the phase-shift criteria; analytically, it corresponds to a pole in the first Riemann sheet, which leads to a violation of causality with a negative width. Also, whenever our code is not able to calculate the width over the profile of the "resonance" because it is too wide and the half maximum surpasses the HEFT validity, we include the symbol ${\bf\Gamma}$, knowing that such a BP can never represent a physical resonance.\myspace
From the Tables \ref{table_s_differences_coupled_gamma}-\ref{table_s_differences_coupled_eta} above we can see a really different scenario from the one in the vector-isovector case. The location of the pole changes $15\%-20\%$ when we use reasonable values of $\gamma$ and $\delta$ ($\sim 10^{-4}$) and softer variations of around $4\%-8\%$ for values of $\eta$ of the same order. The lesson thus is clear: we cannot give a good description of the resonant scalar states from $WW$ scattering without paying attention to the coupled channels. In the first table, we have also included a big $\gamma$ value ($\sim 10^{-2}$) to make evident that the pole position in that case is very similar to that obtained using the single-channel formalism neglecting the extra $I=0$ intermediate states. In fact, for very non-natural values of $\gamma$ ($\sim 1$), the single-channel resonance is reproduced exactly.\myspace
The importance of the mixing parameters in determining the properties of the scalar resonances is now evident.\myspace
In the Tables \ref{table_s_differences_coupled_gamma}-\ref{table_s_differences_coupled_eta}, we have studied the effect in the resonance properties of the different couplings separately. However, this may not be the general case since they are all independent and they are not strongly constrained (or even constrained at all) by the experiment, especially the ones belonging to the Higgs sector, so they could all differ from zero. Hence, it is not the individual effects but the simultaneous contribution of them all that we are interested in. In Figure~ \ref{fig_s_swept_gamma}, we show for the BPs in Table \ref{table_s_BPs} the space parameter in the $\delta - \eta$ plane where physical resonances with scalar masses heavier that $1.8$ TeV are allowed for different values of $\gamma$.
\begin{figure}[t]
\centering
\includegraphics[clip,width=7.0cm,height=6.5cm]{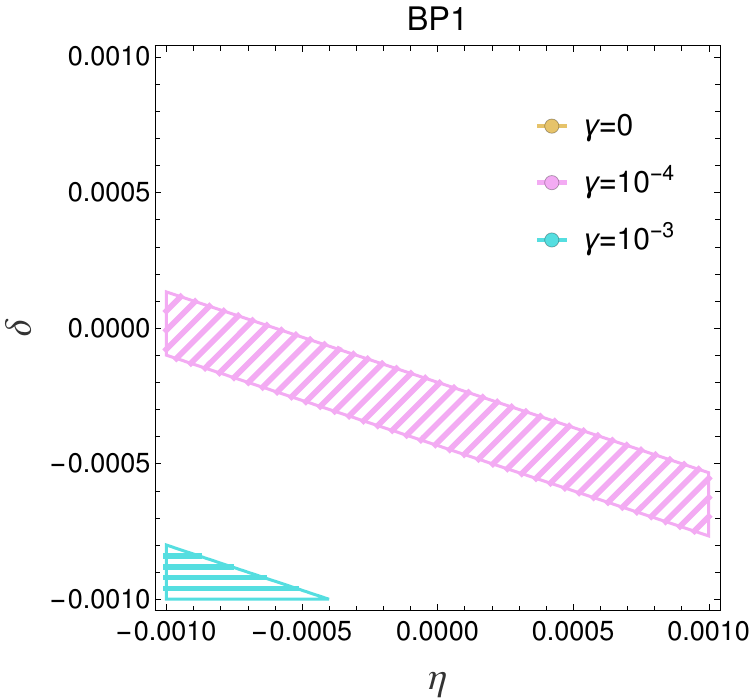} %
\includegraphics[clip,width=7.0cm,height=6.5cm]{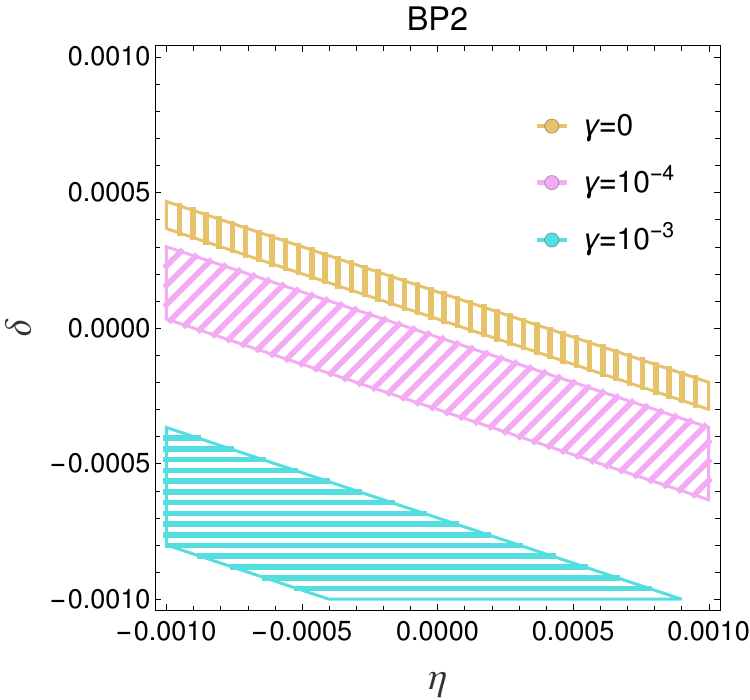} %
\includegraphics[clip,width=7.0cm,height=6.5cm]{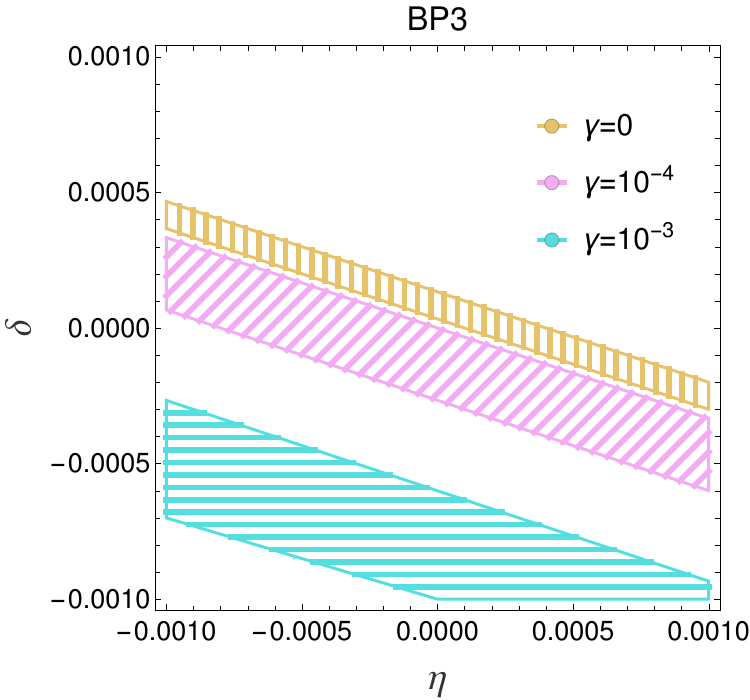} %
\caption{\small{Regions in the $\delta-\eta$ plane where physical resonances satisfying $M_S>1.8$ TeV and for the benchmark points in Table \ref{table_s_BPs} appear for different values of $\gamma$: $\gamma=0$ (golden vertical lines), $\gamma=10^{-4}$ (pink tilted lines) and $\gamma=10^{-3}$ (blue horizontal lines). For all the values of $\gamma$, the region above the bands are excluded by the presence of a non-physical pole. Below the bands we find a non-resonant scenario.}}%
\label{fig_s_swept_gamma}
\end{figure}\myspace
No matter the value of $\gamma$ or the benchmark point selected, the presence of an unphysical pole appearing in the first Riemann sheet leads us to exclude the parameter
space above the bands. This whole range of parameters cannot describe any physical
extension of the SM. We also find that the greater the value of $\gamma$ is, the more
restriction we find (there are more excluded space above the band), especially for BP1. \myspace
Below the bands, we find a nonresonant scenario: we do not find any zero in the determinant of the unitarized amplitude. \myspace
Because the above benchmark points correspond to relatively large masses, the amplitudes are to a large extent dominated by the NLO [i.e. $\mathcal{O}(p^4)$] contributions. Those appearing in the $Wh$ mixed channel vanish when $\delta=\eta=0$, so the decoupling limit results should be retrieved then.\myspace
The question of whether these resonances could be visible in the experiment requires a much more detailed study with Monte Carlo techniques that is beyond the scope of this first study of scalar resonances. However, from the parton level processes studied here, and by looking at the relative size of the residues of the corresponding poles in every channel, we can say whether it is more likely to be a bound system of two $W'$s or a $hh$ composite state. Once the pole structure is factorized from the unitarized amplitude, we are left with function which is a mixture of the other dynamical variables of the system; momentum structures and couplings of the Lagrangian.\myspace
As an example, we show in Figure~\ref{fig_s_relative_residues} the amplitude of two unitarized amplitudes that show a broad (left panel) and a narrow (right panel) resonances. In both cases they correspond to zeros of the determinant of the IAM amplitude. We observe that the bigger the $\gamma$ parameter, the stronger the coupling to a $hh$ final state is, although the $WW$ channel is strongly favored always. In any case, even if the dynamical resonances have a strong admixture of Higgs, they will be easier to spot in the $WW$ elastic channel. This is a very clear prediction.
\begin{figure}[tb]
\centering
\includegraphics[width=0.45\textwidth]{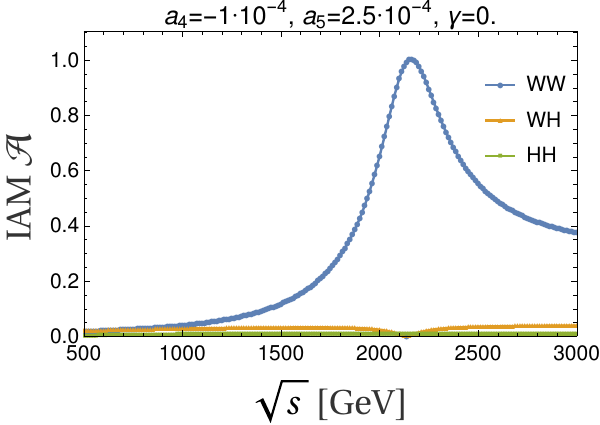} %
\includegraphics[width=0.45\textwidth]{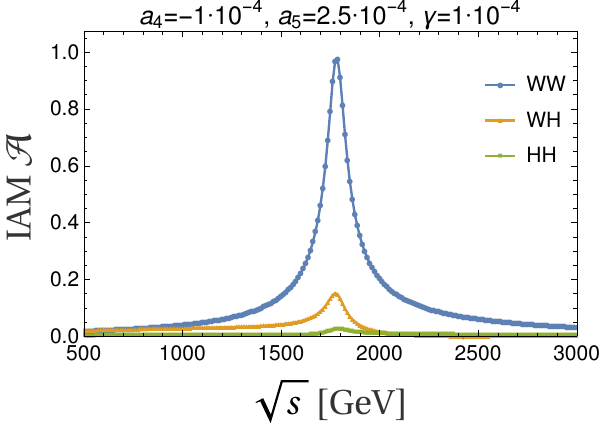} %
\caption{\small{Profile of the unitarized amplitude showing a zero in the determinant for the chiral couplings specified in the title and with the rest of the parameters set to the corresponding SM values.}}%
\label{fig_s_relative_residues}
\end{figure}\myspace
Let us consider the case where some of the $\mathcal{O}(p^2)$ parameters differ from the SM limit. We shall still keep $d_3=d_4=1$, but let us take $a=0.95$ and $b=0.805$. These values correspond to a minimal composite Higgs model living in the subgroup $H=SO(5)/SO(4)$, which presents a symmetry in the Higgs function $\mathcal{F}(h)$ with the relation $b=2a^2-1$. The interested reader may find in Ref.~\cite{Dobado:2019fxe} a complete review for the different realizations of the HEFT, including this minimal extension.\myspace
With these values of $a$ and $b$, and given that the $\mathcal{O}(p^4)$ terms are expected to have a maximum size of $\sim 10^{-3}$, we have not found any resonant state meeting all the criteria of this study, although some scenarios exhibit broad resonance profiles. All physical resonances for this choice of $a$ and $b$ occur when $a_4$ and $a_5$ are of order $10^{-2}$.\myspace
Not much can be concluded in this case.\myspace
Let us now to invesitgate the effects of the Higgs potential. For this analysis we will take the $\mathcal {O}(p^2)$ couplings $a,b$ to be equal to their SM value $a=b=1$, and explore how the resonance scene depends on the triple and quartic Higgs couplings.
\subsection{Triple Higgs self coupling: $d_3$}
The issue of determining the triple Higgs self-coupling is of utmost importance because it would help us to explore the properties of the Higgs potential, crucial to understanding the nature of the Higgs boson itself. However, such a measurement is quite involved at the LCH because it relies on the ability of the experiment to find a double Higgs final state (through its decay products) coming from the fusion of two radiated (off-shell) electroweak gauge bosons or, alternatively, from top pairs. Up to now, not enough statistics have been collected from the experiment, which translates into a very wide range in the experimental bound for this coupling: $-0.4 < d_3 < 6.3$. The upper limit of this interval would make the interaction of $\mathcal{O}(1)$ since the BSM self-interaction is described by $\lambda_3=d_3\lambda_{SM}$ with $\lambda_{SM}\sim 0.13$.\myspace
The fact that this coupling $d_3$ enters now at tree level in the calculation of the $I=0$ processes $Wh$ and $hh$ makes the resonant scalar states in the spectrum of $WW$ scattering more sensitive to it and, hence, a good approach to the problem of investigating the Higgs potential.\myspace
We start by analyzing the effect of this coupling separately, when the rest of the chiral parameters are set to their SM values, and for the benchmark points in Table \ref{table_s_BPs}. The results are gathered in Table \ref{table_s_differences_coupled_d3}.
\begin{table}[h!]
\centering
\resizebox{16cm}{!}{
\begin{tabular}{|c|c|c|c|c|c|c|}
\hline
    $M_S\ih \Gamma_S$  & $d_3=0.5 $ & $d_3=1$ & $d_3=2$ & $d_3=3 $ & $d_3=4 $ & $d_3=5 $  \\ \hline
    $ BP1 $     & \parbox[c][1.5cm][c]{2cm}{$ \bf 2006\ih \Gamma $} & $\bf 1884\ih 487$ & $ 1681\ih 187 $  & $\doublerow{994\ih25}{1756\ih 65}$  & $\doublerow{1044\ih 38}{2069\ih 26}$ & $\doublerow{993 \ih 23}{2444\ih 25}$ \\ \hline
    $ BP2 $     & \parbox[c][1.5cm][c]{2cm}{$ \bf 2369\ih \Gamma$} & $\bf 2156\ih 637 $ &$1906\ih 237 $  & $ \doublerow{1119\ih 27}{1869\ih 75} $ & $\doublerow{1219\ih 37}{2094\ih 31}$ & $ \doublerow{1181\ih 21}{2444\ih 25}  $  \\ \hline
 $ BP3 $ & \parbox[c][1.5cm][c]{2cm}{$ \bf 2468\ih \Gamma$} & $\bf 2244\ih 675 $ & $1969\ih 250 $  & $\doublerow{1131\ih 19}{1894\ih 75}$ & $\doublerow{1269\ih 37}{2094\ih 20}$ & $\doublerow{1231\ih 23}{2444\ih 25}$  \\ \hline
\end{tabular}
}
\caption{{\small Values of the pole position of the benchmark points in Table \ref{table_s_BPs} changing $d_3$. The rest of the parameters are set to their SM values. The cells with two complex numbers indicate the pole position of the two physical Breit-Wigner poles in the denominator of the unitarized amplitude.}}\label{table_s_differences_coupled_d3}
\end{table}\myspace
We find that for $d_3\gtrsim 2.5$ a second pole clearly appears (notation pole1 over pole2) in the low-energy region around $\sim 1$ TeV and it is also physical because it is found in the second Riemann sheet of the complex $s$ plane. However, one of the physical poles is located at energy scales much lower than our preestablished bound of $1.8$ TeV, so, in principle, the corresponding set of parameters should be discarded. The results are shown in Table \ref{table_s_differences_coupled_d3}. In fact, there are already hints of this first resonance at $d_3= 1.7$. \myspace 
Of course, the possibility of a light scalar resonance ($\lesssim 1.8$ TeV) being very weakly coupled to $WW$ channel and, hence, viable but hard to detect yet due to limited statistics remains a logical possibility to be further studied. However, if we discard such possibility, the bound on $d_3$ becomes very stringent.\myspace
We have checked that the inclusion of a natural value of $\gamma$, does not alter the fact that one of the states is too light, making the restriction on $d_3$ not significantly modified as it can be seen in Table \ref{table_s_differences_coupled_d3_gamma}. In this table, we reproduce the same analysis that we have just presented but set the value $\gamma=0.5\cdot 10^{-4}$.
\begin{table}[h!]
\centering
\resizebox{16cm}{!}{
\begin{tabular}{|c|c|c|c|c|c|c|}
\hline
    $M_S\ih \Gamma_S$ & $d_3=0.5 $ & $d_3=1$ & $d_3=2$ & $d_3=3 $ & $d_3=4 $ & $d_3=5 $  \\ \hline
  $ BP1 $  &  \parbox[c][1.5cm][c]{2cm}{ $1769\ih 275 $}  & $1668\ih 212 $ & $ 1544\ih 112 $  & $\doublerow{994\ih 23}{1569\ih 25}$ & $\doublerow{1044\ih 37}{1769\ih 34}$ & $\doublerow{994\ih 27}{1994\ih 54}$  \\ \hline
    $ BP2 $  &  \parbox[c][1.5cm][c]{2cm}{$1981\ih 262 $} & $1881\ih 212 $  & $1719\ih 125  $  & $\doublerow{1106\ih 27}{1656\ih 50}$ & $\doublerow{1219\ih 37}{1781\ih 34}$ & $\doublerow{1118\ih 26}{1994\ih 50}  $  \\ \hline
 $ BP3 $ &  \parbox[c][1,5cm][c]{2cm}{$ 2031\ih 250$} & $1931\ih 200 $ & $1769\ih 125  $  & $\doublerow{1131\ih 37}{1681\ih 38}$ & $\doublerow{1269\ih 37}{1781\ih 27}$ & $\doublerow{1231\ih 23}{1994 \ih 53}$  \\ \hline
\end{tabular}
}
\caption{{\small Values of the pole position of the benchmark points in Table \ref{table_s_BPs} with $\gamma=0.5\cdot 10^{-4}$ changing $d_3$. The rest of the parameters are set to their SM values. The cells with two complex numbers indicate the pole position of the two physical Breit-Wigner poles in the denominator of the unitarized amplitude.}}\label{table_s_differences_coupled_d3_gamma}
\end{table}\myspace
The next step is to check the impact of the crossed channels by varying $\eta$ and $\delta$ in the phenomenological constraint found.\myspace
By doing so, we have not found any resonant state fulfilling all the criteria that we have imposed. For the three selected benchmark points of Table \ref{table_s_BPs}, the behavior is quite similar and can be summed up in following three situations depending on the region in the $\eta-\delta$ plane: the firs scenario (1) with a single light resonance ($\sim 1$ TeV), another scenario (2) where two physical resonances appear but one is too light and a third new scenario (3) where a chain of three resonances emerge but the more massive one is classified as unphysical by the phase shift criteria. With all this, the bound $d_3\lesssim 2.5$ is not modified.
\subsection{Quartic Higgs self coupling: $d_4$}
The coupling $d_4$ parametrizes the strength of the self-interaction of four Higgses, and as it happens with $d_3$, it enters now at the lowest order in chiral perturbation theory and contributes at tree level in the $hh$ process. From experiment, it is extremely poorly constrained because of the difficulty of measuring the pointlike coupling of four Higgses. For this study and in the absence of any relevant experimental bounds up to date, we will be considering values up to $d_4\lesssim 10$, which would make the interaction of order $\mathcal{O}(1)$. Negative values of $d_4$ are to be excluded outright due to vacuum-stability reasons.\myspace
To start the analysis, we select the benchmark points from the tables above and see how the value of $d_4$ affects the properties of the poles. In particular, we focus on the case where $\gamma=0.5\cdot 10^{-4}$ which for all scenarios allowed the presence of resonances satisfying $\Gamma<M/4$.
\begin{table}[h!]
\centering
\resizebox{16cm}{!}{
\begin{tabular}{|c|c|c|c|c|c|c|c|}
\hline
    $M_S\ih \Gamma_S$ & $ d_4=0.5$ & $d_4=1$  & $d_4=2$ & $d_4=3 $ & $d_4=4 $ & $d_4=5 $ & $d_4=8 $ \\ \hline
    $ BP1 $ & \parbox[c][1.2cm][c]{2cm}{$1794\ih 250 $} & $1668\ih 212  $    & $1494\ih 137 $  & $1381\ih 112 $ & $1306\ih 87  $ & $1256\ih 75 $ & $1169\ih 50 $ \\ \hline
    $ BP2 $ & \parbox[c][1.2cm][c]{2cm}{$1981\ih 225 $} & $1881\ih 212  $    & $1719\ih 175 $  & $1606\ih 125  $ & $1531\ih 112  $ & $1481\ih 87 $ & $1381\ih 75 $ \\ \hline
 $ BP3 $ & \parbox[c][1.2cm][c]{2cm}{$2031\ih 225 $} & $1931\ih 200  $    & $1781\ih 162 $  & $1669\ih 137  $ & $1594\ih 112  $ & $1544\ih 100 $ & $1444\ih 75$ \\ \hline
\end{tabular}
}
\caption{{\small Values of the pole position of the benchmark points in Table \ref{table_s_BPs} changing $d_4$ with $\gamma=0.5\cdot 10^{-4}$. The rest of the parameters are set to their SM values.}}\label{table_s_differences_coupled_d4}
\end{table}\myspace
From Table~\ref{table_s_differences_coupled_d4} we can say that, if all the rest of parameters are set to their SM values, we could exclude values of $d_4\gtrsim 2$ for BP2 and BP3 and BP1 would be excluded since these parameters lead to light resonances that should have already been seen. As always we assume (rightly or wrongly) that any scalar resonance above 1.8 TeV should have been observed. And as always, we also force the vector resonances,  if present, to be heavier than that scale.\myspace
The question whether the crossed channel (with the parameters $\delta$ and $\eta$ leading at high energies) could affect this result is depicted in the following graphs, where, for different values of $d_4$, we show the regions in the $\delta-\eta$ plane where resonances $M_S>1.8$ TeV can appear.\myspace
In Figure~\ref{fig_s_swept_BP2y3}, we see how, in fact, some regions that were nonresonant, show resonances after activating the crossed-channels parameters from the values in the Table \ref{table_s_differences_coupled_d4}. The more we depart from the
SM value $d_4=1$, the more restriction we get. In fact, for $d_4=5$ and the values of Table \ref{table_s_differences_coupled_d4}, we only find resonances in the lines $\delta=-\frac{1}{3000}-\frac{\eta}{3}$ for BP2 and $\delta=-\frac{8}{3000}-\frac{\eta}{3}$ for BP3.
\begin{figure}[t]
\centering
\includegraphics[clip,width=7.0cm,height=6.5cm]{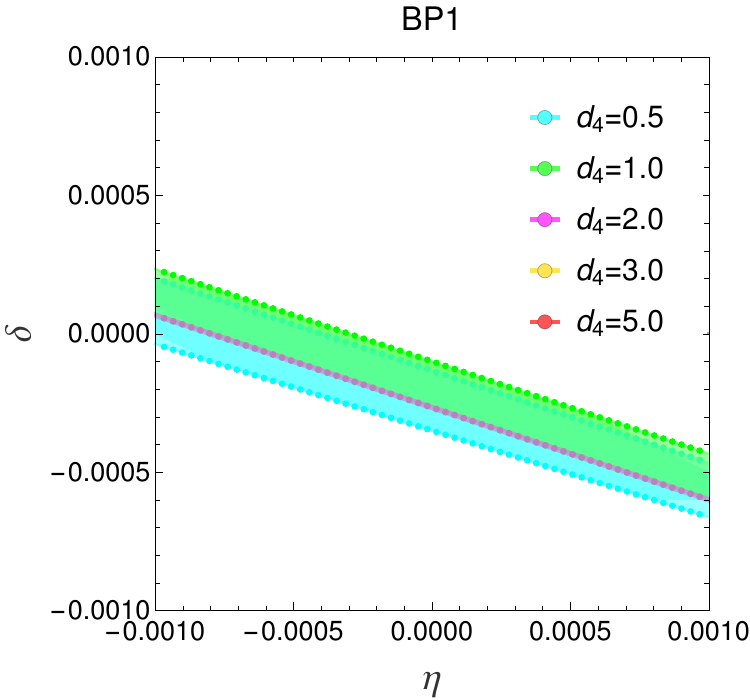} %
\includegraphics[clip,width=7.0cm,height=6.5cm]{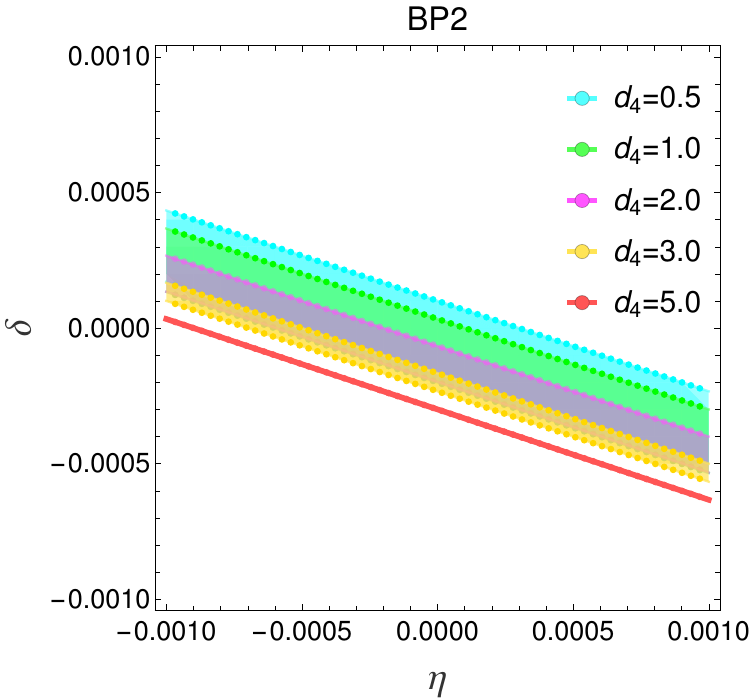} %
\includegraphics[clip,width=7.0cm,height=6.5cm]{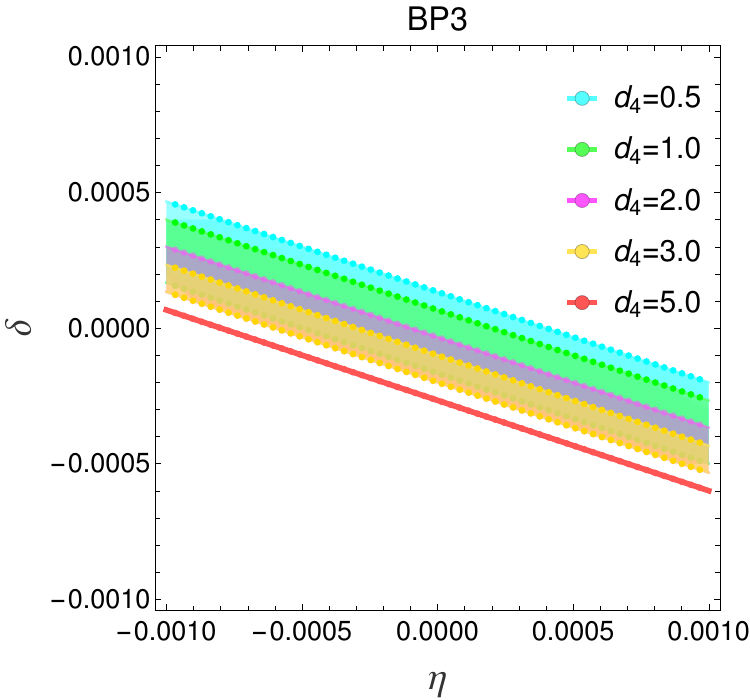} %
\caption{\small{Regions in the $\delta-\eta$ plane where physical resonances satisfying $M_S>1.8$ TeV appear for different values of $d_4$ and setting $\gamma=0.5\cdot 10^{-4}$ for specific values of $a_4$ and $a_5$ corresponding to (top left) BP1, (top right) BP2 and (bottom) BP3 in Table. \ref{table_s_BPs}. For all the values of $d_4\lesssim 6$ and for all the benchmark points, the region above the bands are excluded by the presence of a non-physical pole. Below the bands we find a non-resonant scenario.}}%
\label{fig_s_swept_BP2y3}
\end{figure}\myspace
We do not find any physical resonant state with $M_S\gtrsim 1.8$ TeV and $d_4\gtrsim 6$.\myspace
In this case varying $d_4$, the same behavior that has been observed varying $\gamma$ (see Figure~\ref{fig_s_swept_gamma}) is reproduced: above the color bands, we get excluded regions by the appearance of a second nonphyiscal pole (again using the phase-shift criteria), and below the bands, we get a nonresonant scenario with an absence of any zeros in the determinant of the unitarized amplitude.
\section{Light resonances: the $\text{H}(650)$ in the HEFT}
In the previous sections, we explored scalar resonances at the Terascale, guided by phenomenological constraints on vector resonances. These heavy scalar resonances, dynamically generated within the IAM framework with the coupled channel formalism, are part of the particle spectrum of a theory with strong interactions whose low-energy signatures manifest as operators that induce unitarity violations.\myspace
The gap free of new particles between the electroweak and TeV scales represents a constraint that must be considered in any study of New Physics?. This raises the question: is there room in the theory to accommodate new states with "light" masses---below the Terascale? The answer is yes, supported both by theoretical considerations and experimental evidence.\myspace
Theoretically, two plausible scenarios arise in this context. First, these new light states might have negligible or no impact on the electroweak symmetry-breaking process. Alternatively, light scalars could emerge as pseudo-Goldstones, arising from the vacuum-breaking dynamics of the strongly interacting theory. Analogous to low-energy QCD, where pions have small but nonzero masses on the order of $\sim 100$MeV, the symmetry governing the UV dyanmics must be approximate.\myspace
In the context of the second scenario, there is an obvious interest in the appearance of (relatively) light scalar companions of the $h(125)$ as they may arise in composite Higgs models (see, e.g., Ref.~\cite{Dobado:2019fxe} and the references therein) as (pseudo) Goldstone-like spare states following the spontaneous symmetry breaking (SSB) of the vacuum of a theory possessing a larger global symmetry group. Other models try to explain such triggering of the \textit{electroweak symmetry breaking sector} (EWSBS) with more than one scalar such as the Georgi-Machacek~\cite{Georgi:1985nv} and Chanowitz-Golden~\cite{Chanowitz:1985ug} and two-Higgs-doublet (2HDM)~\cite{Gunion:2002zf} models. In Ref.~\cite{Kundu:2022bpy}, a list of scalar resonances such as H$(650),\ A(400),\, h(151)$ and $h(95)$ and their statistical significances is presented with conclusive results for the particular case of H$(650)$ claiming a 7 standard deviations global significance coming from combined analysis in various channels. Another scalar state $h^{\prime}(515)$ is treated in Ref.~\cite{Afonin:2022qkl} assuming an holographic description of a strong sector beyond-the-SM (BSM).\myspace
Recently, some interest has emerged on a possible signal for a Higgs-like state around 600 GeV, much below the region just mentioned. On one hand, searches in CMS (Ref.~\cite{CMS:2020tkr}) and ATLAS (Ref.~\cite{ATLAS:2021kog}), see Refs. \cite{Cea:2018tmm, Cea:2022zgs} for a combined analysis, have yielded some evidence for the production of this resonance through the clear four-leptonic final state: H$(650)\to ZZ\to 4l$. In particular, they suggest a scalar state peaking at $\sim$ 650 GeV with a total width of approximately 100 GeV,  with a $3.75\sigma$ significance using an integrated luminosity of 139 fb$^{-1}$. The corresponding cross section for the subprocess $pp\to ZZ+X$ is $90\pm 25$ fb. After applying sequential cuts, an ATLAS analysis for vector boson fusion in Ref.~\cite{ATLAS:2021kog} reduces the significance of this resonance to $2.1\sigma$ and the cross section to $30\pm15$ fb, significantly below the inclusive one before the cuts. On the other hand, searches of leptonic decays
from $WW$ ($2l+$missing energy) enhances the production rate for this scalar to more than five times the $ZZ$ one, resulting in a cross section of $160\pm50$ fb. This scenario of unbalanced production rates between channels is actually reproduced in our HEFT description, as we will see.\myspace
Some previous works have already studied models with states similar to H$(650)$~\cite{Kundu:2022bpy, Afonin:2022qkl,Georgi:1985nv,Chanowitz:1985ug,Cea:2018tmm,Cea:2022zgs,Gunion:2002zf}.
The question that naturally emerges is: is such a light resonance compatible with existing bounds on the low energy coefficients of the HEFT? This is not obvious at all, because strict bounds already exist on many of these coefficients, gathered in Table \ref{tab_ET_chiralparams}. These bounds place various such coefficients in the $10^{-4}$ range, which typically provide resonances above the TeV scale, but several other couplings are poorly bounded or not bounded at all. Can therefore the $\text{H(650)}$ be accommodated in the HEFT without violating  any existing bounds? This seems a relevant question because a negative answer -taking into account the generality of the HEFT approach- would most likely rest credibility to the experimental hints. As said in the Introduction, the analysis of this section correspondos to our work in Ref.~{\cite{Asiain:2023myt}}.\myspace
In particular we will admit values of the couplings in $\mathcal{L}_4$ in the range (see Ref.~\cite{CMS:2019uys})
\begin{equation}\label{eq_s_a4a5_bounds}
a_4\in (-0.0061,\,0.0063)  \hspace{0.5cm} a_5\in (-0.0094,\,0.0098) \,
\end{equation} 
that have been obtained using $13$ TeV LHC data in four leptons final states from $WW/WZ$ scattering. In Ref.~\cite{ATLAS:2016nmw}, the bounds are directly for $a_{4,5}$ couplings with K-matrix unitarization method. However, the analysis doesn't fully leverage yet the available LHC statistics. Consequently, the $a_4$ and $a_5$ bounds are about four times weaker than the ones shown in Eq.~(\ref{eq_s_a4a5_bounds}). There is a more strict bound (by a factor 10, see Ref.~\cite{Sirunyan:2019der}) for $a_5$ coming from an SMEFT analysis with $2l2j$ final states, much lesser clear channel. However, it should be noted that in the scalar-isoscalar channel the couplings $a_4$ and $a_5$ always appear in the combination $5a_4 +8a_5$ and therefore the error in $a_4$ amply dominates anyway.
As said in previous sections, the rest of the $\alpha_{p^4}$ couplings relevant for the present discussion, namely $\delta,\,\eta$ and $\gamma$, remain unconstrained experimentally, but taking into account the fact that they are absent in the SM, we will allow these to have a maximum (absolute) value of $10^{-3}$.\myspace
The exercise we want to do in this section is to search for a  set of $\alpha_{p^4}$ HEFT parameters that lead to the presence of a resonance with the properties of the H$(650)$ in $WW$ scattering that are tentatively claimed in Ref.~\cite{Kundu:2022bpy}. Experimentally, this resonance appears to have a total width of $\sim 100$ GeV so, so we focus on the production of a scalar resonance whose mass lies within the $600-700$ GeV range.\myspace
The anomalous parameters of chiral order two are all set to their SM values: $\alpha_{p^2}=1$. This leaves us with free $a_4,\,a_5,\,\delta,\,\eta$ and $\gamma$. All of them intervene at the NLO (formally tree level) contributions for different processes: the first two to $WW$ elastic scattering, $\gamma$ to elastic $hh$ and $\delta$ and $\eta$ to the crosses channel $WW\to hh$. However, these separated contributions mix among themselves along the unitarization process.\myspace
Following previous results, when we set $\alpha_{p^2}=1$ there are only two physical situations for any choice of the $\alpha_{p^4}$ chiral parameters: a nonresonant scenario with the absence of any complex pole in the unitarized amplitude, or
a resonant one with only one such pole. Whenever there are two poles, one is identified as non physical by the phase shift criteria (lies on the first Riemann sheet in the complex $s$ plane). Secondly, the resonances emerging are much more visible in the $WW$ channel than in the coupled ones so the $a_4-a_5$ plane is the most sensible parameter space to represent the results. As already mentioned in a previous section, this plane is restricted by the experimental bounds quoted in Eq.~(\ref{eq_s_a4a5_bounds}).\myspace
In Figure~ \ref{fig_s_map_a4a5_h650} we show the regions in $a_4-a_5$ parameter space where a resonance with mass between $600-700$ GeV appears using different selection of the $\alpha_{p^4}$ chiral parameters. The different areas are obtained by activating different sets of the NLO HEFT coefficients besides $a_4$ and $a_5$, with maximum values of $|10^{-3}|$, following the explanation in the legend. It should be clarified that all the regions overlap with each other: the red one includes the rest of the areas, and the blue one the smaller one in green.
\begin{figure}
\centering
\includegraphics[clip,width=8.5cm,height=8.5cm]{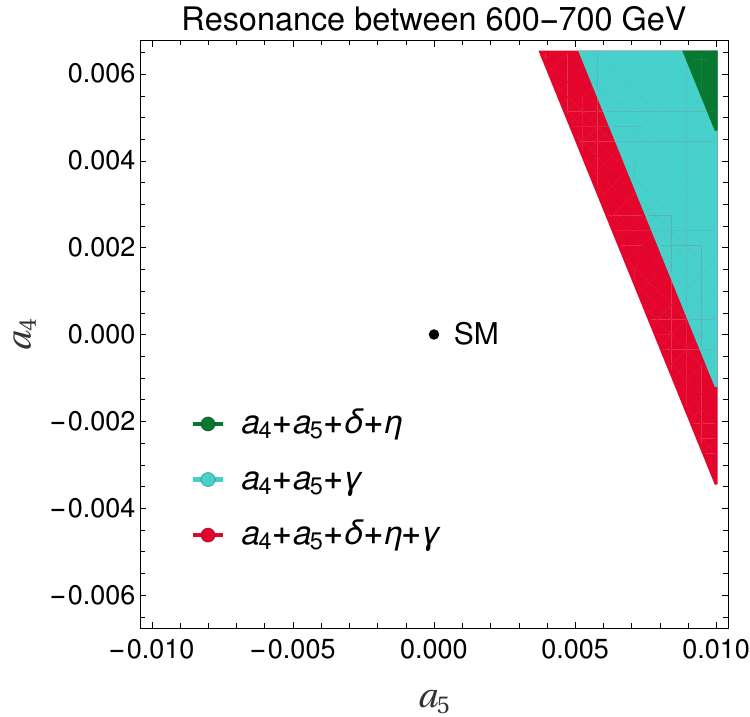}
\caption{\small{Regions in the $a_4-a_5$ plane allowed by experimental constraints where resonances between $600-700$ GeV appear when activating different NLO chiral couplings. The more parameters one activates, the less restriction in the plane to achieve the desired scalar light resonance. The LO parameters are set to the corresponding SM values.}}
\label{fig_s_map_a4a5_h650}
\end{figure}\myspace
In fact, no bound state with the expected characteristics appears assuming non-zero values for $a_4$ and $a_5$ only. One needs the help of at least
one more anomalous coupling.\myspace
The Figure~\ref{fig_s_map_a4a5_h650} above shows regions in the main parameter space where resonances with masses in the range 600-700 GeV are allowed but it says nothing about their properties. One thing that we can indeed extract from Figure~ \ref{fig_s_map_a4a5_h650} is that the inclusion of $\delta$ and $\eta$ does not affect very much the results when looking for light resonances. With these, we now investigate the physical properties of the resonances using only $a_4,\,a_5$ and $\gamma$. Actually, these \textit{a priori} independent three parameters are reduced to two when studying scalar resonances since, as said before, the lines $5a_4+8a_5=k$ contain resonances with the same properties for a fixed $k$. We choose $k\in (0.055,0.11)$ so we lie within the pure scalar region, not vector nor tensor states appear, and the experimental bounds in Eq.~(\ref{eq_s_a4a5_bounds}) are satisfied.\myspace
In Figure~ \ref{fig_s_map_Kgamma} we show the results. As one can observe, no scenario  with negative values of $\gamma$ exhibits physical resonances: the dashed region is forbidden by the emergence of a second pole identified as non-physical using the phase-shift criteria and below this area, a nonresonant region appears. Positive values of $\gamma$ that are not colored in Figure~ \ref{fig_s_map_Kgamma} also lead to physical resonances heavier than $700$ GeV, though, so they are excluded from our analysis. Another interesting thing we have observed in the right panel is that for a specific value of $\gamma$ the width remains constant along the lines $5a_4+8a_5=k$ in the region of interest and they are relatively small (with respect to the masses) leading to quite stable intermediate states emerging in $WW$ scattering. This feature can be explained with the information provided in Ref.~\cite{paper2_doi}, where we observed that for large (yet natural) values of $\gamma\left(\sim 10^{-3}\right)$, the single-channel resonance, this is ignoring coupled-channels, was recovered.
\begin{figure}
\centering
\includegraphics[clip,width=7.0cm,height=7.5cm]{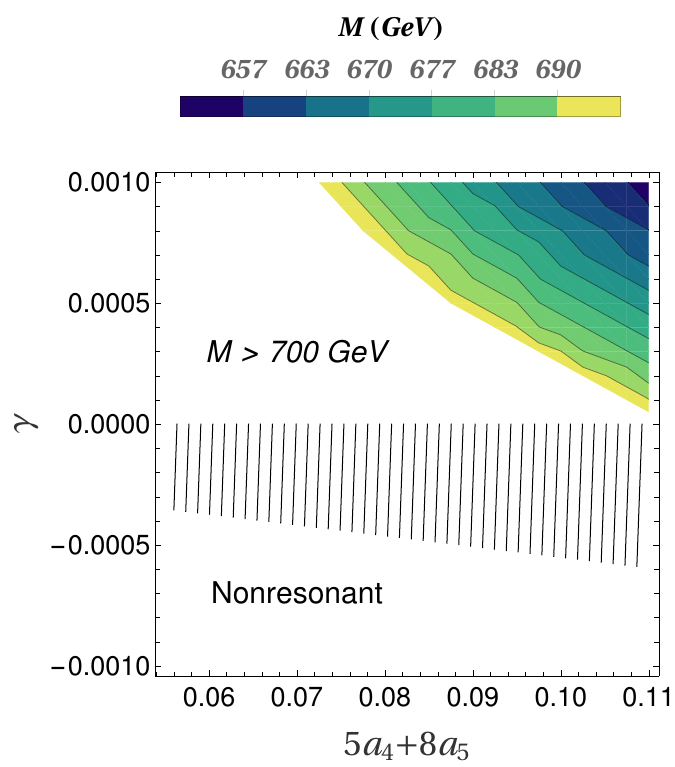}\qquad\quad
\includegraphics[clip,width=7.0cm,height=7.5cm]{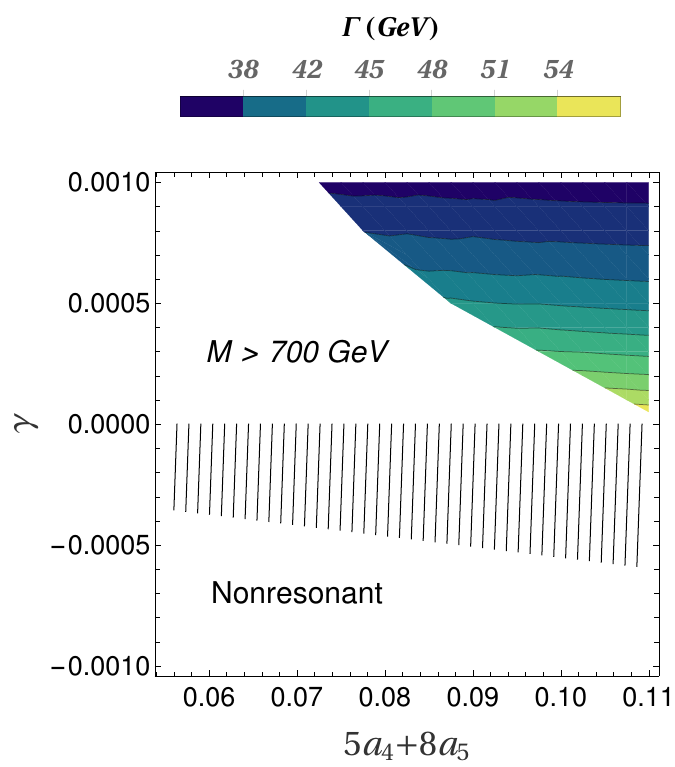}
\caption{\small{Masses (left panel) between 600 and 700 GeV and widths (right panel) of scalar resonances in the plane described by the set of lines $5a_4+8a_5=k$ and $\gamma$. All negative values of $\gamma$ are non-physical (stripped area) by the phase-shift criteria or nonresonant. Positive values of $\gamma$ below the colorful region exhibit physical resonances, too, but they are heavier than $700$ GeV so they are of no interest for this work. The right panel shows how for a fixed value of $\gamma$, the widths of the corresponding resonances are independent of the combination $5a_4+8a_5$.}}
\label{fig_s_map_Kgamma}
\end{figure}\myspace
All in all, we have been able to reproduce a scalar resonance with mass around 650 GeV in the HEFT using SM values of the LO Lagrangian and deviations at the next-to-leading order in chiral counting that are within the existing experimental bounds up to date. The more relevant couplings to describe such a resonance happen to be those driving the elastic NLO processes, $a_4$, $a_5$ and $\gamma$ with subleading effects with the off-diagonal ones, $\delta$ and $\eta$, along the coupled-channel unitarization process. The widths obtained for those new states are quite small compared to their masses, 30-65 GeV.\myspace
However, up to now, the more data the experiment collects the more compatible the anomalous couplings are with the successful SM. The BSM H(650)-like resonances in this study appear close to the upper limit of the experimental bounds in Eq.~(\ref{eq_s_a4a5_bounds}), see Figure~ \ref{fig_s_map_a4a5_h650},  meaning that a possible future improvement in these bounds pointing towards consolidation of the SM values would imply their exclusion.\myspace
As said, detailed information regarding the H$(650)$ can be found in Ref.~\cite{Kundu:2022bpy}. Now, our task if to connect these tentative results with the ones extracted from our analysis.\myspace
We will work under the assumption that the resonant profile obtained, if any, is produced by a single resonance, neglecting the possibility of two overlapping resonances. Besides, we also ignore the decay mode H$(650)\to h(125)h(95)$, being $h(125)$ the Higgs described in the minimal SM. One way to search for H$(650)$ in vector boson fusion is via the decay to $b\overline{b}\gamma\gamma$, but there is yet not enough resolution in the experiment to distinguish between a $b\overline{b}$ pair decayed from a $Z$ or a hypothetical h$(95)$. In other words, we ignore the possible presence of the $h(95)$ that seems less motivated. The interested reader may find this scalar decay in Ref.~\cite{Banik:2023ecr} in the context of 2HDM. This is why for this work we assume the decay mode of H$(650)$ to be exclusively via gauge bosons (however, do keep in mind that other channels contribute in the unitarization procedure).\myspace
In Ref.~\cite{Kundu:2022bpy}, the authors gave a total width to gauge bosons of $\Gamma=90\pm 28$ GeV, of the order of the widths presented in the right panel of Figure~ \ref{fig_s_map_Kgamma}.\myspace 
To get a first estimate of the cross section for the production of such resonance we will use the effective $W$ approximation (EWA), see Ref.~\cite{Dawson:1984gx}, which takes $W$s and $Z$s as proton constituents and it is approximately valid for energies well above the EW scale. Within the EWA approach the differential cross-section is given by 
\begin{equation}\label{eq_s_dxsdMww}
\frac{d\sigma}{dM_{WW}^2}=\sum_{i,j}\int_{M_{WW}^2/s_{}}^{1}\int_{M_{WW}^2/(x_1s_{})}^{1}\frac{dx_1dx_2}{x_1x_2s_{}}f_i(x_1,\mu_F)f_j(x_2,\mu_F)\frac{dL_{WW}}{d\tau}\int_{-1}^{1}\frac{d\sigma_{WW}}{d\cos\theta}d\cos\theta,
\end{equation}
with $s_{}$ is the centre of mass energy of the two opposite protons at the LHC and $M_{WW}$ is the invariant mass of the two $W$s. Here the "partonic" differential cross section in the $WW$ rest frame is
\begin{equation}\label{eq_s_dxsdcos}
\frac{d\sigma_{WW}}{d\cos\theta}=\frac{|A(M_{WW}^2,\cos\theta)|^2}{32\pi M_{WW}^2},
\end{equation}
This expression factorizes both energy scales: the one for the long-distance non-perturbative part describing the dynamics inside the proton and a perturbative one for the $WW$ hard scattering.\myspace
The amplitude $A(M_{WW}^2,\cos\theta)$ appearing in Eq.~(\ref{eq_s_dxsdMww}) describes the amplitude of a $WW$ scattering in the charged (physical) basis that is detected at the LHC and not the amplitude in the $IJ$ basis that we rendered unitary. Thus, by moving backwards along the  process we used for unitarization, we have to recover the "unitary" physical amplitude that would produce such a unitary partial wave. From now on, a superindex $\cal U$ will refer to unitarized quantities, in our case obtained thorough IAM. This is done by reversing the unitarization procedure
\begin{equation}\label{eq_s_TIIAM}
T^{\cal U}_{I}=32\pi\sum_{J=0}^{\infty}(2J+1)\,t^{\cal U}_{IJ}P_{J}(\cos\theta)
\end{equation}
and we truncate the infinite series at the leading order (LO) for every isospin channel assuming that is a good approximation close to the
resonance mass where the peak dominates the amplitude:
\begin{equation}\label{eq_s_TsIAM}
\begin{split}
&T_0^{\cal U}\approx 32\pi\, t^{\cal U}_{00}\\
&T_1^{\cal U}\approx 32\pi\left(3t_{11}^{\cal U}\cos\theta\right)\\
&T_2^{\cal U}\approx 32\pi t_{20}^{\cal U}.
\end{split}
\end{equation}
Now, we simply use the isospin relation to build the "unitary" physical amplitudes. We will be using for comparison with the literature those with $WW$ and $ZZ$ in the final states after VBS:
\begin{equation}\label{eq_s_Aphysical}
\begin{split}
A^{\cal U}\left(\wpwm \to \wpwm\right)&=\frac{1}{3}T_0^{\cal U}+\frac{1}{2}T_1^{\cal U}+\frac{1}{6}T_2^{\cal U}\\
A^{\cal U}\left(\wpwm \to ZZ\right)&=\frac{1}{3}T_0^{\cal U}-\frac{1}{3}T_2^{\cal U}\\
A^{\cal U}\left(ZZ\to ZZ\right)&=\frac{1}{3}T_0^{\cal U}+\frac{2}{3}T_2^{\cal U}
\end{split}
\end{equation}
and the symmetric processes under time reversion.\myspace
The EWA consists in convoluting the probability for a quark inside a proton to radiate a gauge boson with the actual parton distribution function (pdf) for the constituent quarks $q$ at some energy scale, $f_q(x,\mu)$, using the effective luminosity in Ref.~\cite{Espriu:2012ih}
\begin{equation}\label{eq_s_luminosity_WW}
\!\!\!\!\frac{dL_{W W}}{d\tau}=\left(\frac{g}{4\pi}\right)^4\left[\left(\frac{1}{\tau}+1\right)\ln\left(\frac{1}{\tau}\right)-2\left(\frac{1}{\tau}-1\right)\right]
\end{equation}
where $\tau=M_{WW}^2/(x_1x_2s_{})$ connects both energy scales. A factor $1/2$ must be added for $ZZ$ final states accounting for their indistinguishability.\myspace
Other luminosity functions are available in the literature (see for example Ref.~\cite{Alboteanu:2008my}). As soon as the experimental evidence of H$(650)$ gets consolidated, it would be good to consider all them for a  deeper and more complete analysis.\myspace
After performing the convolution of these functions we are in disposition to compute the integral in Eq.~(\ref{eq_s_dxsdMww}) to obtain the differential cross section of the process with respect to the invariant mass of the $WW$ system. Moreover, the total cross section of the process is obtained assuming that the peak indeed dominates the amplitude in such a way that
\begin{equation}\label{eq_s_total_xs}
\sigma=\int_{M-2\Gamma}^{M+2\Gamma}dM_{WW}\,\frac{d\sigma}{dM_{WW}}
\end{equation}
where $M$ and $\Gamma$ are the characteristic parameters of the resonance obtained from the unitarized amplitudes.\myspace
We are now in disposition to compare our results from the unitarized analysis of the cross section next to the experimental ones. But first, two aspects need to be taken into account. Firstly, our analysis  only says something about longitudinally polarized gauge bosons in the external states; any contribution coming from different polarization combinations is to be computed separately and with the corresponding effective luminosity. However, we expect the purely longitudinal process to dominate at high energies when we separate, even just a little, from the SM. If that is the case we should not saturate the experimental value which is unpolarized. Secondly, we can easily include in our calculation kinematical cuts on the pseudo-rapidity and the invariant mass of the outgoing gauge bosons but we can not demand restrictions in the kinematics of the radiated light jets suitable for VBS detection that are usually included in experimental analysis.\myspace
Taking into account all the machinery developed in this section to compute a theoretical cross section in $pp$ collisions and the comments in the above paragraph, we can now compare with experimental results. In Figure~ \ref{fig_s_xsections} we show values of the cross sections obtained using Eq.~(\ref{eq_s_total_xs}) for a subset of the parameter space in Figure~ \ref{fig_s_map_Kgamma} ($k\times\gamma= \left[0.1,0.11\right]\times\left[0.0007,0.001\right]$) where resonances with masses close to 650 GeV appear.
\begin{figure}
\centering
\includegraphics[clip,width=7.0cm,height=7.5cm]{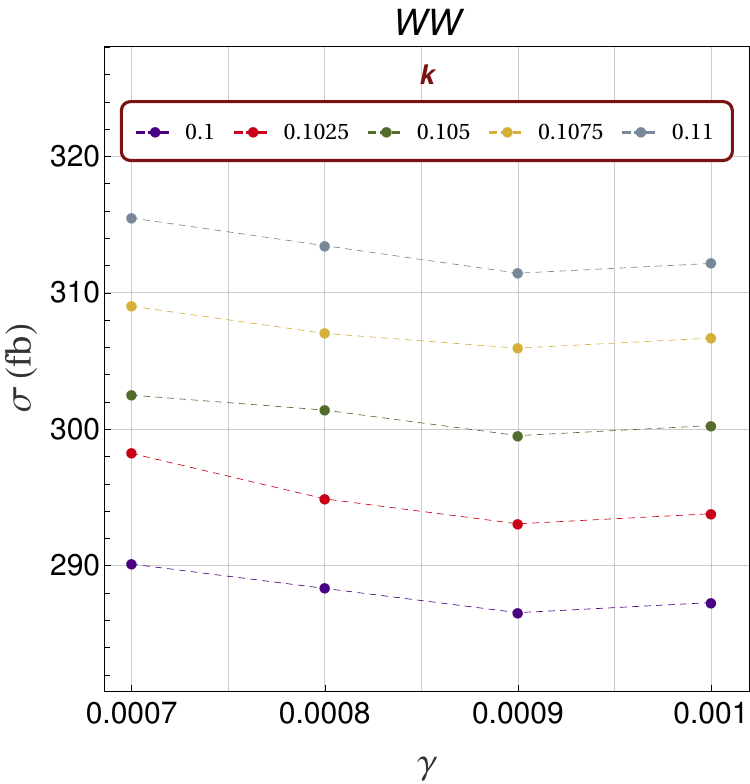}\qquad\qquad
\includegraphics[clip,width=7.0cm,height=7.5cm]{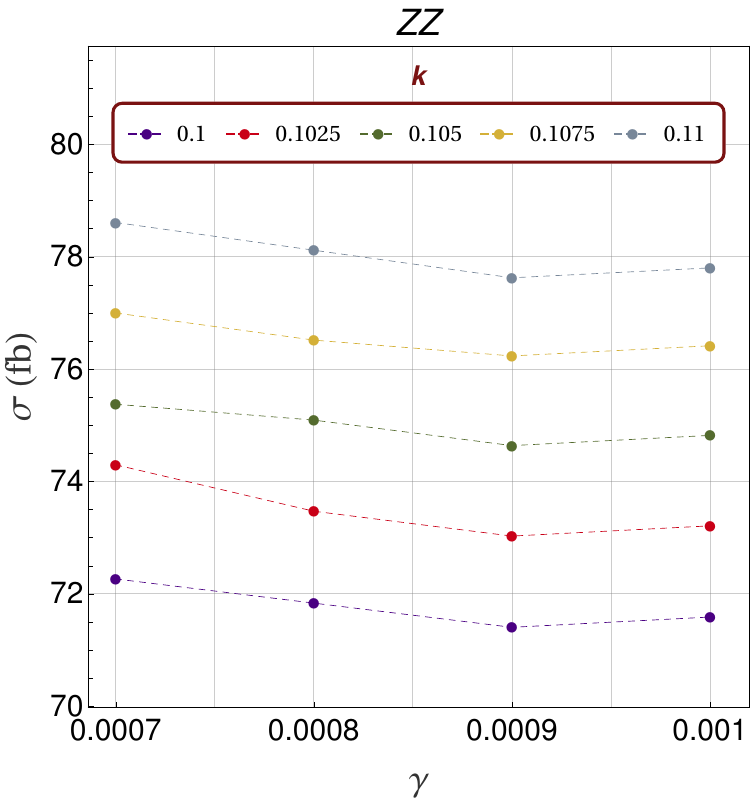}
\caption{\small{Values for the VBS cross section in Eq.~(\ref{eq_s_total_xs}) with $WW$ (left) and $ZZ$ (right) final states versus the NLO chiral parameter $\gamma$ and for different values of $k=5a_4+8a_5$ in the legend. The combination of the values of $k$ and $\gamma$ present in the figure make up the top-right region of Figure~ \ref{fig_s_map_Kgamma} where resonances close to 650 GeV appear. The centre of mass energy for this calculation is set to $\sqrt{s}=13$ TeV.}}
\label{fig_s_xsections}
\end{figure}\myspace
In Figure~ \ref{fig_s_xsections}, we show the values of the cross sections obtained for $WW$ (left panel) and $ZZ$ (right panel) final states at $\sqrt{s}=13$ TeV $pp$ collision energy after VBS using Eq.~(\ref{eq_s_Aphysical}). The cross sections for WW result to be of order $\sim300$ fb and $\sim$75 fb for $ZZ$ for all values of $k$ and $\gamma$. The measured cross sections from VBS~\cite{Kundu:2022bpy} of $\text{H(650)}\to WW$ and $\text{H(650)}\to ZZ$ are $160\pm 50\,\text{fb}$ and $30\pm 15\, \text{fb}$, respectively, close to SM values. These measurements really favor a $WW$ scenario after VBF rather than a $ZZ$ final state, with a ratio between cross sections $\sigma_{WW}/\sigma_{ZZ}\sim 5$. Our calculation implies $\sigma_{WW}/\sigma_{ZZ}\sim 4$, relatively close to the ATLAS and CMS analysis but really distant from the prediction using the Georgi-Machacek model, which infers the inverse situation with a $ZZ$ final state dominating over the $WW$ one with $\sigma_{WW}/\sigma_{ZZ}\sim 0.5$ (Refs.~\cite{Georgi:1985nv,Kundu:2022bpy}).\myspace
We obtain with our calculation, thus, two times the measured central values. As explained before, no cuts, and actually no kinematical condition, are imposed in our calculation in Figure~ \ref{fig_s_xsections} for a better comparison with the experiment so we would expect our computed cross section to exceed the measured one. We can easily introduce a cut in the pseudo-rapitidies of the final state gauge bosons that favors identification of VBS events. In particular if we impose $|\eta_W|<2$ following Ref.~\cite{Delgado:2017cls}, the cross sections are reduced to $\sim 275$ fb and $\sim$70 fb for $WW$ and $ZZ$ processes, respectively, getting closer to the experimental data. Presumably, further cuts on the kinematical variables of the light jets produced after radiation of the gauge bosons triggering the VBS would point towards even closer cross sections.\myspace
We present now a test of our calculation by making a comparison in the number of events obtained for a process using MonteCarlo (MC) techniques in Ref.~\cite{Delgado:2017cls}. In this work the authors reproduced the signal expected at the LHC for vector charged resonances emerging in the subprocess $WZ\to WZ$. The range of chiral parameters used lead to resonances in the mass range 1.5-2.5 TeV. For the event simulation at centre of mass energy of $\sqrt{s}=14$ TeV, the authors used a series of kinematical cuts both in the $WZ$ bound state and in the radiated light jets. On one hand, we also introduce for a more reliable comparison the cut on the pseudo-rapidity $|\eta_W|<2$ by integrating the partonic amplitude in the corresponding values of $\cos\theta$. On the other hand, we can not apply any cut on the light jets easily. \myspace 
The number of events obtained from our calculation of the VBS cross section for a specific value of the integrated luminosity is $N_{\mathcal{L}}=\sigma\cdot\mathcal{L}$, where $\sigma$ is the total cross section in Eq.~(\ref{eq_s_total_xs}). The results for both number of events, the MC simulation in Ref.~\cite{Delgado:2017cls} and our theoretical prediction, are gathered in Table \ref{table_s_events}:
\begin{table}[h!]
\centering
\begin{tabular}{|c|c|c|c|}
\hline
 & $M_V\ih \Gamma_V$ & $N^{\text{MC}}_{1000}$ & $N_{1000} $ \\ \hline
$ \text{BP1}$ &$1510\ih 13 $ & $488 $ & $904 $  \\ [0.6ex]\hline
$ \text{BP2}$ &$2092\ih 20 $ & $82 $ & $121 $  \\ [0.6ex]\hline
$ \text{BP3}$ & $2541\ih 27 $ & $30 $ & $31 $  \\ [0.6ex]\hline
\end{tabular}
\caption{{\small Number of MC events for three benchmark points BP1, BP2 and BP3 in Ref.~\cite{Delgado:2017cls} that produce vector resonances for different values of the integrated luminosity $\mathcal{L}$ in units of fb$^{-1}$, $N^{MC}_{\mathcal{L}}$. The number of events obtained with our calculation is $N_{\mathcal{L}}=\sigma\cdot\mathcal{L}$.}}\label{table_s_events}
\end{table}\myspace
As expected and argued before, all our number of events exceed the ones obtained using a MC simulation due to the lack of extra kinematical cuts. We also observe that the heavier the vector resonance is the more accurate the calculation with respect the MC.\myspace
More recently, and building upon the study in Ref.~\cite{Kundu:2022bpy}, the authors of Ref.~\cite{Yaouanc:2024xqg} extended the predictions of the earlier work regarding the $H$650 to propose it as a tensor resonance---which they refer to as $T$650. They resolve the apparent contradiction adducing an overly strict signal-versus-background event selection, particularly in the angular distribution. By relaxing those kinematic cuts, the existence of $T$650 should not be ruled out, as it could also be well accommodated in $WW/ZZ$ or $h(95)h(125)$ decays (via $bb\bar{b}\bar{b}$). According to their analysis, the mere existence of $T$650 does not exclude $H$650, as both could coexist as degenerate states: H650 belonging to an isotriplet and T650 to a tensor isospin quintuplet. Theoretically speaking, this coexistance is allowed by the Randall-Sundrum model involving extra 'warped' dimensions, see Ref.~\cite{Randall:1999ee}.

\thispagestyle{empty}

\lhead{Chapter 6}
\rhead{Resonances in Gravity}

\chapter{Resonances in Gravity (?)}
\label{chp:graviton}
The reader who has followed along this far might share some of the perplexity I felt when my supervisor first proposed this topic for my thesis. However, with time, I have come to find it immensely interesting and perfectly fitting as a conclusion, tying together the themes of effective theories and unitarization techniques. If the machinery we have developed in the previous chapters rests on a solid foundation, it should apply meaningfully regardless of the phenomena under investigation.\myspace
I had the privilege of beginning my physics studies during a particularly momentous time. At the end of my first year, the Higgs boson---the central focus of this doctoral thesis---was discovered at the LHC. A few years later, shortly before completing my degree in 2016, the LIGO collaboration (Laser Interferometer Gravitational-Wave Observatory) announced the first-ever detection of gravitational waves. This final chapter addresses gravitational waves in the context of quantum field theory. The physical description of these waves at the quantum level is realized by means of a tensor quantum field whose excitations correspond to a massless particle: the graviton.\myspace
This chapter does not aim to serve as a formal review of general relativity. Instead, it seeks to demonstrate the power of effective theories and unitarization techniques within a framework that extends beyond the traditional scope of quantum field theory, typically focused on studies of the Standard Model and its extensions.
\section{The graviton}
From the perspective of quantum field theory, the graviton mediates the gravitational interaction, analogous to how the photon mediates the electromagnetic interaction. The key difference is that the graviton must have spin $j=2$ to satisfy both theoretical and experimental results, making it a tensor particle. This result can also be checked by analyzing the properties of the graviton field under Poincaré transformations. Furthermore, the graviton is massless, which further distinguishes it from the carriers of the weak interaction in the SM, due to the infinite range of the gravitational interaction.\myspace
A massless tensor particle has two degrees of freedom, corresponding to the two physical helicity states of the graviton, $\lambda_{\pm} = \pm 2$. This result is analogous to the well-known situation in quantum electrodynamics (QED), where the photon, a massless spin-1 particle, also has two physical helicity states, $\lambda^\gamma_{\pm} = \pm 1$. To illustrate this analogy, we start from the QED Lagrangian
\begin{equation}\label{eq_g_qedlag}
\mathcal{L} = -\frac{1}{4}F_{\mu\nu}F^{\mu\nu},
\end{equation}
where the field strength tensor $F_{\mu\nu}$ is
\begin{equation}
F_{\mu\nu}=\partial_\mu A_{\nu}-\partial_\nu A_\mu,
\end{equation}
and $A_\mu$ is the gauge field representing the photon.\myspace
Naively, the photon field $A_\mu$ has four degrees of freedom. However, the invariance of the QED Lagrangian under the gauge transformation
\begin{equation}\label{eq_g_qedgauge}
A_\mu \to A_\mu - \partial_\mu \lambda(x),
\end{equation}
where $\lambda(x)$ is an arbitrary function, allows us to eliminate non-physical components of $A_\mu$. For instance, in the Lorenz gauge,
\begin{equation}\label{eq_g_qedlorentz}
\partial_\mu A^\mu = 0,
\end{equation}
the gauge freedom is used to reduce the degrees of freedom from four to three. Additionally, the equation of motion in the Lorentz gauge, $\Box A_\mu = 0$, allows for a further gauge fixing by imposing $\Box \lambda = 0$, leaving only two physical degrees of freedom corresponding to the transverse polarizations of the photon. These correspond to the physical helicity states $\lambda^\gamma_{\pm} = \pm 1$.\myspace
As we will see, just as the photon in QED has two physical helicity states due to gauge invariance, the graviton in general relativity has two helicity states, characterized by $\lambda_{\pm} = \pm 2$, as a result of the invariance of the Einstein-Hilbert action under diffeomorphisms.\myspace
What we aim to construct is a Lagrangian that describes the dynamics of the gravitational waves on top of a flat space, as well as its gravitational interactions with matter. We start from the Einstein-Hilbert action in natural units
\begin{equation}\label{eq_g_einsteinhilbert}
S_{EH} = \frac{1}{16\pi G} \int d^4x \sqrt{-g} R,  
\end{equation}  
where $R = g^{\mu\nu} R_{\mu\nu}$ is the Ricci scalar and $R_{\mu\nu}$ is the Ricci tensor. Both quantities are defined from the Rimeann tensor and the Christoffel symbols
\begin{equation}\label{eq_g_RandChris}
\begin{aligned}
&R^{\mu}_{\nu\sigma\rho}=\partial_{\rho}\Gamma^{\mu}_{\nu\sigma}-\partial_{\sigma}\Gamma^{\mu}_{\nu\rho}+\Gamma^\mu_{\alpha\rho}\Gamma^\alpha_{\nu\sigma}-\Gamma^\mu_{\alpha\sigma}\Gamma^\alpha_{\nu\rho}\\
&\Gamma^{\rho}_{\mu\nu}=\frac{1}{2}g^{\rho\sigma}\left(\partial_\mu g_{\sigma\nu}+\partial_{\nu}g_{\sigma\mu}-\partial_{\sigma}g_{\mu\nu}\right),
\end{aligned}
\end{equation}
contracting one of the indices in the first line of Eq.~(\ref{eq_g_RandChris}) such that $ R_{\mu\nu}=R_{\nu}\, ^\sigma \, _{\sigma\nu}$.\myspace
From the linearization of the action in Eq.~(\ref{eq_g_einsteinhilbert}) through an expansion around flat space-time, i.e., Minkowski space-time,
\begin{equation}\label{eq_g_metricexpansion}
\begin{aligned}
&g_{\mu\nu}=\eta_{\mu\nu}+ h_{\mu\nu},\\
&g^{\mu\nu}=\eta^{\mu\nu}- h^{\mu\nu}+ h^{\mu\lambda}h_{\lambda}^\nu+\cdots
\end{aligned}
\end{equation}
with $\left|h^{\mu\nu}\right|\ll 1$, the interpretation of  $h^{\mu\nu}$ as a gravitational wave inducing perturbations in the spacetime geometry is quite direct.\myspace
The expansion of the metric in Eq.~(\ref{eq_g_metricexpansion}) induces an expansion of the Ricci tensor in terms of $R^{(n)}_{\mu\nu}$, where $R^{(n)}_{\mu\nu}$ is of order $h^n$ (see Eq.~(\ref{eq_g_RandChris})), and the expanded Ricci scalar is
\begin{equation}
R=g^{\mu\nu}R_{\mu\nu}=\left(\eta^{\mu\nu}-h^{\mu\nu}+\cdots\right)\left(R^{(1)}_{\mu\nu}+R^{(2)}_{\mu\nu}+\cdots\right).
\end{equation}
With this, the following action for the Einstein gravity in its linearized version is found
\begin{equation}\label{eq_g_SEH}
S_{EH}=-\frac{c^3}{64\pi G}\int d^4x\left[\partial_\mu h_{\alpha\beta}\partial^\mu h^{\alpha\beta}-\partial^\mu h \partial_\mu h+2\partial_\mu h^{\mu\nu}\partial_\nu h-2\partial_{\mu}h^{\mu\nu}\partial_\rho h^{\rho}_{\nu}\right],
\end{equation}
where we have defined $h\equiv h^\mu_\mu$, the trace of $h^{\mu\nu}$, and it has been used that $\sqrt{-g}\approx 1+\frac{1}{2}h+\mathcal{O}(h^2)$.\myspace
Inspired in this linearized version of the Einstein gravity, our goal is to identify a Lagrangian that includes all possible quadratic terms with two derivatives of a massless tensor field $h^{\mu\nu}$ and that respects the local symmetry. The tensor field $h^{\mu\nu}$ is what we will call the graviton.\myspace
To do so, first we need to identify the symmetry group governing general relativity. Indeed, general relativity possesses a huge symmetry group, the group of all coordinate transformations $x^{\mu}\to x^{\prime\mu}(x)$ given that is a sufficiently smooth function of $x$. The variation of the metric under these transformations is what is known as the gauge symmetry of general relativity.\myspace
Using this symmetry and after choosing a reference frame where the expansion in Eq.~(\ref{eq_g_metricexpansion}) meaningfully holds, there remains an extra residual symmetry
\begin{equation}\label{eq_g_residualgauge}
x^\mu\to x^{\prime\mu}=x^\mu+\xi^\mu(x)
\end{equation}
that, in the linearized version of gravity implies a symmetry under the graviton transformation
\begin{equation}\label{eq_g_gauge}
h_{\mu\nu} \to h_{\mu\nu} - \left( \partial_\mu\xi_\nu + \partial_\nu\xi_\mu \right). 
\end{equation}
This gauge transformation in Eq.~(\ref{eq_g_gauge}) for the theory describing gravity, is a natural generalization of Eq.~(\ref{eq_g_qedgauge}) to the tensor case by solely introducing an additional Lorentz index into the function generating the redundant states, the gauge orbits.\myspace
After choosing a suitable normalization for the coefficients of the theory, we find the Pauli-Fierz action as the most general action fulfilling the requirements described above
\begin{equation}\label{eq_g_SPF}
S_2=\frac{1}{2}\int d^4x\left[-\partial_\rho h_{\mu\nu}\partial^\rho h^{\mu\nu}+2\partial_\rho h_{\mu\nu}\partial^\nu h^{\mu\rho}-2\partial_\nu h^{\mu\nu}\partial_\mu h+\partial^\mu h\partial_\mu h\right],
\end{equation}
that can be proven to be the same as the $S_{EH}$ action for linear gravity in Eq.~(\ref{eq_g_SEH}) after some integration by parts and the field rescaling $h_{\mu\nu}\to \left(32\pi G\right)^{-1/2}h_{\mu\nu}$. This rescaling means that now $h_{\mu\nu}$ in the action in Eq.~(\ref{eq_g_SPF}) must be thought of as a classical field with canonical mass dimensions.\myspace
To be fully consistent regarding dimension analysis, from now on the expansion of the metric on top of a Minkowski background must be, without lost of generality, understood to be
\begin{equation}\label{eq_g_metricexpansionkappa}
\begin{aligned}
&g_{\mu\nu}=\eta_{\mu\nu}+\kappa h_{\mu\nu},\\
&g^{\mu\nu}=\eta^{\mu\nu}-\kappa h^{\mu\nu}+\kappa^2 h^{\mu\lambda}h_{\lambda}^\nu+\cdots
\end{aligned}
\end{equation}
where the follwing customary definition has been applied
\begin{equation}
\kappa\equiv\left(32\pi G\right)^{1/2}\equiv\frac{\left(32\pi\right)^{1/2}}{M_{\text{pl}}}
\end{equation}
and the energy dimension of $\kappa$ is shown explicitely to be an inverse power of mass.\myspace
Of the ten degrees of freedom in the symmetric tensor $h_{\mu\nu}$, which represents the graviton field, only two physical degrees of freedom remain after imposing the local symmetries. By applying the gauge invariance described in Eq.~(\ref{eq_g_gauge}), one can adopt the Lorentz\footnote{According to Ref.~\cite{Maggiore:2007ulw}, the designation of Eq.~(\ref{eq_g_lorentz}) as the "Lorentz gauge" is inspired by its analogy with QED (see Eq.~(\ref{eq_g_qedgauge})). However, this terminology originates from a transcription error, as it was actually the physicist Lorenz (without the "t") who proposed it. Nevertheless, due to its widespread use in the literature, this text will adopt the term "Lorentz gauge."} gauge (also called the Hilbert gauge, or the harmonic gauge or the De Donder gauge)
\begin{equation}\label{eq_g_lorentz}
\partial^\nu \tilde{h}_{\mu\nu} = 0 \longleftrightarrow \partial^\nu h_{\mu\nu}=\frac{1}{2}\partial_\mu h,
\end{equation}
where $\tilde{h}_{\mu\nu} = h_{\mu\nu} - \frac{1}{2} \eta_{\mu\nu} h$. These four equations reduce the degrees of freedom to six by choosing four independent functions $\xi_\mu$ satisfying $\Box \xi_\mu = f_\mu (x)$ for any arbitrary $f_\mu (x)$ initial field configuration. In this very same gauge, the equations of motion for the graviton in the linearized theory reduce to
\begin{equation}\label{eq_g_eomlinearnomatter}
\Box \tilde{h}_{\mu\nu}=0,
\end{equation}
which allows us to use four functions $\xi_\mu$ satisfying $\Box \xi^\mu = 0$ to further reduce the number of independent degrees of freedom for the gaviton to two. These two physical degrees of freedom correspond to two independent helicity states, $\lambda_{\pm} = \pm 2$ that we simply characterize as $\pm$.\myspace
Just in the very same way as in electrodynamics, a gauge fixing is required to find the propagator of the graviton field. The Lorenz gauge can be selected by introducing the following piece
\begin{equation}\label{eq_g_gaugefixing}
S_{gf}=-\int d^4x \left(\partial^\nu\tilde{h}_{\mu\nu}\right)^2,
\end{equation}
resulting in a propagator for the graviton $h_{\mu\nu}$ with momentum $k$
\begin{equation}
D_{\mu\nu\rho\sigma}(k)=\left(\eta_{\mu\rho}\eta_{\nu\sigma}+\eta_{\mu\sigma}\eta_{\nu\rho}-\eta_{\mu\nu}\eta_{\rho\sigma}\right)\left(\frac{-i}{k^2-i\epsilon}\right).
\end{equation}
After writing the gauge-fixing piece in Eq.~(\ref{eq_g_gaugefixing}) in terms of $h_{\mu\nu}$ and some integration by parts, one finds the action
\begin{equation}\label{eq_g_fullS2}
S=S_2+S_{gf}=\int d^4x \left[-\frac{1}{2}\partial_\rho h_{\mu\nu}\partial^\rho h^{\mu\nu}+\frac{1}{4}	\partial^\mu h \partial_\mu h\right].
\end{equation}
However, this cannot be the end of the story. The action above describes free gravitons, but has little to do with the gravitational interaction.\myspace
The interaction with matter is added by means of the energy-momentum tensor 
\begin{equation}
S_{\text{matter}}=\frac{\kappa}{2}\int d^4x\,h_{\mu\nu}T^{\mu\nu},
\end{equation}
that changes the equation of motion in Eq.~(\ref{eq_g_eomlinearnomatter}) to  
\begin{equation}\label{eq_g_eomlinearmatter}  
\Box \tilde{h}_{\mu\nu} = -\frac{\kappa}{2} T_{\mu\nu}.  
\end{equation}  
Then, Eqs.~(\ref{eq_g_lorentz}) and (\ref{eq_g_eomlinearmatter}) together imply 
\begin{equation}\label{eq_g_conservationtmunu}
\partial_\mu T^{\mu\nu} = 0,
\end{equation}
 which, written in integral form for the $\nu=0$ component, becomes  
\begin{equation}  
\frac{d}{dt} \int_{V} d^3x \, T^{00} = -\int_{V} d^3x \, \partial_i T^{0i},  
\end{equation}  
indicating that changes in the system's energy within a volume $V$ arise solely from the energy flux across a surface enclosing this volume. This situation does not represent energy exchange between matter and the gravitational field, a requirement for a dynamical theory of gravity, so Eq.~(\ref{eq_g_conservationtmunu}) cannot be exact for curved space-time.\myspace
To address this, the idea is to modify the right-hand side of Eq.~(\ref{eq_g_eomlinearmatter}) so that \(T_{\mu\nu} \to T_{\mu\nu} + t^{(2)}_{\mu\nu}\), where \(t^{(2)}_{\mu\nu}\) is the graviton energy-momentum tensor derived from the Pauli-Fierz action in Eq.~(\ref{eq_g_SPF}), which must therefore be quadratic in \(h_{\mu\nu}\)---hence the superscript \((2)\). With this modification, new equations of motion and conservation laws are obtained:  
\begin{equation} \label{eq_g_SPF3}
\Box \tilde{h}_{\mu\nu} = -\frac{\kappa}{2} \left(T_{\mu\nu} + t^{(2)}_{\mu\nu}\right), \quad \partial^{\mu} \left(T_{\mu\nu} + t^{(2)}_{\mu\nu}\right) = 0.  
\end{equation}
For consistency, the action in Eq.~(\ref{eq_g_SPF}) must now be modified to produce the additional term $-\kappa/2 \, t^{(2)}_{\mu\nu}$, which, as noted, is quadratic in $h_{\mu\nu}$ and dimensionally requires the form $t^{(2)} \sim \partial h \partial h$. To achieve this, one must include in the action a term of the form $\kappa h \partial h \partial h$. Including indices,  
\begin{equation}  
S_3\equiv\kappa \bar{S}_3 = \frac{\kappa}{2} \int d^4x \, h_{\mu\nu} S^{\mu\nu}(\partial h \partial h),  
\end{equation}  
which introduces a triple graviton interaction.\myspace
But we don’t need to stop there. We can now apply the same procedure iteratively to produce vertices with successive insertions of gravitons. At the next order, the equations of motion would be modified, as well as the associated conservation law, Eq.~(\ref{eq_g_SPF3}), to consistently include the next additional term $t^{(4)}_{\mu\nu}$, which is quartic in $h_{\mu\nu}$. Dimensional analysis reveals that such a term must be accompanied by $\kappa^2$. At this point, we would determine what contribution needs to be included in the action to generate this additional term at the level of eoms, and we would find that it corresponds to a four-graviton operator, contributing to a generic action $S_4$, which produces a four-graviton contact vertex.\myspace
All in all, an action representing gravity is of the form
\begin{equation}
S=\int d^4x \left[-\frac{1}{2}\partial_\rho h_{\mu\nu}\partial^\rho h^{\mu\nu}+\frac{1}{4}	\partial^\mu h \partial_\mu h+\frac{\kappa}{2}h_{\mu\nu}T^{\mu\nu}+\kappa \bar{S}_3+\kappa^2 \bar{S}_4+\cdots\right].
\end{equation}
The above discussion gives meaningful context to the expansion in Eq.~(\ref{eq_g_metricexpansionkappa}), as the graviton's triple and quartic self-interaction vertices are automatically generated in the action when performing an expansion in $\kappa \sim 1/\masspl$.\myspace
With all this, the efforts to develop a theory that describes gravity and its interactions from a quantum field-theoretical perspective have resulted in a theory with a gauge symmetry and that contains non-linear interactions among the gravitons. This appears strikingly similar to a non-Abelian gauge theory, such as the one that explains electroweak interactions in the Standard Model. 
\section{An Effective Field Theory for gravitons}
The fact that the parameter $\kappa$ has dimensions of inverse energy, and that the action itself can be constructed as an expansion in this parameter, $\kappa \sim 1/\masspl$, is a clear indication that the quantization of the theory of gravity leads to a non-renormalizable theory.\myspace
This non-renormalizability entails a loss of predictive power, as, in principle, an infinite comparison of amplitudes with experimental data would be required to fix the infinite number of coefficients explicitly appearing in the theory. However, this situation is not unkwnown for us. If our interest is genuinely limited to a low-energy regime, it suffices to introduce an appropriate cutoff in the expansion in $\kappa$. This is evident in this case, given that $\masspl \sim 10^{19} \, \mathrm{GeV}$; practically, effects suppressed by inverse powers of this scale simply vanish.\myspace
The resulting theory from this expansion can be rendered finite order by order in $\kappa$, requiring only a finite, though not necessarily small, set of counterterms. From this perspective, we are dealing with an effective theory that describes the low-energy effects of a more fundamental theory governed by an unknown dynamics, and whose leading-order term is the Einstein-Hilbert action in Eq.~\ref{eq_g_einsteinhilbert} after the expansion of the metric.\myspace
But there is more. In the previous section, we argued for the necessity of nonlinear self-interactions of gravitons—namely, trilinear and quartic vertices—arising from the gauge structure of the theory itself. From the perspective of an effective theory constructed as an expansion in $\kappa$, we encounter a familiar situation. For instance, to include the trilinear graviton coupling at leading order in the operator expansion, one might initially think of a term that can be schematically written—omitting the tensor structure—as $\kappa hhh$. However, a simple dimensional analysis immediately shows that this term is not viable.\myspace
The term $\kappa hhh$ has mass dimension two. To obtain a dimensionless action, two derivatives must be introduced. The term $\kappa h\partial h\partial h$ represents a trilinear coupling and is dimensionally consistent. The same reasoning applies to the construction of the quartic self-coupling. The only dimensionally viable option to generate it at order $\kappa^2$ in the expansion of effective operators is $\kappa^2 hh\partial h\partial h$.\myspace
To produce vertices with arbitrary graviton insertions, the same procedure is followed by consecutively adding derivatives to construct dimensionless terms in the action. This hierarchy of operators, organized by the number of derivatives, directly mirrors the structure of the chiral Lagrangian.
\subsection{Resemblance between Einstein-Hilbert and the Chiral Lagrangian}
From the previous section it can be seen that the Einstein-Hilbert gravity shares some features with the non-linear chiral Lagrangian presented in Chapter \ref{chp:effective_theories}, that describes the dynamics of massless Goldstones with strong interactions at low energies. A nice discussion about the possibility for the gravitons to be understood as Goldstones coming from a dynamical breaking of some UV theory can be found in Refs.~\cite{Alfaro:2010ui, Alfaro:2012fs}.\myspace
First, both theories are non-renormalizable, in the traditional sense. When one chooses a cut-off in the operator expansion and the precision of the calculation is fixed, a finite number of parameter redefinitions are needeed to get rid of all the ultraviolet divergences appearing in the different processes.\myspace
Second, the interactions of both theories possess a high-energy scale weighting the arbitrary-high momenta insertions of the various fields; Goldstones or gravitons depending on the theory we are referring to. The counterpart of the pion constant $f_\pi$ that appears in the chiral Lagrangian, in the gravity side is the planck Mass, $M_{\text{pl}}\sim 10^{19}$ GeV, \myspace
Third, the power counting. As discussed before, the interactions among gravitons in the Einstein-Hilbert theory happen to be naturally organised as a derivative expansion, which in the chiral Lagrangian language would represent the different $\mathcal{O}(p^n)$ contributions to a process. With this chiral counting, the Lagrangian extracted from the action in Eq.~(\ref{eq_g_fullS2}) represents $\lag_2$, the leading order in the chiral expansion. For consistency $T_{\mu\nu}$ must be assigned "chiral order" two, which is also consistent to the equation of motion in Eq.~(\ref{eq_g_eomlinearmatter}).\myspace
Apart from these similarities, there is a notorious difference between these two theories and it is that unlike the chiral Lagrangian that, is built upon a global symmetry, the chiral symmetry, the Einstein-Hilbert theory has a local gauge symmetry, that is responsible for the reduction of the number of degrees of freedom in the low-energy spectrum.
\subsection{Higher orders and the scalaron}
As in the case of the chiral Lagrangian, the next order in the expansion contributes to processes at $\mathcal{O}(p^4)$, and the set of operators forming $\lag_4$ contains terms with exactly four derivatives. Using the building blocks with which we construct the Einstein-Hilbert action—representing the lowest order in the effective operator expansion—it is also possible to obtain terms with four derivatives.\myspace
To see this, recall that schematically $R \sim \partial \partial g$. In pure gravity---i.e., in the absence of matter and cosmological constant ($\Lambda=0$)--- at the next order in the chiral counting, we can only construct the following action
\begin{equation}\label{eq_g_S4}
S_{\text{NLO}} = \int d^4x \sqrt{-g} \left[ \alpha_1 R^2 + \alpha_2 \left( R_{\mu\nu} \right)^2 + \alpha_3 \left(R_{\mu\nu\alpha\beta} \right)^2 \right].
\end{equation}
In fact, one can reduce these three operators to two, since they are related by topology through the Gaus-Bonnet term 
\begin{equation}\label{eq_g_GBterm}
GB=R_{\mu\nu\rho\sigma}R^{\mu\nu\rho\sigma}-4R_{\mu\nu}R^{\mu\nu}+R^2.
\end{equation}
Since the integral of GB leads to no contribution except for topologically non-trivial solutions, see for instance Ref.~\cite{Alvarez-Gaume:2015rwa}, the three operators in Eq.~(\ref{eq_g_S4}) are not independent. Consequently, the NLO basis of pure gravity reduces to two independent operators.\myspace
Interestingly enough, the theory in Eq.~(\ref{eq_g_S4}) is free from ultraviolet divergences when, assuming the Lorentz gauge, the equations of motion in the absence of matter and a cosmological constant ($R_{\mu\nu} = 0$) are applied. For more details, the interested reader is referred to Ref.~\cite{Espriu:2009ju}.\myspace
Considering the structure of quadratic corrections in gravity, Eq.~(\ref{eq_g_S4}), these types of theories are generically referred to as $f(R)$ theories, as the action representing the dynamics of pure gravity is generally written as  
\begin{equation}
S=\frac{1}{2\kappa^2}\int d^4x \sqrt{-g}f(R).
\end{equation}  
The idea is that each possible theory of gravity is determined by a specific, generally nonlinear, function $f(R)$. For instance, $f(R)=R$ corresponds to the Einstein-Hilbert action, the leading-order term in the effective theory. See Refs.~\cite{Sotiriou:2008rp, DeFelice:2010aj} for more information.\myspace
A particularly interesting case from the perspective of inflation, and the one chosen for our study, is given by  
\begin{equation}\label{eq_g_starobinski}
\begin{aligned}
&f(R)=R+\alpha R^2,\\
&S=\frac{1}{2\kappa^2}\int d^4x \sqrt{-g}\left(R+\alpha R^2\right),
\end{aligned}
\end{equation}  
known as the Starobinsky model, Ref.~\cite{STAROBINSKY198099}.\myspace
From the viewpoint of effective theories, this model, that is obtained by setting $\alpha_1=\alpha$ and $\alpha_{i=2,3}=0$ in Eq.~(\ref{eq_g_S4}),  represents the $R^2$-order corrections to the Einstein-Hilbert theory.\myspace
The equations of motion for $R$ with $f(R) = R + \alpha R^2$ are  
\begin{equation}\label{eq_g_eoms}
R+6\alpha\Box R = 0,
\end{equation}  
which, for nonvanishing values of $\alpha$, immediately resembles the Klein-Gordon equation
\begin{equation}
\left(\Box+m^2\right) R=0,
\end{equation}
describing a free scalar field with a mass given by  
\begin{equation}\label{eq_g_scalaronmass}
 m^2 = \frac{1}{6\alpha}.
\end{equation}  
Indeed, by performing a conformal transformation  
\begin{equation}\label{eq_g_conformaltrans}
\tilde{g}_{\mu\nu} = \Omega^2 g_{\mu\nu}, \qquad \Omega^2 = \exp\left(\beta\varphi\right), \qquad \beta = \frac{1}{\masspl} \sqrt{\frac{16\pi}{3}},
\end{equation}  
to rewrite the action in Eq.~(\ref{eq_g_starobinski}) in the Einstein frame (instead of the Jordan frame, see Ref.~\cite{Maggiore:2007ulw}), expressed in terms of quantities denoted with tildes, it is found that  
\begin{equation}
R = \Omega^2 \tilde{R} - 6 \Omega^{-1} \Box \Omega.
\end{equation}  
In this frame, the action becomes  
\begin{equation}\label{eq_g_actioneinsteinframe}
\begin{aligned}
&S = \int d^4x \sqrt{-\tilde{g}} \left( \frac{1}{2\kappa^2} \tilde{R} + \frac{1}{2} \partial^\mu \varphi \partial_\mu \varphi - V\left(\varphi\right) \right), \\
&V\left(\varphi\right) = \frac{3}{4\kappa^2} m^2 \left( \exp\left(-\beta\varphi\right) - 1 \right)^2,
\end{aligned}
\end{equation}  
where $m$ and $\beta$ are defined in Eqs.~(\ref{eq_g_scalaronmass}) and (\ref{eq_g_conformaltrans}), respectively.\myspace
Expanding the potential $V\left(\varphi\right)$ for $ \varphi / \masspl \ll 1 $,  
\begin{equation}\label{eq_g_potential}
V\left(\varphi\right) \approx \frac{1}{2}m^2\varphi^2 - 2\sqrt{\frac{\pi}{3}}\frac{m^2}{\masspl}\varphi^3 + \frac{14\pi}{9}\frac{m^2}{\masspl^2}\varphi^4 + \cdots,
\end{equation}  
we find that the action at next to leading order in Eq.~(\ref{eq_g_starobinski}), rewrites as the leading-order Einstein-Hilbert term plus a massive, dynamical scalar field with self-interactions. This scalar field receives the name of \textit{scalaron}.\myspace
A natural expansion requires that the contributions $\alpha R^2$ in Eq.~(\ref{eq_g_starobinski}) be much smaller than those of $R$, or at least not dominant. However, in this situation, the equation of motion at the lowest order, $R=0$ in pure gravity, could be applied perturbatively, canceling the $R^2$ term and preventing any curvature corrections. Thus, the chiral counting that organizes the effective expansion cannot be strictly understood as being carried out in this theory with corrections. This fact imposes theoretical constraints on the parameter $\alpha$ and, consequently, on the mass of the scalaron itself.\myspace
Recalling that schematically $R\sim \partial\partial g\sim p^2$, there will be an equivalence between the $R^2$ corrections and the scalar for an energy regime such that  
\begin{equation}  
p^2\sim \alpha p^4 \Rightarrow s\sim \frac{1}{\alpha}=6m^2.  
\end{equation}
The above result is truly promising since, naively, we can derive insights into corrections to the Einstein-Hilbert theory in pure gravity—calculations that are a priori highly complex due to the proliferation of indices—at the cost of introducing a scalar field. With the subsequent expansion of the metric, $g_{\mu\nu} = \eta_{\mu\nu} + \kappa h_{\mu\nu}$, the scalaron couples non-linearly to the graviton.
\section{The scalar spectrum of the Einstein-Hilbert theory}
We saw in the previous section how adding perturbative corrections proportional to $R^2$ to the Einstein-Hilbert theory in pure gravity is equivalent to introducing a massive scalar field coupled to gravity.\myspace
This scalar field, the scalaron, has a squared mass inversely proportional to the coupling constant $\alpha$ in the next-to-leading order effective theory of pure gravity, as given by Eq.~(\ref{eq_g_scalaronmass}). The more significant the corrections to Einstein-Hilbert, the smaller the scalaron's mass. Conversely, the smaller the coupling constant $\alpha$ associated with the higher-order term, the larger the scalaron's mass, leading to greater decoupling from the pure gravity modes. The limit $\alpha \to 0$, which one might expect to recover as the mass tends to infinity ($m \to \infty$), can be pathological due to the structure of the potential. In its expanded form, as shown in Eq.~(\ref{eq_g_potential}), the potential includes interactions proportional to $m$.\myspace
Thus, in a regime where the scalar contribution is non-negligible, we find that the dynamics of the linearized Einstein theory up to the next order, as an effective theory, is described by two types of particles: gravitons, which are spin-2 particles, and the scalaron, a spin-0 particle.\myspace  
Again, this situation is not unfamiliar. Revisiting earlier chapters, we addressed the chiral Lagrangian with a light Higgs to effectively describe electroweak interactions up to the Terascale, the EChL or HEFT. This theory quickly violates unitarity for non-standard values of the couplings, and the relevant degrees of freedom are the spin-1 gauge bosons and the spin-0 Higgs boson.\myspace
As we also noticed when studying $2\to 2$ processes, the $WW$ pair shares quantum numbers with the $hh$ state; in particular, both can be found in states with $IJ = 00 $. This leads to mixing during the unitarization process of the scalar partial waves, necessitating a coupled-channel formalism. Scalar resonances emerge from this unitarization process, which are understood to be dynamically generated by an ultraviolet theory with strong interactions.\myspace
Similarly, we can expect such mixing in the $f(R) = R + \alpha R^2$ theory. By studying the relevant $2 \to 2$ processes projected onto the scalar channel, the quantum numbers of the graviton-graviton and the scalaron-scalaron pairs mix, and scalar resonances may emerge from these interactions, smoothing out the non-unitary behavior of a theory built as an expansion in derivatives.\myspace
Previous works, such as those referenced in Refs.~\cite{Oller:2022ozd, Delgado:2022qnh, Oller:2024neq}, have explored the presence of scalar resonances---and for higher values of $J$ in the case of Ref.~\cite{Delgado:2022qnh}---. These studies perform a detailed analysis of the $++++$ channel at tree level within the IAM framework, searching for resonances in the poles of the unitarized amplitude. Their results do not reveal genuine scalar resonances of the theory; instead, they find a pole that vanishes in the soft limit of the calculation, identifying it as a mathematical artifact, the \textit{graviball} that will be presented in the following sections.\myspace
One of the motivations for this project is precisely to determine, based on the experience with the chiral Lagrangian, whether the inclusion of higher orders in the effective theory gives rise to scalar resonances intrinsic to the theory.
\subsection{Relevant amplitudes}
To study the scalar projection of $2 \to 2$ processes, three relevant amplitudes come into play: $h^{\lambda_1}_{\mu_1\nu_1} h^{\lambda_2}_{\mu_2\nu_2} \to h^{\lambda_3}_{\mu_3\nu_3} h^{\lambda_4}_{\mu_4\nu_4}$, $h^{\lambda_1}_{\mu_1\nu_1} h^{\lambda_2}_{\mu_2\nu_2} \to \varphi \varphi$, and $\varphi \varphi \to \varphi \varphi$. These are schematically represented as $\mathcal{A}^{hhhh}$, $\mathcal{A}^{hh\varphi\varphi}$, and $\mathcal{A}^{\varphi\varphi\varphi\varphi}$, respectively.\myspace
In general, we define the momenta and helicities---for the gravitons---of the incoming particles as $p_1$, $\lambda_1$, $p_2$, $\lambda_2$, and the outgoing particles as $p_3$, $\lambda_3$, $p_4$, $\lambda_4$. The helicity amplitudes are then expressed as 
\begin{equation}
\begin{aligned}
&\mathcal{A}^{hhhh}_{\lambda_1, \lambda_2, \lambda_3, \lambda_4} = \matelemgen{p_1, \lambda_1; p_2, \lambda_2}{p_3, \lambda_3; p_4, \lambda_4}{\mathcal{A}^{hhhh}}, \\
&\mathcal{A}^{hh\varphi\varphi}_{\lambda_1, \lambda_2} = \matelemgen{p_1, \lambda_1; p_2, \lambda_2}{p_3; p_4}{\mathcal{A}^{hh\varphi\varphi}}, \\
&\mathcal{A}^{\varphi\varphi\varphi\varphi} = \matelemgen{p_1; p_2}{p_3; p_4}{\mathcal{A}^{\varphi\varphi\varphi\varphi}},
\end{aligned}
\end{equation}
where $\lambda_i$ takes values $\pm 2$, denoted as $\pm$. The usual Mandelstam relations are defined as $s = (p_1 + p_2)^2$, $t = (p_1 - p_3)^2$, and $u = (p_1 - p_4)^2$.\myspace
Now, we present the explicit amplitudes for the three processes under study in this section. All of them are computed at tree level using the FeynGrav program (Ref.~\cite{Latosh:2023zsi}) and our private Mathematica license.\myspace
For each helicity combination, the diagrams involved in the elastic graviton scattering process include the $s$-, $t$-, and $u$-channel diagrams with a graviton exchange and a four-graviton contact diagram and are represented in Figure~\ref{fig_g_diags_gggg}. The explicit sixteen decomposed helicity amplitudes $\mathcal{A}^{hhhh}_{\lambda_1, \lambda_2, \lambda_3, \lambda_4}$ are gathered in Table~\ref{tab_g_elasticgraviton}. 
\begin{figure}
\centering
\includegraphics[clip,width=6cm,height=3.5cm]{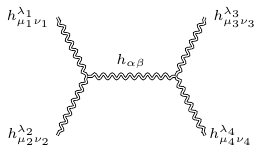}
\includegraphics[clip,width=6cm,height=3.5cm]{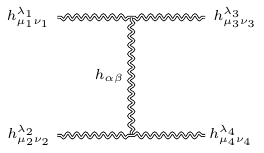}
\includegraphics[clip,width=6cm,height=3.5cm]{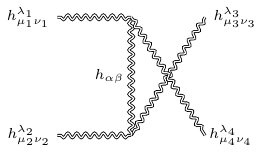}
\includegraphics[clip,width=6cm,height=3.5cm]{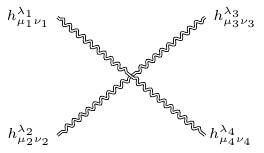}
\caption{\small{Diagrams contributing to the elastic scattering of gravitons in the helicity states $\lambda_i=\pm 2$.}}
\label{fig_g_diags_gggg}
\end{figure}
\begin{table}[h!]
\centering
\renewcommand{\arraystretch}{1.5}
\begin{tabular}{|c|c|c|c|c|}
\hline
  & \(\ket{+; +}\) & \(\ket{+; -}\) & \(\ket{-; +}\) & \(\ket{-; -}\) \\ \hline
\(\bra{+; +}\) & \(\frac{\kappa^2}{4} \frac{s^3}{tu}\) & \(0\) & \(0\) & \(0\) \\ \hline
\(\bra{+; -}\) & \(0\) & \(\frac{\kappa^2}{4} \frac{u^3}{st}\) & \(\frac{\kappa^2}{4} \frac{t^3}{su}\) & \(0\) \\ \hline
\(\bra{-; +}\) & \(0\) & \(\frac{\kappa^2}{4} \frac{t^3}{su}\) & \(\frac{\kappa^2}{4} \frac{u^3}{st}\) & \(0\) \\ \hline
\(\bra{-; -}\) & \(0\) & \(0\) & \(0\) & \(\frac{\kappa^2}{4} \frac{s^3}{tu}\) \\ \hline
\end{tabular}
\caption{Table summarizing the sixteen helicity combinations for elastic graviton scattering.}
\label{tab_g_elasticgraviton}
\end{table}\myspace
Parity ($T_{\lambda_1,\lambda_2,\lambda_3,\lambda_4}=T_{-\lambda_1,-\lambda_2,-\lambda_3,-\lambda_4}$) and time-reversal invariance ($T_{\lambda_1,\lambda_2,\lambda_3,\lambda_4}=T_{\lambda_3,\lambda_4,\lambda_1,\lambda_2}$), corresponding to $PT$ symmetry, are explicitly shown in Table~\ref{tab_g_elasticgraviton}, whose entries are fully symmetric.\myspace
The crossed channel, i.e., $h^{\lambda_1}_{\mu_1\nu_1}h^{\lambda_2}_{\mu_2\nu_2}\to \varphi \varphi$, has contributions from four diagrams that are depicted in Figure~\ref{fig_g_diags_ggss}. First, there is a contact vertex involving two gravitons and two scalarons. Next, the $s$-channel includes a graviton exchange, followed by two diagrams in the $t$- and $u$-channels with a scalar exchange. Four helicity combinations are possible, summarized as  
\begin{figure}
\centering
\includegraphics[clip,width=6.0cm,height=3.5cm]{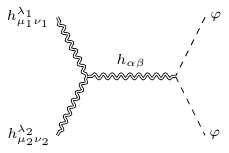}
\includegraphics[clip,width=6.0cm,height=3.5cm]{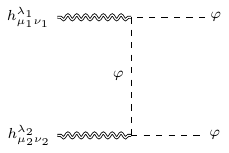}
\includegraphics[clip,width=6.0cm,height=3.5cm]{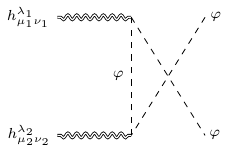}
\includegraphics[clip,width=6.0cm,height=3.5cm]{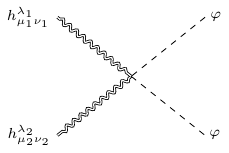}
\caption{\small{Diagrams contributing to the crossed process $hh\to \varphi\varphi$ where the gravitons are in the helicity states $\lambda_i=\pm 2$.}}
\label{fig_g_diags_ggss}
\end{figure}
{\small
\begin{equation}
\mathcal{A}^{hh\varphi\varphi}=-\frac{\kappa^2}{64s^2}\left(2 s^3 (h_1 h_2 + 1) + 2 s (h_1 h_2 + 1) (t - u)^2 - \frac{s \left(s^2 + 2 s (t + u) + (t - u)^2\right)^2}{\left(m^2 - t\right) \left(m^2 - u\right)}\right),
\end{equation}}
where $h_{1,2}$ represents the helicity with a normalization such that $h_{1,2}=\pm 1$.\myspace
For elastic scattering of scalarons, seven types of diagrams are involved. Four of these, namely the $s$-, $t$-, $u$-channel, and contact diagrams with scalar exchange, are identical to the equivalent Higgs process in Eq.~(\ref{eq_ET_tree_hh_2}) with the replacements  
\begin{equation}
\lambda_3 \to -2 \sqrt{\frac{\pi}{3}} \frac{m^2}{\masspl v}, \qquad \lambda_4 \to \frac{56 \pi}{9} \frac{m^2}{\masspl^2}.
\end{equation}
The remaining three diagrams correspond to the $s$-, $t$-, and $u$-channels with graviton exchange. All of them together are shown in Figure~\ref{fig_g_diags_ssss}.
\begin{figure}
\centering
\includegraphics[clip,width=6.0cm,height=3.5cm]{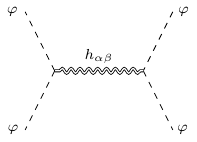}
\includegraphics[clip,width=6.0cm,height=3.5cm]{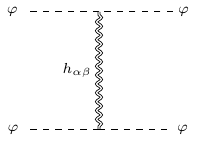}
\includegraphics[clip,width=6.0cm,height=3.5cm]{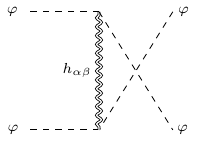}
\includegraphics[clip,width=6.0cm,height=3.5cm]{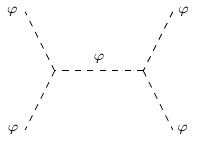}
\includegraphics[clip,width=6.0cm,height=3.5cm]{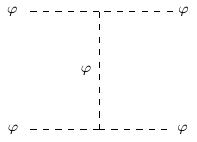}
\includegraphics[clip,width=6.0cm,height=3.5cm]{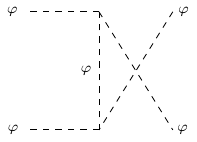}
\includegraphics[clip,width=6.0cm,height=3.5cm]{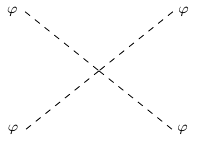}
\caption{\small{Diagrams contributing to the elastic scattering of scalarons.}}
\label{fig_g_diags_ssss}
\end{figure}
The total amplitude for the elastic scattering of scalarons is
{\small
\begin{equation}
\begin{aligned}
&\mathcal{A}^{\varphi\varphi\varphi\varphi}=\frac{1}{12} \kappa ^2 \left(18 m^4 \left(\frac{1}{m^2-s}+\frac{1}{m^2-t}+\frac{1}{m^2-u}\right)-14 m^2-\right.\\
&\left.\frac{3}{8stu}\left(-s^2 \left(6 t^2+7 t u+6 u^2\right)+s^3 (t+u)+s (t+u) \left(t^2-8 t u+u^2\right)+t u \left(t^2-6 t u+u^2\right)\right)\right).
\end{aligned}
\end{equation}}
\myspace
With these amplitudes, we can construct the associated partial waves using Eq.~\ref{eq_un_pw_wigner}, where $\lambda = \lambda_1 - \lambda_2$ and $\lambda^\prime = \lambda_3 - \lambda_4$. As mentioned earlier, we will focus on the $++++$ amplitudes, so $\lambda = \lambda^\prime = 0$, and the Wigner functions reduce to Legendre polynomials, which for the $J = 0$ projection is $P_0\left(\cos\theta\right) = 1$.\myspace
For the elastic graviton-graviton case, the partial waves are not well-defined even at tree level, as can be directly seen in Table~\ref{tab_g_elasticgraviton}, since they diverge for $t \to 0$ ($\cos\theta \to \pm 1$), requiring a regularization.\myspace
Thus, the integral in Eq.~\ref{eq_un_pw_wigner} is redefined over $\cos\theta \in \left[-1 + \eta, 1 - \eta\right]$, with $\eta > 0$, and the physical region where the partial wave is defined is $s > \mu^2$. These two free parameters are related because crossing symmetry requires $t < -\mu^2$, as noted by the authors in Ref.~\cite{Delgado:2022uzu}, taking into account that for massless external states $t=-s/2(1-\cos\theta)$ so
\begin{equation}
\int_{-1 + \eta}^{1 - \eta} d\cos\theta = \frac{2}{s} \int_{-s + \mu^2}^{-\mu^2} dt,
\end{equation}
from which it follows that
\begin{equation}
\mu^2 = \eta \frac{s}{2}.
\end{equation}\myspace
The result of this integration for the three processes are
\begin{equation}
t_0^{hhhh}=\frac{s}{2\masspl^2}\log\left(\frac{2}{\eta}\right)
\end{equation}
\begin{equation}
t_0^{hh\varphi\varphi}=-\frac{s+8 m^2}{24 \masspl^2}+\frac{m^4}{2\sigma\masspl^2 s}\log\left(\frac{1+\sigma}{1-\sigma}\right)
\end{equation}
{\small
\begin{equation}
\begin{aligned}
t_0^{\varphi\varphi\varphi\varphi}=& \frac{1}{24\masspl^2 s \left(s-m^2\right)\left(s-4m^2\right)}\left[s^4\left(12 \log \left(\frac{2}{\eta }\right)-11\right)+m^2 s^3 \left(55-60 \log \left(\frac{2}{\eta }\right)\right)\right.\\
&\left.-4 m^4 s^2 \left(19-18 \log \left(\frac{2}{\eta }\right)-18 \log \left(\frac{s-3 m^2}{m^2}\right)\right)-4m^6s\left(-31+6\log\left(\frac{2}{\eta}\right)\right.\right.\\
&\left.\left. -18\log\left(\frac{s-3m^2}{m^2}\right)\right)+16m^8\right],
\end{aligned}
\end{equation}}
where $\sigma$ is the two-body phase space $\sigma=\sqrt{1-\frac{4m^2}{s}}$.\myspace
All of them are gathered in a $t_0$ scalar wave matrix defined as
\begin{equation}\label{eq_g_t0matrix}
t_0=\begin{pmatrix}
t_0^{hh} & t_0^{h\varphi}\\
t_0^{h\varphi} & t_0^{\varphi\varphi}
\end{pmatrix},
\end{equation}
that is suitable for the unitarization of the processes in the context of the coupled-channels formalism, similar to Eq.~(\ref{eq_un_wavematrix}) in the context of the HEFT.
\subsection{Unitarization and Riemann sheets}
The unitarization method chosen for the previous partial waves is the K-matrix in its improved version, the IK-matrix, described in Chapter \ref{chp:unitarization}. This unitarization method admits amplitudes calculated solely at tree level, adding by hand a physical cut for the relevant continuations.\myspace
In contrast, as we have seen, the IAM requires the calculation of the amplitude at next-to-leading order for all three amplitudes. These amplitudes are not currently available and are exceedingly laborious to compute due to the proliferation of indices. To the best of our knowledge, no literature exists to date that includes these amplitudes with the scalar contribution at the one-loop level.\myspace
As previously mentioned in Chapter \ref{chp:unitarization}, the presence of both massive particles, such as the scalaron, and massless particles, such as the graviton, in the asymptotic states of the processes for the unitarization of the scalar channel leads to an analytical structure of the amplitudes with different cuts on the real axis. Through these cuts, the different partial waves can be analytically continued to access the different Riemann sheets. In general, with $n$ coupled channels, $2^n$ Riemann sheets open up.\myspace
In our case, we have two cuts: the first at the origin and the second at $\text{Re}\,s = 4m^2$, where $m$ is the mass of the scalaron, as defined in Equation (\ref{eq_g_scalaronmass}), resulting in four Riemann sheets.
\subsection{The Graviball}
According to previous bibliography, the presence of the graviball should be recovered in our calculation, see for instance Ref.~\cite{Guiot:2020pku}. The graviball is a pole that appears at very low energies in the $S$-wave of graviton-graviton scattering in pure gravity, i.e. it is a 2-graviton compound by gravitational forces. Whether it constitutes a genuine resonance or not is unclear.\myspace
For example, the authors of Refs.~\cite{Oller:2022ozd,Oller:2024neq} indeed find it and they argue for the presence of a resonance, whose position depends on a parameter $a $ that regulates infrared divergences appearing for the propagation of massless modes. This parameter redefines the energy scale $Q^2 = \pi G^{-1}/\log a $, which is characteristic of the elastic graviton process. The authors find the position of the graviball at $s = \left(0.22 - i 0.63\right)(G \log a)^{-1} $. Under the criterion used throughout this work, this cannot be classified as a resonance because it does not satisfy a natural relationship between its mass and width \textit{à la Wigner}.\myspace
Instead, the authors of Refs.~\cite{Delgado:2022qnh,Delgado:2022uzu} also identify this structure but explain it as an artifact of introducing the parameter $\mu^2$, which regulates the infrared misbehavior of partial waves, analogous to $a$ in the previous case. Using their unitarized analysis with the IAM at next-to-leading order, they determine that the position of this pole is consistent with $s = 0$ as $\mu\to 0$, distancing it from any physical significance as a dynamical state.\myspace
Our work aligns with this latter scenario. First, we verified that, indeed, using a single channel for unitarization, the IK-matrix reproduces the graviball. Furthermore, as the infrared cut-off tends to zero, $\mu \to 0$, the pole moves closer and closer to the origin. This single-channel calculation has no scalar contribution; it represents a perturbative calculation of pure gravity at leading order.\myspace 
In Figure~\ref{fig_g_graviball}, the position we obtain for the graviball (blue points) is shown for values of $\mu \in (0.001, 0.5)\masspl$ and two UV cut-offs, $\Lambda = 0.8\masspl$ and $\Lambda=1.0\masspl$. Recall that quantities with a bar are in units of the Planck mass and are therefore dimensionless. We also show, for comparison, the position of the pole from Ref.~\cite{Oller:2022ozd} with a red point.
\begin{figure}
\centering
\includegraphics[width=0.47\textwidth]{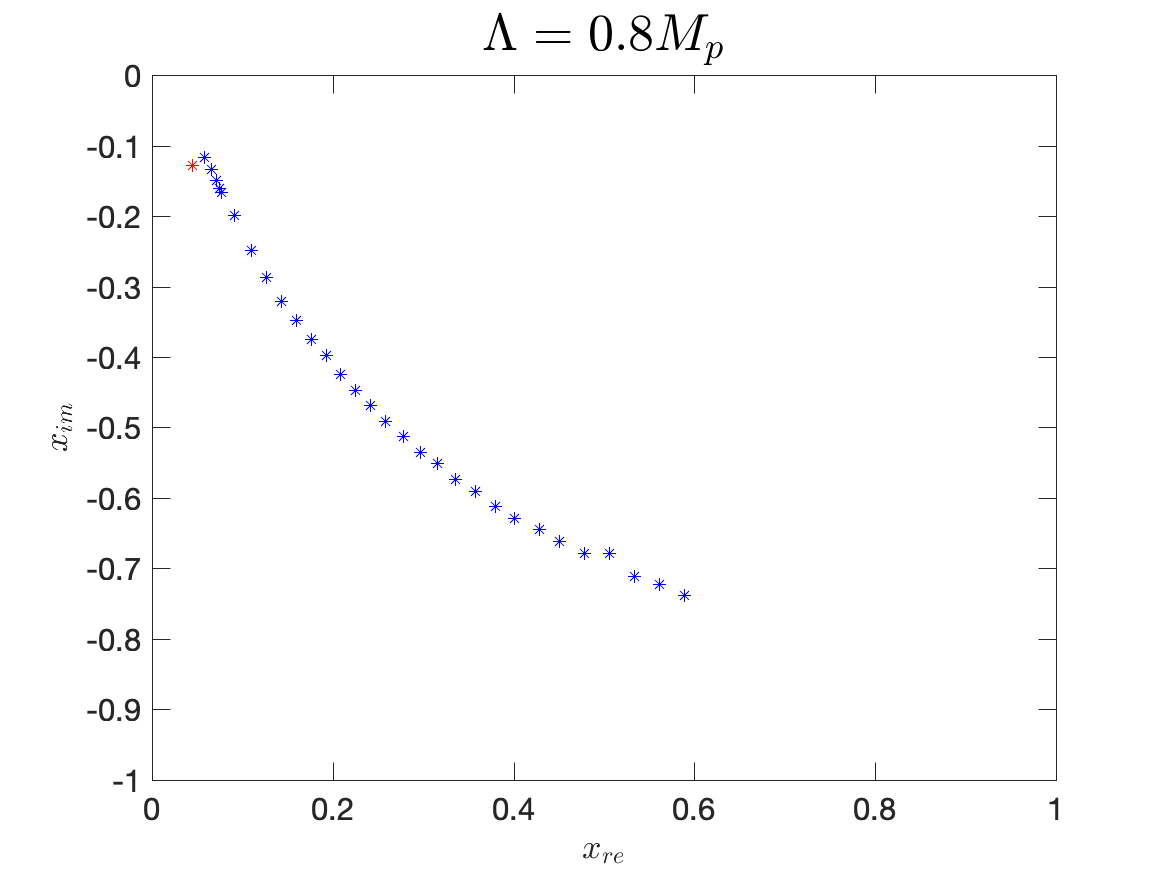}
\includegraphics[width=0.47\textwidth]{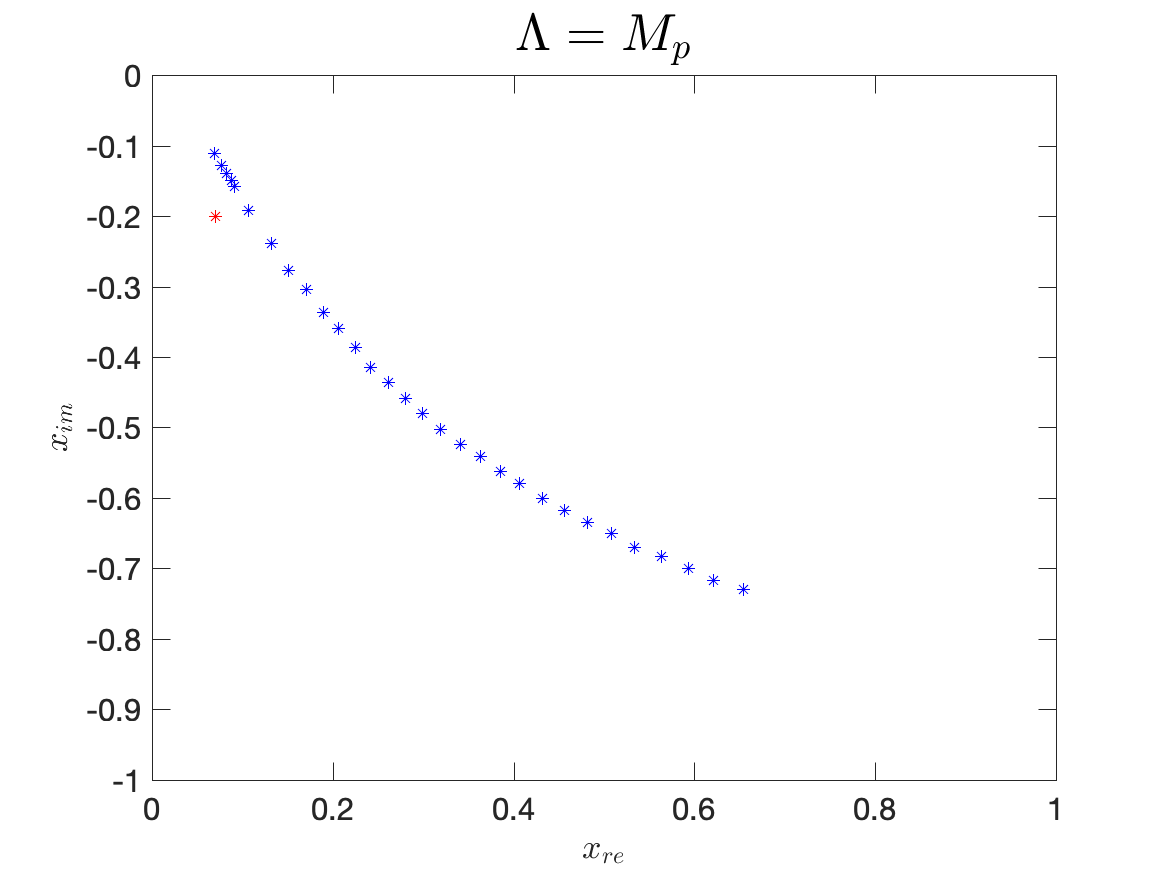} 
\caption{\small{Position of the graviball (blue points) for $\bar{\mu} \in (0.001, 0.5)$ in pure gravity at leading order and two values of $\Lambda = 0.8 \masspl$ (left) and $\Lambda = \masspl$ (right). The results show a trend toward the origin, ruling out the possibility of it being a physical resonance. The pole from Ref.~\cite{Oller:2022ozd} is also indicated with a red point.}}
\label{fig_g_graviball}
\end{figure}
\subsection{Resonances at the Planck scale}
In this section, we present the results obtained for scalar resonances that are generated dynamically.\myspace
All dimensional quantities and plots are shown in units of the Planck mass. Thus, we define:
\begin{equation}
\bar{m} = \frac{m}{\masspl}, \quad \bar{\mu} = \frac{\mu}{\masspl}, \quad \bar{\Lambda} = \frac{\Lambda}{\masspl}, \quad \bar{s} = \frac{s}{\masspl^2},
\end{equation}
where $m$ is the mass of the scalaron, $\mu$ is the IR cut-off regulating the bad behavior of the partial wave in the extremes, $\Lambda$ is the UV regulator that has to be included in the unitarization process and $s$ is the Mandelstam variable that is promoted to a complex quantity.\myspace
For this work we select values of $\bar{m}>0.5$ so the threshold opened for the scalaron, situated at $\bar{s}=4\bar{m}^2$, is pushed away up to the postplanckian scale and we do not need to care about extra analytical continuations of the partial waves across that high-energy cut.\myspace
We did not find significant variations in the results when changing the ultraviolet cut-off $\bar{\Lambda}$, so our analysis is restricted to two parameters: the infrared cut-off $\bar{\mu}$, and the scalaron mass $\bar{m}$, which is related to the parameter $\alpha$ through Eq.~(\ref{eq_g_scalaronmass}). This equation indicates the importance of the $R^2$ corrections to Einstein-Hilbert gravity. From this point forward, for the remainder of the analysis, we fix $\bar{\Lambda} = 0.9$.\myspace 
On the other hand, our Mathematica code allows us to reach extremely small values of the infrared cut-off, on the order of $\bar{\mu} \sim 10^{-80}$. However, although it may be tempting to approach the $\bar{\mu} \to 0$ limit as realistically as possible, the appearance of logarithms makes us cautious with these values. Specifically, the results presented in this section should be considered tentative, as we explore values of $\bar{\mu}$ down to $10^{-10}$ and focus on the observed trend. A more detailed analysis of the logarithmic divergences in the unitarized waves is required to access smaller values of $\bar{\mu}$.\myspace
We focus separately on two different values of the dimensionless scalaron mass, in particular $\bar{m}=0.8$ and $\bar{m}=0.6$.
\begin{center}
$\underline{\bar{m}=0.8}$
\end{center}
We begin with $\bar{m} = 0.8$. Fig~\ref{fig_g_scalaron_m08} shows the imaginary part of the $(2,2)$ element of the unitarized scalar wave on the second Riemann sheet. This matrix element represents the unitarized scattering process $\varphi\varphi \to \varphi\varphi$. The different panels display various values of $\bar{\mu} = 10^{-2},\,10^{-4},\,10^{-8}$, and $10^{-10}$, following a clockwise order. The parameters $\bar{\Lambda} = 0.9$ and $\bar{m} = 0.8$ are kept fixed.
\begin{figure}
\centering
\includegraphics[clip,width=7.0cm,height=6.5cm]{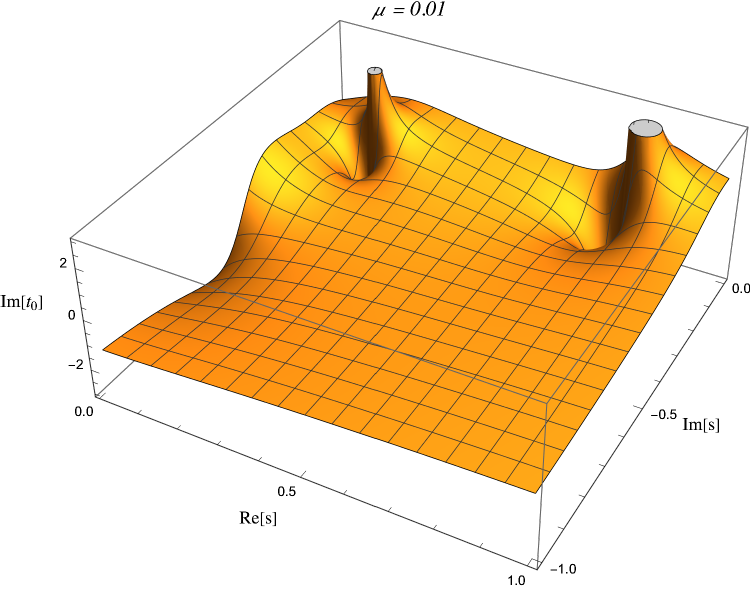}
\includegraphics[clip,width=7.0cm,height=6.5cm]{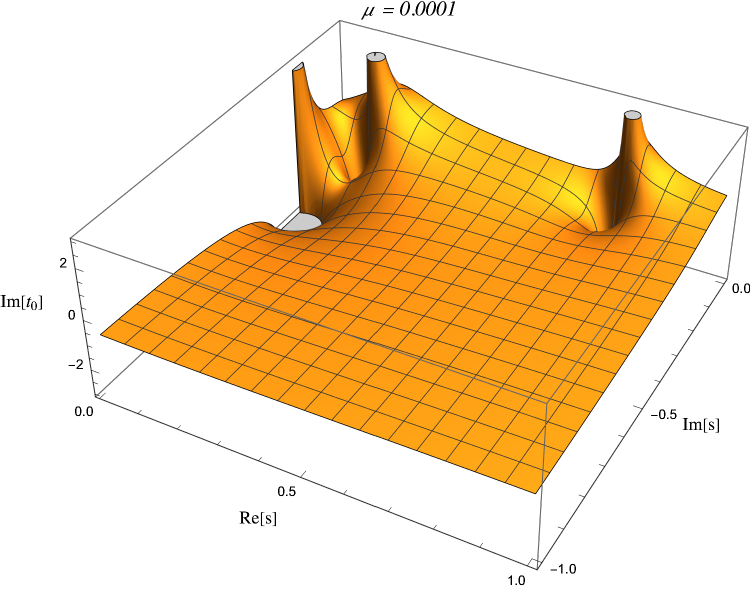}
\includegraphics[clip,width=7.0cm,height=6.5cm]{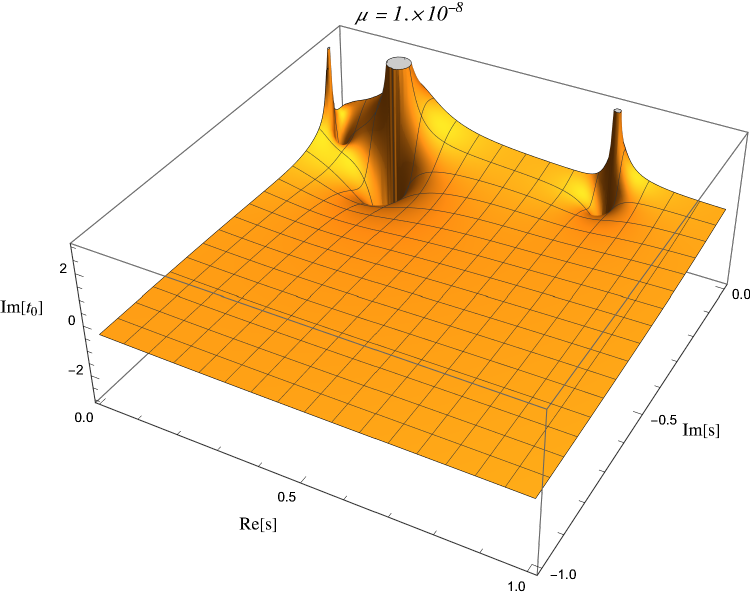}
\includegraphics[clip,width=7.0cm,height=6.5cm]{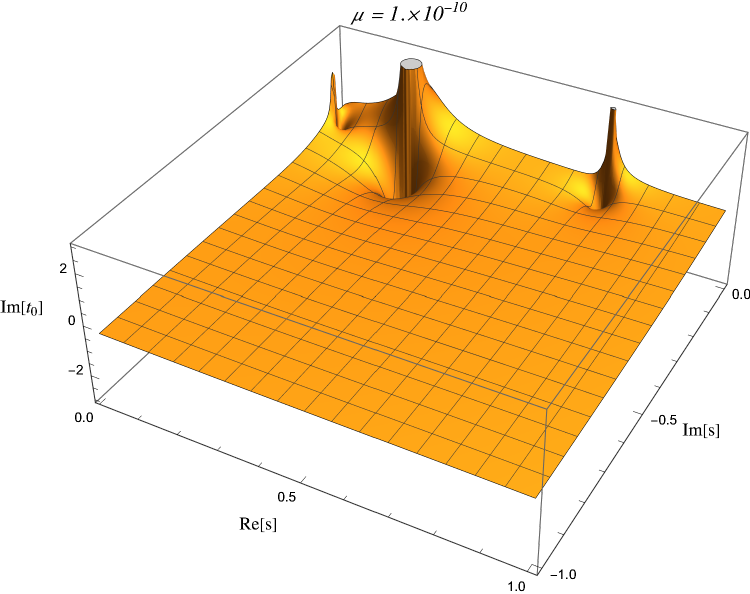}
\caption{\small{Figure showing the imaginary part of the $(2,2)$ element of the unitarized scalar partial wave representing the unitarized scattering process $\varphi\varphi \to \varphi\varphi$. The fixed parameters are $\bar{\Lambda} = 0.9$, $\bar{m} = 0.8$, and on top of each panel there is a text indicating the specific value of $\bar{\mu} = 10^{-2},\, 10^{-4},\, 10^{-8},$ and $10^{-10}$.}}
\label{fig_g_scalaron_m08}
\end{figure}
Three structures can be observed, all of them become more stable as $\mu \to 0$. In particular, we identify the one that gets closer to the origin for smaller values of $\bar{\mu}$ as the graviball. The trend observed in the single-channel analysis persists for coupled channels: the graviball moves closer to the origin as $\bar{\mu} \to 0$.\myspace
We identify the scalaron as the structure located near the real axis for all values of $\bar{\mu}$, at energy values around $\text{Re}\,\bar{s} = \bar{m}^2$. Both the scalaron and the other structure, distinct from the graviball, are strongly coupled to this channel, as evidenced by the clearly emerging structures.\myspace
We also observe that as $\bar{\mu}$ tends to smaller and smaller values, the third structure that we have not yet identified originates from negative $\text{Re}\,s$ values and eventually positions itself near the real axis, between the graviball and the scalaron. We associate this structure with a dynamic scalar resonance that arises from the unitarization process, similar to what we found in the coupled-channel study of the EWSBS.\myspace
Furthermore, we find a relationship between the position of the dynamical resonance and that of the scalaron. In particular, for the present $\bar{m} = 0.8$ and other values we have tested, the position of the dynamical resonance close to the real axis (its mass) is found to be approximately 0.55 times the squared mass of the scalaron. A more detailed study is required to fully understand this situation.\myspace
These three structures are confirmed when examining the $(1,1)$ element in Figure~\ref{fig_g_graviball_m08}, which represents the unitarized scattering $h_{\mu_1\nu_1}h_{\mu_2\nu_2} \to h_{\mu_3\nu_3}h_{\mu_4\nu_4}$---in the $++++$ helicity combination. Recall that, due to the structure of the unitarized matrix in Eq.~(\ref{eq_un_IK_cc}), particularly from $\left(t^{(2)}_{00}-t^{(4)}_{00}\right)^{-1}$, any emergent poles must appear across all channels, as is the case here. Indeed, the three poles are observed in the different pannels of Figure~\ref{fig_g_graviball_m08}, even though the scalaron and the dynamical scalar resonance become almost invisible for decreasing values of $\bar{\mu}$, showing a relatively small coupling to this channel.
\begin{figure}
\centering
\includegraphics[clip,width=7.0cm,height=6.5cm]{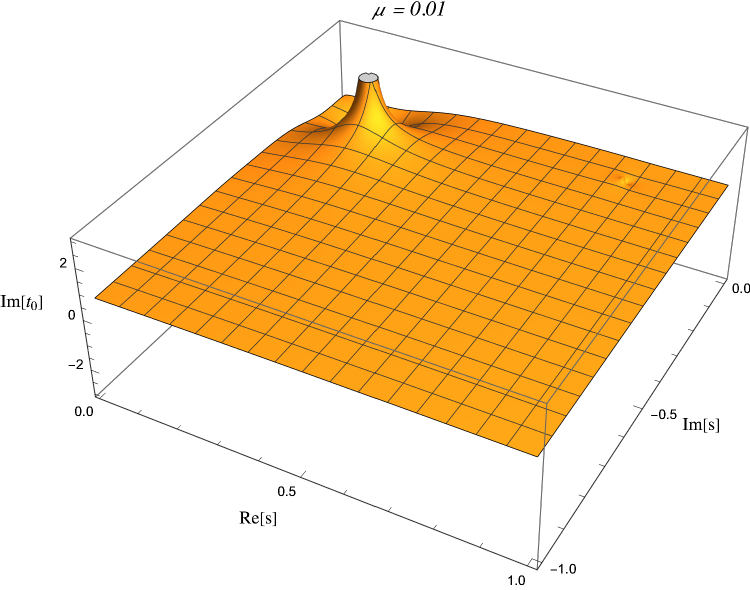}
\includegraphics[clip,width=7.0cm,height=6.5cm]{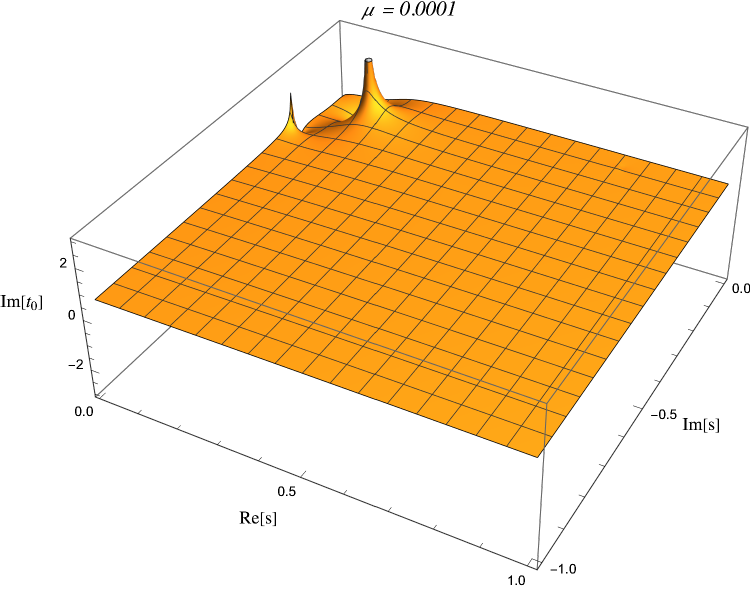}
\includegraphics[clip,width=7.0cm,height=6.5cm]{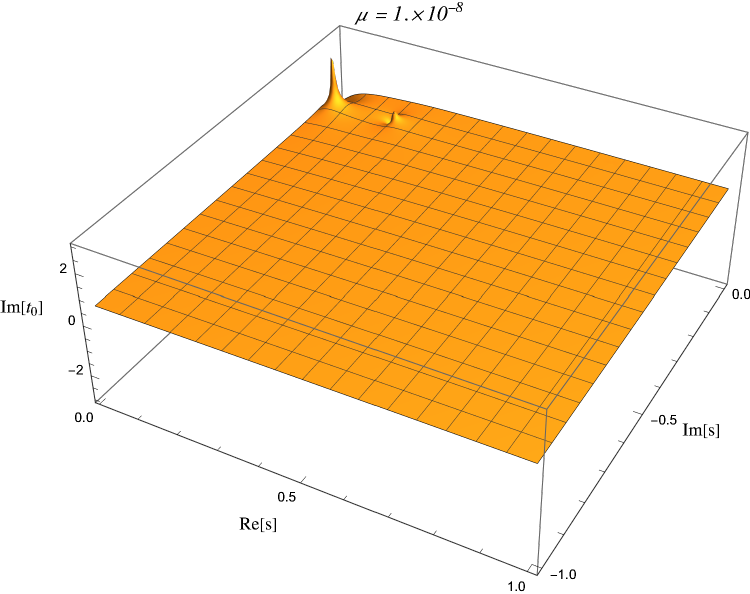}
\includegraphics[clip,width=7.0cm,height=6.5cm]{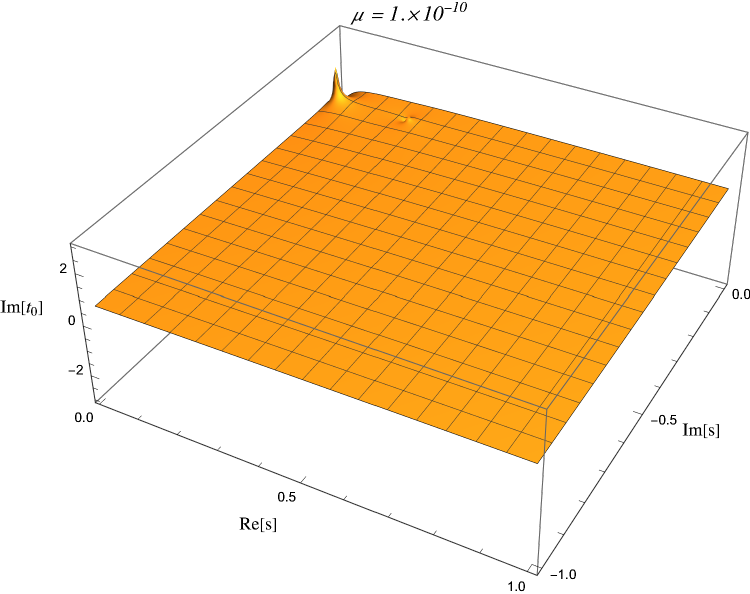}
\caption{\small{Figure showing the imaginary part of the $(1,1)$ element of the unitarized scalar partial wave representing the unitarized scattering process $h_{\mu_1\nu_1}h_{\mu_2\nu_2} \to h_{\mu_3\nu_3}h_{\mu_4\nu_4}$. The fixed parameters are $\bar{\Lambda} = 0.9$, $\bar{m} = 0.8$, and on top of each panel there is a text indicating the specific value of $\bar{\mu} = 10^{-2},\, 10^{-4},\, 10^{-8},$ and $10^{-10}$.}}
\label{fig_g_graviball_m08}
\end{figure}\myspace
\newpage
\begin{center}
$\underline{\bar{m}=0.6}$
\end{center}
Repeating the same analysis but now with $\bar{m} = 0.6$, we obtain the same tentative results.\myspace
We find a total of three structures. One of them is the scalaron , which is more strongly coupled to the $(2,2)$ channel (see Figure~(\ref{fig_g_scalaron_m06})) and located near $\text{Re}\bar{s} \approx \bar{m}^2$.\myspace
\begin{figure}
\centering
\includegraphics[clip,width=7.0cm,height=6.5cm]{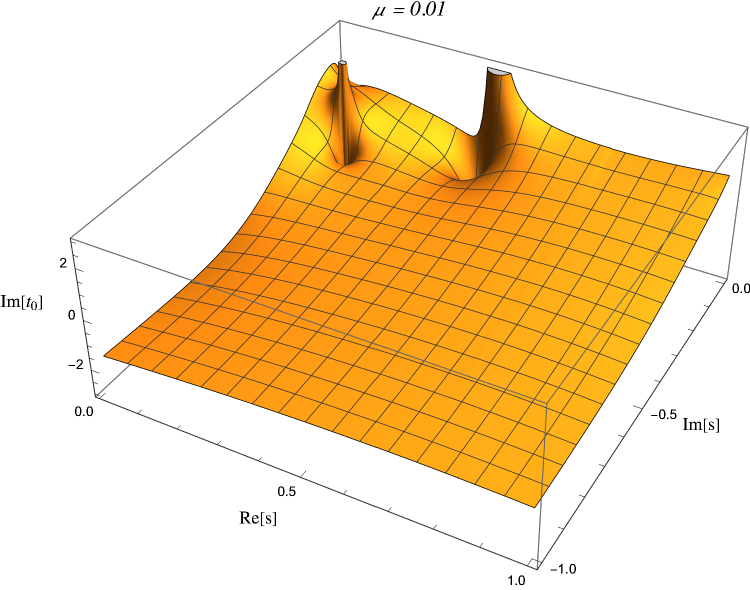}
\includegraphics[clip,width=7.0cm,height=6.5cm]{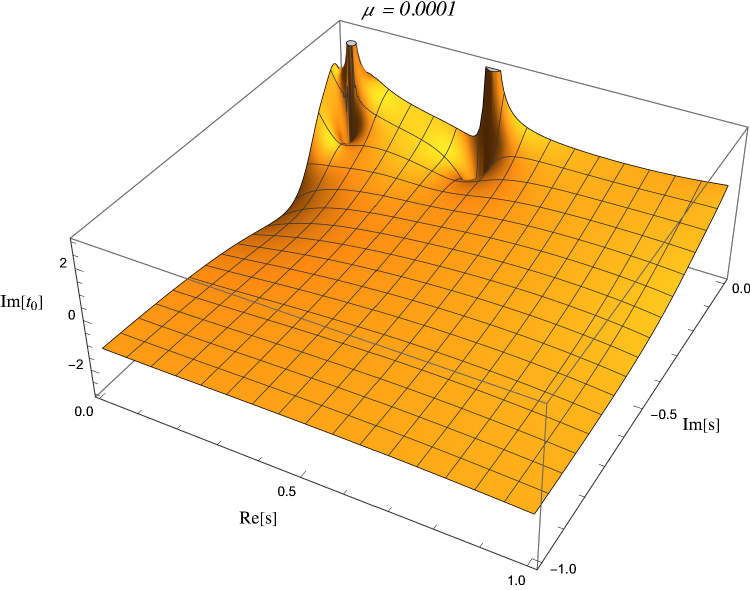}
\includegraphics[clip,width=7.0cm,height=6.5cm]{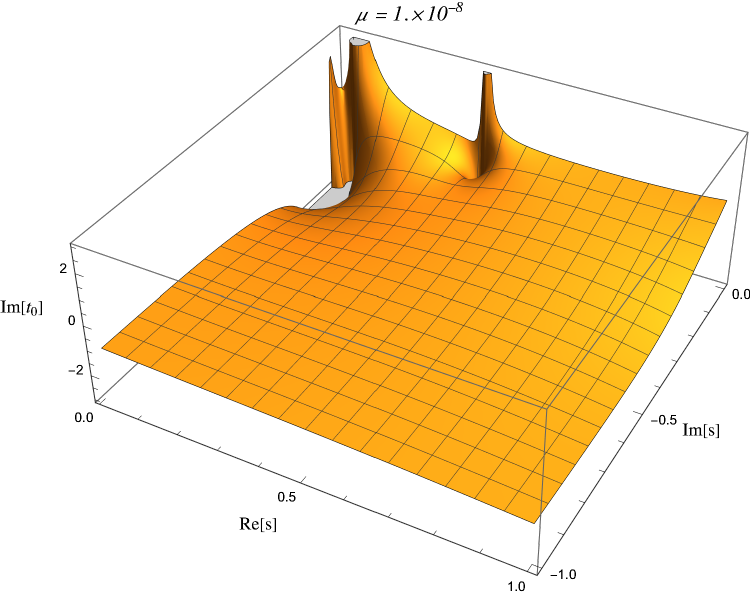}
\includegraphics[clip,width=7.0cm,height=6.5cm]{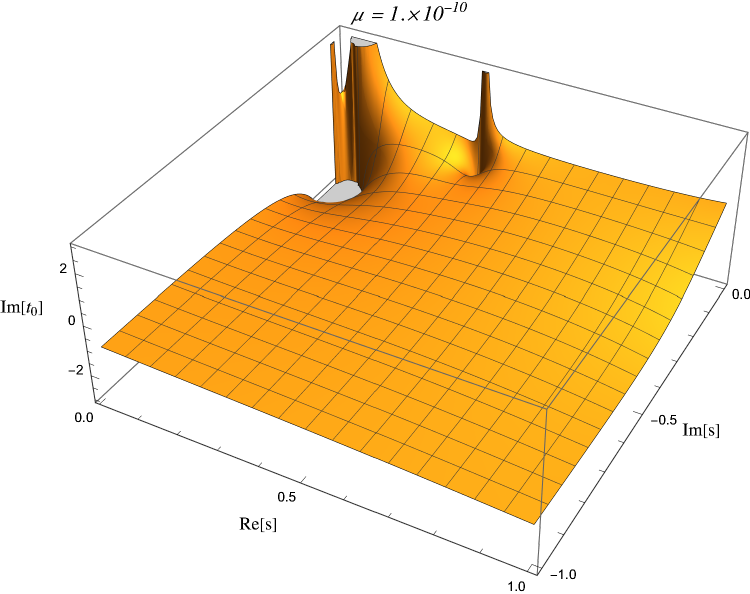}
\caption{\small{Figure showing the imaginary part of the $(2,2)$ element of the unitarized scalar partial wave representing the unitarized scattering process $\varphi\varphi \to \varphi\varphi$. The fixed parameters are $\bar{\Lambda} = 0.9$, $\bar{m} = 0.6$, and on top of each panel there is a text indicating the specific value of $\bar{\mu} = 10^{-2},\, 10^{-4},\, 10^{-8},$ and $10^{-10}$.}}
\label{fig_g_scalaron_m06}
\end{figure}
The second structure, which is also more strongly coupled to the $(2,2)$ channel, is identified as a dynamically generated resonance as a result of the unitarization process. Its position is approximately 0.55 times that of the scalaron, just as it happened in the case studied previously with $\bar{m}=0.8$.\myspace
For this value of the dimensionless mass, we also identify the graviball, that it is more noticeable in the $(1,1)$ channel represented in Figure~(\ref{fig_g_graviball_m06}) and, furthermore, tends to the origin as $\bar{\mu}$ takes increasingly smaller values.
\begin{figure}
\centering
\includegraphics[clip,width=7.0cm,height=6.5cm]{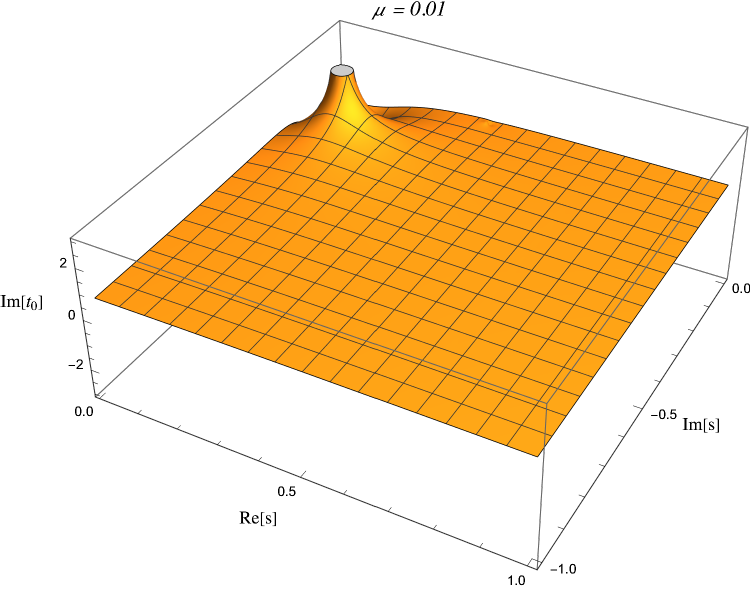}
\includegraphics[clip,width=7.0cm,height=6.5cm]{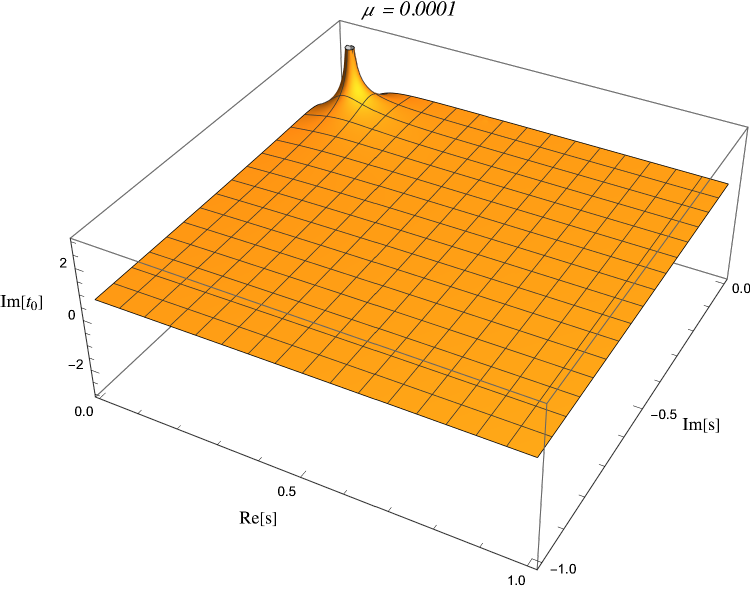}
\includegraphics[clip,width=7.0cm,height=6.5cm]{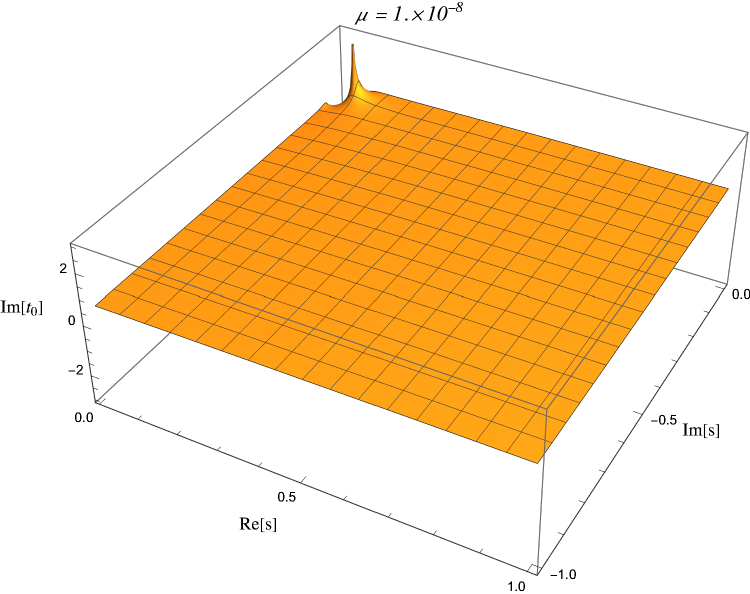}
\includegraphics[clip,width=7.0cm,height=6.5cm]{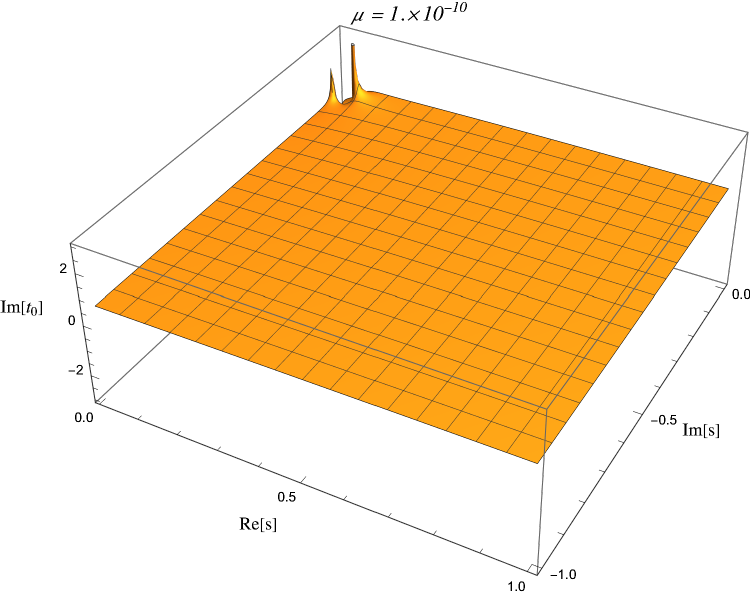}
\caption{\small{Figure showing the imaginary part of the $(1,1)$ element of the unitarized scalar partial wave representing the unitarized scattering process $h_{\mu\nu}h_{\mu\nu} \to h_{\mu\nu}h_{\mu\nu}$. The fixed parameters are $\bar{\Lambda} = 0.9$, $\bar{m} = 0.6$, and on top of each panel there is a text indicating the specific value of $\bar{\mu} = 10^{-2},\, 10^{-4},\, 10^{-8},$ and $10^{-10}$.}}
\label{fig_g_graviball_m06}
\end{figure}\myspace

\thispagestyle{empty}

\lhead{Chapter 7}
\rhead{Conclusions}

\chapter{Summary and conclusions}
\label{chp:conclusions}
\section*{\centering \underline{CHAPTERS 1, 2 AND }3}
It is well established that the Higgs provides mass to elementary particles via the Higgs mechanism through spontaneous symmetry breaking. However, it is also true that, to this day, we know very little about the potential responsible for this breaking. The Higgs potential (which encompasses both the mass and self-interactions of the Higgs boson) remains largely unknown.\myspace
Due to the very nature of the spontaneous symmetry breaking process, the scattering of longitudinally polarized $W$ bosons is one of the channels that can provide the most direct information about the mechanism. However, this channel is subdominant in LHC processes, where strong interactions dominate. Moreover, current experimental results on the Higgs are not very precise. In particular, its self-interactions remain poorly constrained, and their detailed measurement would ideally require studying processes involving two or more Higgs bosons in the final state.\myspace
Thus, given the difficulty of extracting experimental information about the interactions of the Higgs with Standard Model particles, and particularly about its self-couplings, it is reasonable to explore alternative approaches to the problem.\myspace
In the context of beyond the Standard Model, both resonant and non-resonant phenomenological studies are valuable for constraining the parameters of the theory they represent. On the one hand, non-resonant studies aim to identify deviations from the predictions of the effective theory in amplitudes, cross sections, kinematic distributions, etc. On the other hand, scenarios where new resonances emerge focus on their properties and the parameter spaces in which they appear. Thus, the absence or presence of these resonances is expected to provide insights into the couplings. The latter approach is the one we adopt, as will be shown later.\myspace
Our main theoretical assumption is that physics at energies higher than the electroweak symmetry breaking scale involves strong interactions. To study this situation, we use an analogous description to the chiral theory used to study the low-energy regime of QCD, the chiral Lagrangian, but within the framework of electroweak theories: the electroweak chiral Lagrangian with a light Higgs (EChL) or simply Higgs Effective Field Theory (HEFT).\myspace
The amplitudes of processes calculated in the HEFT are, by construction, expansions in powers of the external particle momenta. As such, they are unable to describe resonances, which appear as poles in the amplitude. Furthermore, this expansion in external momenta quickly leads to unitarity violations for values outside the Standard Model.\myspace
Throughout this thesis, we argue that these violations are not significant as long as we remain within the effective framework. However, if we want to make realistic predictions for experiments—predictions that do not absurdly overestimate the signal events for unknown physics—we must extrapolate our knowledge of the low-energy region, which is well-described within the effective framework, to the high-energy regime. This allows us to produce well-behaved amplitudes across the entire range of validity of the effective theory. For this purpose, unitarization techniques are required.\myspace
One of the main results of our work is the determination of the one-loop quantum corrections to all the relevant $2 \to 2$ processes that are significant for two-Higgs production via the scattering of electroweak gauge bosons in the HEFT. The calculation has been explicitly performed in the 't Hooft-Landau gauge, although physical amplitudes are gauge-independent.\myspace
A diagrammatic computation of all the on-shell $2 \to 2$ processes relevant for two-Higgs production is presented. In the one-loop calculation, both transverse and longitudinal polarization modes are included. In the on-shell scheme, this necessarily involves considering the physical values for the Higgs and weak gauge boson masses.\myspace
To maintain the analytical properties of the amplitudes at low energies, the unitarization methods chosen in this thesis are the Inverse Amplitude Method (IAM) and the K-matrix in its improved version, the IK-matrix. These methods are employed in different contexts. Specifically, the IAM is the method of choice for studying the EWSBS, while the IK-matrix is applied exclusively to unitarize graviton scattering, whose full one-loop expression at the next-to-leading order of the effective theory for gravity is not yet available. 
\section*{\centering 	\underline{CHAPTER 4}}
The resulting HEFT amplitudes are unitarized in the isovector-vector projection and we analyze the characteristics of the dynamical resonances appearing. An interesting result is that, after unitarization of the partial waves, the effect of including the gauge boson masses is small but significant increasing the mass of the vector resonances typically in the range 2-3$\%$. The widths are unchanged.\myspace
While traditionally the effective couplings $a_4$ and $a_5$ have been regarded as driving the masses of dynamical resonances, it turns out that the coupling $a_3$ (that plays a role only if the a priori subdominant transverse modes are included) is relevant too. It should also be mentioned that while $a_4$ and $a_5$ are by now fairly constrained by LHC analysis, the bounds on $a_3$ are still rather loose. We believe this makes the investigation of Chapter \ref{chp:vector} particularly relevant.
\section*{\centering \underline{CHAPTER 5}}
In this Chapter we have seen that when considering the $IJ=00$ case, where the formalism of coupled channels is unavoidable when transverse gauge degrees of freedom are included, scalar resonances become substantially broader.\myspace
Assuming that no resonances exist below the scales that have already been experimentally probed, the next-to-leading $\mathcal{O}(p^4)$ coefficients in the HEFT should be at most of order $10^{-3}$ and probably of order $10^{-4}$. In this work we have assumed that no resonance, vector or scalar, exists below 1.8 TeV, and from that, we derive bounds on the HEFT couplings. But there is another way of restricting the HEFT; namely, if in the unitarization process one encounters acausal or unphysical resonances, the corresponding set of parameters in the effective theory can be ruled out.\myspace
First, we have verified  that not all ${\cal O}(p^4)$ coefficients are equally important. Those determining the appearance of resonances correspond to operators that survive in the nET limit. This, which is in agreement with previous studies in the vector case, and it is quite useful as it tells us where to look for resonances in the vast space of HEFT. Note that the inclusion  of transverse modes becomes relevant (unlike in the vector case) in the scalar case.\myspace
Second, besides $a_4$ and $a_5$, three new parameters appear at next-to-leading order.
Taking all the ${\cal O}(p^2)$ couplings to be identical to the SM, we have found that, for specific values of $a_4$ and $a_5$, the resonance spectrum that could be observed is more restricted in the $\delta-\eta$ plane the lower the value of $\gamma$ is. For $\gamma=0$, resonances live in a narrow band of values of $\delta$ and $\eta$, and the greater the value of $\gamma$, the broader the band is. The region above these narrow bands can be excluded on causality grounds. So, this places a very strong restriction in parameter space. It is also seen almost immediately that if a resonance is present in  one channel it is present in all, but they always couple more strongly to $WW$ final states.\myspace
In this study, and making use of the arguments explained above, we have also set encouraging theoretical bounds on the self-interactions of the Higgs, especially in the case of the triple self-coupling whose BSM deviations are parametrized via $d_3$ (in units of $\lambda_{SM}$). We have found for this coupling that whenever it exceeds $d_3\sim 2.5$ a second very light pole appears, which we assume it would have already been detected in the experiment. The emergence of this light pole becomes noticeable even before, from $d_3\sim 1.7$. The absence of such a resonance make us exclude all the values above this threshold. This behavior is not significantly modified when considering nonzero values for the parameters $\delta$, $\gamma$, and $\eta$, so from this study, we could set a bound $d_3\lesssim 2$, much more restrictive than current
experimental bounds (assuming, of course, that indeed no scalar resonance exists below 1.8 TeV). This is an important prediction; even if a resonance is more likely to be observed in the $WW$ elastic channel, the Higgs self-coupling enters in the determination of its properties via the coupled-channel formalism,  and it should not be dramatically different from its SM value.\myspace
For the case of the four-Higgs coupling, parametrized by $d_4$ (again in units of $\lambda_{SM}$), there are no experimental bounds in the literature to our knowledge. From this study of the single resonance spectrum, we have set an overall phenomenological bound $d_4\lesssim 6$ with regions of the parameter space where it could be more restrictive, in particular for the point with both scalar and vector contribution (our BP1).\myspace
In conclusion, somewhat unexpectedly, the study of possible scalar resonances in $WW$ fusion places very interesting restrictions on the space of Higgs couplings,
a region that is hard to experimentally study. We have presented here some, we believe, relevant results, but certainly this line of research deserves further
more systematic studies.\myspace
We finished this Chapter by seeing that the H(650) can be accommodated in the HEFT but it requires the cooperation of at least one more next-to-leading coupling (for which no relevant bounds exist) when the coefficients $a_4$ and $a_5$ are pushed to the limit of the experimentally allowed region. Further restrictions on them derived from experiment would probably hinder the viability of the tentative resonance H(650). The prediction for the width of this resonance also fits well with the preliminary experimental observations.\myspace
We computed within the EWA, the cross section for the production of a 650 GeV resonance in the vector boson fusion channel via $pp\to WW+X$ and compared it with the preexisting bibliography, assuming that the scalar state only decays in gauge bosons and not to any other two scalars.\myspace
The results we find are encouraging in the sense that the predicted cross section is relatively close to the experimental analyses performed by ATLAS and CMS. First, without taking into account any event selection cut we obtained cross sections of $\sim 300$ fb and $\sim 75$ fb for $WW$ and $ZZ$ final states, respectively. These results are obtained within the EWA and assuming that the peak dominates the cross section. After applying cuts on the pseudorapidty of the di-boson state, the values of the cross sections are reduced to $\sim 275$ fb and $\sim 70$ fb for $WW$ and $ZZ$ states, respectively.

\section*{\centering \underline{CHAPTER 6}}
In this chapter, we have investigated the phenomenon of gravitational waves from the perspective of quantum field theory. Gravitational waves can be represented as a tensor field whose excitations give rise to the graviton.\myspace 
To carry out this study, we started with the Einstein-Hilbert theory and added a next-to-leading order contribution, $\alpha R^2$. Through a conformal transformation, this higher-order contribution in the effective operator expansion transforms into a contribution from a scalar field---the scalaron---whose mass is inversely related to the Wilson coefficient $\alpha$. We believe this is significant, as it allows us to access information about the next order by performing a first-order calculation, at the cost of including a scalar in the spectrum. The unitarization of the theory, therefore, requires the use of coupled channels.\myspace 
The tentative results hold across the entire parameter space explored. Three structures emerge: the graviball, understood as a composite of two gravitons, although it is not identified as a physical resonance; the scalaron, which appears explicitly in the scalar spectrum of the theory; and a third resonance, which we interpret as dynamically generated during the unitarization process. Furthermore, the position of the dynamically generated scalar resonance always seems to maintain a direct relationship with the position of the scalaron, with the former consistently located at approximately $0.55$ times the scalaron's mass.

\thispagestyle{empty}




\rhead{}
\lhead{}
\addcontentsline{toc}{chapter}{Bibliography}

\begin{small}
\printbibliography

@article{PhysRevLett.13.508,
  title = {Broken Symmetries and the Masses of Gauge Bosons},
  author = {Higgs, Peter W.},
  journal = {Phys. Rev. Lett.},
  volume = {13},
  issue = {16},
  pages = {508--509},
  numpages = {0},
  year = {1964},
  month = {Oct},
  publisher = {American Physical Society},
  doi = {10.1103/PhysRevLett.13.508},
  url = {https://link.aps.org/doi/10.1103/PhysRevLett.13.508}
}

@article{Longhitano:1980iz,
    author = "Longhitano, Anthony C.",
    title = "{Heavy Higgs Bosons in the Weinberg-Salam Model}",
    reportNumber = "YTP-80-04",
    doi = "10.1103/PhysRevD.22.1166",
    journal = "Phys. Rev. D",
    volume = "22",
    pages = "1166",
    year = "1980"
}

@article{Urbano:2013aoa,
    author = "Urbano, Alfredo",
    title = "{Remarks on analyticity and unitarity in the presence of a Strongly Interacting Light Higgs}",
    eprint = "1310.5733",
    archivePrefix = "arXiv",
    primaryClass = "hep-ph",
    doi = "10.1007/JHEP06(2014)060",
    journal = "JHEP",
    volume = "06",
    pages = "060",
    year = "2014"
}

@article{Arnan:2015csa,
    author = "Arnan, Pere and Espriu, Domenec and Mescia, Federico",
    title = "{Interpreting a 2 TeV resonance in WW scattering}",
    eprint = "1508.00174",
    archivePrefix = "arXiv",
    primaryClass = "hep-ph",
    doi = "10.1103/PhysRevD.93.015020",
    journal = "Phys. Rev. D",
    volume = "93",
    number = "1",
    pages = "015020",
    year = "2016"
}

@article{Eboli:2016kko,
    author = "\'Eboli, O. J. P. and Gonzalez-Garcia, M. C.",
    title = "{Classifying the bosonic quartic couplings}",
    eprint = "1604.03555",
    archivePrefix = "arXiv",
    primaryClass = "hep-ph",
    reportNumber = "YITP-SB-16-09",
    doi = "10.1103/PhysRevD.93.093013",
    journal = "Phys. Rev. D",
    volume = "93",
    number = "9",
    pages = "093013",
    year = "2016"
}

@article{Eboli:2006wa,
    author = "Eboli, O. J. P. and Gonzalez-Garcia, M. C. and Mizukoshi, J. K.",
    title = "{p p ---\ensuremath{>} j j e+- mu+- nu nu and j j e+- mu-+ nu nu at O( alpha(em)**6) and O(alpha(em)**4 alpha(s)**2) for the study of the quartic electroweak gauge boson vertex at CERN LHC}",
    eprint = "hep-ph/0606118",
    archivePrefix = "arXiv",
    reportNumber = "YITP-SB-06-10, IFUSP-1620-2006",
    doi = "10.1103/PhysRevD.74.073005",
    journal = "Phys. Rev. D",
    volume = "74",
    pages = "073005",
    year = "2006"
}

@article{Contino:2013kra,
    author = "Contino, Roberto and Ghezzi, Margherita and Grojean, Christophe and Muhlleitner, Margarete and Spira, Michael",
    title = "{Effective Lagrangian for a light Higgs-like scalar}",
    eprint = "1303.3876",
    archivePrefix = "arXiv",
    primaryClass = "hep-ph",
    reportNumber = "CERN-PH-TH-2013-047, KA-TP-06-2013, PSI-PR-13-04",
    doi = "10.1007/JHEP07(2013)035",
    journal = "JHEP",
    volume = "07",
    pages = "035",
    year = "2013"
}

@article{Weinberg:1967kj,
    author = "Weinberg, Steven",
    title = "{Precise relations between the spectra of vector and axial vector mesons}",
    doi = "10.1103/PhysRevLett.18.507",
    journal = "Phys. Rev. Lett.",
    volume = "18",
    pages = "507--509",
    year = "1967"
}

@article{PhysRevD.55.4193,
  title = {Resonance spectrum of the strongly interacting symmetry-breaking sector},
  author = {Pel\'aez, J. R.},
  journal = {Phys. Rev. D},
  volume = {55},
  issue = {7},
  pages = {4193--4202},
  numpages = {0},
  year = {1997},
  month = {Apr},
  publisher = {American Physical Society},
  doi = {10.1103/PhysRevD.55.4193},
  url = {https://link.aps.org/doi/10.1103/PhysRevD.55.4193}
}

@article{Rosell:2020iub,
    author = "Rosell, Ignasi and Pich, Antonio and Sanz-Cillero, Juan Jos\'e",
    title = "{Constraining resonances by using the electroweak effective theory}",
    eprint = "2010.08271",
    archivePrefix = "arXiv",
    primaryClass = "hep-ph",
    doi = "10.22323/1.390.0077",
    journal = "PoS",
    volume = "ICHEP2020",
    pages = "077",
    year = "2021"
}

@article{Herrero:2022krh,
    author = "Herrero, M. J. and Morales, R. A.",
    title = "{One-loop corrections for WW to HH in Higgs EFT with the electroweak chiral Lagrangian}",
    eprint = "2208.05900",
    archivePrefix = "arXiv",
    primaryClass = "hep-ph",
    reportNumber = "IFT-UAM/CSIC-22-80",
    doi = "10.1103/PhysRevD.106.073008",
    journal = "Phys. Rev. D",
    volume = "106",
    number = "7",
    pages = "073008",
    year = "2022"
}

@article{Herrero:2020dtv,
    author = "Herrero, Maria and Morales, Roberto A.",
    title = "{Anatomy of Higgs boson decays into $\gamma \gamma$ and $\gamma Z$ within the electroweak chiral Lagrangian in the $R_\xi$ gauges}",
    eprint = "2005.03537",
    archivePrefix = "arXiv",
    primaryClass = "hep-ph",
    doi = "10.1103/PhysRevD.102.075040",
    journal = "Phys. Rev. D",
    volume = "102",
    number = "7",
    pages = "075040",
    year = "2020"
}

@article{Herrero:1993nc,
    author = "Herrero, Maria J. and Ruiz Morales, Ester",
    title = "{The Electroweak chiral Lagrangian for the Standard Model with a heavy Higgs}",
    eprint = "hep-ph/9308276",
    archivePrefix = "arXiv",
    reportNumber = "FTUAM-93-24",
    doi = "10.1016/0550-3213(94)90525-8",
    journal = "Nucl. Phys. B",
    volume = "418",
    pages = "431--455",
    year = "1994"
}

@article{Garcia-Garcia:2019oig,
    author = "Garcia-Garcia, Claudia and Herrero, Maria and Morales, Roberto A.",
    title = "{Unitarization effects in EFT predictions of WZ scattering at the LHC}",
    eprint = "1907.06668",
    archivePrefix = "arXiv",
    primaryClass = "hep-ph",
    reportNumber = "IFT-UAM/CSIC-19-81, FTUAM-19-13",
    doi = "10.1103/PhysRevD.100.096003",
    journal = "Phys. Rev. D",
    volume = "100",
    number = "9",
    pages = "096003",
    year = "2019"
}

@article{Domenech:2022uud,
    author = "Domenech, D. and Herrero, M. J. and Morales, R. A. and Ramos, M.",
    title = "{Double Higgs boson production at TeV e+e- colliders with effective field theories: Sensitivity to BSM Higgs couplings}",
    eprint = "2208.05452",
    archivePrefix = "arXiv",
    primaryClass = "hep-ph",
    reportNumber = "IFT-UAM/CSIC-22-79",
    doi = "10.1103/PhysRevD.106.115027",
    journal = "Phys. Rev. D",
    volume = "106",
    number = "11",
    pages = "115027",
    year = "2022"
}

@article{Anisha:2024ryj,
    author = "Anisha and Domenech, Daniel and Englert, Christoph and Herrero, Maria J. and Morales, Roberto A.",
    title = "{HEFT's appraisal of triple (versus double) Higgs weak boson fusion}",
    eprint = "2407.20706",
    archivePrefix = "arXiv",
    primaryClass = "hep-ph",
    reportNumber = "IFT-UAM/CSIC-24-111",
    month = "7",
    year = "2024"
}

@article{Rauch:2016pai,
    author = "Rauch, Michael",
    title = "{Vector-Boson Fusion and Vector-Boson Scattering}",
    eprint = "1610.08420",
    archivePrefix = "arXiv",
    primaryClass = "hep-ph",
    reportNumber = "KA-TP-35-2016",
    month = "10",
    year = "2016"
}

@article{PhysRevD.91.096007,
  title = {High-energy vector boson scattering after the Higgs boson discovery},
  author = {Kilian, Wolfgang and Sekulla, Marco and Ohl, Thorsten and Reuter, J\"urgen},
  journal = {Phys. Rev. D},
  volume = {91},
  issue = {9},
  pages = {096007},
  numpages = {23},
  year = {2015},
  month = {May},
  publisher = {American Physical Society},
  doi = {10.1103/PhysRevD.91.096007},
  url = {https://link.aps.org/doi/10.1103/PhysRevD.91.096007}
}

@article{Gavela:2014uta,
    author = "Gavela, M. B. and Kanshin, K. and Machado, P. A. N. and Saa, S.",
    title = "{On the renormalization of the electroweak chiral Lagrangian with a Higgs}",
    eprint = "1409.1571",
    archivePrefix = "arXiv",
    primaryClass = "hep-ph",
    reportNumber = "FTUAM-14-29, IFT-UAM-CSIC-14-071, DFPD2014-TH-15",
    doi = "10.1007/JHEP03(2015)043",
    journal = "JHEP",
    volume = "03",
    pages = "043",
    year = "2015"
}

@inproceedings{Grozin:2005yg,
    author = "Grozin, Andrey",
    title = "{Lectures on QED and QCD}",
    booktitle = "{3rd Dubna International Advanced School of Theoretical Physics}",
    eprint = "hep-ph/0508242",
    archivePrefix = "arXiv",
    reportNumber = "TTP05-15",
    month = "8",
    year = "2005"
}

@article{Espriu:1994ep,
    author = "Espriu, D. and Matias, J.",
    title = "{Renormalization and the equivalence theorem: On-shell scheme}",
    eprint = "hep-ph/9501279",
    archivePrefix = "arXiv",
    reportNumber = "UB-ECM-PF-94-16",
    doi = "10.1103/PhysRevD.52.6530",
    journal = "Phys. Rev. D",
    volume = "52",
    pages = "6530--6552",
    year = "1995"
}

@article{CMS:2020tkr,
    author = "Sirunyan, Albert M and others",
    collaboration = "CMS",
    title = "{Search for nonresonant Higgs boson pair production in final states with two bottom quarks and two photons in proton-proton collisions at $ \sqrt{s} $ = 13 TeV}",
    eprint = "2011.12373",
    archivePrefix = "arXiv",
    primaryClass = "hep-ex",
    reportNumber = "CMS-HIG-19-018, CERN-EP-2020-222",
    doi = "10.1007/JHEP03(2021)257",
    journal = "JHEP",
    volume = "03",
    pages = "257",
    year = "2021"
}

@article{ATLAS:2019nkf,
    author = "Aad, Georges and others",
    collaboration = "ATLAS",
    title = "{Combined measurements of Higgs boson production and decay using up to $80$ fb$^{-1}$ of proton-proton collision data at $\sqrt{s}=$ 13 TeV collected with the ATLAS experiment}",
    eprint = "1909.02845",
    archivePrefix = "arXiv",
    primaryClass = "hep-ex",
    reportNumber = "CERN-EP-2019-097",
    doi = "10.1103/PhysRevD.101.012002",
    journal = "Phys. Rev. D",
    volume = "101",
    number = "1",
    pages = "012002",
    year = "2020"
}

@article{Passarino:1978jh,
    author = "Passarino, G. and Veltman, M. J. G.",
    title = "{One Loop Corrections for e+ e- Annihilation Into mu+ mu- in the Weinberg Model}",
    reportNumber = "Print-79-0284 (UTRECHT)",
    doi = "10.1016/0550-3213(79)90234-7",
    journal = "Nucl. Phys. B",
    volume = "160",
    pages = "151--207",
    year = "1979"
}

@article{Shtabovenko:2020gxv,
    author = "Shtabovenko, Vladyslav and Mertig, Rolf and Orellana, Frederik",
    title = "{FeynCalc 9.3: New features and improvements}",
    eprint = "2001.04407",
    archivePrefix = "arXiv",
    primaryClass = "hep-ph",
    reportNumber = "P3H-20-002, TTP19-020, TUM-EFT 130/19",
    doi = "10.1016/j.cpc.2020.107478",
    journal = "Comput. Phys. Commun.",
    volume = "256",
    pages = "107478",
    year = "2020"
}

@article{Shtabovenko:2016whf,
    author = "Shtabovenko, Vladyslav",
    title = "{FeynHelpers: Connecting FeynCalc to FIRE and Package-X}",
    eprint = "1611.06793",
    archivePrefix = "arXiv",
    primaryClass = "physics.comp-ph",
    reportNumber = "TUM-EFT-75-15",
    doi = "10.1016/j.cpc.2017.04.014",
    journal = "Comput. Phys. Commun.",
    volume = "218",
    pages = "48--65",
    year = "2017"
}

@article{Hahn:2000kx,
    author = "Hahn, Thomas",
    title = "{Generating Feynman diagrams and amplitudes with FeynArts 3}",
    eprint = "hep-ph/0012260",
    archivePrefix = "arXiv",
    reportNumber = "KA-TP-23-2000",
    doi = "10.1016/S0010-4655(01)00290-9",
    journal = "Comput. Phys. Commun.",
    volume = "140",
    pages = "418--431",
    year = "2001"
}

@article{Sirunyan:2019der,
    author = "Sirunyan, Albert M and others",
    collaboration = "CMS",
    title = "{Search for anomalous electroweak production of vector boson pairs in association with two jets in proton-proton collisions at 13 TeV}",
    eprint = "1905.07445",
    archivePrefix = "arXiv",
    primaryClass = "hep-ex",
    reportNumber = "CMS-SMP-18-006, CERN-EP-2019-089",
    doi = "10.1016/j.physletb.2019.134985",
    journal = "Phys. Lett. B",
    volume = "798",
    pages = "134985",
    year = "2019"
}

@article{CMS:2019uys,
    author = "Sirunyan, Albert M and others",
    collaboration = "CMS",
    title = "{Measurement of electroweak WZ boson production and search for new physics in WZ + two jets events in pp collisions at $\sqrt{s} =$ 13TeV}",
    eprint = "1901.04060",
    archivePrefix = "arXiv",
    primaryClass = "hep-ex",
    reportNumber = "CMS-SMP-18-001, CERN-EP-2018-333",
    doi = "10.1016/j.physletb.2019.05.042",
    journal = "Phys. Lett. B",
    volume = "795",
    pages = "281--307",
    year = "2019"
}

@article{ATLAS:2016nmw,
    author = "Aaboud, Morad and others",
    collaboration = "ATLAS",
    title = "{Search for anomalous electroweak production of $WW/WZ$ in association with a high-mass dijet system in $pp$ collisions at $\sqrt{s}=8$ TeV with the ATLAS detector}",
    eprint = "1609.05122",
    archivePrefix = "arXiv",
    primaryClass = "hep-ex",
    reportNumber = "CERN-EP-2016-171",
    doi = "10.1103/PhysRevD.95.032001",
    journal = "Phys. Rev. D",
    volume = "95",
    number = "3",
    pages = "032001",
    year = "2017"
}

@article{PhysRev.177.2239,
  title = {Structure of Phenomenological Lagrangians. I},
  author = {Coleman, S. and Wess, J. and Zumino, Bruno},
  journal = {Phys. Rev.},
  volume = {177},
  issue = {5},
  pages = {2239--2247},
  numpages = {0},
  year = {1969},
  month = {Jan},
  publisher = {American Physical Society},
  doi = {10.1103/PhysRev.177.2239},
  url = {https://link.aps.org/doi/10.1103/PhysRev.177.2239}
}

@article{Delgado:2015kxa,
    author = "Delgado, Rafael L. and Dobado, Antonio and Llanes-Estrada, Felipe J.",
    title = "{Unitarity, analyticity, dispersion relations, and resonances in strongly interacting $W_LW_L$, $Z_LZ_L$, and hh scattering}",
    eprint = "1502.04841",
    archivePrefix = "arXiv",
    primaryClass = "hep-ph",
    doi = "10.1103/PhysRevD.91.075017",
    journal = "Phys. Rev. D",
    volume = "91",
    number = "7",
    pages = "075017",
    year = "2015"
}

@article{Delgado:2017cls,
    author = "Delgado, R. L. and Dobado, A. and Espriu, D. and Garcia-Garcia, C. and Herrero, M. J. and Marcano, X. and Sanz-Cillero, J. J.",
    title = "{Production of vector resonances at the LHC via WZ-scattering: a unitarized EChL analysis}",
    eprint = "1707.04580",
    archivePrefix = "arXiv",
    primaryClass = "hep-ph",
    reportNumber = "IFT-UAM-CSIC-17-048, FTUAM-17-09",
    doi = "10.1007/JHEP11(2017)098",
    journal = "JHEP",
    volume = "11",
    pages = "098",
    year = "2017"
}

@article{Delgado:2014dxa,
    author = "Delgado, Rafael L. and Dobado, Antonio and Llanes-Estrada, Felipe J.",
    title = "{Possible new resonance from $W_L W_L$-$hh$ interchannel coupling}",
    eprint = "1408.1193",
    archivePrefix = "arXiv",
    primaryClass = "hep-ph",
    doi = "10.1103/PhysRevLett.114.221803",
    journal = "Phys. Rev. Lett.",
    volume = "114",
    number = "22",
    pages = "221803",
    year = "2015"
}

@article{Alonso:2012px,
    author = "Alonso, R. and Gavela, M. B. and Merlo, L. and Rigolin, S. and Yepes, J.",
    title = "{The Effective Chiral Lagrangian for a Light Dynamical ''Higgs Particle''}",
    eprint = "1212.3305",
    archivePrefix = "arXiv",
    primaryClass = "hep-ph",
    reportNumber = "FTUAM-12-115, IFT-UAM-CSIC-12-113, CERN-PH-TH-2012-335, DFPD-2012-TH-23",
    doi = "10.1016/j.physletb.2013.04.037",
    journal = "Phys. Lett. B",
    volume = "722",
    pages = "330--335",
    year = "2013",
    note = "[Erratum: Phys.Lett.B 726, 926 (2013)]"
}

@article{Buchalla:2013rka,
    author = "Buchalla, Gerhard and Cat\`a, Oscar and Krause, Claudius",
    title = "{Complete Electroweak Chiral Lagrangian with a Light Higgs at NLO}",
    eprint = "1307.5017",
    archivePrefix = "arXiv",
    primaryClass = "hep-ph",
    reportNumber = "LMU-ASC-42-13, LMU-ASC\textasciitilde{}42-13",
    doi = "10.1016/j.nuclphysb.2014.01.018",
    journal = "Nucl. Phys. B",
    volume = "880",
    pages = "552--573",
    year = "2014",
    note = "[Erratum: Nucl.Phys.B 913, 475--478 (2016)]"
}

@article{Pelaez:1996wk,
    author = "Pelaez, J. R.",
    title = "{Resonance spectrum of the strongly interacting symmetry breaking sector}",
    eprint = "hep-ph/9609427",
    archivePrefix = "arXiv",
    reportNumber = "LBL-39378, LBNL-39378",
    doi = "10.1103/PhysRevD.55.4193",
    journal = "Phys. Rev. D",
    volume = "55",
    pages = "4193--4202",
    year = "1997"
}

@article{Salas-Bernardez:2020hua,
    author = "Salas-Bern\'ardez, Alexandre and Llanes-Estrada, Felipe J. and Escudero-Pedrosa, Juan and Oller, Jose Antonio",
    title = "{Systematizing and addressing theory uncertainties of unitarization with the Inverse Amplitude Method}",
    eprint = "2010.13709",
    archivePrefix = "arXiv",
    primaryClass = "hep-ph",
    reportNumber = "doi: 10.21468/SciPostPhys.11.2.020",
    doi = "10.21468/SciPostPhys.11.2.020",
    journal = "SciPost Phys.",
    volume = "11",
    number = "2",
    pages = "020",
    year = "2021"
}

@article{Oller:1998hw,
    author = "Oller, J. A. and Oset, E. and Pelaez, J. R.",
    title = "{Meson meson interaction in a nonperturbative chiral approach}",
    eprint = "hep-ph/9804209",
    archivePrefix = "arXiv",
    reportNumber = "SLAC-PUB-7787",
    doi = "10.1103/PhysRevD.59.074001",
    journal = "Phys. Rev. D",
    volume = "59",
    pages = "074001",
    year = "1999",
    note = "[Erratum: Phys.Rev.D 60, 099906 (1999), Erratum: Phys.Rev.D 75, 099903 (2007)]"
}

@article{Oller:2000ma,
    author = "Oller, J. A. and Oset, E. and Ramos, A.",
    title = "{Chiral unitary approach to meson meson and meson - baryon interactions and nuclear applications}",
    eprint = "hep-ph/0002193",
    archivePrefix = "arXiv",
    reportNumber = "FZJ-IKP-TH-1999-37, FTUV-99-1215, IFIC-99-1215",
    doi = "10.1016/S0146-6410(00)00104-6",
    journal = "Prog. Part. Nucl. Phys.",
    volume = "45",
    pages = "157--242",
    year = "2000"
}

@article{Delgado:2013hxa,
    author = "Delgado, Rafael L. and Dobado, Antonio and Llanes-Estrada, Felipe J.",
    title = "{One-loop $W_LW_L$ and $Z_LZ_L$ scattering from the electroweak Chiral Lagrangian with a light Higgs-like scalar}",
    eprint = "1311.5993",
    archivePrefix = "arXiv",
    primaryClass = "hep-ph",
    doi = "10.1007/JHEP02(2014)121",
    journal = "JHEP",
    volume = "02",
    pages = "121",
    year = "2014"
}

@article{Dobado:2019fxe,
    author = "Dobado, Antonio and Espriu, Dom\`enec",
    title = "{Strongly coupled theories beyond the Standard Model}",
    eprint = "1911.06844",
    archivePrefix = "arXiv",
    primaryClass = "hep-ph",
    doi = "10.1016/j.ppnp.2020.103813",
    journal = "Prog. Part. Nucl. Phys.",
    volume = "115",
    pages = "103813",
    year = "2020"
}

@article{Espriu:2014jya,
    author = "Espriu, Domenec and Mescia, Federico",
    title = "{Unitarity and causality constraints in composite Higgs models}",
    eprint = "1403.7386",
    archivePrefix = "arXiv",
    primaryClass = "hep-ph",
    doi = "10.1103/PhysRevD.90.015035",
    journal = "Phys. Rev. D",
    volume = "90",
    number = "1",
    pages = "015035",
    year = "2014"
}

@article{Espriu:2013fia,
    author = "Espriu, Domenec and Mescia, Federico and Yencho, Brian",
    title = "{Radiative corrections to WL WL scattering in composite Higgs models}",
    eprint = "1307.2400",
    archivePrefix = "arXiv",
    primaryClass = "hep-ph",
    reportNumber = "ICCUB-13-221, UB-ECM-PF-13-92",
    doi = "10.1103/PhysRevD.88.055002",
    journal = "Phys. Rev. D",
    volume = "88",
    pages = "055002",
    year = "2013"
}

@article{Espriu:2012ih,
    author = "Espriu, Dom\`enec and Yencho, Brian",
    title = "{Longitudinal WW scattering in light of the \textquotedblleft{}Higgs boson\textquotedblright{} discovery}",
    eprint = "1212.4158",
    archivePrefix = "arXiv",
    primaryClass = "hep-ph",
    reportNumber = "ICCUB-12-480, UB-ECM-PF-84-12",
    doi = "10.1103/PhysRevD.87.055017",
    journal = "Phys. Rev. D",
    volume = "87",
    number = "5",
    pages = "055017",
    year = "2013"
}

@article{Dobado:1990zh,
    author = "Dobado, Antonio and Espriu, Domenec and Herrero, Maria J.",
    title = "{Chiral Lagrangians as a tool to probe the symmetry breaking sector of the SM at LEP}",
    reportNumber = "CERN-TH-5785-90",
    doi = "10.1016/0370-2693(91)90786-P",
    journal = "Phys. Lett. B",
    volume = "255",
    pages = "405--414",
    year = "1991"
}

@article{Herrero:2021iqt,
    author = "Herrero, Maria J. and Morales, Roberto A.",
    title = "{One-loop renormalization of vector boson scattering with the electroweak chiral Lagrangian in covariant gauges}",
    eprint = "2107.07890",
    archivePrefix = "arXiv",
    primaryClass = "hep-ph",
    reportNumber = "IFT-UAM/CSIC-21-30",
    doi = "10.1103/PhysRevD.104.075013",
    journal = "Phys. Rev. D",
    volume = "104",
    number = "7",
    pages = "075013",
    year = "2021"
}

@article{Grzadkowski:2010es,
      author         = "Grzadkowski, B. and Iskrzynski, M. and Misiak, M. and
                        Rosiek, J.",
      title          = "{Dimension-Six Terms in the Standard Model Lagrangian}",
      journal        = "JHEP",
      volume         = "10",
      year           = "2010",
      pages          = "085",
      doi            = "10.1007/JHEP10(2010)085",
      eprint         = "1008.4884",
      archivePrefix  = "arXiv",
      primaryClass   = "hep-ph",
      reportNumber   = "IFT-9-2010, TTP10-35",
      SLACcitation   = "%%CITATION = ARXIV:1008.4884;%%"
}

@inproceedings{Kundu:2022bpy,
    author = "Kundu, Anirban and Le Yaouanc, Alain and Mondal, Poulami and Richard, Fran\c{c}ois",
    title = "{Searches for scalars at LHC and interpretation of the findings}",
    booktitle = "{2022 ECFA Workshop on e+e- Higgs/EW/Top factories}",
    eprint = "2211.11723",
    archivePrefix = "arXiv",
    primaryClass = "hep-ph",
    reportNumber = "IJCLab22-002",
    month = "11",
    year = "2022"
}

@inproceedings{Yaouanc:2024xqg,
    author = "Yaouanc, Alain Le and Richard, Fran\c{c}ois",
    title = "{X650-\ensuremath{>}ZZ/WW/H125H95/A450Z -- scalar, tensor or both ?}",
    booktitle = "{3rd ECFA workshop on e+e- Higgs, Electroweak and Top Factories}",
    eprint = "2408.12178",
    archivePrefix = "arXiv",
    primaryClass = "hep-ph",
    month = "8",
    year = "2024"
}

@article{Randall:1999ee,
    author = "Randall, Lisa and Sundrum, Raman",
    title = "{A Large mass hierarchy from a small extra dimension}",
    eprint = "hep-ph/9905221",
    archivePrefix = "arXiv",
    reportNumber = "MIT-CTP-2860, PUPT-1860, BUHEP-99-9",
    doi = "10.1103/PhysRevLett.83.3370",
    journal = "Phys. Rev. Lett.",
    volume = "83",
    pages = "3370--3373",
    year = "1999"
}

@article{Afonin:2022qkl,
    author = "Afonin, Sergey",
    title = "{Mass ratio of composite Higgs bosons from bottom-up holographic Wilson loop modeling of beyond the Standard Model strong sector}",
    eprint = "2211.07500",
    archivePrefix = "arXiv",
    primaryClass = "hep-ph",
    month = "11",
    year = "2022"
}

@article{Dawson:1984gx,
    author = "Dawson, Sally",
    title = "{The Effective W Approximation}",
    reportNumber = "LBL-17497",
    doi = "10.1016/0550-3213(85)90038-0",
    journal = "Nucl. Phys. B",
    volume = "249",
    pages = "42--60",
    year = "1985"
}

@article{ATLAS:2021kog,
    author = "Aad, Georges and others",
    collaboration = "ATLAS",
    title = "{Measurements of differential cross-sections in four-lepton events in 13 TeV proton-proton collisions with the ATLAS detector}",
    eprint = "2103.01918",
    archivePrefix = "arXiv",
    primaryClass = "hep-ex",
    reportNumber = "CERN-EP-2021-019",
    doi = "10.1007/JHEP07(2021)005",
    journal = "JHEP",
    volume = "07",
    pages = "005",
    year = "2021"
}

@article{Georgi:1985nv,
    author = "Georgi, Howard and Machacek, Marie",
    title = "{DOUBLY CHARGED HIGGS BOSONS}",
    reportNumber = "HUTP-85/A051",
    doi = "10.1016/0550-3213(85)90325-6",
    journal = "Nucl. Phys. B",
    volume = "262",
    pages = "463--477",
    year = "1985"
}

@article{Chanowitz:1985ug,
    author = "Chanowitz, Michael S. and Golden, Mitchell",
    title = "{Higgs Boson Triplets With M ($W$) = M ($Z$) $\cos \theta \omega$}",
    reportNumber = "LBL-20269",
    doi = "10.1016/0370-2693(85)90700-2",
    journal = "Phys. Lett. B",
    volume = "165",
    pages = "105--108",
    year = "1985"
}

@article{Cea:2022zgs,
    author = "Cea, Paolo",
    title = "{The Higgs condensate as a quantum liquid: A critical comparison with observations}",
    eprint = "2210.01579",
    archivePrefix = "arXiv",
    primaryClass = "hep-ph",
    month = "10",
    year = "2022"
}

@article{Cea:2018tmm,
    author = "Cea, Paolo",
    title = "{Evidence of the true Higgs boson $H_T$ at the LHC Run 2}",
    eprint = "1806.04529",
    archivePrefix = "arXiv",
    primaryClass = "hep-ph",
    doi = "10.1142/S0217732319501372",
    journal = "Mod. Phys. Lett. A",
    volume = "34",
    number = "18",
    pages = "1950137",
    year = "2019"
}

@article{Gunion:2002zf,
    author = "Gunion, John F. and Haber, Howard E.",
    title = "{The CP conserving two Higgs doublet model: The Approach to the decoupling limit}",
    eprint = "hep-ph/0207010",
    archivePrefix = "arXiv",
    reportNumber = "SCIPP-02-10",
    doi = "10.1103/PhysRevD.67.075019",
    journal = "Phys. Rev. D",
    volume = "67",
    pages = "075019",
    year = "2003"
}

@article{Banik:2023ecr,
    author = "Banik, Sumit and Crivellin, Andreas and Iguro, Syuhei and Kitahara, Teppei",
    title = "{Asymmetric Di-Higgs Signals of the N2HDM-$U(1)$}",
    eprint = "2303.11351",
    archivePrefix = "arXiv",
    primaryClass = "hep-ph",
    reportNumber = "PSI-PR-23-7, ZU-TH 15/23, P3H-23-015, TTP23-010, KEK-TH-2506",
    month = "3",
    year = "2023"
}

@article{Alonso:2023upf,
    author = "Alonso, Rodrigo",
    title = "{A primer on Higgs Effective Field Theory with Geometry}",
    eprint = "2307.14301",
    archivePrefix = "arXiv",
    primaryClass = "hep-ph",
    reportNumber = "IPPP/23/38",
    month = "7",
    year = "2023"
}

@phdthesis{Brivio:2016uid,
    author = "Brivio, Ilaria",
    title = "{Hunting a dynamical Higgs}",
    school = "U. Autonoma, Madrid (main)",
    year = "2016"
}

@article{Alboteanu:2008my,
    author = "Alboteanu, Ana and Kilian, Wolfgang and Reuter, Juergen",
    title = "{Resonances and Unitarity in Weak Boson Scattering at the LHC}",
    eprint = "0806.4145",
    archivePrefix = "arXiv",
    primaryClass = "hep-ph",
    reportNumber = "SI-HEP-2008-11",
    doi = "10.1088/1126-6708/2008/11/010",
    journal = "JHEP",
    volume = "11",
    pages = "010",
    year = "2008"
}

@article{Alfaro:2010ui,
    author = "Alfaro, Jorge and Espriu, Domene and Puigdomenech, Daniel",
    title = "{The emergence of geometry: a two-dimensional toy model}",
    eprint = "1004.3664",
    archivePrefix = "arXiv",
    primaryClass = "hep-th",
    doi = "10.1103/PhysRevD.82.045018",
    journal = "Phys. Rev. D",
    volume = "82",
    pages = "045018",
    year = "2010"
}

@article{Alfaro:2012fs,
    author = "Alfaro, Jorge and Espriu, Domenec and Puigdomenech, Daniel",
    title = "{Spontaneous generation of geometry in four dimensions}",
    eprint = "1201.4697",
    archivePrefix = "arXiv",
    primaryClass = "hep-th",
    doi = "10.1103/PhysRevD.86.025015",
    journal = "Phys. Rev. D",
    volume = "86",
    pages = "025015",
    year = "2012"
}

@article{Espriu:2009ju,
    author = "Espriu, Domenec and Puigdomenech, Daniel",
    editor = "Praszalowicz, Michal",
    title = "{Gravity as an effective theory}",
    eprint = "0910.4110",
    archivePrefix = "arXiv",
    primaryClass = "hep-th",
    reportNumber = "UB-ECM-FP-28-09, ICCUB-09-239",
    journal = "Acta Phys. Polon. B",
    volume = "40",
    pages = "3409--3437",
    year = "2009"
}

@article{DeFelice:2010aj,
    author = "De Felice, Antonio and Tsujikawa, Shinji",
    title = "{f(R) theories}",
    eprint = "1002.4928",
    archivePrefix = "arXiv",
    primaryClass = "gr-qc",
    doi = "10.12942/lrr-2010-3",
    journal = "Living Rev. Rel.",
    volume = "13",
    pages = "3",
    year = "2010"
}

@article{STAROBINSKY198099,
title = {A new type of isotropic cosmological models without singularity},
journal = {Physics Letters B},
volume = {91},
number = {1},
pages = {99-102},
year = {1980},
issn = {0370-2693},
doi = {https://doi.org/10.1016/0370-2693(80)90670-X},
url = {https://www.sciencedirect.com/science/article/pii/037026938090670X},
author = {A.A. Starobinsky}
}

@article{Sotiriou:2008rp,
    author = "Sotiriou, Thomas P. and Faraoni, Valerio",
    title = "{f(R) Theories Of Gravity}",
    eprint = "0805.1726",
    archivePrefix = "arXiv",
    primaryClass = "gr-qc",
    doi = "10.1103/RevModPhys.82.451",
    journal = "Rev. Mod. Phys.",
    volume = "82",
    pages = "451--497",
    year = "2010"
}

@article{Latosh:2023zsi,
    author = "Latosh, Boris",
    title = "{FeynGrav 2.0}",
    eprint = "2302.14310",
    archivePrefix = "arXiv",
    primaryClass = "hep-th",
    reportNumber = "CTPU-PTC-23-06",
    doi = "10.1016/j.cpc.2023.108871",
    journal = "Comput. Phys. Commun.",
    volume = "292",
    pages = "108871",
    year = "2023"
}

@article{Delgado:2022uzu,
    author = "Delgado, Rafael L. and Dobado, Antonio and Espriu, Dom\`enec",
    title = "{Unitarized one-loop graviton-graviton scattering}",
    eprint = "2211.10406",
    archivePrefix = "arXiv",
    primaryClass = "hep-th",
    doi = "10.1051/epjconf/202227408010",
    journal = "EPJ Web Conf.",
    volume = "274",
    pages = "08010",
    year = "2022"
}

@article{Delgado:2022qnh,
    author = "Delgado, Rafael L. and Dobado, Antonio and Espriu, Dom\`enec",
    title = "{Seeking for resonances in unitarized one-loop graviton-graviton scattering}",
    eprint = "2207.06070",
    archivePrefix = "arXiv",
    primaryClass = "hep-th",
    doi = "10.1103/PhysRevD.107.044073",
    journal = "Phys. Rev. D",
    volume = "107",
    number = "4",
    pages = "044073",
    year = "2023"
}

@article{Oller:2024neq,
    author = "Oller, J. A. and Pel\'aez, Marcela",
    title = "{Unitarization of the one-loop graviton-graviton scattering amplitudes and study of the graviball}",
    eprint = "2407.16538",
    archivePrefix = "arXiv",
    primaryClass = "hep-th",
    month = "7",
    year = "2024"
}

@article{Oller:2022ozd,
    author = "Oller, Jos\'e Antonio",
    title = "{Unitarizing infinite-range forces: Graviton-graviton scattering, the graviball, and Coulomb scattering}",
    eprint = "2211.02084",
    archivePrefix = "arXiv",
    primaryClass = "hep-th",
    doi = "10.1051/epjconf/202227408011",
    journal = "EPJ Web Conf.",
    volume = "274",
    pages = "08011",
    year = "2022"
}

@article{Guiot:2020pku,
    author = "Guiot, B. and Borquez, A. and Deur, A. and Werner, K.",
    title = "{Graviballs and Dark Matter}",
    eprint = "2006.02534",
    archivePrefix = "arXiv",
    primaryClass = "gr-qc",
    doi = "10.1007/JHEP11(2020)159",
    journal = "JHEP",
    volume = "11",
    pages = "159",
    year = "2020"
}

@article{Alvarez-Gaume:2015rwa,
    author = {Alvarez-Gaume, Luis and Kehagias, Alex and Kounnas, Costas and L\"ust, Dieter and Riotto, Antonio},
    title = "{Aspects of Quadratic Gravity}",
    eprint = "1505.07657",
    archivePrefix = "arXiv",
    primaryClass = "hep-th",
    reportNumber = "LMU-ASC-26-15, MPP-2015-95, CERN-PH-TH-2015-099",
    doi = "10.1002/prop.201500100",
    journal = "Fortsch. Phys.",
    volume = "64",
    number = "2-3",
    pages = "176--189",
    year = "2016"
}

@article{Asiain:2021lch,
    author = "Asi\'ain, Inigo and Espriu, Dom\`enec and Mescia, Federico",
    title = "{Introducing tools to test Higgs boson interactions via WW scattering: One-loop calculations and renormalization in the Higgs effective field theory}",
    eprint = "2109.02673",
    archivePrefix = "arXiv",
    primaryClass = "hep-ph",
    doi = "10.1103/PhysRevD.105.015009",
    journal = "Phys. Rev. D",
    volume = "105",
    number = "1",
    pages = "015009",
    year = "2022"
}

@article{paper2_doi,
    author = "Asi\'ain, I\~nigo and Espriu, Dom\`enec and Mescia, Federico",
    title = "{Introducing tools to test Higgs boson interactions via WW scattering. II. The coupled-channel formalism and scalar resonances}",
    eprint = "2301.13030",
    archivePrefix = "arXiv",
    primaryClass = "hep-ph",
    doi = "10.1103/PhysRevD.107.115005",
    journal = "Phys. Rev. D",
    volume = "107",
    number = "11",
    pages = "115005",
    year = "2023"
}

@article{Asiain:2023myt,
    author = "Asi\'ain, I\~nigo and Espriu, Dom\`enec and Mescia, Federico",
    title = "{Accommodating a 650~GeV scalar resonance in HEFT}",
    eprint = "2305.03622",
    archivePrefix = "arXiv",
    primaryClass = "hep-ph",
    doi = "10.1103/PhysRevD.108.055013",
    journal = "Phys. Rev. D",
    volume = "108",
    number = "5",
    pages = "055013",
    year = "2023"
}

@article{LHCHiggsWG,
    author = "LHC Higgs Cross Section Working Group",
    url = {https://twiki.cern.ch/twiki/bin/view/LHCPhysics/LHCHWG"}
}

@article{cernwebpage,
    author = "CERN",
    url = {"https://home.cern/science/accelerators/future-circular-collider"}
}

@article{SMcircle,
author =  "ATLAS",
url = {"https://kaw.wallenberg.org/en/research/measuring-higgs-boson"}
}

@article{aliceexp,
author =  "ALICE",
url = {"https://home.cern/science/experiments/alice"}
}

@article{rhicexp,
author = "RHIC",
url = {"https://www.bnl.gov/rhic/physics.php"}
}

@article{ESPP,
author= "CERN",
URL={"https://europeanstrategy.cern"}
}

@article{Dawson:2023ebe,
    author = "Dawson, Sally and Fontes, Duarte and Quezada-Calonge, Carlos and Sanz-Cillero, Juan Jos\'e",
    title = "{Matching the 2HDM to the HEFT and the SMEFT: Decoupling and perturbativity}",
    eprint = "2305.07689",
    archivePrefix = "arXiv",
    primaryClass = "hep-ph",
    reportNumber = "IPARCOS-UCM-23-034",
    doi = "10.1103/PhysRevD.108.055034",
    journal = "Phys. Rev. D",
    volume = "108",
    number = "5",
    pages = "055034",
    year = "2023"
}

@article{HASERT1973138,
title = {Observation of neutrino-like interactions without muon or electron in the gargamelle neutrino experiment},
journal = {Physics Letters B},
volume = {46},
number = {1},
pages = {138-140},
year = {1973},
issn = {0370-2693},
doi = {https://doi.org/10.1016/0370-2693(73)90499-1},
url = {https://www.sciencedirect.com/science/article/pii/0370269373904991},
author = {F.J. Hasert and S. Kabe and W. Krenz and J. {Von Krogh} and D. Lanske and J. Morfin and K. Schultze and H. Weerts and G.H. Bertrand-Coremans and J. Sacton and W. {Van Doninck} and P. Vilain and U. Camerini and D.C. Cundy and R. Baldi and I. Danilchenko and W.F. Fry and D. Haidt and S. Natali and P. Musset and B. Osculati and R. Palmer and J.B.M. Pattison and D.H. Perkins and A. Pullia and A. Rousset and W. Venus and H. Wachsmuth and V. Brisson and B. Degrange and M. Haguenauer and L. Kluberg and U. Nguyen-Khac and P. Petiau and E. Belotti and S. Bonetti and D. Cavalli and C. Conta and E. Fiorini and M. Rollier and B. Aubert and D. Blum and L.M. Chounet and P. Heusse and A. Lagarrigue and A.M. Lutz and A. Orkin-Lecourtois and J.P. Vialle and F.W. Bullock and M.J. Esten and T.W. Jones and J. McKenzie and A.G. Michette and G. Myatt and W.G. Scott}
}

@inbook{Heisenberg+1979+229+238,
url = {https://doi.org/10.1515/9783112643884-014},
title = {Über den Bau der Atomkerne I},
booktitle = {Das Neutron},
booktitle = {Eine Artikelsammlung},
author = {W Heisenberg},
editor = {B. M. Kedrow},
publisher = {De Gruyter},
address = {Berlin, Boston},
pages = {229--238},
doi = {doi:10.1515/9783112643884-014},
isbn = {9783112643884},
year = {1979},
lastchecked = {2025-01-15}
}

@article{PhysRev.127.965,
  title = {Broken Symmetries},
  author = {Goldstone, Jeffrey and Salam, Abdus and Weinberg, Steven},
  journal = {Phys. Rev.},
  volume = {127},
  issue = {3},
  pages = {965--970},
  numpages = {0},
  year = {1962},
  month = {Aug},
  publisher = {American Physical Society},
  doi = {10.1103/PhysRev.127.965},
  url = {https://link.aps.org/doi/10.1103/PhysRev.127.965}
}

@article{Gell-Mann:1960mvl,
    author = "Gell-Mann, Murray and Levy, M",
    title = "{The axial vector current in beta decay}",
    doi = "10.1007/BF02859738",
    journal = "Nuovo Cim.",
    volume = "16",
    pages = "705",
    year = "1960"
}

@article{Kamefuchi:1961sb,
    author = "Kamefuchi, S. and O'Raifeartaigh, L. and Salam, Abdus",
    title = "{Change of variables and equivalence theorems in quantum field theories}",
    doi = "10.1016/0029-5582(61)90056-6",
    journal = "Nucl. Phys.",
    volume = "28",
    pages = "529--549",
    year = "1961"
}

@article{CHISHOLM1961469,
title = {Change of variables in quantum field theories},
journal = {Nuclear Physics},
volume = {26},
number = {3},
pages = {469-479},
year = {1961},
issn = {0029-5582},
doi = {https://doi.org/10.1016/0029-5582(61)90106-7},
url = {https://www.sciencedirect.com/science/article/pii/0029558261901067},
author = {J.S.R. Chisholm}
}

@article{tHooft:1979rat,
    author = "'t Hooft, Gerard",
    editor = "'t Hooft, Gerard and Itzykson, C. and Jaffe, A. and Lehmann, H. and Mitter, P. K. and Singer, I. M. and Stora, R.",
    title = "{Naturalness, chiral symmetry, and spontaneous chiral symmetry breaking}",
    reportNumber = "PRINT-80-0083 (UTRECHT)",
    doi = "10.1007/978-1-4684-7571-5_9",
    journal = "NATO Sci. Ser. B",
    volume = "59",
    pages = "135--157",
    year = "1980"
}

@article{Espriu:2017mlq,
    author = "Espriu, D. and Katanaeva, A.",
    title = "{Holographic description of $SO(5) \rightarrow SO(4)$ composite Higgs model}",
    eprint = "1706.02651",
    archivePrefix = "arXiv",
    primaryClass = "hep-ph",
    month = "6",
    year = "2017"
}

@article{RevModPhys.96.015006,
  title = {The standard model effective field theory at work},
  author = {Isidori, Gino and Wilsch, Felix and Wyler, Daniel},
  journal = {Rev. Mod. Phys.},
  volume = {96},
  issue = {1},
  pages = {015006},
  numpages = {60},
  year = {2024},
  month = {Mar},
  publisher = {American Physical Society},
  doi = {10.1103/RevModPhys.96.015006},
  url = {https://link.aps.org/doi/10.1103/RevModPhys.96.015006}
}

@article{Brivio:2017vri,
    author = "Brivio, Ilaria and Trott, Michael",
    title = "{The Standard Model as an Effective Field Theory}",
    eprint = "1706.08945",
    archivePrefix = "arXiv",
    primaryClass = "hep-ph",
    doi = "10.1016/j.physrep.2018.11.002",
    journal = "Phys. Rept.",
    volume = "793",
    pages = "1--98",
    year = "2019"
}

@article{Salas-Bernardez:2022hqv,
    author = "Salas-Bernardez, Alexandre and Sanz-Cillero, Juan J. and Llanes-Estrada, Felipe J. and Gomez-Ambrosio, Raquel",
    title = "{SMEFT as a slice of HEFT\textquoteright{}s parameter space}",
    eprint = "2211.09605",
    archivePrefix = "arXiv",
    primaryClass = "hep-ph",
    doi = "10.1051/epjconf/202227408013",
    journal = "EPJ Web Conf.",
    volume = "274",
    pages = "08013",
    year = "2022"
}

@article{Cohen:2020xca,
    author = "Cohen, Timothy and Craig, Nathaniel and Lu, Xiaochuan and Sutherland, Dave",
    title = "{Is SMEFT Enough?}",
    eprint = "2008.08597",
    archivePrefix = "arXiv",
    primaryClass = "hep-ph",
    doi = "10.1007/JHEP03(2021)237",
    journal = "JHEP",
    volume = "03",
    pages = "237",
    year = "2021"
}

@article{PhysRevD.22.200,
  title = {Strongly interacting Higgs bosons},
  author = {Appelquist, Thomas and Bernard, Claude},
  journal = {Phys. Rev. D},
  volume = {22},
  issue = {1},
  pages = {200--213},
  numpages = {0},
  year = {1980},
  month = {Jul},
  publisher = {American Physical Society},
  doi = {10.1103/PhysRevD.22.200},
  url = {https://link.aps.org/doi/10.1103/PhysRevD.22.200}
}

@article{Longhitano:1980tm,
    author = "Longhitano, Anthony C.",
    title = "{Low-Energy Impact of a Heavy Higgs Boson Sector}",
    reportNumber = "YTP-80-27-REV, YTP-80-27",
    doi = "10.1016/0550-3213(81)90109-7",
    journal = "Nucl. Phys. B",
    volume = "188",
    pages = "118--154",
    year = "1981"
}

@article{Feruglio:1992wf,
    author = "Feruglio, F.",
    title = "{The Chiral approach to the electroweak interactions}",
    eprint = "hep-ph/9301281",
    archivePrefix = "arXiv",
    reportNumber = "DFPD-92-TH-50",
    doi = "10.1142/S0217751X93001946",
    journal = "Int. J. Mod. Phys. A",
    volume = "8",
    pages = "4937--4972",
    year = "1993"
}

@article{FADDEEV196729,
title = {Feynman diagrams for the Yang-Mills field},
journal = {Physics Letters B},
volume = {25},
number = {1},
pages = {29-30},
year = {1967},
issn = {0370-2693},
doi = {https://doi.org/10.1016/0370-2693(67)90067-6},
url = {https://www.sciencedirect.com/science/article/pii/0370269367900676},
author = {L.D. Faddeev and V.N. Popov}
}

@article{Becchi:1974md,
    author = "Becchi, C. and Rouet, A. and Stora, R.",
    title = "{Renormalization of the Abelian Higgs-Kibble Model}",
    reportNumber = "CPT-74-P.634-MARSEILLE, CNRS-CPT-74-P634",
    doi = "10.1007/BF01614158",
    journal = "Commun. Math. Phys.",
    volume = "42",
    pages = "127--162",
    year = "1975"
}

@book{Peskin:1995ev,
    author = "Peskin, Michael E. and Schroeder, Daniel V.",
    title = "{An Introduction to quantum field theory}",
    doi = "10.1201/9780429503559",
    isbn = "978-0-201-50397-5, 978-0-429-50355-9, 978-0-429-49417-8",
    publisher = "Addison-Wesley",
    address = "Reading, USA",
    year = "1995"
}

@article{ATLAS:2022jtk,
    author = "Aad, Georges and others",
    collaboration = "ATLAS",
    title = "{Constraints on the Higgs boson self-coupling from single- and double-Higgs production with the ATLAS detector using pp collisions at s=13 TeV}",
    eprint = "2211.01216",
    archivePrefix = "arXiv",
    primaryClass = "hep-ex",
    reportNumber = "CERN-EP-2022-149",
    doi = "10.1016/j.physletb.2023.137745",
    journal = "Phys. Lett. B",
    volume = "843",
    pages = "137745",
    year = "2023"
}

@article{CMS:2024fkb,
    author = "Hayrapetyan, Aram and others",
    collaboration = "CMS",
    title = "{Search for Higgs boson pair production with one associated vector boson in proton-proton collisions at $ \sqrt{s} $ = 13 TeV}",
    eprint = "2404.08462",
    archivePrefix = "arXiv",
    primaryClass = "hep-ex",
    reportNumber = "CMS-HIG-22-006, CERN-EP-2024-064",
    doi = "10.1007/JHEP10(2024)061",
    journal = "JHEP",
    volume = "10",
    pages = "061",
    year = "2024"
}

@article{PhysRevD.10.1145,
  title = {Derivation of gauge invariance from high-energy unitarity bounds on the $S$ matrix},
  author = {Cornwall, John M. and Levin, David N. and Tiktopoulos, George},
  journal = {Phys. Rev. D},
  volume = {10},
  issue = {4},
  pages = {1145--1167},
  numpages = {0},
  year = {1974},
  month = {Aug},
  publisher = {American Physical Society},
  doi = {10.1103/PhysRevD.10.1145},
  url = {https://link.aps.org/doi/10.1103/PhysRevD.10.1145}
}

@article{CHANOWITZ1985379,
title = {The TeV physics of strongly interacting W's and Z's},
journal = {Nuclear Physics B},
volume = {261},
pages = {379-431},
year = {1985},
issn = {0550-3213},
doi = {https://doi.org/10.1016/0550-3213(85)90580-2},
url = {https://www.sciencedirect.com/science/article/pii/0550321385905802},
author = {Michael S. Chanowitz and Mary K. Gaillard}
}

@article{PhysRev.123.1053,
  title = {Asymptotic Behavior and Subtractions in the Mandelstam Representation},
  author = {Froissart, Marcel},
  journal = {Phys. Rev.},
  volume = {123},
  issue = {3},
  pages = {1053--1057},
  numpages = {0},
  year = {1961},
  month = {Aug},
  publisher = {American Physical Society},
  doi = {10.1103/PhysRev.123.1053},
  url = {https://link.aps.org/doi/10.1103/PhysRev.123.1053}
}

@book{book:tesisrafa,
author = {Delgado, Rafael},
year = {2017},
month = {01},
pages = {},
title = {Study of the Electroweak Symmetry Breaking Sector for the LHC},
isbn = {978-3-319-60497-8},
doi = {10.1007/978-3-319-60498-5}
}

@book{Maggiore:2007ulw,
    author = "Maggiore, Michele",
    title = "{Gravitational Waves. Vol. 1: Theory and Experiments}",
    doi = "10.1093/acprof:oso/9780198570745.001.0001",
    isbn = "978-0-19-171766-6, 978-0-19-852074-0",
    publisher = "Oxford University Press",
    year = "2007"
}
\end{small}

\end{document}